%% file: main.tex
\documentclass{jfm}

\input{utility/paper-info}
\input{utility/preamble}%

\begin{document}
\maketitle
\begin{abstract}
We extend unsteady thin aerofoil theory to model aerofoils with generalised chordwise porosity distributions.
The analysis considers a linearised porosity boundary condition where the seepage velocity through the aerofoil is related to the local pressure jump across the aerofoil surface and to the unsteady characteristics of the porous medium.
Application of the Plemelj formulae to the resulting boundary value problem yields a singular Fredholm--Volterra integral equation which does not admit an analytic solution.
Accordingly, we develop a numerical solution scheme by expanding the bound vorticity distribution in terms of appropriate basis functions.
Asymptotic analysis at the leading- and trailing-edges reveals that the appropriate basis functions are weighted Jacobi polynomials whose parameters are related to the porosity distribution.
The Jacobi polynomial basis enables the construction of a numerical scheme that is accurate and rapid, in contrast to the standard choice of Chebyshev basis functions that are shown to {be unsuitable} for porous aerofoils.
Applications of the numerical solution scheme to discontinuous porosity profiles, quasi-static problems, and the separation of circulatory and non-circulatory contributions are presented.
Further	asymptotic analysis of the singular Fredholm--Volterra integral equation corroborates the numerical scheme and elucidates the behaviour of the unsteady solution for small or large reduced frequency in the form of scaling laws.
	At low frequencies, the porous resistance dominates, whereas at high frequencies, an asymptotic inner region develops near the trailing edge and the effective mass of the porous medium dominates.
Analogues to the classical Theodorsen and Sears functions are computed numerically, which show that an effect of trailing-edge porosity is to reduce the amount of vorticity shed into the wake, thereby reducing the magnitude of the circulatory lift. {Fourier transform inversion of these frequency-domain functions produces porous extensions to the Wagner and K\"{u}ssner functions for transient aerofoil motions or gust encounters, respectively.} Results from the present analysis and its underpinning numerical framework aim to enable the unsteady aerodynamic assessment of design strategies using porosity, which has implications for {unsteady gust rejection}, noise-reducing aerofoil design, and biologically-inspired flight.
\end{abstract}
\input{sections/introduction}
\input{sections/derivation}
\input{sections/numerical}
\input{sections/results}
\input{sections/conclusions}
\input{sections/acknowledgement}

\appendix
\input{sections/appendix-jacobi}
\input{sections/large-frequency}
\bibliographystyle{jfm}
\bibliography{bibliography}

\end{document}

%% file: utility/paper-info.tex
\shorttitle{Unsteady aerodynamics of porous aerofoils}
\shortauthor{P. J. Baddoo, R. Hajian, and J. W. Jaworski}

\title{Unsteady aerodynamics of porous aerofoils}

\author{Peter J. Baddoo\aff{1},  
Rozhin Hajian\aff{2},
 \and Justin W. Jaworski\aff{3}}

\affiliation{\aff{1}Department of Mathematics,
	South Kensington Campus,
	Imperial College London,
	London
	SW7 2AZ
\aff{2} School of Engineering and Applied Sciences, Harvard University, Cambridge, Massachusetts 02138
\aff{3} Department of Mechanical Engineering and Mechanics, Lehigh University, Bethlehem, Pennsylvania 18015}



%% file: utility/preamble.tex
\usepackage[utf8]{inputenc}
\usepackage{amsmath,amssymb}
\usepackage{mathtools}
\usepackage{lipsum}

  \usepackage{color,graphicx,amsmath}
 \usepackage{tikz,pgfplots,pstool}
 \pgfplotsset{compat = 1.3}
 
   \usetikzlibrary{plotmarks}
   \usetikzlibrary{arrows.meta}
   \usepgfplotslibrary{patchplots}
 \usepackage[toc,page]{appendix}
 \def\Xint#1{\mathchoice
 {\XXint\displaystyle\textstyle{#1}}%
 {\XXint\textstyle\scriptstyle{#1}}%
 {\XXint\scriptstyle\scriptscriptstyle{#1}}%
 {\XXint\scriptscriptstyle\scriptscriptstyle{#1}}%
 \!\int}
 \def\XXint#1#2#3{{\setbox0=\hbox{$#1{#2#3}{\int}$}
 \vcenter{\hbox{$#2#3$}}\kern-.5\wd0}}
 
\newcommand{\dashint}{\Xint-}

\usepackage{hyperref}
\hypersetup{
    colorlinks,
    linkcolor={red!50!black},
    citecolor={blue!50!black},
    urlcolor={blue!80!black}
}
\usepackage{bigfoot}
\usepackage{subcaption}

\newcommand{\helmNum}{k}
\tikzset{mark size=50}

\input{utility/define-tikz-objects}

\input{utility/define-colours}

\newcommand*\pFqskip{8mu}
\catcode`,\active
\newcommand*\pFq{\begingroup
        \catcode`\,\active
        \def ,{\mskip\pFqskip\relax}%
        \dopFq
}
\catcode`\,12
\def\dopFq#1#2#3{%
        {}_{2}F_{1}\biggl[\genfrac..{0pt}{}{#1}{#2};#3\biggr]%
        \endgroup
}
\newcommand{\weight}[3]{w^{#2,#3}(#1)}

\def\d{\textrm{d}}
\def\i{\textrm{i}}
\def\e{\textrm{e}}

\date{}
\newlength{\fheight}
\newlength{\fwidth}
\usepackage{bm}

\input{utility/define-colours}
\usepackage[normalem]{ulem}
\definecolor{lightblue}{rgb}{.831, .922, .949}
\newcommand\hl{\bgroup\markoverwith
	{\textcolor{lightblue}{\rule[-.5ex]{2pt}{2.5ex}}}\ULon}

\definecolor{lightred}{rgb}{1, .8, .8}
\newcommand\hlt{\bgroup\markoverwith
	{\textcolor{lightred}{\rule[-.5ex]{2pt}{2.5ex}}}\ULon}

\renewcommand{\Re}{\mathbb{R}\textnormal{e}}
\DeclareRobustCommand\drawLine[1]{\tikz[baseline=(current bounding box.base)] \draw[#1] (0,2.5pt)--(.75,2.5pt);}
	

%% file: utility/define-tikz-objects.tex
\usetikzlibrary{calc, 
				decorations.markings, 
				decorations.text,
				decorations.pathreplacing,
				shapes.multipart,
				positioning,
				arrows.meta,
				arrows,
				matrix,
				patterns,
				shadows,
				shadings,
				hobby}

\tikzset{
    side by side/.style 2 args={
        line width=2pt,
        #1,
        postaction={
            clip,postaction={draw,#2}
        }
    },
    circle node/.style={
        circle,
        draw,
        fill=white,
        minimum size=1.3cm
    }
}

\usetikzlibrary{fadings}
\tikzfading[name=fade out,
inner color=transparent!0,
outer color=transparent!100]

\tikzset{test/.style={
    postaction={
        decorate,
        decoration={
            markings,
            mark=at position \pgfdecoratedpathlength-0.5pt with {\arrow[blue,line width=#1] {}; },
            mark=between positions 0 and \pgfdecoratedpathlength-0pt step 0.5pt with {
                \pgfmathsetmacro\myval{multiply(divide(
                    \pgfkeysvalueof{/pgf/decoration/mark info/distance from start}, \pgfdecoratedpathlength),100)};
                \pgfsetfillcolor{white!\myval!black};
                \pgfpathcircle{\pgfpointorigin}{#1};
                \pgfusepath{fill};}
}}}}

\tikzset{
    mark position/.style args={#1(#2)}{
        postaction={
            decorate,
            decoration={
                markings,
                mark=at position #1 with \coordinate (#2);
            }
        }
    }
}

\def\myShiftUp#1{\raisebox{1ex}}
\def\myShiftDown#1{\raisebox{-2.5ex}}

\usetikzlibrary{decorations.text,fit,chains,calc,shapes.geometric,intersections}

\newlength{\myww}
\newlength{\myhh}

%% file: utility/define-colours.tex
\usepackage{xcolor}
\usepackage{colortbl}

\definecolor{themecolour}{rgb}{0.64, 0.76, 0.68}
\definecolor{mygray}{rgb}{0.7, 0.7, 0.7}
\definecolor{tablegray}{rgb}{0.75, 0.75, 0.75}
\definecolor{confcol}{RGB}{255, 255, 255}
\definecolor{camblue}{RGB}{108, 172, 228}
\definecolor{camred}{RGB}{213, 0, 50}
\definecolor{camnavy}{RGB}{0, 60, 113}
\definecolor{camgreen}{RGB}{114, 180, 49}

\definecolor{myblue}{rgb}{0 ,  0.4470 , 0.7410}
\definecolor{myorange}{rgb}{0.8500,    0.3250,    0.0980}
\definecolor{myyellow}{rgb}{0.9290,    0.6940,    0.1250}
\definecolor{mypurple}{rgb}{ 0.4940,    0.1840,    0.5560}
\definecolor{myred}{rgb}{     0.6350 ,   0.0780 ,   0.1840}
\definecolor{mygreen}{rgb}{         0.4660  ,  0.6740   , 0.1880}

\definecolor{branchCol}{rgb}{0, .8, .8}

\definecolor{prsared}{RGB}{ 219,25,73}
\definecolor{prsablue}{RGB}{ 4,146,210}

\definecolor{confblue}{rgb}{0.75, 0.72, 0.95}
\definecolor{confgreen}{rgb}{0.76, 1, 0.74}
\definecolor{confgrey}{rgb}{0.9, .9, 0.9}

\definecolor{cvblue}{rgb}{0.22,0.45,0.70}

\definecolor{matlab1}{rgb}{0,0.4470,0.7410}
\definecolor{matlab2}{rgb}{0.8500,0.3250,0.0980}
\definecolor{matlab3}{rgb}{0.9290,0.6940,0.0980}
\definecolor{matlab4}{rgb}{0.4940,0.1840,0.5560}
\definecolor{matlab5}{rgb}{0.4660,0.6740,0.1880}
\definecolor{matlab6}{rgb}{0.3010,0.7450,0.9330}
\definecolor{matlab7}{rgb}{0.6350,0.0780,0.1840}

%% file: sections/introduction.tex
\section{Introduction}
The seminal works of \citet{Theodorsen1935} and \citet{Sears1941} continue to ground the modern understanding of unsteady aerodynamic phenomena experienced by lifting bodies.
Their analyses considered the unsteady potential flow about an impermeable aerofoil: whilst Theodorsen considered the effect of unsteady (harmonic) aerofoil motions, Sears was concerned with the fluctuating pressure response of the aerofoil to an incident harmonic gust.
Both authors were able to derive closed-form, analytic expressions for the unsteady lift in terms of Hankel functions, {which \mbox{\citet{Garrick1938}} showed could be connected to the indicial lift functions of \mbox{\citet{Wagner1925}} and \mbox{\citet{Kussner1936}} for impulsive airfoil motion or entry into a sharp-edged gust, respectfully.}
The analysis of Theodorsen was originally motivated by the need to predict flutter instability but has been used \emph{inter alia} to form the basis for predicting and comparing unsteady forces on flapping foils~\citep{garrick1936,jaworski2012high,jaworski2015thrust,floryan2017scaling} and to develop load prediction methods relevant to rotorcraft blades \citep{loewy1957two,peters2008two}.
The work of Sears relates directly to aerodynamic gust responses and enables the prediction of acoustic radiation from aerofoils encountering vortical sources \citep[see][]{glegg2017aeroacoustics}, where extensions to Sears's analysis have included the distortion of the incoming gust by the aerofoil \citep{goldstein1976complete}, as well as the effects of mean aerofoil loading \citep{scott1993high}, aerofoil shape \citep{kerschen1993influence}, and finite Mach number \citep{graham1970similarity,leishman1997unsteady}.

Understanding, exploiting, and extending the analyses of Theodorsen and Sears remains a vibrant area of research.
Recent work by \citet{Cordes2017} explored the limitations of {these aerodynamic} transfer functions and found that, whilst the Theodorsen function performed well against experimental data, the Sears function required the second-order correction for gust distortion by the aerofoil provided by \citet{goldstein1976complete} and \citet{Atassi1984}.
The discrepancies between these models were investigated in greater detail by \citet{Wei2019}, who concluded that the original Sears function may even be used when there are considerable fluctuations in the streamwise velocity component.
Of particular note is the recent extension of unsteady potential flow to include viscosity  via triple deck analysis at the trailing edge by \citet{Taha2019}.
This work presented a viscous extension of the Theodorsen function to elucidate the role of viscosity-induced lag that becomes increasingly important at large reduced frequencies.
Extension of the {aerodynamic} transfer functions from two-dimensional aerofoils  to three-dimensional wings is another popular research direction, which has been pursued with a variety of possible methods \citep{Bird2019,Yang2019}.
However, the original analyses by Theodorsen and Sears and these subsequent investigations they have inspired involve impermeable lifting surfaces that do not permit any flow seepage through the aerofoil or wing.
In the present work, we extend these classical unsteady analyses to consider aerofoils with chordwise porosity gradients.
In particular, we consider a {linearised, unsteady} porosity law where the seepage velocity {depends on the local values of the flow resistance and effective fluid inertia of the porous medium, and on the local pressure gradient across the aerofoil}.

Porous aerofoils have received considerable attention over recent years due to their apparent ability to reduce acoustic emissions \citep{Geyer1,JWJ,Ayton2016a,Kisil2018}.
It is generally believed that porosity at the trailing edge weakens the scattering of turbulence there and  therefore reduces sound production in a manner similar to turbulence noise suppression by an edgeless perforated sheet \citep{paper:ffowcswilliams:1972,nelson1982noise}.
However, the fluid loads on perforated aerofoils are also affected by porosity and are expected to be aerodynamically poorer in comparison to impermeable aerofoils \citep{Geyer1,gil,gil2,Hajian2017}.
Recent experiments by \citet{hanna2019aerodynamic} demonstrate that porosity can also be aerodynamically beneficial by suppressing unwanted flow phenomena that are dependent upon the Reynolds number of the configuration.
Consequently, aircraft designers seeking to use porosity as a noise mitigation strategy are faced with the difficult task of balancing the aeroacoustic advantages of porous aerofoils with their aerodynamic disadvantages.

With the goal to assess these aerodynamic effects, \citet{Hajian2017} developed an analytic formulation and solution for the steady aerodynamic loads on {aerofoils} with arbitrary, realistic (specifically, H{\"o}lder continuous) porosity distributions to investigate the impact of a chordwise variation in  porosity.
This analysis was later extended to determine the unsteady forces on an arbitrarily deforming panel with generalised porosity distributions~\citep{Hajian2018}.
An analytical expression for the unsteady pressure distribution was presented and evaluated for the special cases of uniform and variable-porosity panels undergoing harmonic deformations, where the effect of the panel end conditions was also investigated. Subsequent research used these unsteady loads to show generally that the primary instability of porous panels with fixed ends is aeroelastic divergence~\citep{hajian2020porous}.

A comprehensive unsteady aerodynamic theory for lifting porous bodies is also essential to predict aeroelastic stability and aeroacoustic emissions.
The classical theory of \citet{Theodorsen1935} and its later extensions~\citep{JWJ3} developed closed-form expressions for the unsteady aerodynamic forces on a piecewise-continuous  impermeable {aerofoil} undergoing small-amplitude harmonic motions in a uniform incompressible flow.
These analyses separated the total fluid forces or moments into circulatory and non-circulatory parts, which correspond respectively to the contribution of the unsteady shedding of vorticity into the wake and the hydrodynamic reaction of the fluid to aerofoil motion.
These unsteady fluid forces also contribute fundamentally to the aerofoil gust response problem \citep[cf.][pp.~281-293]{A} and to the aerodynamic noise generation from gust encounters \citep{Atassi} and vortex-structure interactions \citep{howebook}.
Therefore, an extension of the classical unsteady aerodynamic response models to include the effects of porosity distributions is desired. 

The classical aerodynamic functions for impermeable aerofoils depend on the solution of a singular integral equation for the vorticity or pressure distribution on the aerofoil, which may be integrated to furnish the aerodynamic loads of interest.
\citet{schwarz1940berechnung} employed the integral inversion of \citet{Sohngen} to produce an exact solution for the pressure distribution and fluid loads on unsteady impermeable aerofoils.
\citet{Hajian2017} determined an exact solution for steady aerofoils with chordwise porosity gradients using conventional analysis methods \citep[see][]{NIM}, as noted above.
However, the singular integral equation describing the generalized aerodynamics of unsteady porous {aerofoils} with a wake cannot be treated by conventional analysis, and a different mathematical approach is required.
A new approach to circumvent the analytical challenges of unsteady porous aerofoil modelling is the focus of the present research.

In complement to standard analytical approaches, there are many methods available for the numerical solution of singular integral equations \citep{Erdogan1973}.
Numerical solutions in terms of orthogonal polynomials were first considered by \citet{Erdogan1972}, who expressed the solution function as a series of weighted Chebyshev polynomials.
However, this numerical approach was limited to particular endpoint behaviours until the generalisation by  \citet{Krenk1975} to Jacobi polynomials allowed a broader class of endpoint zeros and singularities to be examined.
In the present research, we adapt the approach of \citet{Krenk1975} to a broader class of singular integral equations, including the generalisation to discontinuous coefficients.

The expansion of the jump in surface pressure across the aerofoil into a series of weighted Chebyshev polynomials has previously been applied to aerodynamic problems for impermeable \citep{Rienstra1992} and permeable \citep{Weidenfeld2016} aerofoils.
The weighted Chebyshev expansion (also referred to as a Glauert--Fourier series) is an essential feature of many reduced-order discrete-vortex models \citep{Ramesh2014,SureshBabu2019}.
These models require a detailed understanding of the pressure at the leading and trailing edges to predict the vortex shedding behaviour correctly.
In particular, the leading-edge suction parameter must be accurately computed \citep{Ramesh2014}.
In the present work, we show that the Chebyshev expansion is unsuitable for porous aerofoils, and an expansion in terms of weighted Jacobi polynomials is essential to capture the subtle {solution} behaviour at the endpoints.



Further details relevant to the unsteady forces on porous aerofoils can be educed from the asymptotic examination of the model equations in the low- and high-frequency limits. The low-frequency limit produces a regular perturbation correction to the steady analysis of \cite{Hajian2017}, where, for given reduced frequency $k$, the magnitudes of the aerofoil circulation and unsteady lift coefficient each scale as a constant with ${\it O}(k)$ correction. However, the high-frequency limit yields a singular perturbation problem, where two asymptotic expansions are sought in two overlapping regions about the aerofoil that are matched in the spirit of \citet{VanDyke1964}: an outer region along most of the 
aerofoil, and an inner region confined to the vicinity of the trailing edge.
{We show that, unintuitively, the scaling laws for unsteady aerofoil lift and circulation with respect to reduced frequency are independent of the porosity: in the high-frequency limit, the unsteady lift scales like $k^2$, whereas the aerofoil circulation scales like $\sqrt{k}$.
This fractional scaling law arises from the singular nature of the asymptotic inner region near the trailing edge.}

The remainder of this paper is structured in the following manner.
Section \ref{MM} presents the mathematical model for a porous aerofoil undergoing unsteady motions, and a numerical solution of the ensuing singular integral equation is presented in \S\ref{Sec:numSol}.
{Numerical and asymptotic solutions are then used in \S\ref{Sec:Results} to draw physical insights regarding porosity from a range of canonical aerofoil motions, and \S\ref{sec:scaling} provides numerical confirmation of the asymptotic scaling behaviours of the unsteady lift and aerofoil circulation with respect to reduced frequency. Section \ref{sec:standard} develops and discusses porous analogues of the classical Theodorsen, Sears, Wagner, and K{\"u}ssner functions that must be computed numerically.
Finally, \S\ref{Sec:Conclusions} summarises the main findings of the research and outlines a number of possible directions of future work.
Useful identities for the Jacobi polynomials are catalogued in appendix \ref{Ap:Jacobi}, and the low- and high-frequency solution behaviours of porous aerofoils are analysed asymptotically in appendix \ref{Sec:asymp}.
All results in this paper can be reproduced using the computer codes that are publicly available at} \textcolor{blue}{\url{https://github.com/baddoo/unsteady-porous-aerofoils}}.

%% file: sections/derivation.tex
\def\airfoilLength{5}
\def\airfoilHeight{1.5}
\def\ang{20}
\def\axisx{-4}
\def\axisy{-2}

\begin{figure}
	\centering
	\begin{subfigure}{\linewidth}
%
%
\hspace{.5cm}
\includegraphics[height = 3.5cm]{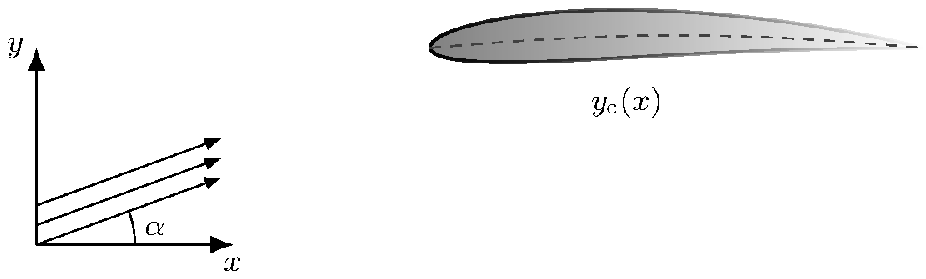}
	\caption{Steady case}
	\label{Fig:steadySchematic}
		\end{subfigure}
		
\begin{subfigure}{\linewidth}

\tikzset{test/.style={
    postaction={
        decorate,
        decoration={
            markings,
            mark=between positions 0 and \pgfdecoratedpathlength-0pt step 0.5pt with {
                \pgfmathsetmacro\myval{multiply(divide(
                    \pgfkeysvalueof{/pgf/decoration/mark info/distance from start}, \pgfdecoratedpathlength),100)};
                \pgfsetfillcolor{blue!\myval!white};
                \pgfpathcircle{\pgfpointorigin}{#1};
                \pgfusepath{fill};}
}}}}

\centering
\includegraphics[height = 4cm]{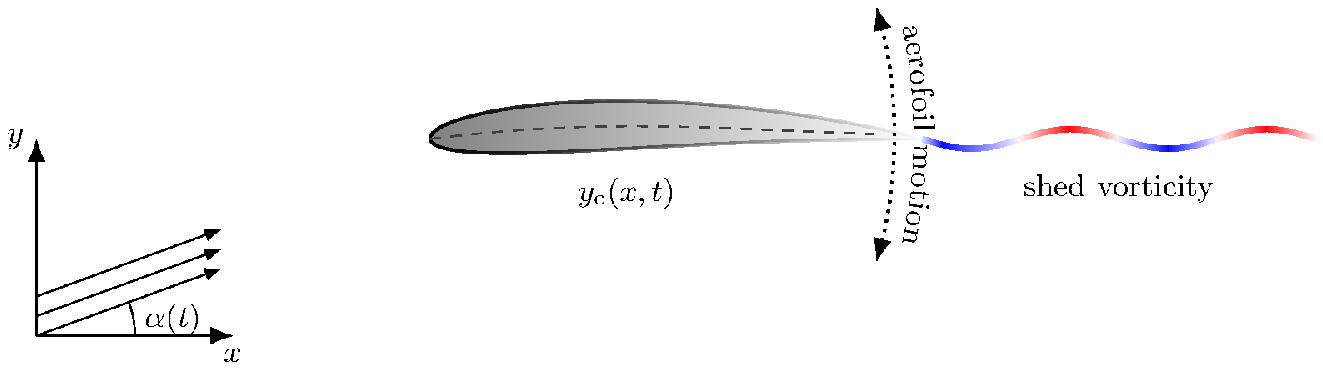}
	\caption{Unsteady case}
	\label{Fig:unsteadySchematic}
	\end{subfigure}
\caption{Schematic diagrams of a porous aerofoil with mean camber profile $y_{\rm c}$ (dashed lines) at angle of attack $\alpha$ for (a) steady and (b) unsteady scenarios. Aerofoil porosity is illustrated by the colour gradient on the aerofoils. In the unsteady case, the angle of attack and aerofoil surface profile may vary with time. A pitching motion is illustrated as an example, and the unsteady wake shed from the trailing edge is illustrated by the wavy line in (b).}
\label{Fig:aerofoil}
\end{figure}

\section{Mathematical model}\label{MM}
We consider a thin {aerofoil} immersed in a uniform, two-dimensional incompressible flow. In the steady case, the aerofoil and incident flow are stationary, whereas in the unsteady case the aerofoil and  flow {velocity field} may be time-dependent, as illustrated in figure \ref{Fig:aerofoil}. In the latter case, the aerofoil sheds vorticity into a wake whose strength is unknown. Additionally, the flow is irrotational away from the aerofoil and wake. Supposing a semi-chord length $l$, mean flow speed $U$, and fluid density $\rho$, all {physical terms in the ensuing analysis} are nondimensionalized using $l$, $l/U$, and $\frac{1}{2} \rho U^2$ as the length, time, and pressure scales, respectively.

%
%
%
\subsection{Porous boundary condition}
%
Along a porous {aerofoil}, the perturbation flow velocity on the aerofoil surface, $w$, is related to the local seepage flow rate directed along the unit normal to the wing surface, $w_{\rm s}$, by
\begin{eqnarray}\label{Eq:w01}
w(x,t)=w_{\rm s}(x,t)+\frac{\partial y_{\rm a}}{\partial x}(x,t) + \frac{\partial y_{\rm a}}{\partial t}(x,t),
\end{eqnarray}
where the function $y_{\rm a}(x,t) =  y_{\rm c}(x,t) - \alpha(t)$ defines the mean surface of the aerofoil relative to the angle of attack.
This linearised boundary condition assumes that the deformation of the aerofoil camber line is small: $|y_{\rm a}|\ll 1$.
{Additionally, the aerofoil thickness is assumed to be small.
Aerofoils with small but non-zero thickness can typically be handeled with thin aerofoil theory (see \S 4.2 of \mbox{\cite{VanDyke1964}} or \mbox{\cite{Baddoo2018c}}) but
in this problem aerofoil thickness does not contribution to the pressure jump across the chord so thickness decouples from the hydrodynamic effects of porosity,
Thus, we do not consider the effects of thickness further.
}
{%
The seepage velocity depends on the pressure jump across the aerofoil and on the local porous structure.
Porosity imparts nondimensional hydrodynamic inductance $\rho_{\rm e}$ and resistance $\Phi$ 
due to the viscous fluid-solid interactions of the flow within the porous medium.
Combining these effects yields the unsteady porosity boundary condition~\mbox{\citep[p.~254]{Morse1986}} }
\begin{align}
  2 \rho_{\rm e}(x) \frac{\partial w_{\rm s}}{\partial t}(x,t) + \Phi(x) w_{\rm s}(x,t) = - \Delta p(x,t), \label{Eq:morseBC}
\end{align}
{where $\Delta p$ is the local difference of the surface pressure (upper minus lower).
The dimensional scalings used in this work identify the} {so-called} {effective density $\rho_{\rm e} = m_{\rm e}/V > 1$, 
where $m_{\rm e} \ge 1$ is the effective mass or structure factor of the porous medium,
and $V$ is the fluid-to-solid volumetric fraction of the porous matrix~\mbox{\citep{bliss1982study,attenborough1983acoustical}}.
Similarly, the dimensionless flow resistance $\Phi$ is the pressure drop per 
unit length per unit mean flow velocity through the porous aerofoil scaled by $\rho U/(2 l)$.
Strictly speaking, $\rho_{\rm e}$ and $\Phi$ are also functions of frequency, although we do not consider that detail here.}

The linearised Bernoulli equation for unsteady flow enables the pressure to be expressed as a function of the velocity potential $\phi$ as
\begin{align}
p(x,y,t) & = -\left(\frac{\partial \phi}{\partial x} + \frac{\partial \phi}{\partial t}\right).\label{Eq:pres1}
\end{align}
Applying \eqref{Eq:pres1} to $y = 0^\pm$, $x>-1$, and taking the difference from the upper and lower sides {yields expressions} for the pressure jump along the {aerofoil} and the wake:
\begin{align}
\addtocounter{equation}{1}
&&\Delta p(x,t) &= -2 \left(\gamma_{\rm a}(x,t) + \frac{\partial }{\partial t} \int_{-1}^x \gamma_{\rm a}(\xi,t)\d \xi \right), & -1<x<1, &&\tag{\theequation.a} \label{Eq:pres2a}\\
&&\Delta p(x,t) &= -2 \left(\gamma_{\rm w}(x,t) + \frac{\partial }{\partial t} \int_{1}^x \gamma_{\rm w}(\xi,t) \d \xi + \frac{\d \Gamma }{\d t}(t)\right), & \phantom{-}1<x,\phantom{<1,} \tag{\theequation.b}\label{Eq:pres2b}&&
\end{align}
where $\Gamma$ represents the circulation around the aerofoil. 
%
%
%
On the other hand, application of the Plemelj formula \citep{Fokas2003} to the Biot--Savart law shows that the {fluid normal velocity} on the wing, $w$, is {related to} the vorticity distributions on the aerofoil ($\gamma_{\rm a}$) and in the wake ($\gamma_{\rm w}$) through the following singular integral equation \cite[(5-313a)]{A}: 
\begin{align}\label{Eq:ww}
   w(x,t)&= \frac{1}{2\pi}\dashint_{-1}^{1}\frac{\gamma_{\rm a}(\xi,t)}{\xi-x}\d\xi +\frac{1}{2\pi}\int_{1}^{\infty}\frac{\gamma_{\rm w}(\xi,t)}{\xi-x} \d\xi, \qquad -1<x<1,
\end{align}
{where the bar on the integral denotes the Cauchy principal value.
Note that once $\gamma_a$ and $\gamma_w$ have been determined,
the full complex velocity field can be determined
from the Biot--Savart law.
Combining, \eqref{Eq:w01}, \eqref{Eq:morseBC}, \eqref{Eq:pres2a} and \eqref{Eq:ww} yields the integral equation 
}
\begin{align}\label{Eq:SIE1}
  \left[ 2 \rho_{\rm e}(x) \frac{\partial}{\partial t} + \Phi(x) \right]
  \left\{ \frac{1}{2\pi}\dashint_{-1}^{1}\frac{\gamma_{\rm a}(\xi,t)}{\xi-x}\d\xi 
  +  \frac{1}{2\pi} \int_{1}^{\infty}\frac{\gamma_{\rm w}(\xi,t)}{\xi-x} \d\xi 
  - \frac{\partial y_{\rm a}}{\partial t} - \frac{\partial y_{\rm a}}{\partial x}  \right\}= \notag \\
 2 \left(\gamma_{\rm a}(x,t) + \frac{\partial }{\partial t} \int_{-1}^x \gamma_{\rm a}(\xi,t)\d \xi \right) 
\end{align}
%
%
%
%
%
%
%
%
for $-1<x<1$.

We further assume harmonic motions of reduced frequency $k$, so that we may write the vorticity distributions and mean camber line as
\begin{align}
{\gamma}_{\rm{a}}(x,t) = \hat{\gamma}_{\rm{a}}(x) \e^{\i k t}, \quad  {\gamma}_{\rm{w}}(x,t) = \hat{\gamma}_{\rm{w}}(x) \e^{\i k t}, \quad \textnormal{and} \quad y_{\rm a}(x,t) = \hat{y}_{\rm a}(x) \e^{\i k t}, \label{Eq:harmonic}
\end{align}
where the real part is assumed.
 Since we do not allow any pressure jump across the wake, we require the {expression on the right-hand side of} \eqref{Eq:pres2b} to vanish. Solving the associated integral equation yields%
\begin{align}
	\hat{\gamma}_{\rm{w}}(x) = - \i \helmNum \e^{\i \helmNum (1-x)} 
	\int_{-1}^1 \hat{\gamma}_{\rm a}(\xi) \d \xi  =
	-\i k \hat{\Gamma}\e^{\i k (1-x)} , \label{Eq:wakeRep}
\end{align}
where Kelvin's circulation theorem has been applied to enforce that the net circulation of the aerofoil and its wake vanishes,
and $\hat{\Gamma}$ is the circulation around the aerofoil with the time dependence factored out.
We also enforce the Kutta condition, namely that the pressure jump vanishes at the trailing edge:
\begin{align}
    \Delta p(1) = 0.
    \label{eq:kutta}
\end{align}
Substitution of \eqref{Eq:harmonic} and \eqref{Eq:wakeRep} into \eqref{Eq:SIE1} yields
\begin{align}
\mathcal{L}\hat{\gamma}_{\rm{a}} = f_{\rm a}(x) + \hat{\Gamma} f_{\rm w}(x) \label{Eq:SIEfin}
\end{align}
for $-1<x<1$, where we use the notation $\mathcal{L}$ to represent the {singular linear} operator
\begin{align}
\mathcal{L}f &\triangleq - {\i \helmNum}\psi(x,k) \int_{-1}^x f(\xi) \d \xi
+ \frac{1}{\pi}\dashint_{-1}^{1}\frac{f(\xi)}{\xi-x}\d \xi
- \psi(x,k) f(x)
, \label{Eq:Ldef}
\end{align}
and
\begin{align}
    \psi (x,k) \triangleq \frac{4}{2 \i k  \rho_{\rm e}(x) + \Phi(x)}.
\label{Eq:psidef}
\end{align}
%
%
The forcing functions $f_{\rm a}$ and $f_{\rm w}$ are defined as
\begin{align}
f_{\rm a}(x) & \triangleq 2 \left( \frac{\d \hat{y}_{\rm a}}{\d x}(x)  + \i \helmNum \hat{y}_{\rm a}(x) \right),  \\
f_{\rm w}(x) & \triangleq \frac{\i \helmNum}{\pi} \int_{1}^\infty \frac{\e^{\i k (1-\xi)}}{\xi - x} \d \xi.
\end{align}
The subscript notations ``${\rm a}$'' and ``${\rm w}$'' are again employed here to symbolise that $f_{\rm a}$ corresponds to contributions from the mean aerofoil profile and its motions, whereas $f_{\rm w}$  corresponds to contributions from the unsteady wake. The problem is now to determine {the function $\hat{\gamma}_{\rm a}$, from which the constant $\hat{\Gamma}$ can be found via integration per \eqref{Eq:wakeRep}}.

The operator $\mathcal{L}$ consists of two parts: a Volterra part (the first term in \eqref{Eq:Ldef}), and a singular Fredholm integral part (the second and third terms in \eqref{Eq:Ldef}). 
Accordingly, we refer to \eqref{Eq:SIEfin} as a singular \mbox{Fredholm--Volterra} integral equation (SF--VIE).
The literature on these types of integral equations is apparently non-existent: the closest comparisons that could be found {by the authors}  considered only the case where the kernel is weakly singular \citep{Darwish1999,Abdou2003} and not the Cauchy principal value considered in the present work.
In particular, it is the presence of the Volterra part of $\mathcal{L}$ that precludes the possibility of a solution using the classical singular integral equation methods of \citet{NIM}.
Consequently, we now seek a numerical solution by expanding $\hat{\gamma}_{\rm{a}}$ into an appropriate series of basis functions. 

%% file: sections/numerical.tex
\section{Numerical solution} \label{Sec:numSol}

We now introduce our numerical solution for the SF--VIE \eqref{Eq:SIEfin} that is central to the unsteady aerodynamics of porous aerofoils. We motivate our approach to the unsteady problem by first examining a numerical solution of the steady case. 

\subsection{Motivation -- the steady case} \label{Sec:steady}
We first consider the case where the field is steady {($k=0$)} {and the wake vanishes}, as illustrated in figure \ref{Fig:steadySchematic}. The SF--VIE \eqref{Eq:SIEfin} for the bound vorticity distribution becomes
\begin{align}\label{Eq:p_steady}
 \frac{1}{\pi}\dashint_{-1}^{1}\frac{\hat{\gamma}_{\rm a}(\xi)}{\xi-x}\d \xi
-\psi(x,0) \hat{\gamma}_{\rm a}(x) 
~= 2\frac{\d \hat{y}_{\rm a}}{\d x}(x), \qquad \qquad -1<x<1.
\end{align}
In the impermeable case ($\psi \equiv 0$), the typical {solution} approach is to expand $\hat{\gamma}_{\rm a}$ in terms of weighted Chebyshev polynomials \citep{Rienstra1992}. {However, this approach dictates the behaviour of the vorticity distribution at the endpoints.} In particular, the vorticity distribution is usually written as
\begin{align}
\hat{\gamma}_{\rm a}(x) = \hat{\gamma}_{0} \sqrt{\frac{1-x}{1+x}} + \sqrt{1-x^2} \sum_{n=1}^N \hat{\gamma}_{n} U_{n-1} (x), \label{Eq:ChebRig}
\end{align}
where $U_n$ are the Chebyshev polynomials of the second kind and $\hat{\gamma}_{n}$ are coefficients to be determined. Consequently, $\hat{\gamma}_{\rm a}$ possesses a square-root singularity at the leading edge and a square-root zero at the trailing edge. This series necessarily satisfies the steady Kutta condition at the trailing edge. However, as we will now show, this choice of basis expansion leads to invalid results at the endpoints when the aerofoil is permeable.

By sending $x \rightarrow -1$, we obtain the  asymptotic behaviours
{\setlength{\jot}{12pt}
\begin{align}
\addtocounter{equation}{1}
\psi(x,0)\hat{\gamma}_{\rm a}(x) & \sim \psi(-1,0) \hat{\gamma}_{0} \sqrt{\frac{2}{1+x}}, \tag{\theequation.a}\label{Eq:asympA}\\
\frac{1}{\pi}\dashint_{-1}^{1}\frac{\hat{\gamma}_{\rm a}(\xi)}{\xi-x}\d \xi&\sim \Phi^\ast(x),\tag{\theequation.b} \label{Eq:asympB} \\
2 \frac{\d y_{\rm a}}{\d x}(x) & \sim 2 \frac{\d y_{\rm a}}{\d x}(-1), \tag{\theequation.c} \label{Eq:asympC}
\end{align}
where $\Phi^\ast(x)= o\left((1+x)^{-1/2}\right)$ according to \citet[(29.8)]{NIM}. Substitution of these limits into \eqref{Eq:p_steady} results in a contradictory equation where the left-hand side scales like $(1+x)^{-1/2}$ whereas the right-hand side tends to a constant as $x \rightarrow -1$. Asymptotic analysis at the trailing edge generates similar contradictions. Consequently, the Chebyshev expansion generates spurious results at both endpoints, and the $\hat{\gamma}_{n}$ coefficients for $n>1$ must account for the contradiction, resulting in a slowly converging series. The modification of the square-root behaviour at the endpoints due to porosity is embedded in the partially-porous aerofoil solution by \citet{gil,gil2} and is detailed in the generalised porous aerofoil solution by \citet{Hajian2017}.


Suppose we do not explicitly enforce the square-root behaviour of $\hat{\gamma}_{\rm a}$ at the endpoints and instead express $\hat{\gamma}_{\rm a}$ in the form
\begin{align}
\hat{\gamma}_{\rm a}(x) =\weight{x}{\alpha}{-\beta} \hat{\gamma}_{\rm a}^\ast(x), \label{Eq:presExp1}
\end{align}
where $w^{a,b}$ represents the weight function
\begin{align}
\weight{x}{a}{b} \triangleq \left(1-x\right)^a \left(1+x\right)^b. \label{Eq:jacobiWeight}
\end{align}
{The function} $\hat{\gamma}_{\rm a}^\ast(x)$ is H\"older continuous on $x \in [-1,1]$ and is finite and \mbox{non-zero} at $x=\pm1$. The constants $\alpha$ and $\beta$ in \eqref{Eq:presExp1} are unknown and must be found as a part of the solution. Using the new expansion \eqref{Eq:presExp1}, the limits \eqref{Eq:asympA} and \eqref{Eq:asympB} instead become
%
%
\begin{align}
\addtocounter{equation}{1}
\psi(x,0)\hat{\gamma}_{\rm a}(x) & \sim \psi(-1,0) \hat{\gamma}_{\rm a}^\ast(-1) \frac{2^\alpha}{(1+x)^{\beta}}, \tag{\theequation.a}\label{Eq:asympJA}\\
\frac{1}{\pi}\dashint_{-1}^{1}\frac{\hat{\gamma}_{\rm a}(\xi)}{\xi-x}\d \xi&\sim  \hat{\gamma}_{\rm a}^\ast(-1) \frac{2^\alpha \cot(\pi \beta)}{(1+x)^{\beta}}+ \Phi^\ast(x),\tag{\theequation.b} \label{Eq:asympJB}
\end{align}%
where now $\Phi^*(x) = o((1+x)^{-\beta})$. Accordingly, by matching the singularities in the above two terms through \eqref{Eq:p_steady}, we obtain the following expression for $\beta$:
\begin{align}
\beta & = \frac{1}{\pi} \cot^{-1} \left(\psi(-1,0)\right) + n_\beta, \label{Eq:beta}
\end{align}
for $n_\beta \in \mathbb{Z}$. A similar procedure at the endpoint $x = 1$ yields a similar expression for $\alpha$:
\begin{align}
\alpha & = \frac{1}{\pi} \cot^{-1} \left(\psi(1,0)\right) + n_\alpha, \label{Eq:alpha}
\end{align}
for $n_\alpha \in \mathbb{Z}$.
{Physically speaking, we require a finite force when integrating the surface pressure round the leading edge, and that the Kutta condition holds at the trailing edge.  
Accordingly, we restrict $n_\alpha\geq 0$ and $n_\beta\geq -1$.}
Consequently, we seek an expansion of the vorticity distribution as a sequence of weighted Jacobi polynomials of the form
\begin{align}
\hat{\gamma}_{\rm a}(x) = \hat{\gamma}_{0} \weight{x}{\alpha}{-\beta}+ \weight{x}{\alpha}{1 - \beta} \sum_{n=1}^\infty \hat{\gamma}_{n} P_{n-1}^{\alpha,1-\beta} (x), \label{Eq:JacExpan}
\end{align}
where $P_n^{a,b}$ represents the $n^{\textnormal{th}}$ Jacobi polynomial with parameters $a$ and $b$. The Jacobi polynomials are a classical family of orthogonal polynomials \citep{Szego1939} and represent a generalisation of Chebyshev polynomials. Some important properties of the Jacobi polynomials are catalogued in appendix \ref{Ap:Jacobi}. 
{In the unsteady case, the presence of an effective density in the porous boundary condition implies that the parameters $\alpha$ and $\beta$ may be complex-valued.
However, this scenario is not an issue, as \mbox{\citet{Kuijlaars2005a}} have established a theory for Jacobi polynomials with generalised parameters.
}

Note that the inverse cotangent function in \eqref{Eq:beta} and \eqref{Eq:alpha} decreases monotonically for positive arguments. Therefore, the effect of porosity is to decrease the strength of the both leading-edge singularity and the trailing-edge zero. In the large porosity limit, $\psi \rightarrow \infty$, the singularity and zero vanish and we have $\alpha = \beta = 0$. Accordingly, the pressure jump along the chord also vanishes in this limit. 

We may now substitute our Jacobi polynomials expansion \eqref{Eq:JacExpan} into the singular integral equation \eqref{Eq:p_steady} and collocate at the Jacobi nodes to determine the coefficients $\hat{\gamma}_{n}$ following the procedure of \citet{Baddoo2019e} to furnish a solution to the full unsteady problem. 
This numerical technique is an example of a \textit{spectral method} \citep{Trefethen2000}, where the unknown function is expanded globally in terms of basis functions whose coefficients are chosen by collocation.

Before we apply this strategy to the unsteady problem, we point out that an alternative weight function should be used for higher-accuracy solutions. For uniformly porous aerofoils with constant $\psi$, the remainder terms in the asymptotic expansions \eqref{Eq:asympJA} and \eqref{Eq:asympJB} are regular. Accordingly, the Jacobi weight function \eqref{Eq:jacobiWeight} precisely captures the behaviour at the endpoints. However, in the general case of non-constant porosity profiles, the weight function should be written more generally as
\begin{align}
   \weight{x}{\alpha}{\beta}= (1-x)^{\alpha(x)} (1+x)^{\beta(x)}, \label{Eq:modWeight}
\end{align}
where $\alpha$ and $\beta$ are regular at $x = \pm1$ \citep{Hajian2018a}. Although a set of orthogonal polynomials could in principle be constructed with the general weight function \eqref{Eq:modWeight} via Gram--Schmidt orthogonalisation, we find it more appropriate to use the weight function \eqref{Eq:jacobiWeight} with the Jacobi polynomials due to the availability of many useful identities (see appendix \ref{Ap:Jacobi}) and practical ease of computation. Whilst our choice of weight function precludes the possibility of spectral accuracy, only a few polynomials are usually required to obtain a degree of accuracy that is finer than the size of other physical quantities that are being ignored {within the assumptions of the theoretical model}.

\input{sections/unsteady}

%

%% file: sections/unsteady.tex
\input{sections/alternate}
\input{sections/discontinuous}

\input{sections/noncirculatory}
\input{sections/quasi-static}


%% file: sections/alternate.tex
\subsection{Unsteady solution} \label{Sec:numUnsteady}
We adapt the steady solution {\eqref{Eq:JacExpan}} to the full unsteady problem \eqref{Eq:SIEfin}. The SF--VIE in the unsteady case is distinct from the singular integral equation of the steady case in a number of ways. Firstly, the forcing term $f_{\rm w}$ is not regular but possesses a logarithmic singularity at $x=1$. Secondly, the coefficient $\hat{\Gamma}$ multiplying $f_{\rm w}$ is unknown \emph{a priori} because it is proportional to the aerofoil circulation. Thirdly, the integral equation \eqref{Eq:SIEfin} now contains {a term} of Volterra type. We will now adapt the  solution approach in  \S\ref{Sec:steady} to address these issues simultaneously.

We first address the fact that the forcing term $f_{\textrm w}$ is not regular as $x\rightarrow 1$ in \eqref{Eq:SIEfin}. 
In particular, we have the logarithmic behaviour \citep{AS}
\begin{align}
f_{\rm w}(x) \sim -\frac{\i k}{\pi} \log(1-x) , \qquad \textrm{as } x \rightarrow 1.
\end{align}
 For the asymptotic matching procedure at the endpoints,
 we therefore require that the left side of \eqref{Eq:SIEfin} possesses a logarithmic singularity of the same strength of $f_{\rm w}$ at $x = 1$.
 If $\hat{\gamma}_{\rm a}$ itself is to be regular, the only way to generate this logarithmic singularity is through the principal value part of the operator $\mathcal{L}$ \citep{NIM}.
 In particular, we require
\begin{align}
	\hat{\gamma}_{\rm a}(1) = -\i k \hat{\Gamma}.
\end{align}
This behaviour may be alternatively derived by enforcing the Kutta condition and requiring the pressure to vanish at the trailing edge in \eqref{Eq:pres2a}. Accordingly, we adapt the expansion in \eqref{Eq:presExp1} and seek a solution of the form
\begin{align}
\hat{\gamma}_{\rm a}(x) =\weight{x}{\alpha}{-\beta} \hat{\gamma}_{\rm a}^\ast(x) - \i k  2^{\beta-1} \hat{\Gamma} 
\weight{x}{0}{1-\beta} , \label{Eq:presExp2}
\end{align}
where $\hat{\gamma}^\ast_{\rm a}$ is a smooth function. We now note the leading-order asymptotic behaviours as $x\rightarrow -1$:
{ \setlength{\jot}{12pt}%
\begin{align}
\addtocounter{equation}{1}
\int_{-1}^x \hat{\gamma}_{\rm a}(\xi) \d \xi &\sim \frac{(1+x)^{1-\beta}2^{\alpha} }{1-\beta} \hat{\gamma}^\ast_{\rm a}(-1) , \tag{\theequation.a} \\
\psi(x,k)\hat{\gamma}_{\rm a}(x) &\sim \frac{\psi(-1,k)2^{\alpha} }{(1+x)^\beta} \hat{\gamma}^\ast_{\rm a}(-1), \tag{\theequation.b} \\
\frac{1}{\pi}\dashint_{-1}^1 \frac{\hat{\gamma}_{\rm a}(\xi)}{\xi - x} \d \xi &
\sim \frac{\cot(\beta \pi)2^{\alpha} }{(1+x)^\beta }\hat{\gamma}^\ast_{\rm a}(-1) +\Phi^\ast(x), \tag{\theequation.c} \\
f_{\rm a}(x) + \hat{\Gamma} f_{\rm w}(x) &\sim f_{\rm a}(-1) + \hat{\Gamma} f_{\rm w}(-1). \tag{\theequation.d} 
\end{align}%
}%
Consequently, we see that the Volterra part of the SF--VIE does not contribute to the asymptotic behaviour at the leading edge, and the expression for $\beta$ is the same as for the steady case \eqref{Eq:beta} with $\psi(-1,0)$ replaced with $\psi(-1,k)$. 

We now inspect the behaviour as $x \rightarrow 1$ and track terms at higher orders to ensure the correct asymptotic matching. In this limit, we obtain the behaviours
{\setlength{\jot}{12pt}
\begin{align}
\addtocounter{equation}{1}
	\int_{-1}^x \hat{\gamma}_{\rm a}(\xi) \d \xi &\sim \hat{\Gamma}, \tag{\theequation.a} \\
		\psi(x,k) \hat{\gamma}_{\rm a}(x) &
	\sim \psi(1,k) \left(- \i k \hat{\Gamma} + \frac{(1-x)^\alpha}{2^\beta} \hat{\gamma}_{\rm a}^\ast(1) \right), \tag{\theequation.b} \\
\frac{1}{\pi}\dashint_{-1}^1 \frac{\hat{\gamma}_{\rm a}(\xi)}{\xi - x} \d \xi & \sim -\frac{\i k \hat{\Gamma
}}{\pi} \log(1-x) + \textnormal{const. } + \cot(\alpha \pi)2^{-\beta}(1-x)^\alpha \hat{\gamma}_{\rm a}(1), \tag{\theequation.c} \\
f_{\rm a}(x) + \hat{\Gamma} f_{\rm w}(x) &\sim f_{\rm a}(1) - \frac{\i k \hat{\Gamma}}{\pi} \log(1-x). \tag{\theequation.d} 
\end{align}
}%
We note that the logarithmic singularities on the third and fourth lines cancel by virtue of the Kutta condition. 
The unknown constant $\hat{\Gamma}$ will be found through the collocation procedure, so we must choose $\alpha$ such that the leading-order zero (proportional to $(1-x)^\alpha$) vanishes. 
Consequently, we again obtain the same expression for $\alpha$ as in the steady case \eqref{Eq:alpha}
with $\psi(1,0)$ replaced with $\psi(1,k)$. 

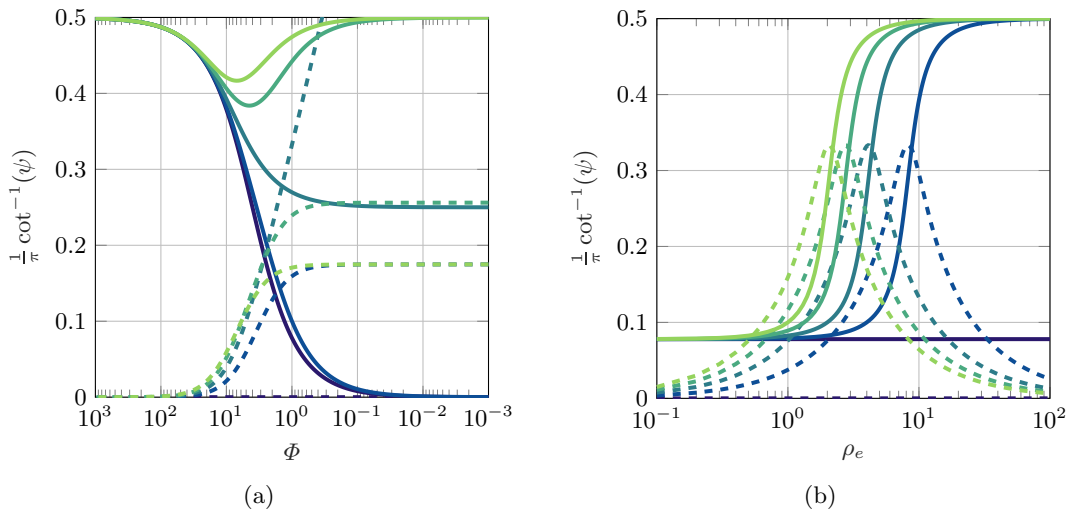
\begin{figure}
	\begin{subfigure}[b]{.45\linewidth}
		\setlength{\fheight}{5cm}
		\setlength{\fwidth}{.9\linewidth}
		\centering
		\input{images/alphVarPsi.tex}
		\caption{}
	\label{Fig:alphPlotA}
	\end{subfigure}
	\hfill
	\begin{subfigure}[b]{.45\linewidth}
		\setlength{\fheight}{5cm}
		\setlength{\fwidth}{.9\linewidth}
		\centering
		\input{images/alphVarRho.tex}
		\caption{}
		\label{Fig:alphPlotB}
	\end{subfigure}
	\caption{The real (\drawLine{ultra thick}) and imaginary (\drawLine{ultra thick, dashed}) parts of the first term of the Jacobi parameters $\alpha$ and $\beta$ as functions of porous medium resistivity 
		$\Phi$ and effective density $\rho_{\rm e}$ for a range of frequencies.
		The values of $\Phi$ and $\rho_{\rm e}$ here are taken to be constant with respect to chordwise position so $\alpha$ and $\beta$ values can be determined by a value of $\psi$ at the aerofoil endpoints in an application.
	In figure (a), $\rho_{\rm e} = 1$ and $k$ is 0 (purple), 1, 2, 3, 4 (green) and in figure (b) $\Phi = 1$ and $k=$0, 0.25, 0.5, 0.75, 1.}
	\label{Fig:jacobiParams}
\end{figure}

{Increasing the flow resistance $\Phi$ can have different effects on the Jacobi parameters $\alpha$ and $\beta$, as illustrated in figure \ref{Fig:alphPlotA}.
For very large values of $\Phi$, the Jacobi parameters are close to $1/2$ and the aerofoil is almost impermeable.
As $\Phi$ is decreased, the real part of the Jacobi parameters tend to one of three values:
if $k \rho_{\rm e}<2$ the real part tends to 0, if $k \rho_{\rm e} = 2$ it tends to $1/4$,
and if $k \rho_{\rm e}>2$ it tends to $1/2$.
Similar behaviour can be observed as the effective density $\rho_{\rm e}$ is increased in figure \mbox{\ref{Fig:alphPlotB}}.
The parameters transition from a lower value of around 0.1 to the impermeable limit of 1/2.
This behaviour is due to the branch point of $\cot^{-1}(z)$ at $z = \i$.}

{The behaviour at the endpoints} motivates an expansion of $\hat{\gamma}_{\rm a}$ of the form
\begin{align}
	\hat{\gamma}_{\rm a}(x) &=  - \i k \hat{\Gamma} \weight{x}{0}{1-\beta}2^{\beta-1} + \weight{x}{\alpha}{-\beta} \hat{\gamma}_0 
+ \weight{x}{\alpha}{1-\beta} \sum_{n=1}^\infty \hat{\gamma}_n P_{n-1}^{\alpha,1-\beta}(x). \label{Eq:newExp}
\end{align}
Recall that we do not know the circulation $\hat{\Gamma}$ \emph{a priori} because it must be determined as part of the solution. 
However, integrating \eqref{Eq:newExp} yields 
\begin{align}
	\hat{\Gamma} & = -2 \i k \hat{\Gamma} B(2-\beta,1)+ \hat{\gamma}_0 2^{1+\alpha-\beta} B(1-\beta,1+\alpha) + \hat{\gamma}_1 2^{2+\alpha - \beta} B(2-\beta,1+\alpha), \label{Eq:gam2}
\end{align}
where we have used \eqref{Eq:JacobiOrthogonality0} to express the integral in terms of the Beta function, $B$.
Rearranging \eqref{Eq:gam2} then yields an equation for $\hat{\Gamma}$ in terms of the first two coefficients of the Jacobi expansion:
\begin{align}
\hat{\Gamma}&= \frac{\hat{\gamma}_0 2^{1+\alpha-\beta} B(1-\beta,1+\alpha) + \hat{\gamma}_1 2^{2+\alpha - \beta} B(2-\beta,1+\alpha)}{1+2\i k B(2- \beta,1)}.
\end{align}
It proves convenient to express
\begin{align}
\hat{\Gamma} &= \hat{\gamma}_0 \hat{\Gamma}_0 + \hat{\gamma}_1 \hat{\Gamma}_1,
\end{align}
so that the SF--VIE equation \eqref{Eq:SIEfin} may be expressed in the new form
\begin{align}
\mathcal{H}\hat{\gamma}_{\rm a}&= f_{\textrm{a}}(x), \label{Eq:Heq}
\end{align}
where the new, regularised operator $\mathcal{H}$ is defined as
\begin{align}
\mathcal{H}\hat{\gamma}_{\rm a} & \triangleq \mathcal{L}\hat{\gamma}_{\rm a}(x)-\left(\hat{\gamma}_0 \hat{\Gamma}_0 + \hat{\gamma}_1 \hat{\Gamma}_1\right) f_{\textrm{w}}(x).
\end{align}
{It is straightforward to show that $\mathcal{H}\hat{\gamma}_{\rm a}(x)$ is bounded at the endpoints.} 
This regularisation improves the conditioning of the collocation matrix below.

{We may now find approximate solutions for the coefficients $\hat{\gamma}_n$ by a collocation procedure. 
In particular, we truncate the infinite sum in \eqref{Eq:newExp} at $N$ and collocate the resulting equation \eqref{Eq:Heq} at the zeros of the Jacobi polynomial $P_{N+1}^{\alpha,-\beta}$.
The result is an $(N+1) \times (N+1)$ system of linear equations for the coefficients $\hat{\gamma}_n$. When constructing the system of equations, it is useful to note that the effect of the operator $\mathcal{L}$ on each individual weighted Jacobi polynomial can be computed using standard functions. For example,}
\begin{align}
	\mathcal{L} w^{a,b} P_n^{a,b}(x) &= 
\frac{\weight{x}{a}{b}}{\pi} Q_n^{a,b}(x)
	-\i k \psi(x,k) I_n^{a,b}(x)
-\psi(x,k) \weight{x}{a}{b}P_n^{a,b}(x),
\end{align}
{%
where $I_n^{a,b}$ is the integral of the weighted Jacobi (see \eqref{Eq:WeightedJacobiIntegral}) and $Q_n^{a,b}$ is the associated Jacobi function of the second kind (see \eqref{Eq:JacobiHilbertTransform}). 
This result allows us to rapidly evaluate the collocation matrix below.
The system of equations for the coefficients $\hat{\gamma}_n$ is given by $\boldsymbol{A} \hat{\boldsymbol{\gamma}} = \boldsymbol{f}_{\rm a}$, where}
\begin{align}
        \boldsymbol{A}^T=
    \begin{bmatrix}
    L_{0,0}^{\alpha,-\beta}+\hat{\Gamma}_0 (C L_{0,0}^{0,1-\beta}-f_{\rm w}(x_0)) 
    & \cdots &
    L_{0,N}^{\alpha,-\beta}+\hat{\Gamma}_0 (C L_{0,N}^{0,1-\beta}-f_{\rm w}(x_N))\\
    L_{0,0}^{\alpha,1-\beta}+\hat{\Gamma}_1 (C L_{0,0}^{0,1-\beta}-f_{\rm w}(x_0)) 
    & \cdots &
    L_{0,N}^{\alpha,1-\beta}+\hat{\Gamma}_1 (C L_{0,N}^{0,1-\beta}-f_{\rm w}(x_N))\\
    L_{1,0}^{\alpha,1-\beta} 
    & \cdots &
    L_{1,N}^{\alpha,1-\beta}\\
    \vdots & \ddots & \vdots\\
    L_{N,0}^{\alpha,1-\beta} 
    & \cdots &
    L_{N,N}^{\alpha,1-\beta}
    \end{bmatrix},
    \end{align}
    \begin{align}
    \hat{\boldsymbol{\gamma}} =
    \begin{bmatrix}
    \hat{\gamma_0}\\
    \vdots\\
    \hat{\gamma}_N
    \end{bmatrix},
    \qquad \qquad 
    \boldsymbol{f}_{\rm a} =
    \begin{bmatrix}
    f_{\rm a}(x_0)\\
    \vdots\\
    f_{\rm a}(x_N)
    \end{bmatrix},
\end{align}
{%
and where $ C =- \i k 2^{\beta - 1}$ and $L_{i,j}^{a,b} = \mathcal{L} w^{a,b}P^{a,b}_{i}(x_j)$.
We solve this system in the least-squares sense using, for example, the backslash operation in \textsc{Matlab}.}

For low reduced frequencies that are typically of interest ($k<3$), fewer than 10 Jacobi polynomials are usually required to resolve the vorticity distribution. As the reduced frequency increases, more polynomials are required to resolve an asymptotic inner region near the trailing edge; we comment on this scenario further in appendix \mbox{\ref{Sec:asymp}}.

We now present several extensions to our method, including the case where the porosity distribution is discontinuous. The numerical method verifies the solution for quasi-steady aerodynamics and establishes both the circulatory and non-circulatory vorticity distributions for generalized unsteady aerodynamics of porous aerofoils.

%% file: images/alphVarPsi.tex
%
%
\definecolor{mycolor1}{rgb}{0.16295,0.09522,0.42257}%
\definecolor{mycolor2}{rgb}{0.05084,0.30506,0.58943}%
\definecolor{mycolor3}{rgb}{0.17572,0.48809,0.53686}%
\definecolor{mycolor4}{rgb}{0.29112,0.66634,0.50553}%
\definecolor{mycolor5}{rgb}{0.58106,0.82876,0.36638}%
\begin{tikzpicture}[%
trim axis left, trim axis right
]

\begin{axis}[%
width=0.951\fwidth,
height=\fheight,
at={(0\fwidth,0\fheight)},
scale only axis,
x dir=reverse,
xmode=log,
xmin=0.001,
xmax=1000,
xminorticks=true,
xlabel style={font=\color{white!15!black}},
xlabel={$\Phi$},
ymin=0,
ymax=0.5,
ylabel style={font=\color{white!15!black}},
ylabel={$\frac{1}{\pi} \cot^{-1}(\psi)$},
axis background/.style={fill=white},
xmajorgrids,
ymajorgrids
]
\addplot [color=mycolor1, line width=1.5pt, forget plot]
  table[row sep=crcr]{%
0.001	7.95774698882923e-05\\
0.00421332174384729	0.000335285367184746\\
0.00939664831495469	0.000747760138392284\\
0.0163385387780986	0.00130017237395252\\
0.0250841505927754	0.00199610711408038\\
0.0359381366380463	0.00285978909822537\\
0.0487178021879463	0.00387664783893271\\
0.0633580499265825	0.00504145182656579\\
0.080150069615654	0.00637728647762881\\
0.0986265846131282	0.0078468643208649\\
0.11805165285688	0.00939152596846382\\
0.139361927422414	0.0110855858183547\\
0.162259528707809	0.0129051276281116\\
0.188919277620767	0.0150225550259666\\
0.216938351838518	0.0172465091906906\\
0.245691646298279	0.0195269876788777\\
0.278255940220712	0.0221072899170038\\
0.310808217386906	0.0246837348893134\\
0.347168681892656	0.0275577479480638\\
0.3824569722467	0.0303427179531925\\
0.421332174384729	0.0334053677958703\\
0.464158883361278	0.036772130595299\\
0.504315948717136	0.0399215475360273\\
0.547947233690029	0.0433345355612809\\
0.595353313081437	0.0470314459814003\\
0.646860766154633	0.0510337302050656\\
0.702824426430835	0.0553638641460692\\
0.753142016597437	0.0592395924840554\\
0.80706201411495	0.0633731369873649\\
0.864842327573173	0.0677786684548325\\
0.926759330114688	0.0724704795146667\\
0.99310918137498	0.0774628237391357\\
1.06420924406472	0.0827697172063755\\
1.14039960197003	0.0884046981789535\\
1.22204468663149	0.0943805411388934\\
1.30953502048267	0.100708922490223\\
1.38401609657313	0.106032366633491\\
1.46273335620113	0.111592356197832\\
1.54592773641948	0.117392313885647\\
1.63385387780986	0.123434561592959\\
1.72678090388436	0.129720149566397\\
1.82499324481615	0.136248682204163\\
1.92879150802078	0.143018145075406\\
2.06688024962908	0.151812811088829\\
2.21485523372636	0.160966136298629\\
2.37342425002387	0.170461287914681\\
2.54334576130465	0.180276041751315\\
2.72543253128103	0.190382662229959\\
2.96122543798803	0.202848786698768\\
3.21741815067637	0.215620097343145\\
3.59381366380463	0.232990097308154\\
5.07815211232768	0.287627313559944\\
5.51749237612913	0.300328482064094\\
5.99484250318941	0.312706599752356\\
6.42403365939419	0.32272809790272\\
6.8839520696455	0.332448749282297\\
7.37679760252773	0.341842489211144\\
7.90492762269643	0.350888746489324\\
8.47086826655741	0.359572272864885\\
8.9526571259964	0.366250889159664\\
9.461848194722	0.37268761155701\\
10	0.378881058409156\\
10.5687597118481	0.384831401899936\\
11.1698681846782	0.390540193999965\\
11.8051652856881	0.39601019305493\\
12.650337203959	0.402517758042691\\
13.5560178532937	0.408667504837176\\
14.5265392594678	0.414470424156538\\
15.5665435927106	0.419938789387449\\
16.6810053720006	0.425085797757363\\
17.8752552590424	0.429925260584866\\
19.1550055557353	0.434471340681013\\
20.5263775270925	0.438738333602879\\
21.9959306803007	0.442740488658054\\
23.5706941399673	0.446491865191719\\
25.6099310025846	0.450681831814651\\
27.8255940220713	0.454553484249139\\
30.2329468440578	0.458128881108547\\
32.8485736603005	0.461429019718976\\
35.6904934567523	0.464473777135002\\
39.3182875570577	0.467728145830413\\
43.314832233764	0.470688133942104\\
47.7176094893875	0.473379434380245\\
53.2999408084409	0.476156497472136\\
59.5353313081437	0.478645808315348\\
67.4262224177834	0.48113865811875\\
76.3629826128224	0.483341706872141\\
87.6885609458743	0.485490039916715\\
100.693863147603	0.487361986087375\\
117.23818032866	0.489143930448512\\
138.401609657313	0.490802959466847\\
165.660595894991	0.49231566092049\\
201.04964162605	0.493667874376443\\
250.841505927754	0.494924557503059\\
321.741815067637	0.49604287073453\\
424.255643071778	0.496998974933099\\
591.250841383188	0.497846565229237\\
870.843149769072	0.498537933506233\\
1000	0.49872676724581\\
};
\addplot [color=mycolor1, dashed, line width=1.5pt, forget plot]
  table[row sep=crcr]{%
0.001	0\\
1000	0\\
};
\addplot [color=mycolor2, line width=1.5pt, forget plot]
  table[row sep=crcr]{%
0.001	0.000106103288517456\\
0.00361874981241128	0.000383960954410068\\
0.00763629826128224	0.000810233347835254\\
0.0130953502048267	0.00138944436765076\\
0.0198288394912707	0.00210385160085691\\
0.028018665564592	0.0029727214987525\\
0.0374605003274899	0.00397432109257334\\
0.0487178021879463	0.00516832444910253\\
0.0607832312829723	0.00644775759081329\\
0.0747952251562182	0.00793314464198724\\
0.0895265712599639	0.00949413529206389\\
0.10568759711848	0.0112056969916141\\
0.123052400435926	0.0130434796295069\\
0.141302599059953	0.0149733251559567\\
0.162259528707809	0.0171870037395245\\
0.183765620038817	0.0194556687310263\\
0.208122156998634	0.0220208385245497\\
0.232469705998565	0.0245800758522958\\
0.259665597293487	0.0274321352042097\\
0.286059553517574	0.0301927668889443\\
0.315136348486648	0.0332248277054354\\
0.347168681892656	0.0365529165672878\\
0.3824569722467	0.0402031861081693\\
0.415545533471887	0.0436092904689676\\
0.45149677720361	0.0472904348948897\\
0.490558370636505	0.0512652250999102\\
0.532999408084409	0.0555525826333949\\
0.579112264764176	0.0601714858612561\\
0.62057288067765	0.0642873491854177\\
0.665001803043112	0.0686569434302249\\
0.712611543011175	0.0732902420180723\\
0.763629826128224	0.0781964867266138\\
0.818300681586739	0.0833839471146534\\
0.876885609458743	0.0888596605372953\\
0.939664831495469	0.0946291594893673\\
1.00693863147603	0.10069619576494\\
1.07902879151618	0.107062473703794\\
1.15628013120738	0.113727407273248\\
1.23906215694792	0.120687917467056\\
1.32777082935543	0.127938286994769\\
1.42283045721435	0.135470088006313\\
1.52469572701757	0.143272195295544\\
1.63385387780986	0.151330891952872\\
1.77520801171763	0.161317189222471\\
1.92879150802078	0.171617871793983\\
2.09566239948043	0.182197643403415\\
2.30867799418717	0.194842878378926\\
2.5787628875938	0.209616128059598\\
2.96122543798803	0.228425370910732\\
3.85110700232557	0.264583789074238\\
4.5462954695324	0.287239446623284\\
5.07815211232768	0.302073588407476\\
5.59432570616938	0.31478851103702\\
6.16296625513294	0.327190001299812\\
6.69616005485322	0.33752432308893\\
7.27548352919623	0.347548122190793\\
7.90492762269643	0.357229160918853\\
8.47086826655741	0.365014547460856\\
9.07732652521023	0.37252884551269\\
9.72720319245055	0.379761098370453\\
10.423606739764	0.386703050221841\\
11.1698681846782	0.393349153667215\\
11.9695570235904	0.399696492409088\\
12.8264983052806	0.405744631827595\\
13.7447909267754	0.411495413323156\\
14.728827239075	0.416952709414958\\
15.7833140565212	0.422122155998771\\
16.9132951702965	0.427010876372798\\
18.1241754737424	0.431627209131267\\
19.4217468148903	0.435980449235717\\
20.8122156998634	0.440080608836112\\
22.6128006633728	0.444681561749709\\
24.5691646298279	0.448951965927669\\
26.6947849403432	0.452910804300034\\
29.0043049386399	0.456576987660169\\
31.5136348486648	0.459969099085\\
34.2400613797143	0.463105199158037\\
37.72042493417	0.466463578185225\\
41.5545533471888	0.469523441909262\\
45.7784053837662	0.472309544483567\\
51.1338753841433	0.475188061082048\\
57.1158647812643	0.477771151007659\\
63.7976680860628	0.480088069076134\\
72.2534949178721	0.482409538493611\\
82.9695852083491	0.484674884231418\\
95.2750047242729	0.486649873094728\\
110.928986489522	0.488530721405076\\
130.953502048267	0.490282448338343\\
156.74554102056	0.491880111743173\\
190.230118866894	0.493308570939382\\
237.342425002387	0.494636320645417\\
300.246170908555	0.495759786883445\\
395.911026646846	0.496784217526769\\
544.171428686589	0.497660297197037\\
790.492762269642	0.498389333098614\\
1000	0.498726772338586\\
};
\addplot [color=mycolor2, dashed, line width=1.5pt, forget plot]
  table[row sep=crcr]{%
0.001	0.17484955859915\\
0.0895265712599639	0.174707997408724\\
0.157833140565212	0.174410565591705\\
0.226128006633728	0.173951697909293\\
0.294082017058706	0.17333830622245\\
0.361874981241128	0.17257502183711\\
0.427199396630678	0.171701649721059\\
0.497389595879006	0.17061898264227\\
0.563314267060136	0.169472903802402\\
0.637976680860628	0.168032927059132\\
0.712611543011175	0.166453291343711\\
0.785045620020451	0.164797034913057\\
0.864842327573173	0.162844877467031\\
0.939664831495469	0.160905351946489\\
1.02096066230605	0.158692421378166\\
1.10928986489522	0.156180138061102\\
1.20526093687084	0.153343808120571\\
1.29154966501488	0.150716511190465\\
1.38401609657313	0.14783774508123\\
1.4831025143361	0.144698995675038\\
1.58928286562298	0.141294738242713\\
1.70306502925284	0.137622991573656\\
1.82499324481615	0.13368581011465\\
1.95565071586595	0.129489679565988\\
2.09566239948043	0.125045784454073\\
2.27697025538168	0.119408922737905\\
2.47396410088681	0.113475215256861\\
2.72543253128103	0.106239031177513\\
3.0442722120643	0.0976627991069785\\
4.60960448682843	0.0651449608815051\\
5.07815211232768	0.0580605186563599\\
5.51749237612913	0.0522857839576001\\
5.99484250318941	0.0468293417017587\\
6.42403365939419	0.0425482923070271\\
6.8839520696455	0.0385229595637377\\
7.37679760252773	0.0347613887332794\\
7.90492762269643	0.0312670880120378\\
8.47086826655741	0.028039384284567\\
9.07732652521023	0.0250738865274456\\
9.72720319245055	0.0223630229763656\\
10.423606739764	0.0198966172817214\\
11.3254131515281	0.0172423305940255\\
12.3052400435926	0.0148986637714192\\
13.3698374182495	0.0128400940800577\\
14.5265392594678	0.0110404233368615\\
16.003103137387	0.00923340008908058\\
17.6297537528721	0.00770376131383088\\
19.6922025547917	0.00624790161375133\\
22.3022329796594	0.0049233523982628\\
25.6099310025846	0.00376841752829993\\
29.8177229001967	0.00280139874073182\\
35.6904934567523	0.00196830006691817\\
43.9180089259609	0.00130672588560499\\
57.1158647812643	0.0007758461350309\\
80.7062014114951	0.000389758584043332\\
134.626057929891	0.000140346953390935\\
344.776405473446	2.14186442710762e-05\\
1000	2.54642816122086e-06\\
};
\addplot [color=mycolor3, line width=1.5pt, forget plot]
  table[row sep=crcr]{%
0.001	0.250019894367924\\
0.0104959323055823	0.250208809818803\\
0.028018665564592	0.250557411361306\\
0.0529326605836056	0.251053046455875\\
0.0858882855954625	0.251708627505924\\
0.12650337203959	0.252516494887641\\
0.173876240021625	0.253458613349748\\
0.229276931286565	0.254560071384811\\
0.290043049386399	0.255767696891096\\
0.361874981241128	0.257194369764042\\
0.439180089259609	0.258728448884231\\
0.525679112201842	0.260443040631597\\
0.62057288067765	0.262321230981409\\
0.722534949178721	0.264335481689367\\
0.829695852083491	0.266447472085134\\
0.939664831495469	0.268608772377136\\
1.06420924406472	0.271048194732551\\
1.1887076977119	0.273476816026633\\
1.32777082935543	0.276176544608227\\
1.4831025143361	0.279174170715339\\
1.63385387780986	0.282063546541711\\
1.79992850678248	0.285221933697296\\
1.98288394912707	0.28866892137033\\
2.18443607114943	0.292423801331\\
2.40647515001542	0.29650494577959\\
2.61467321180109	0.300275443978098\\
2.8408836901833	0.304306896545274\\
3.08666494333727	0.308606294687837\\
3.35371015200293	0.313178061462339\\
3.64385898376354	0.318023412536787\\
3.95911026646846	0.32313969891566\\
4.3016357581068	0.328519772607824\\
4.73887960971766	0.335113338511482\\
5.22056752784697	0.342020274673689\\
5.75121707184161	0.349201217957769\\
6.42403365939419	0.357679474715157\\
7.47952251562182	0.369650650974917\\
10.1392540755882	0.393683645617758\\
11.4831241454351	0.403155711608967\\
12.650337203959	0.410252712438928\\
13.9361927422414	0.417061783924223\\
15.3527502878042	0.423546983102903\\
16.6810053720006	0.428828363094863\\
18.1241754737424	0.433842683208341\\
19.6922025547917	0.438584569408761\\
21.3958887134342	0.44305291239281\\
23.2469705998565	0.447250218940896\\
25.2582002696278	0.451181957342004\\
27.4434330322837	0.45485593304725\\
30.2329468440578	0.458829299782751\\
33.3060034362459	0.46248221692546\\
36.6914237840249	0.465833433293867\\
40.4209583979631	0.468902397593598\\
44.5295850994266	0.471708687546212\\
49.7389595879007	0.474618939250972\\
55.5577622239888	0.47723904919233\\
62.9214610961034	0.479872613209685\\
71.2611543011175	0.482207266211278\\
81.8300681586739	0.484489732725069\\
93.966483149547	0.486482663255603\\
109.405470720574	0.488382870665325\\
129.154966501488	0.490154347357236\\
154.592773641948	0.491771253064082\\
187.617469143912	0.493217749190333\\
230.867799418717	0.494487190072046\\
292.055551218275	0.495641509859458\\
379.821530619074	0.496648290816002\\
514.886745013749	0.49752734535505\\
737.679760252773	0.49827406186041\\
1000	0.498726787616666\\
};
\addplot [color=mycolor3, dashed, line width=1.5pt, forget plot]
  table[row sep=crcr]{%
0.342400613797143	0.501677145436691\\
0.61204983724767	0.40955360253031\\
0.876885609458743	0.352813360519821\\
1.14039960197003	0.311644668747896\\
1.40328908478587	0.279440319424739\\
1.65660595894991	0.253957792155028\\
1.92879150802078	0.230901813340326\\
2.18443607114943	0.212318885283424\\
2.43998629725955	0.196066663796453\\
2.72543253128103	0.180118467643508\\
3.00246170908555	0.166459899709017\\
3.30764978074424	0.153124505889092\\
3.59381366380463	0.141988450505627\\
3.90473523688556	0.13115871787125\\
4.24255643071778	0.120669566131949\\
4.5462954695324	0.112213924768643\\
4.87178021879463	0.104039064029329\\
5.22056752784697	0.0961644530090227\\
5.59432570616938	0.0886084992007259\\
5.99484250318941	0.0813880442313919\\
6.42403365939419	0.0745178584794779\\
6.78940681269611	0.0692822132187647\\
7.17556091893693	0.0642835248767133\\
7.58367791499719	0.0595253682361934\\
8.0150069615654	0.0550099026182052\\
8.47086826655741	0.0507378199388611\\
8.9526571259964	0.0467083279185068\\
9.461848194722	0.0429191693022335\\
10	0.0393666761086\\
10.7159339982267	0.0352511480518629\\
11.4831241454351	0.0314851497949125\\
12.3052400435926	0.0280537860541017\\
13.1862140139475	0.0249401172566723\\
14.1302599059953	0.0221256655699471\\
15.1418932530435	0.0195909183571277\\
16.2259528707809	0.017315804774416\\
17.6297537528721	0.0148999598199118\\
19.1550055557353	0.0127945414568291\\
20.8122156998634	0.0109664282495161\\
22.9276931286565	0.00914245975348349\\
25.2582002696278	0.00760748317545001\\
28.2130767593947	0.00615414670264558\\
31.9524750575921	0.00483828803200881\\
36.6914237840249	0.00369587122247417\\
42.7199396630678	0.00274285455604373\\
51.1338753841433	0.00192438051942512\\
62.9214610961034	0.00127610288301572\\
81.8300681586739	0.00075696721542684\\
115.628013120738	0.000380020119540525\\
192.879150802078	0.000136781147864085\\
500.840798984821	2.03009009966948e-05\\
1000	5.09279521132555e-06\\
};
\addplot [color=mycolor4, line width=1.5pt, forget plot]
  table[row sep=crcr]{%
0.001	0.499936338029342\\
0.00477176094893875	0.499696220977841\\
0.0109405470720574	0.499303511755836\\
0.0195565071586595	0.498755043286914\\
0.030442722120643	0.498062141688361\\
0.0436153778920801	0.497223904456043\\
0.0591250841383188	0.49623733919882\\
0.0768928372075831	0.495107838014865\\
0.0972720319245054	0.49381351605982\\
0.119695570235904	0.492391200150022\\
0.145265392594678	0.490772219210614\\
0.171488196987054	0.489115751539759\\
0.20244465099768	0.487166218303089\\
0.235706941399673	0.485079860199332\\
0.270665207003324	0.482897933994866\\
0.306539529505653	0.480671927584412\\
0.347168681892656	0.478168905270905\\
0.393182875570577	0.475359963878837\\
0.439180089259609	0.472582467640034\\
0.490558370636505	0.469519586109722\\
0.547947233690029	0.466152238423232\\
0.61204983724767	0.462463894476623\\
0.674262224177834	0.458963436665013\\
0.742798248256492	0.455203239999604\\
0.818300681586739	0.451183760448804\\
0.914031074875623	0.446281565548817\\
1.02096066230605	0.441072381508612\\
1.15628013120738	0.434893831429962\\
1.34626057929891	0.427004760003265\\
1.90230118866894	0.408902392402677\\
2.09566239948043	0.404160999207264\\
2.27697025538168	0.400339345392025\\
2.47396410088681	0.396801379047803\\
2.65108360190854	0.394109001429954\\
2.8408836901833	0.391680606216485\\
3.0442722120643	0.389542912756238\\
3.26222200971167	0.387719626145629\\
3.49577557436328	0.386231052273032\\
3.74605003274899	0.385093799898443\\
3.95911026646846	0.384445614822283\\
4.18428850790158	0.384035106986162\\
4.4222739805059	0.383865207194551\\
4.67379510799246	0.383937128045791\\
4.93962174387832	0.384250356261554\\
5.22056752784697	0.384802655914739\\
5.51749237612913	0.385590082813809\\
5.91250841383188	0.386896286550414\\
6.33580499265825	0.388546387618966\\
6.78940681269611	0.390521986628801\\
7.27548352919623	0.392801118323812\\
7.79636013040524	0.395358590213645\\
8.47086826655741	0.398755385329803\\
9.20373199661822	0.402459959912136\\
10.1392540755882	0.407096369333662\\
11.3254131515281	0.41269805309937\\
13.3698374182495	0.421425224468992\\
16.6810053720006	0.433041454836874\\
18.8919277620767	0.439304631441873\\
21.102034285686	0.444606870312133\\
23.5706941399673	0.449612876768106\\
25.9665597293487	0.453728107107438\\
28.6059553517574	0.457585764006382\\
31.5136348486648	0.461183207851405\\
34.7168681892656	0.464523197834886\\
38.7782841458946	0.468034017321833\\
43.314832233764	0.471232482536316\\
48.3820966492596	0.474136697678073\\
54.0421642070592	0.476766583339739\\
61.204983724767	0.479423026320463\\
69.3171727615541	0.481787738663286\\
79.5977700231499	0.484107449124391\\
91.4031074875623	0.486138485482924\\
106.420924406472	0.488079248718198\\
125.631660247412	0.489891736689766\\
150.375532129974	0.491548369540732\\
182.499324481615	0.493031961634833\\
224.569799553977	0.494334958397588\\
284.08836901833	0.4955204503687\\
369.46012051993	0.496554826625576\\
500.840798984821	0.497458214675929\\
717.556091893692	0.498225731256641\\
1000	0.498726813079318\\
};
\addplot [color=mycolor4, dashed, line width=1.5pt, forget plot]
  table[row sep=crcr]{%
0.001	0.256149980264798\\
0.0858882855954625	0.25600924784875\\
0.153527502878042	0.255701206259523\\
0.216938351838518	0.255256632884208\\
0.282130767593947	0.254645318643718\\
0.347168681892656	0.253883477675217\\
0.409838367175726	0.25301020043957\\
0.477176094893875	0.251925337763545\\
0.540421642070592	0.250774134713975\\
0.603643850607587	0.249501628328672\\
0.674262224177834	0.247944483802691\\
0.742798248256492	0.246305078915639\\
0.818300681586739	0.244363888197163\\
0.889096598952916	0.242425665101463\\
0.966017479952265	0.240202450799985\\
1.04959323055823	0.237663146748856\\
1.12473717836475	0.235283019790615\\
1.20526093687084	0.23264392453068\\
1.29154966501488	0.229728982275068\\
1.38401609657313	0.226522678086671\\
1.4831025143361	0.223011410878593\\
1.58928286562298	0.21918405354593\\
1.70306502925284	0.215032495005783\\
1.82499324481615	0.21055213367157\\
1.95565071586595	0.205742292442519\\
2.09566239948043	0.200606529065622\\
2.24569799553977	0.195152822526111\\
2.40647515001542	0.189393625188742\\
2.5787628875938	0.183345780538585\\
2.76338529005317	0.177030316120005\\
3.00246170908555	0.169133805553741\\
3.26222200971167	0.160937492359663\\
3.59381366380463	0.151070632614407\\
4.07014245321944	0.138052272717757\\
5.67222897164454	0.103114495270383\\
6.24878807200689	0.0933056845087608\\
6.78940681269611	0.0851735333618362\\
7.37679760252773	0.0773531436537569\\
7.90492762269643	0.0711091340518863\\
8.47086826655741	0.0651389430068932\\
9.07732652521023	0.0594624717255039\\
9.72720319245055	0.0540953686956716\\
10.423606739764	0.0490488555719972\\
11.1698681846782	0.0443296814050309\\
11.9695570235904	0.039940212245646\\
12.8264983052806	0.0358786531565034\\
13.7447909267754	0.0321393896897892\\
14.728827239075	0.0287134273447438\\
15.7833140565212	0.0255889015004978\\
16.9132951702965	0.0227516274354582\\
18.1241754737424	0.0201856603405677\\
19.4217468148903	0.0178738382642347\\
21.102034285686	0.0154099315962677\\
22.9276931286565	0.0132549222013578\\
24.9113002606779	0.0113778705425323\\
27.4434330322837	0.00949957531117507\\
30.2329468440578	0.00791458344601459\\
33.7698031082509	0.00641024011153934\\
37.72042493417	0.00518170499083759\\
42.7199396630678	0.00407092740543114\\
49.0558370636505	0.00310778337688777\\
57.9112264764176	0.00224327462432061\\
70.2824426430835	0.00153049913344017\\
88.9096598952916	0.000960110522692847\\
120.526093687084	0.00052402128485296\\
185.04070195423	0.000222775824791466\\
374.605003274899	5.44193483924538e-05\\
1000	7.63904004408289e-06\\
};
\addplot [color=mycolor5, line width=1.5pt, forget plot]
  table[row sep=crcr]{%
0.001	0.499973474176949\\
0.00852964449974103	0.499773744647867\\
0.0218443607114943	0.499420568655971\\
0.0407014245321944	0.498920415000701\\
0.065134909462728	0.498272463410836\\
0.0946184819472199	0.497490842751172\\
0.130051125217341	0.496552041393321\\
0.171488196987054	0.49545515531468\\
0.216938351838518	0.494253666272526\\
0.270665207003324	0.492836165564928\\
0.328485736603004	0.491314820323446\\
0.393182875570577	0.489618730457887\\
0.464158883361278	0.487766965725809\\
0.540421642070592	0.485789449410032\\
0.62057288067765	0.483726760729609\\
0.712611543011175	0.481380688985285\\
0.80706201411495	0.479001328604729\\
0.914031074875623	0.476345086216187\\
1.03517795563018	0.4733914019719\\
1.1723818032866	0.47012328090439\\
1.30953502048267	0.466945329291387\\
1.46273335620113	0.463508689266938\\
1.65660595894991	0.459341849466719\\
1.87617469143912	0.454881960382419\\
2.15443469003188	0.449641506941032\\
2.65108360190854	0.4414517418143\\
3.21741815067637	0.433881710389106\\
3.59381366380463	0.429816456050477\\
3.95911026646846	0.426537255434193\\
4.3016357581068	0.424000882620078\\
4.67379510799246	0.421772642765304\\
5.00840798984821	0.420185889299391\\
5.36697694554048	0.418871103945419\\
5.75121707184161	0.417849216903774\\
6.16296625513294	0.417136823940738\\
6.6041939623303	0.416745660220522\\
7.07701066118189	0.41668222617946\\
7.58367791499719	0.416947579735317\\
8.12661920009194	0.417537302439785\\
8.70843149769072	0.418441640789787\\
9.33189771573324	0.419645817578057\\
10	0.42113050160727\\
10.7159339982267	0.422872417210173\\
11.6430313292088	0.425264775243085\\
12.650337203959	0.427939050427198\\
13.9361927422414	0.431344447568223\\
15.7833140565212	0.436043255365804\\
19.4217468148903	0.444242167634545\\
23.5706941399673	0.451806977114441\\
27.0665207003324	0.456934299438957\\
30.6539529505653	0.461266027424426\\
34.7168681892656	0.465290646764296\\
38.7782841458946	0.468595827952311\\
43.314832233764	0.471642232025621\\
49.0558370636505	0.474766259087606\\
55.5577622239888	0.477582953158716\\
62.9214610961034	0.480111196138949\\
71.2611543011175	0.482372507381073\\
81.8300681586739	0.484599433227631\\
95.2750047242729	0.48673724446869\\
110.928986489522	0.488586236715519\\
130.953502048267	0.490316269281033\\
156.74554102056	0.491899868340171\\
190.230118866894	0.493319637820715\\
237.342425002387	0.494642024455306\\
304.42722120643	0.495820715661818\\
401.424249049932	0.496829558696664\\
551.749237612913	0.497692883718553\\
801.50069615654	0.498411601946564\\
1000	0.498726848725319\\
};
\addplot [color=mycolor5, dashed, line width=1.5pt, forget plot]
  table[row sep=crcr]{%
0.001	0.17484957186206\\
0.178752552590424	0.174708471841733\\
0.319524750575921	0.174399805807425\\
0.45149677720361	0.173954688943002\\
0.587176639073326	0.173343316451661\\
0.722534949178721	0.172582517516744\\
0.852964449974102	0.171711952637463\\
0.99310918137498	0.170632711647847\\
1.12473717836475	0.169490195233695\\
1.27381132318648	0.168054601775242\\
1.42283045721435	0.16647965732031\\
1.5674554102056	0.164828190009077\\
1.72678090388436	0.162881509573545\\
1.87617469143912	0.160947248963326\\
2.03849339825246	0.158740114361001\\
2.21485523372636	0.156234143705531\\
2.40647515001542	0.153404606489875\\
2.5787628875938	0.150783292601698\\
2.76338529005317	0.147910748750416\\
2.96122543798803	0.144778395375049\\
3.17322963473498	0.141380628359833\\
3.40041193270371	0.137715374235703\\
3.64385898376354	0.133784584240607\\
3.90473523688556	0.129594632653731\\
4.18428850790158	0.125156587809248\\
4.5462954695324	0.11952614798882\\
4.93962174387832	0.11359802593128\\
5.44171428686589	0.106367066011515\\
6.07832312829723	0.0977948203720636\\
7.58367791499719	0.08024582678485\\
8.70843149769072	0.0694609862390601\\
9.59360828709315	0.0621908448573247\\
10.423606739764	0.0562206011609443\\
11.3254131515281	0.0505415640931806\\
12.3052400435926	0.0451916649145221\\
13.1862140139475	0.041005446114855\\
14.1302599059953	0.0370785158281279\\
15.1418932530435	0.0334171811576445\\
16.2259528707809	0.0300233311589548\\
17.3876240021625	0.0268948368177013\\
18.6324631193156	0.0240260462597917\\
19.9664245010979	0.0214083405897938\\
21.6938351838518	0.0185828990170926\\
23.5706941399673	0.0160807337594568\\
25.6099310025846	0.013877092435683\\
27.8255940220713	0.0119460045149822\\
30.6539529505653	0.0100026240749984\\
33.7698031082509	0.00835404187865185\\
37.72042493417	0.0067819232185431\\
42.1332174384729	0.0054926239284212\\
47.7176094893875	0.00432267689618815\\
54.7947233690029	0.00330494953996752\\
64.6860766154633	0.00238883491078035\\
78.5045620020451	0.00163163995933502\\
99.310918137498	0.00102448023146584\\
132.777082935543	0.000575161057697837\\
198.288394912707	0.000258537260094283\\
374.605003274899	7.25446638227822e-05\\
1000	1.01851015634757e-05\\
};
\end{axis}

\begin{axis}[%
width=1.227\fwidth,
height=1.227\fheight,
at={(-0.16\fwidth,-0.135\fheight)},
scale only axis,
xmin=0,
xmax=1,
ymin=0,
ymax=1,
axis line style={draw=none},
ticks=none,
axis x line*=bottom,
axis y line*=left
]
\end{axis}
\end{tikzpicture}%

%% file: images/alphVarRho.tex
%
%
\definecolor{mycolor1}{rgb}{0.16295,0.09522,0.42257}%
\definecolor{mycolor2}{rgb}{0.05084,0.30506,0.58943}%
\definecolor{mycolor3}{rgb}{0.17572,0.48809,0.53686}%
\definecolor{mycolor4}{rgb}{0.29112,0.66634,0.50553}%
\definecolor{mycolor5}{rgb}{0.58106,0.82876,0.36638}%
\begin{tikzpicture}[%
trim axis left, trim axis right
]

\begin{axis}[%
width=0.951\fwidth,
height=\fheight,
at={(0\fwidth,0\fheight)},
scale only axis,
xmode=log,
xmin=0.1,
xmax=100,
xminorticks=true,
xlabel style={font=\color{white!15!black}},
xlabel={$\rho_e$},
ymin=0,
ymax=0.5,
ylabel style={font=\color{white!15!black}},
ylabel={$\frac{1}{\pi} \cot^{-1}(\psi)$},
axis background/.style={fill=white},
xmajorgrids,
ymajorgrids
]
\addplot [color=mycolor1, line width=1.5pt, forget plot]
  table[row sep=crcr]{%
0.1	0.0779791303773694\\
100	0.0779791303773694\\
};
\addplot [color=mycolor1, dashed, line width=1.5pt, forget plot]
  table[row sep=crcr]{%
0.1	0\\
100	0\\
};
\addplot [color=mycolor2, line width=1.5pt, forget plot]
  table[row sep=crcr]{%
0.1	0.0779901459896259\\
0.4	0.0781557238617272\\
0.7	0.0785222766402827\\
1	0.0790950100279844\\
1.3	0.0798821704699564\\
1.6	0.0808953154674499\\
1.8	0.08170365902635\\
2	0.0826246702692841\\
2.2	0.0836648029142406\\
2.4	0.0848315308686445\\
2.6	0.0861334700824048\\
2.8	0.0875805223270927\\
3	0.0891840443197718\\
3.2	0.0909570461206619\\
3.4	0.0929144232330659\\
3.6	0.0950732272561772\\
3.8	0.0974529801710293\\
4	0.100076037180843\\
4.2	0.102968002155683\\
4.4	0.106158197624441\\
4.6	0.109680187085671\\
4.8	0.11357233988501\\
4.9	0.115670676192102\\
5	0.117878416079264\\
5.1	0.120201964201614\\
5.2	0.122648127711496\\
5.3	0.125224126668904\\
5.4	0.127937599748145\\
5.5	0.130796603543252\\
5.6	0.133809603378605\\
5.7	0.136985453075351\\
5.8	0.140333360612673\\
5.9	0.143862836068356\\
6	0.147583617650433\\
6.1	0.151505571083919\\
6.2	0.155638557160663\\
6.3	0.159992261994196\\
6.4	0.16457598458019\\
6.5	0.169398376822595\\
6.6	0.174467132460064\\
6.7	0.179788623556421\\
6.8	0.185367486638731\\
6.9	0.191206165357494\\
7	0.197304422754204\\
7.1	0.203658843668896\\
7.2	0.210262355981606\\
7.3	0.217103807308351\\
7.4	0.224167640071927\\
7.5	0.231433710811016\\
7.6	0.238877297386919\\
7.8	0.254176858060298\\
8	0.269791712080283\\
8.7	0.322553584784408\\
8.9	0.336047000298677\\
9.1	0.348560406714519\\
9.3	0.360051062744513\\
9.5	0.370526942770836\\
9.7	0.380031366533894\\
9.90000000000001	0.388629222945619\\
10.1	0.396396186123792\\
10.3	0.403411092152468\\
10.5	0.40975101900261\\
10.7	0.415488415305399\\
10.9	0.420689656931632\\
11.1	0.425414532780411\\
11.3	0.429716296241436\\
11.5	0.433642034134686\\
11.7	0.437233192215797\\
11.9	0.440526157778629\\
12.2	0.444975210446581\\
12.5	0.448915857859404\\
12.8	0.45242349891413\\
13.1	0.455560368322455\\
13.4	0.458378055483541\\
13.8	0.461711705303919\\
14.2	0.464636633967486\\
14.6	0.467219027124502\\
15.1	0.470045358235881\\
15.6	0.472501215385028\\
16.2	0.475048585258587\\
16.9	0.477575411794104\\
17.6	0.479719083692691\\
18.4	0.481796630357027\\
19.3	0.483758625754334\\
20.4	0.485738657279899\\
21.6	0.48749881957172\\
23.1	0.489262833888038\\
24.8	0.490834306475993\\
26.9	0.492331314823931\\
29.4	0.493671647210222\\
32.5	0.494889628352007\\
36.4	0.495974677316106\\
41.5	0.49693648418348\\
48.4	0.497768897834852\\
58.2	0.498469361645566\\
73.1	0.499036095414101\\
98.2000000000001	0.49946855768506\\
100	0.499487631731933\\
};
\addplot [color=mycolor2, dashed, line width=1.5pt, forget plot]
  table[row sep=crcr]{%
0.1	0.00374496257289891\\
0.2	0.00749076757663225\\
0.3	0.0112382581208879\\
0.4	0.0149882786717748\\
0.5	0.0187416757267722\\
0.6	0.0224992984856431\\
0.7	0.0262619995157607\\
0.8	0.0300306354101063\\
0.9	0.033806067435977\\
1	0.0375891621721269\\
1.1	0.0413807921317302\\
1.2	0.0451818363681125\\
1.3	0.0489931810596915\\
1.4	0.0528157200699879\\
1.5	0.0566503554778364\\
1.6	0.0604979980721532\\
1.7	0.0643595678046238\\
1.8	0.0682359941926132\\
1.9	0.0721282166632902\\
2	0.0760371848284982\\
2.1	0.0799638586781755\\
2.2	0.0839092086781501\\
2.3	0.087874215755833\\
2.4	0.0918598711546679\\
2.5	0.0958671761351289\\
2.6	0.0998971414964847\\
2.7	0.103950786889433\\
2.8	0.108029139884942\\
2.9	0.112133234759118\\
3	0.116264110947545\\
3.1	0.120422811115163\\
3.2	0.124610378779225\\
3.3	0.128827855413052\\
3.4	0.133076276946893\\
3.5	0.137356669569138\\
3.6	0.141670044716001\\
3.8	0.150399677771109\\
4	0.159272717766505\\
4.2	0.168295959856035\\
4.4	0.17747505058797\\
4.6	0.186813939873837\\
4.8	0.196314139887395\\
5	0.205973730718926\\
5.2	0.215786038029513\\
5.4	0.225737893767957\\
5.6	0.235807381698424\\
5.8	0.245960973966175\\
6	0.256149999363388\\
6.2	0.266306474915338\\
6.9	0.300007019211089\\
7.1	0.308448915682856\\
7.2	0.312344866042896\\
7.3	0.315985909818651\\
7.4	0.319343482650015\\
7.5	0.322390086391158\\
7.6	0.325100207251385\\
7.7	0.327451270998785\\
7.8	0.329424570856482\\
7.9	0.331006095411204\\
8	0.332187184191886\\
8.1	0.33296494904828\\
8.2	0.333342419551545\\
8.30000000000001	0.333328397650644\\
8.4	0.33293703635077\\
8.5	0.332187184191886\\
8.6	0.331101557397459\\
8.7	0.329705812035714\\
8.8	0.328027588873755\\
8.9	0.326095595294795\\
9	0.323938774488433\\
9.1	0.321585595339793\\
9.2	0.319063479920027\\
9.4	0.3136144335355\\
9.6	0.307776833756011\\
9.80000000000001	0.301703766859214\\
10.1	0.29241045154583\\
11.4	0.254362596171448\\
11.8	0.244063548039982\\
12.2	0.234481145627701\\
12.6	0.225578937522446\\
13	0.217308444576622\\
13.4	0.209617721601323\\
13.8	0.202455766993488\\
14.2	0.195774643644446\\
14.7	0.188032781293479\\
15.2	0.180896960559725\\
15.7	0.174300038842462\\
16.3	0.167013412555189\\
16.9	0.160332599029624\\
17.5	0.154184431968621\\
18.2	0.14760248631173\\
18.9	0.141578407906382\\
19.7	0.135288465209221\\
20.5	0.129550976550828\\
21.4	0.123668546264117\\
22.4	0.117746112728285\\
23.4	0.112379617630971\\
24.5	0.107028019210726\\
25.7	0.101755346205233\\
27	0.0966117261093888\\
28.4	0.0916347611148884\\
29.9	0.0868511928373454\\
31.6	0.0820090524840826\\
33.4	0.0774458775572264\\
35.4	0.0729439293896434\\
37.6	0.0685666188985952\\
40	0.0643595678046238\\
42.7	0.0602091788804913\\
45.7	0.0561881862655453\\
49	0.052347022372174\\
52.7	0.0486238663519623\\
56.9	0.044994556567945\\
61.6	0.0415287027922955\\
66.9	0.0382119894077544\\
73	0.0349971334413675\\
79.9	0.0319576120728864\\
87.9000000000001	0.0290353308101547\\
97.2	0.0262465035769943\\
100	0.0255090119948052\\
};
\addplot [color=mycolor3, line width=1.5pt, forget plot]
  table[row sep=crcr]{%
0.1	0.0780232099865845\\
0.2	0.0781557238617272\\
0.3	0.0783775010351593\\
0.4	0.0786899362834474\\
0.5	0.0790950100279844\\
0.6	0.0795953168386867\\
0.7	0.0801941032058435\\
0.8	0.0808953154674499\\
0.9	0.08170365902635\\
1	0.0826246702692841\\
1.1	0.0836648029142406\\
1.2	0.0848315308686445\\
1.3	0.0861334700824048\\
1.4	0.0875805223270927\\
1.5	0.0891840443197718\\
1.6	0.0909570461206619\\
1.7	0.0929144232330659\\
1.8	0.0950732272561772\\
1.9	0.0974529801710293\\
2	0.100076037180843\\
2.1	0.102968002155683\\
2.2	0.106158197624441\\
2.3	0.109680187085671\\
2.4	0.11357233988501\\
2.5	0.117878416079264\\
2.6	0.122648127711496\\
2.7	0.127937599748145\\
2.8	0.133809603378605\\
2.9	0.140333360612673\\
3	0.147583617650433\\
3.1	0.155638557160663\\
3.2	0.16457598458019\\
3.3	0.174467132460064\\
3.4	0.18536748663873\\
3.5	0.197304422754204\\
3.6	0.210262355981606\\
3.7	0.224167640071927\\
3.8	0.238877297386919\\
3.9	0.254176858060298\\
4	0.269791712080283\\
4.3	0.315470101506133\\
4.4	0.329417733502196\\
4.5	0.342430244402319\\
4.6	0.354434086520033\\
4.7	0.365413796379783\\
4.8	0.375396894297187\\
4.9	0.384439034766803\\
5	0.392611643738639\\
5.1	0.399992745676132\\
5.2	0.406660768915074\\
5.3	0.412690733112318\\
5.4	0.418152162820093\\
5.5	0.423108162309763\\
5.6	0.427615221794139\\
5.7	0.43172345253769\\
5.8	0.435477049806885\\
5.9	0.43891485638015\\
6	0.44207094989006\\
6.1	0.444975210446581\\
6.2	0.447653845989178\\
6.4	0.45242349891413\\
6.6	0.456532779818199\\
6.8	0.460100714298105\\
7	0.463220778013369\\
7.2	0.465967057266936\\
7.4	0.468398815058336\\
7.7	0.471558875721685\\
8	0.47424277284111\\
8.30000000000001	0.476544660075112\\
8.7	0.479140864182432\\
9.1	0.481309818676306\\
9.6	0.483557459295308\\
10.1	0.485407771350858\\
10.7	0.487229812259934\\
11.4	0.488942326065034\\
12.2	0.490498158654769\\
13.2	0.49201098983343\\
14.4	0.493384766934343\\
15.9	0.494647998086754\\
17.8	0.495782727492185\\
20.2	0.49676098168768\\
23.5	0.497630255092527\\
28.1	0.498356341541613\\
35	0.498947751245531\\
46.4	0.499404463319109\\
68.4	0.499726981469989\\
100	0.499872484799971\\
};
\addplot [color=mycolor3, dashed, line width=1.5pt, forget plot]
  table[row sep=crcr]{%
0.1	0.00749076757663225\\
0.2	0.0149882786717748\\
0.3	0.0224992984856431\\
0.4	0.0300306354101063\\
0.5	0.0375891621721269\\
0.6	0.0451818363681125\\
0.7	0.0528157200699879\\
0.8	0.0604979980721532\\
0.9	0.0682359941926132\\
1	0.0760371848284982\\
1.1	0.0839092086781501\\
1.2	0.0918598711546679\\
1.3	0.0998971414964847\\
1.4	0.108029139884942\\
1.5	0.116264110947545\\
1.6	0.124610378779225\\
1.7	0.133076276946893\\
1.8	0.141670044716001\\
1.9	0.150399677771109\\
2	0.159272717766505\\
2.1	0.168295959856035\\
2.2	0.17747505058797\\
2.3	0.186813939873837\\
2.4	0.196314139887395\\
2.5	0.205973730718926\\
2.6	0.215786038029513\\
2.7	0.225737893767957\\
2.8	0.235807381698424\\
2.9	0.245960973966175\\
3	0.256149999363388\\
3.1	0.266306474915338\\
3.4	0.295516675471988\\
3.5	0.304326858444733\\
3.6	0.312344866042896\\
3.7	0.319343482650015\\
3.8	0.325100207251385\\
3.9	0.329424570856482\\
4	0.332187184191886\\
4.1	0.333342419551545\\
4.2	0.33293703635077\\
4.3	0.331101557397459\\
4.4	0.328027588873755\\
4.5	0.323938774488433\\
4.6	0.319063479920027\\
4.7	0.3136144335355\\
4.8	0.307776833756011\\
4.9	0.301703766859214\\
5.1	0.289308657857577\\
5.6	0.25978526340946\\
5.8	0.249122048956183\\
6	0.239184621928361\\
6.2	0.229947892023218\\
6.4	0.221367958271627\\
6.6	0.213393854397637\\
6.8	0.205973730718926\\
7	0.199057948273538\\
7.2	0.192600482442901\\
7.4	0.186559416713684\\
7.7	0.178196978617605\\
8	0.170575933508998\\
8.30000000000001	0.163602136801133\\
8.6	0.157196050303599\\
8.9	0.151290313399301\\
9.2	0.14582765836027\\
9.6	0.139149843522362\\
10	0.133076276946893\\
10.4	0.127526922333304\\
10.9	0.121227312897006\\
11.4	0.115537412894398\\
11.9	0.110371118317212\\
12.5	0.1047644478713\\
13.1	0.0997120532756197\\
13.8	0.0944126894932467\\
14.5	0.0896582758998301\\
15.3	0.0847885878624068\\
16.2	0.0799152693688976\\
17.2	0.0751265778206771\\
18.3	0.070488519962165\\
19.5	0.0660474105199218\\
20.8	0.0618330722172735\\
22.2	0.0578621286922427\\
23.8	0.0539100561971169\\
25.6	0.0500670999158932\\
27.6	0.0463958866741914\\
29.8	0.0429357360279017\\
32.3	0.0395836373501686\\
35.2	0.0362985558382634\\
38.5	0.0331681429509749\\
42.3	0.0301733122412595\\
46.7	0.0273185184280407\\
51.8	0.0246196784703878\\
57.8	0.0220569781615421\\
64.9	0.0196386888563156\\
73.4	0.0173605521065059\\
83.7	0.0152213575163698\\
96.3	0.013227775622993\\
100	0.0127379153893266\\
};
\addplot [color=mycolor4, line width=1.5pt, forget plot]
  table[row sep=crcr]{%
0.1	0.0780783739093258\\
0.2	0.0783775010351593\\
0.3	0.0788807459726382\\
0.4	0.0795953168386867\\
0.5	0.0805316305796677\\
0.6	0.08170365902635\\
0.7	0.0831294045365474\\
0.8	0.0848315308686445\\
0.9	0.0868381835530863\\
1	0.0891840443197718\\
1.1	0.0919116760289902\\
1.2	0.0950732272561772\\
1.3	0.0987325768037612\\
1.4	0.102968002155683\\
1.5	0.107875439493791\\
1.6	0.11357233988501\\
1.7	0.120201964201614\\
1.8	0.127937599748145\\
1.9	0.136985453075351\\
2	0.147583617650433\\
2.1	0.159992261994196\\
2.2	0.174467132460064\\
2.3	0.191206165357494\\
2.4	0.210262355981606\\
2.5	0.231433710811017\\
2.6	0.254176858060298\\
2.7	0.277621159641029\\
2.8	0.300730249839428\\
2.9	0.322553584784408\\
3	0.34243024440232\\
3.1	0.360051062744513\\
3.2	0.375396894297187\\
3.3	0.388629222945619\\
3.4	0.399992745676132\\
3.5	0.40975101900261\\
3.6	0.418152162820093\\
3.7	0.425414532780412\\
3.8	0.43172345253769\\
3.9	0.437233192215797\\
4	0.44207094989006\\
4.1	0.446341223476989\\
4.2	0.450129865646972\\
4.3	0.453507573629662\\
4.4	0.456532779818199\\
4.5	0.459253997633458\\
4.6	0.461711705303919\\
4.7	0.463939851475976\\
4.9	0.467817577126568\\
5.1	0.471068213641064\\
5.3	0.473824207863909\\
5.5	0.476184273489588\\
5.7	0.478223245819431\\
6	0.480802146891598\\
6.3	0.48292986750473\\
6.6	0.484708836940384\\
7	0.486663955462451\\
7.5	0.488606710468188\\
8	0.490142550072488\\
8.6	0.491598595301531\\
9.4	0.493077435915274\\
10.3	0.494310760153776\\
11.5	0.495495157694418\\
13	0.496514617389886\\
15	0.497408392859192\\
17.8	0.498175666967759\\
21.9	0.498803506590627\\
28.5	0.49929755179357\\
40.6	0.499655305076878\\
69.3000000000001	0.499882004798948\\
100	0.499943373827396\\
};
\addplot [color=mycolor4, dashed, line width=1.5pt, forget plot]
  table[row sep=crcr]{%
0.1	0.0112382581208879\\
0.2	0.0224992984856431\\
0.3	0.033806067435977\\
0.4	0.0451818363681125\\
0.5	0.0566503554778364\\
0.6	0.0682359941926132\\
0.7	0.0799638586781755\\
0.8	0.0918598711546679\\
0.9	0.103950786889433\\
1	0.116264110947545\\
1.1	0.128827855413052\\
1.2	0.141670044716001\\
1.3	0.154817825609426\\
1.4	0.168295959856035\\
1.5	0.18212435851684\\
1.6	0.196314139887395\\
1.7	0.210861441961886\\
1.8	0.225737893767957\\
1.9	0.240876305809852\\
2	0.256149999363388\\
2.1	0.27134485350417\\
2.2	0.286125886168872\\
2.3	0.300007019211089\\
2.4	0.312344866042896\\
2.5	0.322390086391159\\
2.6	0.329424570856482\\
2.7	0.33296494904828\\
2.8	0.33293703635077\\
2.9	0.329705812035714\\
3	0.323938774488433\\
3.1	0.316398371285773\\
3.2	0.307776833756011\\
3.3	0.298618041055164\\
3.5	0.28010341728224\\
3.7	0.262564020342415\\
3.8	0.254362596171448\\
3.9	0.24657015077709\\
4	0.239184621928361\\
4.1	0.232193598819242\\
4.2	0.225578937522446\\
4.3	0.219319677806217\\
4.4	0.213393854397637\\
4.5	0.207779595746305\\
4.7	0.19740232351662\\
4.9	0.188032781293479\\
5.1	0.179536452180447\\
5.3	0.171798812100409\\
5.5	0.164723026848403\\
5.7	0.1582273009287\\
5.9	0.15224239864204\\
6.1	0.146709499688443\\
6.3	0.141578407906382\\
6.6	0.134542654813562\\
6.9	0.128194307700605\\
7.2	0.122435379625047\\
7.5	0.117185825253714\\
7.9	0.110866298533305\\
8.30000000000001	0.105209277343931\\
8.7	0.100113968256396\\
9.2	0.0944126894932467\\
9.7	0.0893372784801181\\
10.2	0.0847885878624068\\
10.8	0.0799152693688976\\
11.4	0.0755790998029395\\
12.1	0.0710865461458345\\
12.9	0.0665713580581402\\
13.8	0.0621378759992193\\
14.8	0.0578621286922427\\
15.9	0.053795298958097\\
17.1	0.0499682069546599\\
18.4	0.0463958866741914\\
19.9	0.0428631129365198\\
21.6	0.0394604507422094\\
23.5	0.0362467077385991\\
25.7	0.0331248714353345\\
28.2	0.0301733122412595\\
31.1	0.0273479155352878\\
34.5	0.0246435426834992\\
38.5	0.0220761247908903\\
43.2	0.0196690580523473\\
48.8	0.0174080619961448\\
55.6	0.0152761787500189\\
63.9	0.0132899363342012\\
74.3	0.0114282988311936\\
87.4	0.00971442141791945\\
100	0.00848989874688177\\
};
\addplot [color=mycolor5, line width=1.5pt, forget plot]
  table[row sep=crcr]{%
0.1	0.0781557238617272\\
0.2	0.0786899362834474\\
0.3	0.0795953168386867\\
0.4	0.0808953154674499\\
0.5	0.0826246702692841\\
0.6	0.0848315308686445\\
0.7	0.0875805223270927\\
0.8	0.0909570461206619\\
0.9	0.0950732272561772\\
1	0.100076037180843\\
1.1	0.106158197624441\\
1.2	0.11357233988501\\
1.3	0.122648127711496\\
1.4	0.133809603378605\\
1.5	0.147583617650433\\
1.6	0.16457598458019\\
1.7	0.18536748663873\\
1.8	0.210262355981606\\
1.9	0.238877297386919\\
2	0.269791712080283\\
2.1	0.300730249839428\\
2.2	0.329417733502196\\
2.3	0.354434086520033\\
2.4	0.375396894297187\\
2.5	0.392611643738639\\
2.6	0.406660768915074\\
2.7	0.418152162820093\\
2.8	0.427615221794139\\
2.9	0.435477049806885\\
3	0.44207094989006\\
3.1	0.447653845989178\\
3.2	0.45242349891413\\
3.3	0.456532779818199\\
3.4	0.460100714298105\\
3.5	0.463220778013369\\
3.6	0.465967057266936\\
3.7	0.468398815058336\\
3.8	0.470563885974708\\
3.9	0.472501215385028\\
4.1	0.475815005793755\\
4.3	0.478536107722262\\
4.5	0.480802146891598\\
4.7	0.482712267466554\\
5	0.485064707144303\\
5.3	0.486951668555287\\
5.7	0.488942326065034\\
6.1	0.490498158654769\\
6.6	0.49201098983343\\
7.2	0.493384766934343\\
7.9	0.4945757478365\\
8.8	0.495681479671426\\
10	0.496693360153865\\
11.6	0.497566813802427\\
13.8	0.498295046816426\\
17.1	0.498897302041023\\
22.5	0.499366550835031\\
32.8	0.499703094407252\\
58.9	0.499908147879115\\
100	0.499968157070708\\
};
\addplot [color=mycolor5, dashed, line width=1.5pt, forget plot]
  table[row sep=crcr]{%
0.1	0.0149882786717748\\
0.2	0.0300306354101063\\
0.3	0.0451818363681125\\
0.4	0.0604979980721532\\
0.5	0.0760371848284982\\
0.6	0.0918598711546679\\
0.7	0.108029139884942\\
0.8	0.124610378779225\\
0.9	0.141670044716001\\
1	0.159272717766505\\
1.1	0.17747505058797\\
1.2	0.196314139887395\\
1.3	0.215786038029513\\
1.4	0.235807381698424\\
1.5	0.256149999363388\\
1.6	0.276338518446211\\
1.7	0.295516675471988\\
1.8	0.312344866042896\\
1.9	0.325100207251385\\
2	0.332187184191886\\
2.1	0.33293703635077\\
2.2	0.328027588873755\\
2.3	0.319063479920027\\
2.4	0.307776833756011\\
2.5	0.295516675471988\\
2.8	0.25978526340946\\
2.9	0.249122048956183\\
3	0.239184621928361\\
3.1	0.229947892023218\\
3.2	0.221367958271627\\
3.3	0.213393854397637\\
3.4	0.205973730718926\\
3.5	0.199057948273538\\
3.6	0.192600482442901\\
3.7	0.186559416713684\\
3.8	0.180896960559725\\
3.9	0.1755792315639\\
4	0.170575933508998\\
4.2	0.161407251178181\\
4.4	0.153207020219799\\
4.6	0.14582765836027\\
4.8	0.139149843522362\\
5	0.133076276946893\\
5.2	0.127526922333304\\
5.4	0.122435379625047\\
5.6	0.117746112728285\\
5.9	0.111366064658768\\
6.2	0.10565799640809\\
6.5	0.100519212307243\\
6.8	0.0958671761351289\\
7.2	0.0903073767960407\\
7.6	0.0853676614030956\\
8	0.0809484071497657\\
8.5	0.0760371848284982\\
9	0.0716949380093896\\
9.6	0.0671037800891727\\
10.2	0.0630707732572642\\
10.9	0.0589431969896888\\
11.7	0.0548461848094197\\
12.6	0.05087265814733\\
13.6	0.0470861654770238\\
14.7	0.0435257462040277\\
15.9	0.0402113398142889\\
17.3	0.0369325470062285\\
18.9	0.0337860722073482\\
20.7	0.0308325447091216\\
22.8	0.0279802385280514\\
25.3	0.025205501585194\\
28.2	0.0226059677516948\\
31.6	0.0201680773211099\\
35.7	0.0178476583161991\\
40.6	0.0156906026227315\\
46.6	0.013668187453221\\
54	0.0117936366607623\\
63.4	0.0100440284058632\\
75.4	0.0084448422818153\\
91.2000000000001	0.00698138928395586\\
100	0.00636688747545877\\
};
\end{axis}

\begin{axis}[%
width=1.227\fwidth,
height=1.227\fheight,
at={(-0.16\fwidth,-0.135\fheight)},
scale only axis,
xmin=0,
xmax=1,
ymin=0,
ymax=1,
axis line style={draw=none},
ticks=none,
axis x line*=bottom,
axis y line*=left
]
\end{axis}
\end{tikzpicture}%

%% file: sections/discontinuous.tex
\subsection{Solution for discontinuous porosity distributions}
\label{Sec:Discont}
\begin{figure}
	\centering
	\begin{subfigure}[b]{.6\linewidth}
\centering
\includegraphics[width = \linewidth]{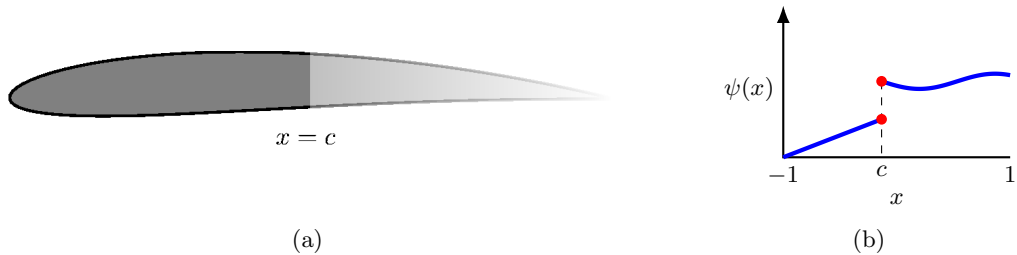}
    \caption{}
		\end{subfigure}
\hfill 		
\begin{subfigure}[b]{.3\linewidth}
\centering
\begin{tikzpicture}
    \draw [Latex-,thick] (0,2) node[pos=0,below] {$-1$}
        |- (3,0) node[right] (xaxis) {} node[pos = 1, below] {$1$} node[pos=.275,left] {$\psi(x)$} node[pos=.75,below=10pt] {$x$};
    \coordinate (c) at (1.3,1);
      \draw[ultra thick,blue] (0,0)--($(c)-(0,0.5)$);
    \draw[dashed] (c) 
        -| (xaxis -| c) node[below] {$c$};
\draw[ultra thick, blue, domain=0:1.7,shift = {(c)},path fading = none,samples =100]   plot (\x,{-.1*sin(180*\x)});
    \fill[red] (c) circle (2pt);
        \fill[red] ($(c)-(0,0.5)$) circle (2pt);
\end{tikzpicture}
    \caption{}
	\end{subfigure}
\caption{Example of an aerofoil with a discontinuous porosity distribution, where the discontinuity occurs at dimensionless chordwise position $x =c$. (a) Illustration of an aerofoil with an discontinuous porosity distribution where the porosity variation along the chord is indicated by at colour gradient. (b) A representative discontinuous porosity distribution. Note that the porosity along the forward section need not be constant.}
\label{Fig:discontSchematic}
\end{figure}
The case of a discontinuous porosity profile is now considered. This scenario is motivated in part by the investigation by \citet{Geyer2}, who showed that, depending on the porous material, aerofoils with porosity at the trailing-edge section only can still lead to a noticeable noise reduction, while maintaining a certain level of aerodynamic performance over a fully-porous aerofoil. A schematic of a partially-porous aerofoil is illustrated in figure \ref{Fig:discontSchematic}.
When the discontinuity is located at $x=c$, the original SF--VIE \eqref{Eq:SIEfin} may be partitioned into two {integral equations}:
%

\begin{align}
\addtocounter{equation}{1}
\frac{1}{\pi}\dashint_{-1}^{1}\frac{\hat{\gamma}_{\rm{a}}(\xi)}{\xi-x}\d \xi
- \psi_{i}(x,k) \left(\hat{\gamma}_{\rm{a}}(x) + \i \helmNum \int_{-1}^x \hat{\gamma}_{\rm{a}}(\xi) \d \xi \right) &= f_{\rm{a}}(x) + \hat{\Gamma} f_{\rm{w}}(x) \tag{\theequation}\label{Eq:SIEdcA},
\end{align}
where
\begin{eqnarray}
\psi_{i}(x,k)=
\begin{cases}
\begin{tabular}{c c}
$\psi_{\rm l}(x,k)$ &\text{for $-1<x<c$},\\
$\psi_{\rm r}(x,k)$ & \text{for $c<x<1$}.
\end{tabular}
\end{cases}
\end{eqnarray}

Note that the subscripts $\rm l$ and $\rm r$ correspond the left and right sides of the  discontinuity so that $\psi_{\rm l}(c^-,k) \neq \psi_{\rm r}(c^+,k)$.

We require the pressure jump across the wing to vanish at \mbox{$x = c$} to ensure that there is no discontinuity in the seepage velocity \eqref{Eq:morseBC}. In particular, asymptotic analysis close to the discontinuity \citep{Baddoo2019e} reveals that, to leading order,
\begin{align}
\left|\hat{\gamma}_{\rm{a}} (x)+ \i k \int_{-1}^x \hat{\gamma}_{\rm{a}}(\xi) \d \xi \right| \sim \left|x-c\right|^\lambda {\hat{\gamma}}^\ast_{\rm{a}}(x), \qquad \quad  \textnormal{as } x \rightarrow c,
\end{align}
where
\begin{align}
\lambda =  \frac{1}{\pi} \left[ \cot^{-1} \left(\psi_{\rm l}(c^-,k)\right)  - \cot^{-1} \left(\psi_{\rm r}(c^+,k)\right) \right],
\end{align}
and where ${\hat{\gamma}}^\ast_{\rm{a}}(x)$ is regular in $-1<x<c$ and $c<x<1$ but may be discontinuous at $x=c$. 

{
These results reflect some physical characteristics of partially porous aerofoils.
On one hand, if the junction transitions from less permeable to more permeable ($\Re[\psi_{\rm l}(c^-,k)] < \Re[\psi_{\rm r}(c^+,k)]$) then the pressure vanishes at $c$ and the junction behaves as a second trailing edge.
On the other hand, if the junction transitions from more permeable to less permeable ($\Re[\psi_{\rm l}(c^-,k)] >  \Re[\psi_{\rm r}(c^+,k)]$) then $\lambda<0$ and the pressure is singular at the junction.
Therefore, in this case the junction behaves like a second leading edge.
}

These observations motivate two separate  expansions for $\hat{\gamma}_{\rm{a}}$ in the left and right regions of the forms
\begin{align}
\addtocounter{equation}{1}
\hat{\gamma}_{\rm l}(\tau_{\rm l}) &= \hat{\gamma}_{\rm l,0} \weight{\tau_{\rm l}}{\lambda}{-\beta}+ 2^{\beta-1} \Pi \weight{\tau_{\rm l}}{0}{1- \beta}+ \weight{\tau_{\rm l}}{\lambda}{1-\beta}\sum_{n=1}^\infty \hat{\gamma}_{{\rm l},n} P_{n-1}^{\lambda,1-\beta} \left( \tau_{\rm l} \right), \tag{\theequation.a}  \label{Eq:impExp}\\
\hat{\gamma}_{\rm r}(\tau_{\rm r}) &= 2^{-\alpha} \Pi \weight{\tau_{\rm r}}{\alpha}{0} + \Lambda \weight{\tau_{\rm r}}{0}{\lambda} + \weight{\tau_{\rm r}}{\alpha}{\lambda} \sum_{n=1}^\infty \hat{\gamma}_{{\rm r},n} P_{n-1}^{\alpha,\lambda} \left( \tau_{\rm r}\right),  \tag{\theequation.b} \label{Eq:perExp}
\end{align}
where $\Pi$ and $\Lambda$ are constants, and we have introduced the rescaled variables
\begin{align}
\addtocounter{equation}{1}
\tau_{\rm l}(x) &= -1+2 \Big(\frac{x+1}{1+c}\Big), \qquad \qquad -1<x<c,    \tag{\theequation.a} \\
\tau_{\rm r}(x) &= \phantom{-} 1+ 2 \Big(\frac{x-1}{1-c}\Big), \qquad \qquad \phantom{-}c<x<1,      \tag{\theequation.b}
\end{align}
so that $-1<\tau_{\rm l},\, \tau_{\rm r} <1$. We now seek to express the constants $\Pi$ and $\Lambda$ in \eqref{Eq:impExp} and \eqref{Eq:perExp} in terms of the unknown coefficients $\hat{\gamma}_{{\rm l},n}$ and $\hat{\gamma}_{{\rm r},n}$. Beginning with the constant $\Pi$, we note the two relations
\begin{align}
\addtocounter{equation}{1}
\hat{\gamma}_{\rm{a}}(c) = \Pi,  \qquad \textrm{ and  }\qquad \hat{\gamma}_{\rm{a}}(c)=-\i k \int_{-1}^c \hat{\gamma}_{\rm{a}}(\xi) \d \xi.     \tag{\theequation.a,b}
\end{align}
The latter expression may be evaluated using the expansion \eqref{Eq:impExp} and the quadrature formula \eqref{Eq:JacobiOrthogonality0}.  A simple rearrangement then allows us to express $\Pi$ in terms of $\hat{\gamma}_{\rm l,0}$ and $\hat{\gamma}_{\rm l,1}$ as
\begin{align}
\Pi &= \Pi_0 \hat{\gamma}_{\rm l,0} + \Pi_1 \hat{\gamma}_{\rm l,1},
\end{align}
where 
\begin{align}
\addtocounter{equation}{1}
\Pi_0 &=  \frac{-\i k (1+c)2^{\lambda -\beta}B(1-\beta,1+\lambda)}{1+ \i k (1+c) B(2- \beta,1)}, \tag{\theequation.a} \\ 
\Pi_1 &=  \frac{-\i k (1+c)2^{\lambda +1 - \beta}B(1-\beta,1+\lambda)}{1+ \i k (1+c) B(2 - \beta,1)}.    \tag{\theequation.b}
\end{align}
We now seek to express $\Lambda$ in terms of the coefficients $\hat{\gamma}_{{\rm l},n}$ and $\hat{\gamma}_{{\rm r},n}$. By employing an approach similar to \S\ref{Sec:numUnsteady}, it is straightforward to show that the new expression for $\Lambda$ is
\begin{align}
\Lambda = \Lambda_{\rm l,0} \hat{\gamma}_{\rm l,0} + \Lambda_{\rm l,1} \hat{\gamma}_{\rm l,1} + \Lambda_{\rm r,1} \hat{\gamma}_{\rm r ,1},
\end{align}
where
\begin{align*}
\addtocounter{equation}{1}
\Lambda_{\rm l,0} &= \Pi_0 \mathcal{M}  \left(\i/k + (1-c)B(1,1+\alpha) \right)  ,  \tag{\theequation.a}\\
\Lambda_{\rm l,0} &= \Pi_1 \mathcal{M}  \left(\i/k + (1-c)B(1,1+\alpha) \right)  ,  \tag{\theequation.b}\\
\Lambda_{\rm r,1} &=  \mathcal{M}  (1-c) 2^{\alpha+\lambda} B(1+\alpha,1+\lambda),  \tag{\theequation.c} \\
\mathcal{M} &=  \frac{ -\i k 2^{-\lambda}}{1+\i k (1-c)B(1,1+\lambda)}. \tag{\theequation.d}
\end{align*}
We may now substitute the expansions \eqref{Eq:impExp} and \eqref{Eq:perExp}  into \eqref{Eq:SIEdcA}. By collocating the Jacobi nodes on the forward and aft sections, we obtain a system of linear equations for the unknown coefficients. During the procedure we encounter the Cauchy integral of the weighted Jacobi polynomials without the principal value, which can be calculated using \eqref{Eq:jacQ2}.

Although only a single discontinuity was considered in this example, any finite number of discontinuities could be modelled using the same approach.  

%% file: sections/noncirculatory.tex
\subsection{Circulatory and non-circulatory solutions} \label{Sec:nc}
The solution to the full unsteady problem in \S\ref{Sec:numUnsteady} may be separated into circulatory and non-circulatory parts by writing
\begin{align}
\hat{\gamma}_{\rm{a}}(x) = \hat{\gamma}_{\rm{a}}^C(x) + \hat{\gamma}_{\rm{a}}^{NC}(x), \label{Eq:CNC}
\end{align}
where the superscripts $C$ and $NC$ denote the circulatory and non-circulatory contributions, respectively. The circulatory part is sometimes referred to the wake-induced component because it contains information about the effect of the downstream wake on the aerofoil. Conversely, the non-circulatory part is sometimes referred to as the added mass component as it represents the effects of the unsteady sloshing of the flow about the aerofoil. Recent research into the origin of added mass led \cite{leonard2001aspects} and \cite{eldredge2010reconciliation} to postulate that its associated force may be represented solely by inviscid theory, even in viscous and separated flows. \cite{corkery2019quantification} confirmed experimentally the ability of inviscid theory to represent added mass effects as a non-circulatory component that depends only on body geometry and its motion, where the circulatory terms in turn measure the viscous effects associated with the bound vorticity and the wake. The circulatory and non-circulatory components combine to give the full vorticity distribution on the aerofoil. 

As its name suggests, the non-circulatory component is the solution to \eqref{Eq:SIEfin} subject to the auxiliary requirement that its net circulation is identically zero:
\begin{align}
\int_{-1}^1 \hat{\gamma}^{NC}_{\rm{a}}(\xi) \d \xi = 0. \label{Eq:ncCirc}
\end{align}
The problem of finding $\hat{\gamma}_{\rm a}^{NC}$ subject to the SF--VIE \eqref{Eq:SIEfin} and to both the non-circulatory condition \eqref{Eq:ncCirc} and the Kutta condition is generally ill-posed. 
{In other words, the Kutta condition \eqref{eq:kutta} cannot be applied to the circulatory and non-circulatory solutions individually.
As such, we permit singularities at the trailing edge in both $\hat{\gamma}_{\rm{a}}^C$ and $\hat{\gamma}_{\rm{a}}^{NC}$.
These singularities are perfectly valid and appear in the impermeable case detailed in, for example, \hbox{\citet[\S\S5-6]{A}}.
The Kutta condition for the full solution is then enforced by specifying that these singularities are equal and opposite and $\Delta p(1)=0$ when the circulatory and non-circulatory solutions are combined according to \eqref{Eq:CNC}, thereby connecting the bound circulation to the wake strength.}

It is simpler to derive the non-circulatory solution and then use the full solution and  \eqref{Eq:CNC} to determine the circulatory solution. Following the analysis of \S\ref{Sec:numUnsteady}, we seek an expansion of the form
\begin{align}
\hat{\gamma}_{\rm{a}}^{NC}(x) &=   \Theta \weight{x}{\alpha-1}{1-\beta} + \weight{x}{\alpha}{-\beta} \hat{\gamma}_0^{NC} + \weight{x}{\alpha}{1-\beta} \sum_{n=1}^N \hat{\gamma}_n^{NC} P_{n-1}^{\alpha,1-\beta}(x),
\end{align}
where the constant $\Theta$ is chosen so that the circulation vanishes:
\begin{align}
\Theta = \frac{-\hat{\gamma}_0^{NC} B(1+\alpha,1-\beta) - \hat{\gamma}_1^{NC} 2B(1+\alpha, 2-\beta)}{B(\alpha,2-\beta)}.
\end{align}
At this point a collocation scheme similar to the procedure described in \S\ref{Sec:numUnsteady} can determine the coefficients ${\hat{\gamma}_n^{NC}}$.

%

%% file: sections/quasi-static.tex
\subsection{Quasi-steady solution} \label{Sec:qs}
The quasi-steady problem is equivalent to the steady problem described in \cite{Hajian2017} with the exception that the kinematic boundary condition is replaced by the  instantaneous unsteady boundary condition \eqref{Eq:w01}. {Specifically, the quasi-steady assumption augments the angle of attack due to aerofoil pitch or camber with the effective angle of attack from heaving motions.  With this modification,} the singular integral equation \eqref{Eq:SIE1} becomes
\begin{align}\label{Eq:qsSIE}
\frac{1}{\pi}\dashint_{-1}^{1}\frac{\hat{\gamma}_{\rm{a}}^Q(\xi)}{\xi-x}\d\xi - \psi(x,0) \hat{\gamma}_{\rm{a}}^Q(x) = 2\left( \i k \hat{y}_{\rm{a}}(x)+ \frac{\d \hat{y}_{\rm{a}}}{\d x} (x)  \right),
\end{align}
where the superscript $Q$ denotes the quasi-steady solution. This singular integral equation may be solved analytically using classical techniques \citep{NIM}. The solution {is rendered unique} by enforcing the Kutta condition, and the corresponding solution is
\begin{align}
\hat{\gamma}_{\rm{a}}^Q(x) = \frac{-2}{1+(\psi(x,0))^2} \Bigg\{&\psi(x,0) \left(\i k \hat{y}_{\rm{a}}(x)+ \frac{\d \hat{y}_{\rm{a}}}{\d x} (x) \right) \notag\\
	+& \frac{Z(x)}{\pi} \dashint_{-1}^1  \left( \i k \hat{y}_{\rm{a}}(\xi)+ \frac{\d \hat{y}_{\rm{a}}}{\d x} (\xi) \right)\frac{\d \xi }{Z(\xi)(\xi - x)} \Bigg\},
	\label{Eq:qsSol}
\end{align}
where
\begin{align}
Z(x) &= \sqrt{1 + (\psi(x,0))^2}  \exp \left[\dashint_{-1}^1 \frac{j(\xi)}{\xi - x}\d \xi\right],
\label{Eq:Zdef}
\end{align}
with $j(x) = (1/\pi) \cot^{-1}(\psi(x,0))$.

In special cases, such as uniformly porous aerofoils with simple geometries, the singular integrals in  \eqref{Eq:qsSol} can be calculated analytically, and the full solution can be expressed in closed form. Otherwise, the solution \eqref{Eq:qsSol} contains nested singular integral equations that can be computed numerically using Gauss--Jacobi quadrature.
%
%

%% file: sections/results.tex
\section{Unsteady pressure distributions on porous aerofoils} \label{Sec:Results}
{This section presents a representative set of unsteady aerodynamic results to showcase the versatility of our numerical scheme detailed in \S\ref{Sec:numUnsteady} for various porosity distributions. Although the numerical scheme is valid for thin aerofoils of arbitrary shape, we focus on symmetric aerofoils undergoing simple pitching or  heaving motions to illustrate the effects of porosity on the pressure difference across the chord. We first consider aerofoils with continuous porosity distributions and then consider aerofoils with discontinuous porosity distributions. Here the porosity distribution \mbox{\eqref{Eq:psidef}} is described by a chordwise function of flow resistivity $\Phi$, and the effective density of the porous medium is held fixed at the representative value of $\rho_{\rm e}=1.2$ \mbox{\citep{bliss1982study}}.}
%

{Physical considerations demand that the analysis in this section is restricted to the case where the leading edge is impermeable, i.e. \mbox{$\psi(-1)=0$}.  Since the pressure gradient possesses a singularity at the leading edge, the seepage velocity also possesses a singularity at the leading edge \emph{unless} $\psi$ vanishes at \mbox{$x=-1$}. Accordingly, in order to obtain physically faithful results, we focus on cases with \mbox{$\psi(-1)=0$}, although the mathematical analysis remains valid in other cases.} 

\input{sections/results-pressure}

\input{sections/asymp-results}

\input{sections/results-theodorsen}

\input{sections/results-gust}
\input{sections/indicial}

%% file: sections/results-pressure.tex
%
%
\subsection{Continuous porosity distributions}
The numerical method in this work is amenable to any H\"older continuous porosity distribution. {Using equations \eqref{Eq:pres2a} and \eqref{Eq:WeightedJacobiIntegral} to construct the aerofoil pressure jump from the numerical scheme,} figure \ref{Fig:contPlots} compares {the magnitudes of the unsteady} pressure distributions for aerofoils with linear and parabolic {reciprocal flow resistance distributions (cf.~\eqref{Eq:psidef}) undergoing pitching or heaving motions of unit amplitude}. The Kutta condition is clearly satisfied at the trailing edge for all cases presented. Figure~\ref{Fig:contPlots} indicates that the introduction of porosity decreases the pressure jump across the aerofoil under both pitching and heaving motions.
As the porosity {parameter $\psi$} increases along the chord, the pressure distribution decreases except in a small region localised to the trailing edge. This reduction corresponds to a significant reduction in the unsteady lift. 
We also note the rapid changes in pressure at the trailing edge $x=1$, which is caused by the reduction in the strength of the trailing-edge zero by \eqref{Eq:alpha}. This behaviour is associated with the reduction in vortex shedding at the porous trailing edge.
%

 \begin{figure}
	\begin{subfigure}[t]{.45\linewidth}
		\hspace{3cm}
		\setlength{\fheight}{5cm}
		\setlength{\fwidth}{\linewidth}
		\centering
		\input{images/largePorCont1.tex}
		\caption{pitching, $k=0.5$}
		\label{Fig:contPlotA}
	\end{subfigure}
	\hfill
	\begin{subfigure}[t]{.45\linewidth}
		\hspace{3cm}
		\setlength{\fheight}{5cm}
		\setlength{\fwidth}{\linewidth}
		\centering
		\input{images/largePorCont2.tex}
		\caption{heaving, $k=0.1$}
		\label{Fig:contPlotB}
	\end{subfigure}
	\caption{{Magnitude of the unsteady pressure} distributions for a porous aerofoil with continuous porosity distributions undergoing harmonic motions with unit amplitude: 
	(a) pitching about the leading edge with linear reciprocal dimensionless flow resistance; 
(b) heaving with parabolic reciprocal dimensionless flow resistance. 
$\rho_{\rm e}=1.2$ in both cases.
The impermeable limit is indicated by the thick black line.}
	\label{Fig:contPlots}
\end{figure}
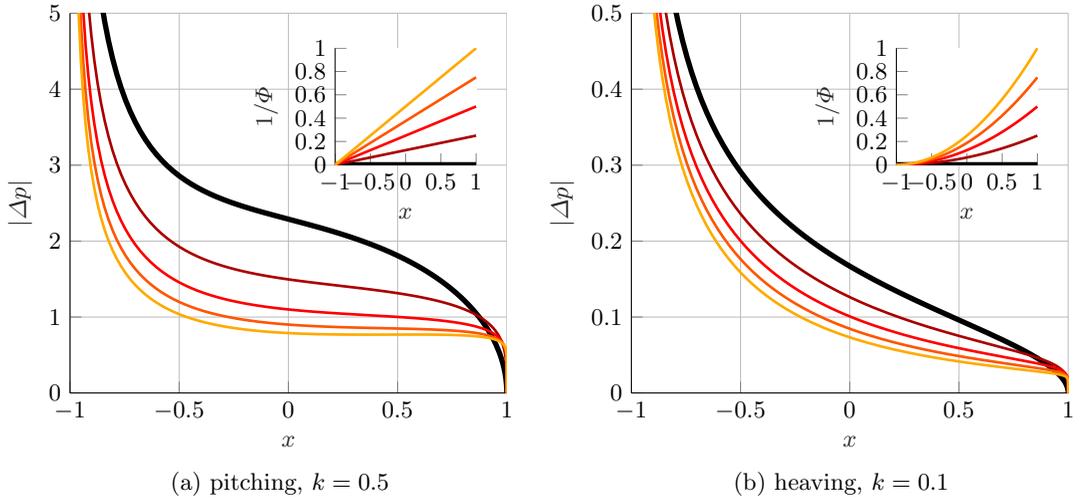

 \begin{figure}
 	\begin{subfigure}[t]{.45\linewidth}
 		\hspace{3cm}
 		\setlength{\fheight}{5cm}
 		\setlength{\fwidth}{\linewidth}
 		\centering
 		\input{images/largePor1.tex}
 		\caption{pitching, $k=0.5$}
 		\label{Fig:ppA}
 	\end{subfigure}
 	\hfill
 	\begin{subfigure}[t]{.45\linewidth}
 		\hspace{3cm}
 		\setlength{\fheight}{5cm}
 		\setlength{\fwidth}{\linewidth}
 		\centering
 		\input{images/largePor2.tex}
 		\caption{heaving, $k=0.1$}
 		\label{Fig:ppB}
 	\end{subfigure}
	\caption{{Magnitude of the unsteady pressure} distributions on a partially-porous aerofoil undergoing pitching or heaving motions with unit amplitude: 
	(a) pitching about the leading edge with linear reciprocal dimensionless flow resistance; 
(b) heaving with parabolic reciprocal dimensionless flow resistance. The porosity distributions are plotted on a logarithmic scale in the figure insets, where $\psi = 0$ for $x<0$. The impermeable case is indicated by the thick black line.  $\rho_{\rm e}=1.2$ in both cases.}
 	\label{Fig:pp}
 \end{figure}
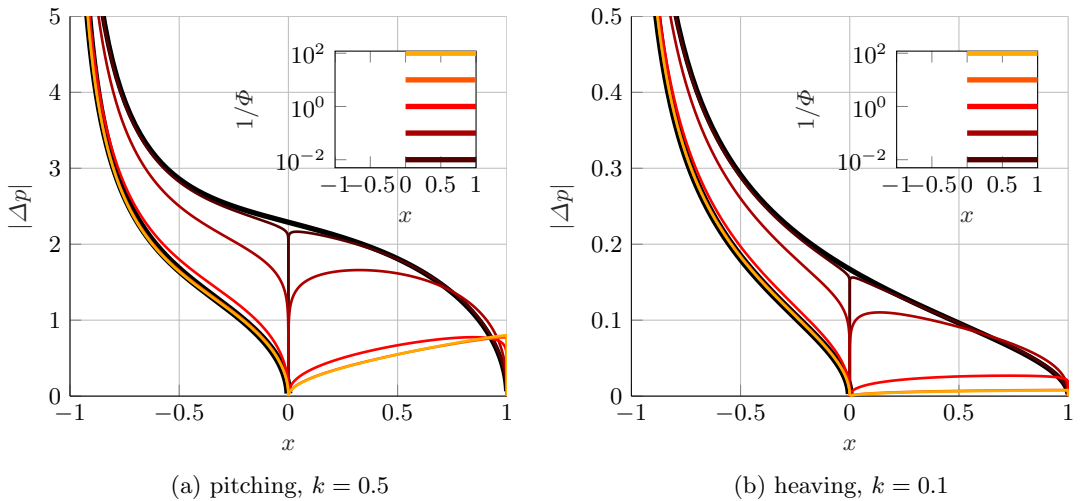
\subsection{Discontinuous porosity distributions}

The numerical method in \S\ref{Sec:Discont} is now applied to aerofoils with discontinuous porosity distributions. Specifically, the numerical scheme is demonstrated for aerofoils with an impermeable leading-edge section for $x<0$ and a constant-porosity section for $x>0$. Figure \ref{Fig:pp} plots the surface pressure jump for pitching motions at $k=0.5$ about the leading edge ($x=-1$) and heaving motions at $k=0.1$. The seepage velocity is continuous across the chord, as evidenced by the vanishing jump at pressure located at the junction $x = 0$. Note that even a small amount of porosity is sufficient to enforce a zero  pressure jump at the junction.

For both the pitching and plunging cases and at different reduced frequencies, the pressure jump along the permeable section vanishes when the porosity is large. Accordingly, the permeable section of the wing behaves effectively as an extension of the wake: although the pressure jump vanishes, the vorticity distribution does not, by \eqref{Eq:pres2a}. For these cases with an impermeable forward section, the aerofoil solution for large porosity in the aft section is the same as the solution of an impermeable aerofoil truncated at $x = c$, i.e., the solution attained as rescaling the characteristic length scale by a factor of $(c+1)/2$, or equivalently, using the mappings
\begin{align}
x \mapsto \frac{(c+1)x+(c-1)}{2}, \qquad y_{\rm a}(x) \mapsto \frac{c+1}{2}y_{\rm a}, \qquad k \mapsto \frac{c+1}{2} k .
\end{align}

%% file: images/largePorCont1.tex
%
%
\definecolor{mycolor1}{rgb}{0.66667,0.00000,0.00000}%
\definecolor{mycolor2}{rgb}{1.00000,0.33333,0.00000}%
\definecolor{mycolor3}{rgb}{1.00000,0.66667,0.00000}%
\begin{tikzpicture}[%
trim axis left, trim axis right
]

\begin{axis}[%
width=0.951\fwidth,
height=\fheight,
at={(0\fwidth,0\fheight)},
scale only axis,
xmin=-1,
xmax=1,
xlabel style={font=\color{white!15!black}},
xlabel={$x$},
ymin=0,
ymax=5,
ylabel style={font=\color{white!15!black}},
ylabel={$|\Delta p|$},
axis background/.style={fill=white},
axis x line*=bottom,
axis y line*=left,
xmajorgrids,
ymajorgrids,
axis on top=false
]
\addplot [color=black, line width=2.0pt, forget plot]
  table[row sep=crcr]{%
-0.857983413234977	5.18166314697026\\
-0.841253532831181	4.89257630846033\\
-0.823676581429833	4.63592825109263\\
-0.805270257531059	4.40709600976295\\
-0.786053094742788	4.20232660112812\\
-0.766044443118978	4.01853596932952\\
-0.745264449675755	3.85316034746179\\
-0.723734038105071	3.7040448211016\\
-0.701474887706321	3.56935882218758\\
-0.678509411557132	3.44753145514034\\
-0.654860733945285	3.3372015946153\\
-0.630552667084523	3.23717904599409\\
-0.605609687137666	3.1464140371877\\
-0.580056909571198	3.06397307721506\\
-0.55392006386611	2.98901980710084\\
-0.527225467610503	2.92079985742409\\
-0.5	2.85862891952047\\
-0.472271074772682	2.80188331053957\\
-0.444066612605774	2.74999238116503\\
-0.415415013001887	2.70243225588529\\
-0.386345125693128	2.65872059669927\\
-0.356886221591872	2.61841225767962\\
-0.327067963317422	2.58109576745627\\
-0.296920375328275	2.54639052879728\\
-0.266473813690035	2.51394453091903\\
-0.235758935509427	2.48343233044139\\
-0.20480666806519	2.45455312298636\\
-0.17364817766693	2.42702886689787\\
-0.142314838273285	2.40060254398654\\
-0.110838199901011	2.37503667090743\\
-0.0475819158237423	2.32562700814649\\
0.0792499568567884	2.22859589225112\\
0.142314838273285	2.17857148131184\\
0.173648177666931	2.15275269330597\\
0.204806668065191	2.12626152557822\\
0.235758935509427	2.09900939458936\\
0.266473813690035	2.07091686166739\\
0.296920375328275	2.04191286722539\\
0.327067963317422	2.01193403935639\\
0.356886221591872	1.98092421220912\\
0.386345125693128	1.94883416160488\\
0.415415013001886	1.91562142952115\\
0.444066612605774	1.88125006978225\\
0.472271074772682	1.845690237882\\
0.5	1.80891769893885\\
0.527225467610503	1.77091341692971\\
0.55392006386611	1.73166333839509\\
0.580056909571199	1.69115833186345\\
0.605609687137667	1.64939411903777\\
0.630552667084523	1.60637105051702\\
0.654860733945285	1.56209373438191\\
0.678509411557132	1.51657068821323\\
0.701474887706321	1.46981419858783\\
0.723734038105071	1.42184040047376\\
0.745264449675755	1.37266936863681\\
0.766044443118978	1.32232495383141\\
0.786053094742788	1.27083429392792\\
0.805270257531059	1.218227249844\\
0.823676581429833	1.16453617168845\\
0.841253532831181	1.10979619311205\\
0.857983413234977	1.0540457833185\\
0.873849377069785	0.997326940444703\\
0.888835448654923	0.939684564671129\\
0.902926538286621	0.881165215643755\\
0.916108457432069	0.821816161534113\\
0.928367933016073	0.761685697947256\\
0.939692620785909	0.700824844641228\\
0.950071117740945	0.639289164019768\\
0.964061863361954	0.544220185979277\\
0.967948701396356	0.514434409082894\\
0.975429786885407	0.451233107716282\\
0.981928697262707	0.387582975562046\\
0.987438888676395	0.323526069545152\\
0.991954812830795	0.259109818654866\\
0.995471922573085	0.194406202093949\\
0.9983318994628	0.117875601290953\\
0.999784556530997	0.0422583026388796\\
0.99999	0\\
};
\addplot [color=mycolor1, line width=1.0pt, forget plot]
  table[row sep=crcr]{%
-0.916108457432069	5.12241145758592\\
-0.902926538286621	4.72542243728688\\
-0.888835448654923	4.38175664183947\\
-0.873849377069785	4.08175642019224\\
-0.857983413234977	3.81811568122356\\
-0.841253532831181	3.58509823129701\\
-0.823676581429833	3.37805024842647\\
-0.805270257531059	3.19315844090259\\
-0.786053094742788	3.02733995456065\\
-0.766044443118978	2.87814364670198\\
-0.745264449675755	2.74360527423056\\
-0.723734038105071	2.62208072521564\\
-0.701474887706321	2.51212049833619\\
-0.678509411557132	2.41242420128556\\
-0.654860733945285	2.32185687513983\\
-0.630552667084523	2.2394712767488\\
-0.605609687137666	2.16449063538604\\
-0.580056909571198	2.096250635792\\
-0.55392006386611	2.03413739500936\\
-0.527225467610503	1.97755999882949\\
-0.5	1.92596518010124\\
-0.472271074772682	1.87886787993511\\
-0.444066612605774	1.8358632146985\\
-0.415415013001887	1.79660649280733\\
-0.386345125693128	1.76077727915632\\
-0.356886221591872	1.72805566862764\\
-0.327067963317422	1.69812589025307\\
-0.296920375328275	1.67069813767905\\
-0.266473813690035	1.64552576580901\\
-0.235758935509427	1.62240198861512\\
-0.20480666806519	1.60113963736097\\
-0.17364817766693	1.58155160382555\\
-0.142314838273285	1.56344709419996\\
-0.110838199901011	1.54664382685548\\
-0.0792499568567884	1.53098293509053\\
-0.0475819158237423	1.5163326772207\\
-0.0158659638348082	1.50257822217863\\
0.0158659638348082	1.4896069896615\\
0.0475819158237423	1.47730179820375\\
0.110838199901012	1.45423489496344\\
0.173648177666931	1.4326092892506\\
0.327067963317422	1.38094115395154\\
0.386345125693128	1.35980696305959\\
0.444066612605774	1.33774953899665\\
0.472271074772682	1.32623459272084\\
0.5	1.31432595037666\\
0.527225467610503	1.30197981469174\\
0.55392006386611	1.28915511435982\\
0.580056909571199	1.27580779778959\\
0.605609687137667	1.26188854128673\\
0.630552667084523	1.24734545329329\\
0.654860733945285	1.23212878371945\\
0.678509411557132	1.2161929750957\\
0.701474887706321	1.1994939623709\\
0.723734038105071	1.18198397944941\\
0.745264449675755	1.16360829783715\\
0.766044443118978	1.14430641054667\\
0.786053094742788	1.12401601106382\\
0.805270257531059	1.10267548965265\\
0.823676581429833	1.08022190118337\\
0.841253532831181	1.0565853264078\\
0.857983413234977	1.0316836865403\\
0.873849377069785	1.00542149945639\\
0.888835448654923	0.977692203534525\\
0.902926538286621	0.948379893379081\\
0.916108457432069	0.917355931677433\\
0.928367933016073	0.884469321833978\\
0.939692620785909	0.849533989793558\\
0.950071117740945	0.81231719691227\\
0.959492973614497	0.772529454117486\\
0.967948701396356	0.729809632621072\\
0.975429786885407	0.683693307870844\\
0.981928697262707	0.633549100074524\\
0.987438888676395	0.578461737957911\\
0.991954812830795	0.517014310883166\\
0.995471922573085	0.446822924343754\\
0.9983318994628	0.346193034759042\\
0.999784556530997	0.204693640505307\\
0.99999	0\\
};
\addplot [color=red, line width=1.0pt, forget plot]
  table[row sep=crcr]{%
-0.939692620785908	5.04230154954396\\
-0.928367933016072	4.57477217152119\\
-0.916108457432069	4.17946145919871\\
-0.902926538286621	3.84083379564051\\
-0.888835448654923	3.547629005671\\
-0.873849377069785	3.29166965600867\\
-0.857983413234977	3.06685451281807\\
-0.841253532831181	2.86836586708672\\
-0.823676581429833	2.69219195477951\\
-0.805270257531059	2.53495908330887\\
-0.786053094742788	2.39393031284014\\
-0.766044443118978	2.2669885721074\\
-0.745264449675755	2.15250790459065\\
-0.723734038105071	2.04914939381811\\
-0.701474887706321	1.95569300914525\\
-0.678509411557132	1.87098647957601\\
-0.654860733945285	1.79399821016816\\
-0.630552667084523	1.72388811309203\\
-0.605609687137666	1.66001425397944\\
-0.580056909571198	1.60186126348179\\
-0.55392006386611	1.54894553989196\\
-0.527225467610503	1.50076651608392\\
-0.5	1.4568284599366\\
-0.472271074772682	1.41669790183064\\
-0.444066612605774	1.38003847826294\\
-0.415415013001887	1.34659234411618\\
-0.386345125693128	1.31612642892085\\
-0.356886221591872	1.28838931947806\\
-0.327067963317422	1.2631103237106\\
-0.296920375328275	1.24003348769233\\
-0.266473813690035	1.21895147453198\\
-0.235758935509427	1.19970914096019\\
-0.20480666806519	1.18217618241058\\
-0.17364817766693	1.16621455204105\\
-0.142314838273285	1.15166767460292\\
-0.110838199901011	1.13837675847441\\
-0.0792499568567884	1.12620600405614\\
-0.0475819158237423	1.11505330634472\\
-0.0158659638348082	1.10483834838043\\
0.0158659638348082	1.09548037604116\\
0.0475819158237423	1.08688540614032\\
0.0792499568567884	1.07895222723766\\
0.110838199901012	1.0715893114132\\
0.173648177666931	1.05830995617627\\
0.235758935509427	1.04660698651604\\
0.327067963317422	1.03092453132243\\
0.472271074772682	1.00652705712224\\
0.527225467610503	0.996354895453281\\
0.580056909571199	0.985501343392019\\
0.630552667084523	0.973663581657251\\
0.654860733945285	0.967264679037628\\
0.678509411557132	0.960496540224399\\
0.701474887706321	0.953328356263895\\
0.723734038105071	0.94572523025042\\
0.745264449675755	0.937642556962436\\
0.766044443118978	0.929027665015655\\
0.786053094742788	0.919826717315186\\
0.805270257531059	0.909990120439562\\
0.823676581429833	0.899470959119289\\
0.841253532831181	0.888217020693225\\
0.857983413234977	0.876162484237035\\
0.873849377069785	0.863225470123808\\
0.888835448654923	0.849312225300053\\
0.902926538286621	0.834322448747654\\
0.916108457432069	0.818148463606162\\
0.928367933016073	0.800665045864622\\
0.939692620785909	0.781713225756719\\
0.950071117740945	0.761084626116139\\
0.959492973614497	0.738509536551925\\
0.967948701396356	0.713643592794747\\
0.975429786885407	0.686039356390119\\
0.981928697262707	0.655082769050353\\
0.987438888676395	0.619865281291042\\
0.991954812830795	0.578929747456544\\
0.995471922573085	0.529703428391563\\
0.9983318994628	0.453381098074749\\
0.999496542383185	0.375820717158832\\
0.999972174405978	0.238506892429646\\
0.99999	0\\
};
\addplot [color=mycolor2, line width=1.0pt, forget plot]
  table[row sep=crcr]{%
-0.959492973614497	5.48782489920189\\
-0.950071117740945	4.87524528263748\\
-0.939692620785908	4.37394082341228\\
-0.928367933016072	3.95685025574841\\
-0.916108457432069	3.60462441996347\\
-0.902926538286621	3.30315863447165\\
-0.888835448654923	3.04226667765562\\
-0.873849377069785	2.81466723031369\\
-0.857983413234977	2.61501467886274\\
-0.841253532831181	2.43907626571808\\
-0.823676581429833	2.28323936402858\\
-0.805270257531059	2.14438408263746\\
-0.786053094742788	2.01996062647734\\
-0.766044443118978	1.90803997167409\\
-0.745264449675755	1.80720419888207\\
-0.723734038105071	1.71631504815358\\
-0.701474887706321	1.63430664741292\\
-0.678509411557132	1.56012017613292\\
-0.654860733945285	1.49277814216429\\
-0.630552667084523	1.43149281813064\\
-0.605609687137666	1.37569705815066\\
-0.580056909571198	1.32496822711726\\
-0.55392006386611	1.27890951983034\\
-0.527225467610503	1.23708088758127\\
-0.5	1.19902045142998\\
-0.472271074772682	1.16431928749378\\
-0.444066612605774	1.13267514154703\\
-0.415415013001887	1.10387913097205\\
-0.386345125693128	1.07775185386751\\
-0.356886221591872	1.05408567681218\\
-0.327067963317422	1.03263762840672\\
-0.296920375328275	1.01316952159339\\
-0.266473813690035	0.995493292608084\\
-0.235758935509427	0.979481098989825\\
-0.20480666806519	0.965034963191457\\
-0.17364817766693	0.95204581061006\\
-0.142314838273285	0.940376975395613\\
-0.110838199901011	0.929881975839202\\
-0.0792499568567884	0.920436042532424\\
-0.0475819158237423	0.911951909999337\\
-0.0158659638348082	0.904367660764101\\
0.0158659638348082	0.897620024133492\\
0.0475819158237423	0.891627501832366\\
0.0792499568567884	0.886296252503823\\
0.110838199901012	0.881540414564144\\
0.173648177666931	0.873522844317836\\
0.235758935509427	0.867208029917813\\
0.296920375328275	0.862145304891659\\
0.386345125693128	0.856014766228967\\
0.55392006386611	0.845313493537278\\
0.605609687137667	0.841155391988996\\
0.654860733945285	0.836234396812845\\
0.701474887706321	0.830284241134494\\
0.745264449675755	0.823089515275075\\
0.766044443118978	0.818927927012374\\
0.786053094742788	0.814329570849488\\
0.805270257531059	0.809255493347757\\
0.823676581429833	0.803671741169762\\
0.841253532831181	0.797540617415629\\
0.857983413234977	0.790811653548603\\
0.873849377069785	0.783419291176991\\
0.888835448654923	0.775288136788404\\
0.902926538286621	0.766339726946678\\
0.916108457432069	0.756492890114532\\
0.928367933016073	0.745654339795202\\
0.939692620785909	0.733703107275881\\
0.950071117740945	0.720475878817584\\
0.959492973614497	0.705756968360644\\
0.967948701396356	0.68926846486356\\
0.975429786885407	0.670647946362198\\
0.981928697262707	0.64939542762945\\
0.987438888676395	0.624762845404788\\
0.991954812830795	0.595529117421777\\
0.995471922573085	0.559487665692073\\
0.9983318994628	0.501494719806569\\
0.999496542383185	0.439453890876881\\
0.999972174405978	0.319044133452048\\
0.99999	0\\
};
\addplot [color=mycolor3, line width=1.0pt, forget plot]
  table[row sep=crcr]{%
-0.966368943466229	5.5\\
-0.959492973614497	4.93065781514541\\
-0.950071117740945	4.36923820834357\\
-0.939692620785908	3.91028305137153\\
-0.928367933016072	3.5290364075864\\
-0.916108457432069	3.20760205127205\\
-0.902926538286621	2.93282743876588\\
-0.888835448654923	2.69523951640533\\
-0.873849377069785	2.48817028049907\\
-0.857983413234977	2.30681088784426\\
-0.841253532831181	2.14735273271317\\
-0.823676581429833	2.00646893306613\\
-0.805270257531059	1.88120982116027\\
-0.786053094742788	1.76914126909318\\
-0.766044443118978	1.66845348395133\\
-0.745264449675755	1.57787185788757\\
-0.723734038105071	1.49640539005321\\
-0.701474887706321	1.42310505141676\\
-0.678509411557132	1.35698151068175\\
-0.654860733945285	1.29709221386728\\
-0.630552667084523	1.2426793214087\\
-0.605609687137666	1.19322258929203\\
-0.580056909571198	1.14836319865977\\
-0.55392006386611	1.10776757217619\\
-0.527225467610503	1.07104087136822\\
-0.5	1.03774534998959\\
-0.472271074772682	1.00748675928152\\
-0.444066612605774	0.979982846557433\\
-0.415415013001887	0.955055856323145\\
-0.386345125693128	0.932562577894663\\
-0.356886221591872	0.912325961850541\\
-0.327067963317422	0.894122448887151\\
-0.296920375328275	0.877725358916383\\
-0.266473813690035	0.862958085111466\\
-0.235758935509427	0.849709361617011\\
-0.20480666806519	0.837901641919044\\
-0.17364817766693	0.827444769251318\\
-0.142314838273285	0.818215410312741\\
-0.110838199901011	0.810075337683336\\
-0.0792499568567884	0.802906789306458\\
-0.0475819158237423	0.796631546885666\\
-0.0158659638348082	0.79119881826898\\
0.0158659638348082	0.786555930971753\\
0.0475819158237423	0.782629059552459\\
0.0792499568567884	0.779329100139535\\
0.110838199901012	0.776574032783418\\
0.173648177666931	0.772487333574799\\
0.235758935509427	0.770047548872506\\
0.296920375328275	0.768831951486651\\
0.356886221591872	0.768434406595387\\
0.444066612605774	0.768829448573561\\
0.580056909571199	0.769607419967294\\
0.630552667084523	0.769261950214219\\
0.678509411557132	0.768136359264985\\
0.723734038105071	0.766046556467225\\
0.766044443118978	0.762768759340821\\
0.786053094742788	0.760586313318931\\
0.805270257531059	0.757996310692205\\
0.823676581429833	0.754973042465943\\
0.841253532831181	0.751487792709987\\
0.857983413234977	0.747499632065109\\
0.873849377069785	0.74295350803733\\
0.888835448654923	0.737786290476386\\
0.902926538286621	0.731934272401132\\
0.916108457432069	0.72533401576298\\
0.928367933016073	0.717913403375159\\
0.939692620785909	0.709576897404117\\
0.950071117740945	0.700192282680439\\
0.959492973614497	0.689582642831953\\
0.967948701396356	0.6775193307083\\
0.975429786885407	0.663704309140874\\
0.981928697262707	0.647725621144247\\
0.987438888676395	0.628962842235896\\
0.991954812830795	0.606392595623753\\
0.995471922573085	0.578140505174329\\
0.9983318994628	0.531689779369779\\
0.999496542383185	0.480532730341032\\
0.999972174405978	0.375966306970669\\
0.99999	0\\
};
\end{axis}

\begin{axis}[%
width=0.307\fwidth,
height=0.307\fheight,
at={(0.577\fwidth,0.601\fheight)},
scale only axis,
xmin=-1,
xmax=1,
xlabel style={font=\color{white!15!black}},
xlabel={$x$},
ymin=0,
ymax=1,
ylabel style={font=\color{white!15!black}},
ylabel={$1/\Phi$},
axis background/.style={fill=white},
axis x line*=bottom,
axis y line*=left,
xminorgrids,
yminorgrids,
axis on top=false
]
\addplot [color=black, line width=2.0pt, forget plot]
  table[row sep=crcr]{%
-0.999496542383185	2.22044604925031e-16\\
0.99999	2.22044604925031e-16\\
};
\addplot [color=mycolor1, line width=1.0pt, forget plot]
  table[row sep=crcr]{%
-0.999496542383185	6.29322021020728e-05\\
0.99999	0.24999875\\
};
\addplot [color=red, line width=1.0pt, forget plot]
  table[row sep=crcr]{%
-0.999496542383185	0.000125864404203924\\
0.99999	0.4999975\\
};
\addplot [color=mycolor2, line width=1.0pt, forget plot]
  table[row sep=crcr]{%
-0.999496542383185	0.000188796606305774\\
0.99999	0.74999625\\
};
\addplot [color=mycolor3, line width=1.0pt, forget plot]
  table[row sep=crcr]{%
-0.999496542383185	0.000251728808407736\\
0.99999	0.999995\\
};
\end{axis}
\end{tikzpicture}%

%% file: images/largePorCont2.tex
%
%
\definecolor{mycolor1}{rgb}{0.66667,0.00000,0.00000}%
\definecolor{mycolor2}{rgb}{1.00000,0.33333,0.00000}%
\definecolor{mycolor3}{rgb}{1.00000,0.66667,0.00000}%
\begin{tikzpicture}[%
trim axis left, trim axis right
]

\begin{axis}[%
width=0.951\fwidth,
height=\fheight,
at={(0\fwidth,0\fheight)},
scale only axis,
xmin=-1,
xmax=1,
xlabel style={font=\color{white!15!black}},
xlabel={$x$},
ymin=0,
ymax=0.5,
ylabel style={font=\color{white!15!black}},
ylabel={$|\Delta p|$},
axis background/.style={fill=white},
axis x line*=bottom,
axis y line*=left,
xmajorgrids,
ymajorgrids,
axis on top=false
]
\addplot [color=black, line width=2.0pt, forget plot]
  table[row sep=crcr]{%
-0.805270257531059	0.515081455688924\\
-0.786053094742787	0.488582484497454\\
-0.766044443118978	0.464403569645838\\
-0.745264449675755	0.442242372603541\\
-0.72373403810507	0.421846933042859\\
-0.701474887706321	0.403005590054676\\
-0.678509411557132	0.385539229788931\\
-0.654860733945285	0.369295256390194\\
-0.630552667084523	0.354142854947872\\
-0.605609687137666	0.339969234026208\\
-0.580056909571198	0.326676619010515\\
-0.55392006386611	0.314179827113582\\
-0.527225467610502	0.302404297490646\\
-0.5	0.291284480361079\\
-0.472271074772683	0.280762511026406\\
-0.444066612605774	0.270787111095921\\
-0.415415013001887	0.26131267207495\\
-0.386345125693129	0.252298486747517\\
-0.356886221591872	0.24370810175521\\
-0.327067963317421	0.235508770477483\\
-0.296920375328275	0.227670989080891\\
-0.266473813690035	0.220168101176331\\
-0.235758935509427	0.21297595877299\\
-0.204806668065191	0.206072629649935\\
-0.17364817766693	0.199438143767769\\
-0.142314838273285	0.193054273377198\\
-0.110838199901011	0.186904342590308\\
-0.0792499568567886	0.180973062400608\\
-0.0475819158237423	0.175246387014469\\
-0.015865963834808	0.169711387551308\\
0.0158659638348082	0.164356139983522\\
0.0475819158237424	0.159169625366985\\
0.110838199901011	0.14926272454109\\
0.173648177666931	0.139917531906278\\
0.235758935509427	0.131070718931889\\
0.296920375328275	0.122667135091303\\
0.386345125693129	0.110788704535648\\
0.5	0.0960968368227054\\
0.630552667084523	0.0792044346387061\\
0.701474887706321	0.0696936133679755\\
0.745264449675755	0.0635666934024315\\
0.786053094742788	0.0575880332735328\\
0.823676581429833	0.0517400541063051\\
0.857983413234977	0.0460065721723435\\
0.888835448654923	0.0403725834687901\\
0.902926538286621	0.037588495378976\\
0.91610845743207	0.0348241177136236\\
0.928367933016073	0.0320778481457223\\
0.939692620785908	0.0293481285981908\\
0.950071117740945	0.0266334361550259\\
0.964061863361954	0.022516350324112\\
0.967948701396356	0.021243194434495\\
0.975429786885407	0.0185647640982846\\
0.981928697262707	0.0158956112021535\\
0.987438888676394	0.0132343992604863\\
0.991954812830795	0.0105798100658003\\
0.995471922573085	0.00793050901363612\\
0.9983318994628	0.00481034871373021\\
0.999496542383185	0.00264212815938636\\
0.999922573631732	0.00103610267722987\\
0.99999	0\\
};
\addplot [color=mycolor1, line width=1.0pt, forget plot]
  table[row sep=crcr]{%
-0.857983413234977	0.532486875522909\\
-0.841253532831181	0.499571464503191\\
-0.823676581429833	0.469952520336808\\
-0.805270257531059	0.44314076861072\\
-0.786053094742787	0.418740713398782\\
-0.766044443118978	0.396429262892761\\
-0.745264449675755	0.375939955765238\\
-0.72373403810507	0.357051150819983\\
-0.701474887706321	0.339577045937199\\
-0.678509411557132	0.323360743202531\\
-0.654860733945285	0.308268829365898\\
-0.630552667084523	0.294187112076959\\
-0.605609687137666	0.281017254214167\\
-0.580056909571198	0.268674102502307\\
-0.55392006386611	0.257083540972161\\
-0.527225467610502	0.246180736220284\\
-0.5	0.235908683986935\\
-0.472271074772683	0.226217004357422\\
-0.444066612605774	0.217060953070793\\
-0.415415013001887	0.208400616995318\\
-0.386345125693129	0.200200253880866\\
-0.356886221591872	0.192427734996628\\
-0.327067963317421	0.185054060978424\\
-0.296920375328275	0.178052940007838\\
-0.266473813690035	0.171400430513471\\
-0.235758935509427	0.165074649827609\\
-0.204806668065191	0.159055539059796\\
-0.17364817766693	0.153324664513536\\
-0.142314838273285	0.147865037061243\\
-0.110838199901011	0.142660942780177\\
-0.0792499568567886	0.137697791205582\\
-0.0475819158237423	0.132961991275277\\
-0.015865963834808	0.128440857053009\\
0.0158659638348082	0.124122533223016\\
0.0475819158237424	0.119995925018634\\
0.0792499568567885	0.116050623496319\\
0.110838199901011	0.112276829644598\\
0.173648177666931	0.105207245875938\\
0.235758935509427	0.0987189481851033\\
0.296920375328275	0.0927507503486281\\
0.356886221591872	0.0872473089130901\\
0.415415013001886	0.0821580070222943\\
0.5	0.0751999849374246\\
0.580056909571198	0.0689271693260499\\
0.766044443118978	0.0545735260703073\\
0.823676581429833	0.0497285606498579\\
0.857983413234977	0.0465562681442918\\
0.888835448654923	0.0433766534135022\\
0.902926538286621	0.0417675990632344\\
0.91610845743207	0.0401354761708845\\
0.928367933016073	0.0384712766723091\\
0.939692620785908	0.0367645872039257\\
0.950071117740945	0.0350032294149296\\
0.964061863361954	0.0321761859425885\\
0.967948701396356	0.031255172785484\\
0.975429786885407	0.0292278727777182\\
0.981928697262707	0.0270602539661156\\
0.987438888676394	0.0247092542478696\\
0.991954812830795	0.0221102779748995\\
0.995471922573085	0.0191572350686476\\
0.9983318994628	0.0149291882155348\\
0.999496542383185	0.0110650447433539\\
0.999922573631732	0.00692644251401009\\
0.999972174405978	0.00536142701406117\\
0.99999	0\\
};
\addplot [color=red, line width=1.0pt, forget plot]
  table[row sep=crcr]{%
-0.873849377069785	0.513838421671385\\
-0.857983413234977	0.47937880571776\\
-0.841253532831181	0.448526519859052\\
-0.823676581429833	0.420721256243122\\
-0.805270257531059	0.395515876893016\\
-0.786053094742787	0.372549312044905\\
-0.766044443118978	0.351526849184194\\
-0.745264449675755	0.332205611138943\\
-0.72373403810507	0.314383706520036\\
-0.701474887706321	0.297891981018867\\
-0.678509411557132	0.282587628391829\\
-0.654860733945285	0.268349175766528\\
-0.630552667084523	0.255072534491428\\
-0.605609687137666	0.242667900799313\\
-0.580056909571198	0.231057322117293\\
-0.55392006386611	0.220172757188238\\
-0.527225467610502	0.209954486788068\\
-0.5	0.20034978322899\\
-0.472271074772683	0.19131179959508\\
-0.444066612605774	0.182798666774727\\
-0.415415013001887	0.17477277961809\\
-0.386345125693129	0.167200230145936\\
-0.356886221591872	0.160050333235217\\
-0.327067963317421	0.153295203733688\\
-0.296920375328275	0.146909374884633\\
-0.266473813690035	0.140869472731061\\
-0.235758935509427	0.135153961938086\\
-0.204806668065191	0.129742957732915\\
-0.17364817766693	0.124618076339543\\
-0.142314838273285	0.119762292405749\\
-0.110838199901011	0.115159789986206\\
-0.0792499568567886	0.1107958184715\\
-0.0475819158237423	0.106656576036877\\
-0.015865963834808	0.102729132386554\\
0.0158659638348082	0.0990013800676686\\
0.0475819158237424	0.0954619890918883\\
0.0792499568567885	0.0921003453234589\\
0.110838199901011	0.0889064739183849\\
0.173648177666931	0.0829849373675565\\
0.235758935509427	0.0776282406931094\\
0.296920375328275	0.0727749803652009\\
0.356886221591872	0.0683704392837714\\
0.415415013001886	0.0643652974509381\\
0.472271074772683	0.0607151638543347\\
0.55392006386611	0.0558190380732406\\
0.630552667084523	0.0515108444040601\\
0.857983413234977	0.0390584727142335\\
0.9	0.0363275324559259\\
0.91610845743207	0.0351373559217626\\
0.939692620785908	0.0331301253434114\\
0.959492973614497	0.0310018338244166\\
0.967948701396356	0.0298603024359735\\
0.975429786885407	0.0286425847098759\\
0.981928697262707	0.0273215499568399\\
0.987438888676394	0.0258574705868652\\
0.991954812830795	0.0241880713964009\\
0.995471922573085	0.022205750700571\\
0.9983318994628	0.0191512486838795\\
0.999496542383185	0.0160400171901753\\
0.999784556530997	0.0141470112449146\\
0.999972174405978	0.0104509200785907\\
0.99999	0\\
};
\addplot [color=mycolor2, line width=1.0pt, forget plot]
  table[row sep=crcr]{%
-0.888835448654923	0.51220568682508\\
-0.873849377069785	0.475302508943672\\
-0.857983413234977	0.442470737645402\\
-0.841253532831181	0.413044497661306\\
-0.823676581429833	0.386499722950805\\
-0.805270257531059	0.362418533661579\\
-0.786053094742787	0.340463694945431\\
-0.766044443118978	0.320360040524728\\
-0.745264449675755	0.301880812237827\\
-0.72373403810507	0.284837481200548\\
-0.701474887706321	0.269072016488101\\
-0.678509411557132	0.254450883179194\\
-0.654860733945285	0.24086031386714\\
-0.630552667084523	0.228202582695883\\
-0.605609687137666	0.216393099705096\\
-0.580056909571198	0.205358157167524\\
-0.55392006386611	0.195033152510972\\
-0.527225467610502	0.185361132850034\\
-0.5	0.176291564278797\\
-0.472271074772683	0.167779295494745\\
-0.444066612605774	0.15978372060801\\
-0.415415013001887	0.152268135166507\\
-0.386345125693129	0.145199243462567\\
-0.356886221591872	0.138546752059392\\
-0.327067963317421	0.132282997484342\\
-0.296920375328275	0.126382596071507\\
-0.266473813690035	0.12082213931073\\
-0.235758935509427	0.115579962162957\\
-0.204806668065191	0.110635984882784\\
-0.17364817766693	0.105971596467533\\
-0.142314838273285	0.10156953836429\\
-0.110838199901011	0.0974137682615248\\
-0.0792499568567886	0.0934893173890883\\
-0.0475819158237423	0.0897821728651741\\
-0.015865963834808	0.0862792053734722\\
0.0158659638348082	0.0829681328750608\\
0.0475819158237424	0.0798374889209067\\
0.0792499568567885	0.076876567898491\\
0.142314838273285	0.0714243968336578\\
0.204806668065191	0.06653833407933\\
0.266473813690035	0.0621536293918412\\
0.327067963317422	0.0582134326699836\\
0.386345125693129	0.0546673797121451\\
0.444066612605774	0.0514705337123736\\
0.527225467610502	0.0472442909874022\\
0.605609687137667	0.0435981748712094\\
0.701474887706321	0.0394667029242548\\
0.888835448654923	0.0315471546619897\\
0.928367933016073	0.0294933374632439\\
0.950071117740945	0.0281010492056025\\
0.967948701396356	0.0266197673915699\\
0.975429786885407	0.0258162727226195\\
0.981928697262707	0.0249462928898578\\
0.987438888676394	0.0239803491952223\\
0.991954812830795	0.0228714054597884\\
0.995471922573085	0.0215364121179421\\
0.9983318994628	0.0194214826876137\\
0.999496542383185	0.0171678955173648\\
0.999784556530997	0.0157333750310547\\
0.999972174405978	0.0127495073918267\\
0.99999	0\\
};
\addplot [color=mycolor3, line width=1.0pt, forget plot]
  table[row sep=crcr]{%
-0.902926538286621	0.522075089758949\\
-0.888835448654923	0.481859461351993\\
-0.873849377069785	0.446350160485532\\
-0.857983413234977	0.414732253141281\\
-0.841253532831181	0.386373969505752\\
-0.823676581429833	0.360778412500326\\
-0.805270257531059	0.337549583760129\\
-0.786053094742787	0.316367995326245\\
-0.766044443118978	0.296972934026693\\
-0.745264449675755	0.279149447445747\\
-0.72373403810507	0.262718677029855\\
-0.701474887706321	0.247530526898242\\
-0.678509411557132	0.233457962515281\\
-0.654860733945285	0.220392504182448\\
-0.630552667084523	0.208240675322329\\
-0.605609687137666	0.19692125194463\\
-0.580056909571198	0.186363157954044\\
-0.55392006386611	0.176503825789684\\
-0.527225467610502	0.167287854465258\\
-0.5	0.158665861660404\\
-0.472271074772683	0.150593506507371\\
-0.444066612605774	0.143030703702778\\
-0.415415013001887	0.135941035563132\\
-0.386345125693129	0.129291320687575\\
-0.356886221591872	0.123051263933797\\
-0.327067963317421	0.117193124169343\\
-0.296920375328275	0.111691384900085\\
-0.266473813690035	0.106522459170951\\
-0.235758935509427	0.101664468604867\\
-0.204806668065191	0.0970971042671541\\
-0.17364817766693	0.0928015344919193\\
-0.142314838273285	0.0887603086198494\\
-0.110838199901011	0.0849572285524504\\
-0.0792499568567886	0.0813772018458243\\
-0.0475819158237423	0.0780061158981668\\
-0.015865963834808	0.0748307626660104\\
0.0158659638348082	0.071838807878409\\
0.0475819158237424	0.0690187688046868\\
0.0792499568567885	0.0663599649805194\\
0.142314838273285	0.0614868475652476\\
0.204806668065191	0.0571469711317166\\
0.266473813690035	0.0532766914538506\\
0.327067963317422	0.049820626423982\\
0.386345125693129	0.0467301816981623\\
0.444066612605774	0.0439623318166815\\
0.527225467610502	0.0403342749008589\\
0.605609687137667	0.037238923475262\\
0.701474887706321	0.0337826496903157\\
0.91610845743207	0.0263637621215151\\
0.950071117740945	0.0248143135194847\\
0.967948701396356	0.0237345254591697\\
0.981928697262707	0.0225329884344223\\
0.987438888676394	0.0218441170137176\\
0.991954812830795	0.0210538165514511\\
0.995471922573085	0.0200989501432625\\
0.9983318994628	0.0185682663961264\\
0.999496542383185	0.0169001606596977\\
0.999784556530997	0.0158128296718419\\
0.999972174405978	0.0134723237852287\\
0.99999	0\\
};
\end{axis}

\begin{axis}[%
width=0.307\fwidth,
height=0.307\fheight,
at={(0.577\fwidth,0.601\fheight)},
scale only axis,
xmin=-1,
xmax=1,
xlabel style={font=\color{white!15!black}},
xlabel={$x$},
ymin=0,
ymax=1,
ylabel style={font=\color{white!15!black}},
ylabel={$1/\Phi$},
axis background/.style={fill=white},
axis x line*=bottom,
axis y line*=left,
xminorgrids,
yminorgrids,
axis on top=false
]
\addplot [color=black, line width=2.0pt, forget plot]
  table[row sep=crcr]{%
-0.999496542383185	2.22044604925031e-16\\
0.99999	2.22044604925031e-16\\
};
\addplot [color=mycolor1, line width=1.0pt, forget plot]
  table[row sep=crcr]{%
-0.999496542383185	1.58418484996758e-08\\
-0.939692620785908	0.0002273112492297\\
-0.873849377069785	0.000994623729105304\\
-0.805270257531059	0.00236997953762652\\
-0.745264449675755	0.00405563753743743\\
-0.678509411557132	0.00645976240358404\\
-0.605609687137666	0.00972148242997828\\
-0.527225467610502	0.0139697349047569\\
-0.472271074772683	0.0174061136575989\\
-0.415415013001887	0.0213587254389742\\
-0.356886221591872	0.025849708248649\\
-0.296920375328275	0.0308950599142837\\
-0.235758935509427	0.0365040252908555\\
-0.17364817766693	0.0426785833920743\\
-0.110838199901011	0.0494130441722049\\
-0.0475819158237423	0.0566937629416235\\
0.0158659638348082	0.0644989785298767\\
0.0792499568567885	0.0727987793359615\\
0.142314838273285	0.0815551993587078\\
0.204806668065191	0.0907224442133969\\
0.266473813690035	0.100247245047662\\
0.327067963317422	0.110069336203966\\
0.386345125693129	0.120122050470819\\
0.444066612605774	0.13033302385267\\
0.527225467610502	0.145776101807383\\
0.605609687137667	0.161123904214395\\
0.678509411557132	0.176087115292867\\
0.745264449675755	0.190371749956376\\
0.823676581429833	0.207862267103475\\
0.9	0.225625\\
0.967948701396356	0.242051380707976\\
0.99999	0.24999750000625\\
};
\addplot [color=red, line width=1.0pt, forget plot]
  table[row sep=crcr]{%
-0.999496542383185	3.16836966662848e-08\\
-0.959492973614497	0.000205102398324719\\
-0.916108457432069	0.000879723864303572\\
-0.873849377069785	0.00198924745821039\\
-0.823676581429833	0.00388624349203404\\
-0.766044443118978	0.00684190032443888\\
-0.72373403810507	0.00954036021271654\\
-0.678509411557132	0.012919524807168\\
-0.630552667084523	0.0170614164747952\\
-0.580056909571198	0.0220440248998618\\
-0.527225467610502	0.0279394698095137\\
-0.472271074772683	0.0348122273151976\\
-0.415415013001887	0.0427174508779483\\
-0.356886221591872	0.0516994164972976\\
-0.296920375328275	0.061790119828567\\
-0.235758935509427	0.0730080505817108\\
-0.17364817766693	0.0853571667841484\\
-0.110838199901011	0.0988260883444095\\
-0.0475819158237423	0.113387525883247\\
0.0158659638348082	0.128997957059753\\
0.0792499568567885	0.145597558671923\\
0.142314838273285	0.163110398717415\\
0.204806668065191	0.181444888426794\\
0.266473813690035	0.200494490095323\\
0.327067963317422	0.220138672407932\\
0.386345125693129	0.240244100941637\\
0.444066612605774	0.26066604770534\\
0.5	0.28125\\
0.55392006386611	0.301833445610707\\
0.605609687137667	0.32224780842879\\
0.654860733945285	0.342320506094241\\
0.701474887706321	0.361877099186905\\
0.745264449675755	0.380743499912752\\
0.805270257531059	0.407375087840782\\
0.857983413234977	0.431512795482037\\
0.91610845743207	0.458933952580338\\
0.967948701396356	0.484102761415951\\
0.99999	0.4999950000125\\
};
\addplot [color=mycolor2, line width=1.0pt, forget plot]
  table[row sep=crcr]{%
-0.999496542383185	4.7525544943916e-08\\
-0.967948701396356	0.000192616076658991\\
-0.928367933016072	0.000962091191323289\\
-0.888835448654923	0.0023170420267038\\
-0.857983413234977	0.00378163329682291\\
-0.823676581429833	0.00582936523805089\\
-0.786053094742787	0.00858248967546371\\
-0.745264449675755	0.012166912612312\\
-0.701474887706321	0.0167094830006165\\
-0.654860733945285	0.0223352086823978\\
-0.605609687137666	0.0291644472899345\\
-0.55392006386611	0.0373101205164781\\
-0.5	0.0468750000000002\\
-0.472271074772683	0.0522183409727964\\
-0.444066612605774	0.0579491121036787\\
-0.415415013001887	0.0640761763169223\\
-0.386345125693129	0.0706073071426094\\
-0.356886221591872	0.0775491247459463\\
-0.327067963317421	0.0849070361238309\\
-0.296920375328275	0.0926851797428504\\
-0.266473813690035	0.100886374875458\\
-0.235758935509427	0.109512075872566\\
-0.204806668065191	0.118562331591297\\
-0.17364817766693	0.128035750176222\\
-0.142314838273285	0.137929469371159\\
-0.110838199901011	0.148239132516614\\
-0.0792499568567886	0.158958870365293\\
-0.0475819158237423	0.17008128882487\\
-0.015865963834808	0.181597462713524\\
0.0158659638348082	0.19349693558963\\
0.0475819158237424	0.205767725692677\\
0.0792499568567885	0.218396338007884\\
0.110838199901011	0.231367782442372\\
0.142314838273285	0.244665598076123\\
0.173648177666931	0.25827188342642\\
0.204806668065191	0.27216733264019\\
0.235758935509427	0.286331277504636\\
0.266473813690035	0.300741735142984\\
0.296920375328275	0.315375461239057\\
0.327067963317422	0.330208008611897\\
0.356886221591872	0.34521379093985\\
0.415415013001886	0.375637436068337\\
0.472271074772683	0.406421647052308\\
0.527225467610502	0.437328305422147\\
0.580056909571198	0.468108719528191\\
0.630552667084523	0.498506625025585\\
0.678509411557132	0.528261345878601\\
0.72373403810507	0.557111068897877\\
0.766044443118978	0.584796182825892\\
0.805270257531059	0.611062631761173\\
0.857983413234977	0.647269193223056\\
0.902926538286621	0.678961764396657\\
0.950071117740945	0.713020755796391\\
0.9983318994628	0.748749446326988\\
0.99999	0.74999250001875\\
};
\addplot [color=mycolor3, line width=1.0pt, forget plot]
  table[row sep=crcr]{%
-0.999496542383185	6.33673932215473e-08\\
-0.967948701396356	0.000256821435545174\\
-0.939692620785908	0.000909244996918246\\
-0.902926538286621	0.00235581424225495\\
-0.873849377069785	0.00397849491642055\\
-0.841253532831181	0.00630011020964549\\
-0.805270257531059	0.00947991815050531\\
-0.766044443118978	0.0136838006488775\\
-0.72373403810507	0.019080720425433\\
-0.678509411557132	0.0258390496143356\\
-0.654860733945285	0.0297802782431971\\
-0.630552667084523	0.0341228329495902\\
-0.605609687137666	0.0388859297199126\\
-0.580056909571198	0.0440880497997235\\
-0.55392006386611	0.049746827355304\\
-0.527225467610502	0.0558789396190272\\
-0.5	0.0625000000000003\\
-0.472271074772683	0.0696244546303951\\
-0.444066612605774	0.0772654828049049\\
-0.415415013001887	0.0854349017558963\\
-0.386345125693129	0.0941430761901458\\
-0.356886221591872	0.103398832994595\\
-0.327067963317421	0.113209381498441\\
-0.296920375328275	0.123580239657134\\
-0.266473813690035	0.134515166500611\\
-0.235758935509427	0.146016101163421\\
-0.204806668065191	0.158083108788396\\
-0.17364817766693	0.170714333568297\\
-0.142314838273285	0.183905959161546\\
-0.110838199901011	0.197652176688819\\
-0.0792499568567886	0.211945160487057\\
-0.0475819158237423	0.226775051766493\\
-0.015865963834808	0.242129950284698\\
0.0158659638348082	0.257995914119506\\
0.0475819158237424	0.274356967590236\\
0.0792499568567885	0.291195117343845\\
0.110838199901011	0.30849037658983\\
0.142314838273285	0.326220797434831\\
0.173648177666931	0.344362511235227\\
0.204806668065191	0.362889776853587\\
0.235758935509427	0.381775036672849\\
0.266473813690035	0.400988980190646\\
0.296920375328275	0.420500614985409\\
0.327067963317422	0.440277344815863\\
0.356886221591872	0.460285054586467\\
0.386345125693129	0.480488201883274\\
0.415415013001886	0.500849914757783\\
0.444066612605774	0.521332095410679\\
0.472271074772683	0.541895529403078\\
0.5	0.5625\\
0.527225467610502	0.58310440722953\\
0.580056909571198	0.624144959370922\\
0.630552667084523	0.664675500034113\\
0.678509411557132	0.704348461171468\\
0.72373403810507	0.742814758530503\\
0.766044443118978	0.779728243767856\\
0.805270257531059	0.814750175681564\\
0.841253532831181	0.847553643040827\\
0.888835448654923	0.891924838023862\\
0.928367933016073	0.92965072127117\\
0.967948701396356	0.968205522831902\\
0.99999	0.999990000025\\
};
\end{axis}
\end{tikzpicture}%

%% file: images/largePor1.tex
%
%
\definecolor{mycolor1}{rgb}{0.33333,0.00000,0.00000}%
\definecolor{mycolor2}{rgb}{1.00000,0.33333,0.00000}%
\definecolor{mycolor3}{rgb}{1.00000,0.66667,0.00000}%
\begin{tikzpicture}[%
trim axis left, trim axis right
]

\begin{axis}[%
width=0.951\fwidth,
height=\fheight,
at={(0\fwidth,0\fheight)},
scale only axis,
xmin=-1,
xmax=1,
xlabel style={font=\color{white!15!black}},
xlabel={$x$},
ymin=0,
ymax=5,
ylabel style={font=\color{white!15!black}},
ylabel={$|\Delta p|$},
axis background/.style={fill=white},
axis x line*=bottom,
axis y line*=left,
xmajorgrids,
ymajorgrids,
axis on top=false
]
\addplot [color=black, line width=2.0pt, forget plot]
  table[row sep=crcr]{%
-0.857983413234977	5.18166220334366\\
-0.841253532831181	4.89257543608171\\
-0.823676581429833	4.63592755095666\\
-0.805270257531059	4.40709558616031\\
-0.786053094742788	4.20232651456647\\
-0.766044443118978	4.01853624553462\\
-0.745264449675755	3.85316098519181\\
-0.723734038105071	3.70404575398806\\
-0.701474887706321	3.56935988480665\\
-0.678509411557132	3.44753243790479\\
-0.654860733945285	3.33720237955821\\
-0.630552667084523	3.23717970107757\\
-0.605609687137667	3.14641475812494\\
-0.580056909571198	3.06397401261378\\
-0.55392006386611	2.98902093595049\\
-0.527225467610503	2.92080105289495\\
-0.5	2.85863013881057\\
-0.472271074772682	2.80188469078263\\
-0.444066612605774	2.74999409746562\\
-0.415415013001886	2.70243427602295\\
-0.386345125693128	2.65872264347064\\
-0.356886221591872	2.61841409806386\\
-0.327067963317422	2.58109753655711\\
-0.296920375328275	2.54639266314046\\
-0.266473813690035	2.51394729267592\\
-0.235758935509427	2.48343544225036\\
-0.20480666806519	2.4545560069675\\
-0.17364817766693	2.4270313005384\\
-0.142314838273285	2.40060498361384\\
-0.110838199901011	2.37503980802379\\
-0.0475819158237423	2.32563144511609\\
0.0792499568567884	2.22860126345516\\
0.142314838273285	2.17857633541377\\
0.17364817766693	2.15275690385626\\
0.20480666806519	2.12626583039828\\
0.235758935509427	2.09901454413117\\
0.266473813690035	2.07092287984593\\
0.296920375328275	2.04191916821201\\
0.327067963317422	2.01194018708739\\
0.356886221591872	1.98093039874135\\
0.386345125693128	1.948840921055\\
0.415415013001886	1.9156290375306\\
0.444066612605774	1.88125833501232\\
0.472271074772682	1.84569879419417\\
0.5	1.80892634098574\\
0.527225467610503	1.77092215130467\\
0.55392006386611	1.73167228434053\\
0.580056909571198	1.69116769029691\\
0.605609687137667	1.64940412802724\\
0.630552667084523	1.60638175937013\\
0.654860733945285	1.56210477035822\\
0.678509411557132	1.51658141874323\\
0.701474887706321	1.4698243253998\\
0.723734038105071	1.42185042556696\\
0.745264449675755	1.37268034459636\\
0.766044443118978	1.3223376309647\\
0.786053094742788	1.27084845944598\\
0.805270257531059	1.21824195534987\\
0.823676581429833	1.16455073557742\\
0.841253532831181	1.10981109766454\\
0.857983413234977	1.05406253863599\\
0.873849377069785	0.997346790990692\\
0.888835448654923	0.939707102769502\\
0.902926538286621	0.881188541075174\\
0.916108457432069	0.821839165371031\\
0.928367933016073	0.761710594897714\\
0.939692620785908	0.700856574168096\\
0.950071117740945	0.639330469730467\\
0.959492973614497	0.577184931636398\\
0.967948701396356	0.514475457371546\\
0.975429786885407	0.451264214813557\\
0.981928697262707	0.387617240971299\\
0.987438888676395	0.323593618070312\\
0.991954812830795	0.259237245692564\\
0.995471922573085	0.194587984366362\\
0.997986676471885	0.129718065119049\\
0.999496542383185	0.064775589947569\\
};
\addplot [color=black, line width=3.0pt, forget plot]
  table[row sep=crcr]{%
-0.920626766415591	5.15499651788821\\
-0.911838290714917	4.86568004913307\\
-0.902635128765529	4.6049243260113\\
-0.893026547371393	4.36871582561481\\
-0.883022221559489	4.15377002609173\\
-0.872632224837878	3.95737237637241\\
-0.861867019052535	3.77725897621194\\
-0.850737443853161	3.61152589827219\\
-0.839254705778566	3.45855948466821\\
-0.827430366972642	3.3169821693474\\
-0.815276333542261	3.18560989344557\\
-0.802804843568834	3.06341828308525\\
-0.790028454785599	2.94951557249558\\
-0.776960031933055	2.84312081492979\\
-0.763612733805251	2.74354626598914\\
-0.75	2.65018303136607\\
-0.736135537386342	2.56248924747141\\
-0.722033306302887	2.47998026423071\\
-0.707707506500943	2.40222048716332\\
-0.693172562846565	2.32881662784916\\
-0.678443110795936	2.2594120949685\\
-0.663533981658711	2.19368223964531\\
-0.648460187664138	2.13133026477884\\
-0.633236906845018	2.07208377620452\\
-0.617879467754713	2.01569200750512\\
-0.602403334032595	1.96192359866081\\
-0.586824088833465	1.91056463123331\\
-0.571157419136642	1.86141669497062\\
-0.555419099950505	1.81429508059594\\
-0.539624978428394	1.76902740533244\\
-0.523790957911872	1.72545278239224\\
-0.507932981917404	1.68342122685357\\
-0.476209042088129	1.60343683366343\\
-0.444580900049495	1.5280583025902\\
-0.413175911166535	1.45639373489171\\
-0.382120532245287	1.38766513759425\\
-0.263864462613658	1.12982045607303\\
-0.236387266194749	1.06725701954528\\
-0.209971545214401	1.00474680499778\\
-0.184723666457739	0.942040742475859\\
-0.160745294221434	0.878940413881253\\
-0.138132980947465	0.815294064537541\\
-0.116977778440511	0.750994460780228\\
-0.097364871234471	0.685969388965245\\
-0.0793732335844091	0.620179971925979\\
-0.0710082933825111	0.586996013701006\\
-0.0630753114651075	0.553621209640741\\
-0.0555822756725384	0.520058615094241\\
-0.0485367308566893	0.486312668535797\\
-0.0419457712839648	0.452389513048886\\
-0.0358160334919635	0.418296997584003\\
-0.0301536896070456	0.384044096203866\\
-0.0249644411295273	0.349640148365367\\
-0.0202535131927517	0.315094819283728\\
-0.016025649301822	0.280419141715412\\
-0.0122851065572966	0.245626522602219\\
-0.00903565136864692	0.210731803755872\\
-0.00628055566180308	0.175748077000897\\
-0.00402259358460277	0.140684321965977\\
-0.00226403871345759	0.105548617671544\\
-0.00100666176405806	0.070358557963619\\
-0.000251728808407847	0.0351536844960254\\
};
\addplot [color=mycolor1, line width=1.0pt, forget plot]
  table[row sep=crcr]{%
0	4.44089209850063e-16\\
0.000251728808407403	2.0560972254957\\
0.00100666176405761	2.09217887329089\\
0.00226403871345759	2.11300980355281\\
0.00402259358460233	2.12730383172738\\
0.00628055566180308	2.13782412009834\\
0.00903565136864648	2.14581271242406\\
0.0122851065572966	2.15193716079069\\
0.016025649301822	2.15659799328103\\
0.0202535131927513	2.16005409017611\\
0.0249644411295273	2.16248214238317\\
0.030153689607046	2.16400813020014\\
0.0358160334919639	2.16472537587205\\
0.0419457712839653	2.16470543465962\\
0.0485367308566893	2.16400480742865\\
0.0555822756725384	2.16266906704712\\
0.0710082933825116	2.15823399853602\\
0.0881617092850835	2.15162118522736\\
0.106973452628606	2.1429717310209\\
0.127367775162122	2.13237233527467\\
0.149262556146839	2.11987497413435\\
0.172569633027357	2.10550558846057\\
0.197195156431166	2.08926410185431\\
0.223039968066945	2.07112499550036\\
0.25	2.05104341022963\\
0.277966693697113	2.02896262158615\\
0.306827437153436	2.00481689731045\\
0.336466018341289	1.97853043646671\\
0.366763093154983	1.95001819023264\\
0.397596665967405	1.91919105923218\\
0.428842580863357	1.88596153367635\\
0.460375021571606	1.85024538767411\\
0.492067018082596	1.81196071379451\\
0.523790957911871	1.77102904777124\\
0.555419099950506	1.72737986179492\\
0.586824088833465	1.68095456199157\\
0.602403334032596	1.65668592996642\\
0.617879467754713	1.63170668617054\\
0.633236906845017	1.60601263785994\\
0.648460187664138	1.57960027016509\\
0.663533981658711	1.55246686844008\\
0.678443110795936	1.52461066432247\\
0.693172562846565	1.49603095629131\\
0.707707506500943	1.46672816427736\\
0.722033306302887	1.43670380302135\\
0.736135537386342	1.40596039038463\\
0.75	1.3745013320732\\
0.763612733805251	1.34233083301546\\
0.776960031933055	1.30945387419385\\
0.790028454785599	1.27587626637453\\
0.802804843568834	1.24160475963996\\
0.815276333542261	1.20664716289111\\
0.827430366972643	1.17101242066383\\
0.839254705778566	1.13471060886757\\
0.850737443853161	1.09775284122493\\
0.861867019052535	1.06015111258259\\
0.872632224837878	1.02191813014754\\
0.883022221559489	0.983067188689889\\
0.893026547371394	0.943612128033482\\
0.902635128765529	0.903567376853811\\
0.911838290714917	0.862948048845071\\
0.92062676641559	0.82177003055298\\
0.928991706617488	0.780049995610105\\
0.936924688534893	0.737805300504332\\
0.944417724327462	0.695053755401094\\
0.951463269143311	0.651813305364384\\
0.958054228716035	0.608101685397489\\
0.964183966508037	0.563936113505613\\
0.969846310392954	0.51933305488379\\
0.975035558870473	0.474308033757312\\
0.979746486807249	0.428875401656302\\
0.983974350698178	0.383047905645797\\
0.987714893442704	0.336835837482916\\
0.990964348631354	0.290245455104942\\
0.993719444338197	0.243276162134365\\
0.995977406415398	0.195915369551749\\
0.997735961286542	0.148128277550183\\
0.998993338235942	0.0998337367314321\\
0.999748271191593	0.050825744129587\\
1	4.44089209850063e-16\\
};
\addplot [color=mycolor1, line width=1.0pt, forget plot]
  table[row sep=crcr]{%
-0.850737443853161	5.01119539485437\\
-0.839254705778566	4.82290913577048\\
-0.827430366972642	4.64971564625618\\
-0.815276333542261	4.49006969904702\\
-0.802804843568834	4.34263554278611\\
-0.790028454785599	4.20625149627079\\
-0.776960031933055	4.07990142534583\\
-0.763612733805251	3.96269160297228\\
-0.75	3.85383181616174\\
-0.736135537386342	3.75261985031952\\
-0.722033306302887	3.65842867962886\\
-0.707707506500943	3.57069584057249\\
-0.693172562846565	3.48891457797571\\
-0.678443110795936	3.41262643861723\\
-0.663533981658711	3.34141505332971\\
-0.648460187664138	3.27490089955171\\
-0.633236906845018	3.21273687611193\\
-0.617879467754713	3.15460455330318\\
-0.602403334032595	3.10021098604294\\
-0.586824088833465	3.04928599764249\\
-0.571157419136643	3.00157985757648\\
-0.555419099950505	2.95686128954194\\
-0.539624978428394	2.91491575667742\\
-0.523790957911872	2.87554397956357\\
-0.507932981917404	2.83856064989277\\
-0.492067018082596	2.8037933087306\\
-0.476209042088129	2.77108136329199\\
-0.460375021571606	2.74027522028587\\
-0.444580900049495	2.7112355173013\\
-0.428842580863358	2.68383243656354\\
-0.413175911166535	2.65794508781368\\
-0.397596665967405	2.63346094916724\\
-0.382120532245287	2.61027535664989\\
-0.366763093154983	2.58829103471853\\
-0.351539812335862	2.56741766144663\\
-0.321556889204064	2.52867483420434\\
-0.292292493499057	2.4934485660326\\
-0.263864462613658	2.46122815886337\\
-0.236387266194749	2.43158025012085\\
-0.197195156431166	2.39114817399958\\
-0.0881617092850835	2.28108379607261\\
-0.071008293382512	2.26221817018508\\
-0.0555822756725384	2.24399778469741\\
-0.0419457712839657	2.22619229791198\\
-0.0301536896070456	2.20847530045921\\
-0.0249644411295273	2.19950091839693\\
-0.0202535131927517	2.19033296184332\\
-0.016025649301822	2.1808457637509\\
-0.0122851065572966	2.17086317254535\\
-0.00903565136864692	2.16012747366105\\
-0.00628055566180308	2.14824003806013\\
-0.00402259358460277	2.1345355124686\\
-0.00226403871345759	2.11777313561043\\
-0.00100666176405806	2.09518762451175\\
-0.000251728808407847	2.0580531877913\\
0	8.88178419700125e-16\\
};
\addplot [color=red!50!mycolor1, line width=1.0pt, forget plot]
  table[row sep=crcr]{%
0	4.44089209850063e-16\\
0.000251728808407403	0.775242727230954\\
0.00100666176405784	0.915124711595508\\
0.00226403871345782	1.0081106506749\\
0.00402259358460233	1.07943408079516\\
0.00628055566180286	1.13786308478601\\
0.0090356513686467	1.1875884291399\\
0.0122851065572964	1.23097122246499\\
0.0160256493018218	1.26948189281006\\
0.0202535131927513	1.30409830481913\\
0.0249644411295273	1.33550146757411\\
0.0301536896070458	1.3641833589729\\
0.0358160334919637	1.39051171972434\\
0.0419457712839653	1.41477114519314\\
0.0485367308566895	1.43718953023732\\
0.0555822756725384	1.45795467574802\\
0.0630753114651075	1.47722415989946\\
0.0710082933825116	1.49513088176374\\
0.0793732335844095	1.51178626845952\\
0.0881617092850837	1.52728267699235\\
0.0973648712344708	1.54169594460639\\
0.106973452628606	1.55508842165669\\
0.116977778440511	1.56751229555768\\
0.127367775162123	1.57901270601712\\
0.138132980947465	1.58963011547889\\
0.149262556146839	1.59940160101862\\
0.160745294221434	1.60836106755564\\
0.172569633027358	1.61653870695948\\
0.184723666457739	1.62396022092987\\
0.197195156431167	1.63064632395334\\
0.209971545214401	1.63661285956278\\
0.223039968066945	1.64187157798702\\
0.236387266194749	1.64643134778774\\
0.25	1.65029940896848\\
0.263864462613659	1.65348227353772\\
0.277966693697113	1.65598603053115\\
0.292292493499057	1.65781604859507\\
0.306827437153436	1.6589762950546\\
0.321556889204064	1.65946862004882\\
0.336466018341289	1.65929234325716\\
0.351539812335863	1.65844434047147\\
0.366763093154983	1.65691961819466\\
0.382120532245287	1.65471217053486\\
0.397596665967405	1.65181580825654\\
0.413175911166535	1.64822467155848\\
0.428842580863357	1.64393327229391\\
0.444580900049494	1.63893610117421\\
0.460375021571606	1.63322700467761\\
0.476209042088129	1.62679861926835\\
0.492067018082596	1.61964211728771\\
0.507932981917404	1.61174738599804\\
0.523790957911871	1.60310358381472\\
0.539624978428394	1.5936998668815\\
0.555419099950506	1.58352601382822\\
0.571157419136643	1.57257272083932\\
0.586824088833465	1.56083147343605\\
0.602403334032595	1.54829407086653\\
0.617879467754714	1.53495201588924\\
0.633236906845017	1.52079603302522\\
0.648460187664137	1.50581592292811\\
0.663533981658711	1.49000082229303\\
0.678443110795936	1.47333977322595\\
0.693172562846564	1.45582237924516\\
0.707707506500943	1.43743928750676\\
0.722033306302887	1.41818230342074\\
0.736135537386341	1.39804408823837\\
0.75	1.37701755518689\\
0.763612733805251	1.35509519961639\\
0.776960031933055	1.33226862515161\\
0.790028454785599	1.30852844931046\\
0.802804843568833	1.28386461820912\\
0.815276333542261	1.25826699070605\\
0.827430366972643	1.23172593447002\\
0.839254705778566	1.20423265762431\\
0.850737443853161	1.17577908929297\\
0.861867019052535	1.14635728751152\\
0.872632224837877	1.11595853007863\\
0.883022221559489	1.08457236317649\\
0.893026547371394	1.05218589471392\\
0.902635128765529	1.01878351513225\\
0.911838290714917	0.984347043048479\\
0.92062676641559	0.948856093549721\\
0.928991706617488	0.912288326845731\\
0.936924688534893	0.87461920650276\\
0.944417724327462	0.835820989564518\\
0.951463269143311	0.795860847571657\\
0.958054228716035	0.75469820340734\\
0.964183966508036	0.712281477068952\\
0.969846310392954	0.668544391098918\\
0.975035558870473	0.623401752046736\\
0.979746486807249	0.576744183313404\\
0.983974350698178	0.528430610800782\\
0.987714893442704	0.478276266091352\\
0.990964348631353	0.426032132849525\\
0.993719444338197	0.37134782895201\\
0.995977406415398	0.313699974246535\\
0.997735961286542	0.252238425447491\\
0.998993338235942	0.185391818725373\\
0.999748271191593	0.109458676726957\\
1	4.44089209850063e-16\\
};
\addplot [color=red!50!mycolor1, line width=1.0pt, forget plot]
  table[row sep=crcr]{%
-0.872632224837878	5.06552699012526\\
-0.861867019052535	4.85172949302214\\
-0.850737443853161	4.65576928425196\\
-0.839254705778566	4.4756590388119\\
-0.827430366972642	4.30970283076758\\
-0.815276333542261	4.15644354538797\\
-0.802804843568834	4.01462118642145\\
-0.790028454785599	3.8831395417666\\
-0.776960031933055	3.76103932588293\\
-0.763612733805251	3.6474763880422\\
-0.75	3.54170391757282\\
-0.736135537386342	3.44305782831531\\
-0.722033306302887	3.35094469059736\\
-0.707707506500943	3.26483171836245\\
-0.693172562846565	3.18423842451169\\
-0.678443110795936	3.10872963815796\\
-0.663533981658711	3.03790963976837\\
-0.648460187664138	2.97141721859882\\
-0.633236906845018	2.90892149460276\\
-0.617879467754713	2.8501183764807\\
-0.602403334032595	2.79472755057446\\
-0.586824088833465	2.74248991343817\\
-0.571157419136643	2.69316537543522\\
-0.555419099950505	2.64653097465503\\
-0.539624978428394	2.60237925054352\\
-0.523790957911872	2.56051683529595\\
-0.507932981917404	2.5207632283915\\
-0.492067018082596	2.48294972562782\\
-0.476209042088129	2.44691847862094\\
-0.460375021571606	2.41252166410044\\
-0.444580900049495	2.37962074477513\\
-0.428842580863358	2.34808580553926\\
-0.397596665967405	2.28863375100704\\
-0.366763093154983	2.23327688019548\\
-0.336466018341289	2.1812303564638\\
-0.292292493499057	2.10786814111154\\
-0.236387266194749	2.0157607823149\\
-0.209971545214401	1.97100788982827\\
-0.184723666457739	1.92651527023039\\
-0.160745294221434	1.88185427434746\\
-0.138132980947465	1.83659839399728\\
-0.127367775162122	1.81360981949955\\
-0.116977778440511	1.79030301756727\\
-0.106973452628607	1.76661514527565\\
-0.097364871234471	1.74247832437202\\
-0.0881617092850835	1.71781818797088\\
-0.07937323358441	1.6925521407768\\
-0.071008293382512	1.66658722580386\\
-0.0630753114651075	1.63981743243226\\
-0.0555822756725384	1.61212020177161\\
-0.0485367308566893	1.58335177773912\\
-0.0419457712839657	1.55334089780177\\
-0.0358160334919635	1.52188007790122\\
-0.0301536896070456	1.48871334642242\\
-0.0249644411295273	1.4535185752042\\
-0.0202535131927517	1.41588124685838\\
-0.016025649301822	1.37525396353908\\
-0.0122851065572966	1.33089082318543\\
-0.00903565136864692	1.28173443319725\\
-0.00628055566180308	1.22620601427845\\
-0.00402259358460277	1.16177471356032\\
-0.00226403871345759	1.08394245134208\\
-0.00100666176405806	0.983282985508323\\
-0.000251728808407847	0.832660564347876\\
0	8.88178419700125e-16\\
};
\addplot [color=red, line width=1.0pt, forget plot]
  table[row sep=crcr]{%
0	2.22044604925031e-16\\
0.000251728808407403	0.0285261236810281\\
0.00100666176405784	0.0506085643264382\\
0.00226403871345782	0.0707534935674192\\
0.00402259358460233	0.0897496619854135\\
0.00628055566180286	0.107950295959644\\
0.0090356513686467	0.125550294841896\\
0.0122851065572964	0.142667756792964\\
0.0160256493018218	0.159378270401282\\
0.0202535131927513	0.175732751679415\\
0.0249644411295273	0.191767866332132\\
0.0301536896070458	0.207512269818422\\
0.0358160334919637	0.222990091110078\\
0.0419457712839653	0.238222522842142\\
0.0485367308566895	0.253228212184522\\
0.0555822756725384	0.268023061040981\\
0.0630753114651075	0.282619928548179\\
0.0710082933825116	0.29702856280759\\
0.0881617092850837	0.32530666389649\\
0.106973452628606	0.352891032301686\\
0.127367775162123	0.379802337005116\\
0.149262556146839	0.406058540418021\\
0.172569633027358	0.43167055019007\\
0.197195156431167	0.456635444682585\\
0.223039968066945	0.480938101862702\\
0.25	0.504558702174573\\
0.277966693697113	0.527476597326982\\
0.306827437153436	0.549667388806679\\
0.336466018341289	0.571099133457129\\
0.366763093154983	0.591733482448508\\
0.397596665967405	0.611530076032641\\
0.428842580863357	0.630448526145874\\
0.460375021571606	0.648446395675284\\
0.492067018082596	0.665477090671032\\
0.523790957911871	0.681491033298563\\
0.555419099950506	0.69643850710512\\
0.586824088833465	0.710270453082157\\
0.617879467754714	0.722936684329615\\
0.648460187664137	0.734384424095575\\
0.678443110795936	0.744559034645069\\
0.707707506500943	0.753405308739271\\
0.736135537386341	0.760866887462683\\
0.763612733805251	0.766883994237124\\
0.790028454785599	0.771391542629784\\
0.815276333542261	0.774318168121238\\
0.839254705778566	0.775584414049254\\
0.861867019052535	0.775098621080885\\
0.872632224837877	0.774165302878545\\
0.883022221559489	0.772751039105309\\
0.893026547371394	0.770838103873413\\
0.902635128765529	0.768406563419955\\
0.911838290714917	0.765433775333094\\
0.92062676641559	0.76189372225666\\
0.928991706617488	0.757756151843282\\
0.936924688534893	0.752985491521713\\
0.944417724327462	0.747539483190488\\
0.951463269143311	0.741367430486522\\
0.958054228716035	0.734407862995178\\
0.964183966508036	0.726585288910498\\
0.969846310392954	0.717805509718785\\
0.975035558870473	0.707948656146134\\
0.979746486807249	0.696858554442514\\
0.983974350698178	0.684325976700016\\
0.987714893442704	0.670061143816629\\
0.990964348631353	0.653645976043191\\
0.993719444338197	0.634444691119197\\
0.995977406415398	0.611418584931473\\
0.997735961286542	0.582683814084127\\
0.998993338235942	0.544198914550872\\
0.999748271191593	0.483949397748894\\
1	2.22044604925031e-16\\
};
\addplot [color=red, line width=1.0pt, forget plot]
  table[row sep=crcr]{%
-0.911838290714917	5.05157521453622\\
-0.902635128765529	4.78355364866255\\
-0.893026547371393	4.54092452504358\\
-0.883022221559489	4.32029499766373\\
-0.872632224837878	4.11886109760771\\
-0.861867019052535	3.93428447222693\\
-0.850737443853161	3.76459870328375\\
-0.839254705778566	3.60813724500566\\
-0.827430366972642	3.46347737910283\\
-0.815276333542261	3.32939618386345\\
-0.802804843568834	3.20483561976878\\
-0.790028454785599	3.08887460922452\\
-0.776960031933055	2.98070653835512\\
-0.763612733805251	2.87962100325574\\
-0.75	2.78498890800515\\
-0.736135537386342	2.69625022938063\\
-0.722033306302887	2.61290391664234\\
-0.707707506500943	2.53449951047631\\
-0.693172562846565	2.46063015458319\\
-0.678443110795936	2.39092674370898\\
-0.663533981658711	2.32505300723409\\
-0.648460187664138	2.26270136994854\\
-0.633236906845018	2.20358946286347\\
-0.617879467754713	2.14745717870805\\
-0.602403334032595	2.094064181786\\
-0.586824088833465	2.0431877933007\\
-0.571157419136643	1.99462118405016\\
-0.555419099950505	1.94817181834135\\
-0.539624978428394	1.90366010614117\\
-0.523790957911872	1.86091823328562\\
-0.507932981917404	1.81978914957406\\
-0.492067018082596	1.78012569971916\\
-0.460375021571606	1.70465221410286\\
-0.428842580863358	1.63349275956768\\
-0.397596665967405	1.56576223751946\\
-0.351539812335862	1.46891574851577\\
-0.277966693697113	1.31486343326336\\
-0.25	1.25429009289297\\
-0.223039968066945	1.19367474804088\\
-0.197195156431166	1.13266181225805\\
-0.172569633027358	1.07093778027387\\
-0.149262556146839	1.00822739159741\\
-0.127367775162122	0.94428849783183\\
-0.116977778440511	0.91178909466394\\
-0.106973452628607	0.878900851450369\\
-0.097364871234471	0.845597222208617\\
-0.0881617092850835	0.811851434698945\\
-0.07937323358441	0.777636424816565\\
-0.071008293382512	0.742924846070706\\
-0.0630753114651075	0.707688979770653\\
-0.0555822756725384	0.671900324149838\\
-0.0485367308566893	0.635528652716182\\
-0.0419457712839657	0.598540408870818\\
-0.0358160334919635	0.560896426414737\\
-0.0301536896070456	0.522549090818713\\
-0.0249644411295273	0.483439122357582\\
-0.0202535131927517	0.44349209695592\\
-0.016025649301822	0.40261454234848\\
-0.0122851065572966	0.360688844743366\\
-0.00903565136864692	0.317565052751923\\
-0.00628055566180308	0.273045381293712\\
-0.00402259358460277	0.226851973874735\\
-0.00226403871345759	0.178553747810562\\
-0.00100666176405806	0.12737462378802\\
-0.000251728808407847	0.0715150021257811\\
0	4.44089209850063e-15\\
};
\addplot [color=mycolor2, line width=1.0pt, forget plot]
  table[row sep=crcr]{%
0	2.22044604925031e-16\\
0.000251728808407403	0.0122567812596437\\
0.00100666176405784	0.0240671559003585\\
0.00226403871345782	0.0356905284560152\\
0.00402259358460233	0.0472229073765569\\
0.00628055566180286	0.0587206748584812\\
0.0090356513686467	0.0702136659827046\\
0.0122851065572964	0.0817121933063647\\
0.0160256493018218	0.0932137935782338\\
0.0249644411295273	0.116191241342114\\
0.0358160334919637	0.139091391519821\\
0.0485367308566895	0.161921223523894\\
0.0630753114651075	0.184723769770129\\
0.0793732335844095	0.207519275251843\\
0.0973648712344708	0.230288545109752\\
0.116977778440511	0.253002454592532\\
0.138132980947465	0.27564964448304\\
0.160745294221434	0.298230736363848\\
0.184723666457739	0.32073520179038\\
0.209971545214401	0.34313354225909\\
0.236387266194749	0.365390824902856\\
0.263864462613659	0.387479693725524\\
0.292292493499057	0.40937630249938\\
0.321556889204064	0.431047876279153\\
0.351539812335863	0.452450004936784\\
0.397596665967405	0.483947300991125\\
0.444580900049494	0.514594890607888\\
0.492067018082596	0.544241511430838\\
0.539624978428394	0.572712039865977\\
0.586824088833465	0.599846369713673\\
0.633236906845017	0.625492249366436\\
0.678443110795936	0.649486102899481\\
0.722033306302887	0.671681786329081\\
0.763612733805251	0.691954126959315\\
0.802804843568833	0.710175355407334\\
0.839254705778566	0.726232899361789\\
0.872632224837877	0.740038325928709\\
0.902635128765529	0.751497475750557\\
0.928991706617488	0.760512650131509\\
0.951463269143311	0.766984873798656\\
0.964183966508036	0.76980859478523\\
0.975035558870473	0.771341445656193\\
0.983974350698178	0.771440304576888\\
0.987714893442704	0.770877361724096\\
0.990964348631353	0.769837092504944\\
0.993719444338197	0.768228870532792\\
0.995977406415398	0.765896337792553\\
0.997735961286542	0.762540010352751\\
0.998993338235942	0.757479115849777\\
0.999748271191593	0.748534212860654\\
1	2.22044604925031e-16\\
};
\addplot [color=mycolor2, line width=1.0pt, forget plot]
  table[row sep=crcr]{%
-0.920626766415591	5.08676177900928\\
-0.911838290714917	4.80122938629303\\
-0.902635128765529	4.54391013137658\\
-0.893026547371393	4.31084328597032\\
-0.883022221559489	4.09878761434133\\
-0.872632224837878	3.90506438977282\\
-0.861867019052535	3.72743964570605\\
-0.850737443853161	3.56403468479711\\
-0.839254705778566	3.41325724370031\\
-0.827430366972642	3.27374796052379\\
-0.815276333542261	3.14433831936745\\
-0.802804843568834	3.02401730212704\\
-0.790028454785599	2.91190471896709\\
-0.776960031933055	2.80722971577771\\
-0.763612733805251	2.7093133348004\\
-0.75	2.61755427727654\\
-0.736135537386342	2.53141721510167\\
-0.722033306302887	2.4504231442656\\
-0.707707506500943	2.37414138249362\\
-0.693172562846565	2.30218289828319\\
-0.678443110795936	2.23419472562027\\
-0.663533981658711	2.16985527195244\\
-0.648460187664138	2.10887036829773\\
-0.633236906845018	2.05096994075553\\
-0.617879467754713	1.99590520365558\\
-0.602403334032595	1.94344628860794\\
-0.586824088833465	1.89338023403965\\
-0.571157419136643	1.84550926957499\\
-0.555419099950505	1.79964934085576\\
-0.539624978428394	1.75562883326915\\
-0.523790957911872	1.71328746588456\\
-0.507932981917404	1.67247533703046\\
-0.492067018082596	1.63305210805011\\
-0.460375021571606	1.55785476262184\\
-0.428842580863358	1.48674011371631\\
-0.397596665967405	1.41887031919897\\
-0.351539812335862	1.32157032126482\\
-0.292292493499057	1.19709579322951\\
-0.263864462613658	1.13600584262809\\
-0.236387266194749	1.07515463102458\\
-0.209971545214401	1.01422369207744\\
-0.184723666457739	0.952946873212263\\
-0.160745294221434	0.891107439418287\\
-0.138132980947465	0.828536869081093\\
-0.116977778440511	0.765109518307639\\
-0.097364871234471	0.700732363089411\\
-0.0881617092850835	0.66816548188191\\
-0.07937323358441	0.635337869751151\\
-0.071008293382512	0.60224568355256\\
-0.0630753114651075	0.568887486476179\\
-0.0555822756725384	0.535264278167384\\
-0.0485367308566893	0.501379100863919\\
-0.0419457712839657	0.467236172849073\\
-0.0358160334919635	0.432839661543112\\
-0.0301536896070456	0.398192374272517\\
-0.0249644411295273	0.363294761018002\\
-0.0202535131927517	0.328144641152743\\
-0.016025649301822	0.292737955972838\\
-0.0122851065572966	0.257070609736365\\
-0.00903565136864692	0.221141119129082\\
-0.00628055566180308	0.18495337547858\\
-0.00402259358460277	0.148518311327443\\
-0.00226403871345759	0.111852367340274\\
-0.00100666176405806	0.0749678486194227\\
-0.000251728808407847	0.0378352044049493\\
0	0\\
};
\addplot [color=mycolor3, line width=1.0pt, forget plot]
  table[row sep=crcr]{%
0	6.66133814775094e-16\\
0.000251728808407403	0.0114462822272738\\
0.00100666176405784	0.0227163392340002\\
0.00226403871345782	0.0338968270754123\\
0.00402259358460233	0.0450482520320548\\
0.00628055566180286	0.0562109219346802\\
0.0090356513686467	0.0674056853086786\\
0.0122851065572964	0.0786371240672035\\
0.0202535131927513	0.101178801401454\\
0.0301536896070458	0.123751810373944\\
0.0419457712839653	0.146315581732105\\
0.0555822756725384	0.16889418347249\\
0.0710082933825116	0.191519937083495\\
0.0881617092850837	0.214188050935879\\
0.106973452628606	0.236866448271871\\
0.127367775162123	0.25953109693386\\
0.149262556146839	0.282178030554589\\
0.172569633027358	0.304804206426563\\
0.197195156431167	0.327388367396004\\
0.223039968066945	0.349895337863251\\
0.25	0.372292843236307\\
0.277966693697113	0.394556642951508\\
0.306827437153436	0.416659670298985\\
0.351539812335863	0.449423977589333\\
0.397596665967405	0.481580563178044\\
0.444580900049494	0.512988854322356\\
0.492067018082596	0.543496748567269\\
0.539624978428394	0.572927677922913\\
0.586824088833465	0.601120745238888\\
0.633236906845017	0.627923799906757\\
0.678443110795936	0.653173828776544\\
0.736135537386341	0.684180571197441\\
0.790028454785599	0.711887793930984\\
0.839254705778566	0.736036231341884\\
0.883022221559489	0.756457192033683\\
0.92062676641559	0.773023244399668\\
0.951463269143311	0.785690019265252\\
0.975035558870473	0.794472171618453\\
0.990964348631353	0.799391910185213\\
0.997735961286542	0.800518987664597\\
0.998993338235942	0.800298214740606\\
0.999748271191593	0.799521329159869\\
1	6.66133814775094e-16\\
};
\addplot [color=mycolor3, line width=1.0pt, forget plot]
  table[row sep=crcr]{%
-0.920626766415591	5.06023083290683\\
-0.911838290714917	4.77591809212297\\
-0.902635128765529	4.51968428072913\\
-0.893026547371393	4.2875874432663\\
-0.883022221559489	4.07640170223281\\
-0.872632224837878	3.88346101826437\\
-0.861867019052535	3.70654199936351\\
-0.850737443853161	3.54377483471291\\
-0.839254705778566	3.3935747869147\\
-0.827430366972642	3.25458891471914\\
-0.815276333542261	3.125654218392\\
-0.802804843568834	3.00576445065582\\
-0.790028454785599	2.89404357400274\\
-0.776960031933055	2.78972436981962\\
-0.763612733805251	2.69213108105837\\
-0.75	2.60066524167732\\
-0.736135537386342	2.51479404325976\\
-0.722033306302887	2.43404073419626\\
-0.707707506500943	2.3579766557543\\
-0.693172562846565	2.28621460357488\\
-0.678443110795936	2.21840326982276\\
-0.663533981658711	2.15422257427121\\
-0.648460187664138	2.09337973380046\\
-0.633236906845018	2.0356059501557\\
-0.617879467754713	1.98065361678121\\
-0.602403334032595	1.92829395955039\\
-0.586824088833465	1.87831503646322\\
-0.571157419136643	1.83052003104397\\
-0.555419099950505	1.78472578528874\\
-0.539624978428394	1.74076153077701\\
-0.523790957911872	1.69846778932614\\
-0.507932981917404	1.65769542466792\\
-0.476209042088129	1.58016524953133\\
-0.444580900049495	1.50715688049164\\
-0.413175911166535	1.43778078623033\\
-0.366763093154983	1.33885004032004\\
-0.277966693697113	1.15183220492881\\
-0.25	1.09091859646611\\
-0.223039968066945	1.03009532531572\\
-0.197195156431166	0.969077592252857\\
-0.172569633027358	0.907631736035753\\
-0.149262556146839	0.845574033579859\\
-0.127367775162122	0.782768244899189\\
-0.106973452628607	0.719117551318686\\
-0.0881617092850835	0.654554927546302\\
-0.07937323358441	0.621918225076274\\
-0.071008293382512	0.589041621048358\\
-0.0630753114651075	0.555926083409368\\
-0.0555822756725384	0.522575262083931\\
-0.0485367308566893	0.488995133951951\\
-0.0419457712839657	0.45519319193681\\
-0.0358160334919635	0.421177299593595\\
-0.0301536896070456	0.386954499690464\\
-0.0249644411295273	0.352530183183139\\
-0.0202535131927517	0.317908046472724\\
-0.016025649301822	0.283091164253218\\
-0.0122851065572966	0.24808428876737\\
-0.00903565136864692	0.212897194048013\\
-0.00628055566180308	0.177548582965822\\
-0.00402259358460277	0.142069844279849\\
-0.00226403871345759	0.106507849415165\\
-0.00100666176405806	0.0709259914716451\\
-0.000251728808407847	0.0354021871634975\\
0	0\\
};
\end{axis}

\begin{axis}[%
width=0.307\fwidth,
height=0.307\fheight,
at={(0.577\fwidth,0.601\fheight)},
scale only axis,
xmin=-1,
xmax=1,
xlabel style={font=\color{white!15!black}},
xlabel={$x$},
ymode=log,
ymin=0.005,
ymax=120,
yminorticks=true,
ylabel style={font=\color{white!15!black}},
ylabel={$1/\Phi$},
axis background/.style={fill=white},
xminorgrids,
yminorgrids,
axis on top=false
]
\addplot [color=mycolor1, line width=2.0pt, forget plot]
  table[row sep=crcr]{%
0.000251728808407403	0.0100000000000002\\
1	0.0100000000000002\\
};
\addplot [color=red!50!mycolor1, line width=2.0pt, forget plot]
  table[row sep=crcr]{%
0.000251728808407403	0.1\\
1	0.1\\
};
\addplot [color=red, line width=2.0pt, forget plot]
  table[row sep=crcr]{%
0.000251728808407403	1\\
1	1\\
};
\addplot [color=mycolor2, line width=2.0pt, forget plot]
  table[row sep=crcr]{%
0.000251728808407403	10\\
1	10\\
};
\addplot [color=mycolor3, line width=2.0pt, forget plot]
  table[row sep=crcr]{%
0.000251728808407403	100\\
1	100\\
};
\end{axis}
\end{tikzpicture}%

%% file: images/largePor2.tex
%
%
\definecolor{mycolor1}{rgb}{0.33333,0.00000,0.00000}%
\definecolor{mycolor2}{rgb}{1.00000,0.33333,0.00000}%
\definecolor{mycolor3}{rgb}{1.00000,0.66667,0.00000}%
\begin{tikzpicture}[%
trim axis left, trim axis right
]

\begin{axis}[%
width=0.951\fwidth,
height=\fheight,
at={(0\fwidth,0\fheight)},
scale only axis,
xmin=-1,
xmax=1,
xlabel style={font=\color{white!15!black}},
xlabel={$x$},
ymin=0,
ymax=0.5,
ylabel style={font=\color{white!15!black}},
ylabel={$|\Delta p|$},
axis background/.style={fill=white},
axis x line*=bottom,
axis y line*=left,
xmajorgrids,
ymajorgrids,
axis on top=false
]
\addplot [color=black, line width=2.0pt, forget plot]
  table[row sep=crcr]{%
-0.805270257531059	0.515081522400168\\
-0.786053094742788	0.488582549509343\\
-0.766044443118978	0.464403633132487\\
-0.745264449675755	0.44224243459437\\
-0.72373403810507	0.421846993187448\\
-0.701474887706321	0.403005647331915\\
-0.678509411557132	0.385539282859673\\
-0.654860733945285	0.369295304568703\\
-0.630552667084523	0.354142898886811\\
-0.605609687137667	0.339969275287237\\
-0.580056909571198	0.326676658796776\\
-0.55392006386611	0.314179865411875\\
-0.527225467610502	0.302404333546688\\
-0.5	0.291284513978898\\
-0.472271074772683	0.280762543162648\\
-0.444066612605774	0.270787142898275\\
-0.415415013001886	0.261312703360968\\
-0.386345125693129	0.252298515856753\\
-0.356886221591872	0.243708127348245\\
-0.327067963317422	0.235508793430812\\
-0.296920375328275	0.227671012035846\\
-0.266473813690035	0.220168125744791\\
-0.235758935509427	0.212975983403348\\
-0.204806668065191	0.206072650951688\\
-0.17364817766693	0.19943816029394\\
-0.142314838273285	0.193054287662221\\
-0.110838199901011	0.18690435880296\\
-0.0792499568567885	0.180973082007874\\
-0.0475819158237423	0.175246407446562\\
-0.015865963834808	0.169711405492087\\
0.015865963834808	0.164356155306594\\
0.0475819158237423	0.15916964082212\\
0.110838199901011	0.14926274042821\\
0.17364817766693	0.139917536459782\\
0.235758935509427	0.131070723874136\\
0.296920375328275	0.122667142634974\\
0.386345125693129	0.110788708694916\\
0.5	0.0960968407165522\\
0.630552667084523	0.0792044318943456\\
0.701474887706321	0.0696935993712245\\
0.745264449675755	0.0635666669442486\\
0.786053094742788	0.0575880190609954\\
0.823676581429833	0.0517400361465109\\
0.857983413234977	0.0460065441927702\\
0.888835448654923	0.0403725602755939\\
0.902926538286621	0.0375884786250249\\
0.91610845743207	0.0348241099881377\\
0.928367933016073	0.0320778473303531\\
0.939692620785908	0.0293481245306844\\
0.950071117740945	0.026633416663108\\
0.959492973614497	0.0239322418194496\\
0.967948701396356	0.0212431550034423\\
0.975429786885407	0.0185647361619655\\
0.981928697262707	0.015895586215629\\
0.987438888676394	0.0132343385327717\\
0.991954812830795	0.0105796706100629\\
0.995471922573085	0.0079302869897232\\
0.997986676471884	0.00528486131148453\\
0.999496542383185	0.00264196131907735\\
};
\addplot [color=black, line width=3.0pt, forget plot]
  table[row sep=crcr]{%
-0.883022221559489	0.503754501165118\\
-0.872632224837877	0.479851722145463\\
-0.861867019052535	0.457854340203842\\
-0.850737443853161	0.437533501823347\\
-0.839254705778566	0.4186955448668\\
-0.827430366972642	0.401175481143289\\
-0.815276333542261	0.384831875359779\\
-0.802804843568833	0.369542783204583\\
-0.790028454785599	0.355202501631591\\
-0.776960031933055	0.341718948335285\\
-0.763612733805251	0.329011532952055\\
-0.75	0.317009415458087\\
-0.736135537386341	0.305650071683636\\
-0.722033306302887	0.294878104431399\\
-0.707707506500943	0.284644252722208\\
-0.693172562846564	0.274904561970085\\
-0.678443110795936	0.265619685334841\\
-0.663533981658711	0.256754292310088\\
-0.648460187664137	0.248276565650848\\
-0.633236906845017	0.240157772011217\\
-0.617879467754714	0.232371894575529\\
-0.602403334032595	0.224895317467094\\
-0.586824088833465	0.217706552794672\\
-0.571157419136643	0.210786003004253\\
-0.555419099950506	0.204115753657585\\
-0.539624978428394	0.197679393450011\\
-0.523790957911871	0.191461858162531\\
-0.507932981917404	0.185449294172925\\
-0.492067018082596	0.179628937163374\\
-0.476209042088129	0.173989003552399\\
-0.444580900049495	0.163207611347786\\
-0.413175911166535	0.15302710376623\\
-0.382120532245286	0.143380105445726\\
-0.351539812335863	0.134208049767809\\
-0.306827437153436	0.121230175139247\\
-0.25	0.105159794673289\\
-0.184723666457739	0.0866642762827922\\
-0.149262556146839	0.0762466769440804\\
-0.127367775162123	0.0695354624804377\\
-0.106973452628606	0.0629872627096939\\
-0.0881617092850836	0.0565832825320822\\
-0.0710082933825116	0.0503061330043557\\
-0.0555822756725383	0.0441396123582957\\
-0.0485367308566894	0.0410930542187443\\
-0.0419457712839653	0.0380685274680441\\
-0.0358160334919637	0.0350642662499794\\
-0.0301536896070459	0.0320785477261237\\
-0.0249644411295273	0.029109689655396\\
-0.0202535131927513	0.0261560481581505\\
-0.0160256493018219	0.0232160125985816\\
-0.0122851065572965	0.0202879988289627\\
-0.0090356513686467	0.0173704460486732\\
-0.00628055566180286	0.0144618196864077\\
-0.00402259358460233	0.011560613949841\\
-0.00226403871345771	0.00866534255848783\\
-0.00100666176405784	0.00577451303650212\\
-0.000251728808407514	0.00288659479331732\\
};
\addplot [color=mycolor1, line width=1.0pt, forget plot]
  table[row sep=crcr]{%
0	0\\
0.000251728808407403	0.150464475910969\\
0.00100666176405784	0.15302673296565\\
0.00226403871345782	0.154419493125978\\
0.00402259358460233	0.155280497104051\\
0.00628055566180286	0.15581204130376\\
0.0090356513686467	0.156105291336999\\
0.0122851065572964	0.156209632906844\\
0.0202535131927513	0.155962084031473\\
0.0301536896070458	0.155213489041215\\
0.0419457712839653	0.154043362918918\\
0.0630753114651075	0.151610915993697\\
0.0973648712344708	0.147302027128195\\
0.336466018341289	0.116512790977169\\
0.460375021571606	0.101123612904945\\
0.602403334032595	0.0834390784008545\\
0.663533981658711	0.0755800494982943\\
0.722033306302887	0.0677540929670037\\
0.763612733805251	0.0619124030031049\\
0.802804843568833	0.0560959703838642\\
0.839254705778566	0.0503038445076871\\
0.861867019052535	0.0464546357559774\\
0.883022221559489	0.042613790513657\\
0.902635128765529	0.0387797665379053\\
0.92062676641559	0.0349507445928419\\
0.936924688534893	0.0311245661055177\\
0.951463269143311	0.0272985991262369\\
0.964183966508036	0.0234696436139381\\
0.969846310392954	0.0215528885236489\\
0.975035558870473	0.0196339218224941\\
0.979746486807249	0.0177121553008057\\
0.983974350698178	0.0157869090287754\\
0.987714893442704	0.0138573538462132\\
0.990964348631353	0.0119224215005713\\
0.993719444338197	0.00998066644668061\\
0.995977406415398	0.00803004088973958\\
0.997735961286542	0.00606747537159369\\
0.998993338235942	0.00408790671045667\\
0.999748271191593	0.00208109718450777\\
1	0\\
};
\addplot [color=mycolor1, line width=1.0pt, forget plot]
  table[row sep=crcr]{%
-0.802804843568833	0.507268378571342\\
-0.790028454785599	0.48966074304182\\
-0.776960031933055	0.473176282176509\\
-0.763612733805251	0.457710699429437\\
-0.75	0.443172376838807\\
-0.736135537386341	0.429480511254502\\
-0.722033306302887	0.416563570125418\\
-0.707707506500943	0.404358004711136\\
-0.693172562846564	0.392807172016407\\
-0.678443110795936	0.381860426999003\\
-0.663533981658711	0.371472354487749\\
-0.648460187664137	0.361602116359589\\
-0.633236906845017	0.352212894295106\\
-0.617879467754714	0.343271412180372\\
-0.602403334032595	0.334747525186899\\
-0.586824088833465	0.326613864918917\\
-0.571157419136643	0.318845531902794\\
-0.555419099950506	0.311419828209739\\
-0.539624978428394	0.304316024228537\\
-0.523790957911871	0.297515154600738\\
-0.507932981917404	0.290999839143515\\
-0.492067018082596	0.284754125252075\\
-0.476209042088129	0.278763348822744\\
-0.460375021571606	0.273014011192035\\
-0.444580900049494	0.267493669964083\\
-0.428842580863357	0.26219084191253\\
-0.413175911166535	0.257094916404914\\
-0.397596665967405	0.252196078016708\\
-0.366763093154983	0.242953967916075\\
-0.336466018341289	0.234399426639505\\
-0.306827437153436	0.226476315812039\\
-0.277966693697113	0.219136033195182\\
-0.25	0.21233634130118\\
-0.223039968066945	0.206040379590142\\
-0.184723666457739	0.197471254357537\\
-0.149262556146839	0.189869743983279\\
-0.106973452628606	0.181102783861583\\
-0.0485367308566894	0.169055138061404\\
-0.0301536896070458	0.16497288086363\\
-0.0202535131927513	0.162524292918583\\
-0.0122851065572964	0.160214057172078\\
-0.0090356513686467	0.159072249621816\\
-0.00628055566180286	0.157902988595022\\
-0.00402259358460233	0.156656953353458\\
-0.00226403871345771	0.155242697287523\\
-0.00100666176405784	0.153457125407174\\
-0.000251728808407403	0.150660667994177\\
0	0\\
};
\addplot [color=red!50!mycolor1, line width=1.0pt, forget plot]
  table[row sep=crcr]{%
0	0\\
0.000251728808407403	0.0580502627104706\\
0.00100666176405784	0.068610717767734\\
0.00226403871345782	0.0756003824926321\\
0.00402259358460233	0.0809219681867548\\
0.00628055566180286	0.0852353006551967\\
0.0090356513686467	0.0888543698410118\\
0.0122851065572964	0.0919552491912645\\
0.0160256493018218	0.0946473989753172\\
0.0202535131927513	0.0970039265800273\\
0.0249644411295273	0.099076340077785\\
0.0301536896070458	0.100902490799461\\
0.0358160334919637	0.102511179957255\\
0.0419457712839653	0.103924987279006\\
0.0485367308566895	0.105162084715877\\
0.0555822756725384	0.106237437564553\\
0.0630753114651075	0.107163619818839\\
0.0710082933825116	0.107951379788978\\
0.0793732335844095	0.108610041878084\\
0.0973648712344708	0.109571941535525\\
0.116977778440511	0.110104979142198\\
0.138132980947465	0.110255019821585\\
0.160745294221434	0.11006128123003\\
0.184723666457739	0.109558122216678\\
0.209971545214401	0.108775987816542\\
0.236387266194749	0.107742031328929\\
0.277966693697113	0.105771737161871\\
0.321556889204064	0.103362357085792\\
0.366763093154983	0.100576669014185\\
0.413175911166535	0.0974679079674727\\
0.460375021571606	0.0940811537541901\\
0.523790957911871	0.089197166651682\\
0.571157419136643	0.0852961597997153\\
0.617879467754714	0.0812166423808576\\
0.663533981658711	0.0769756074774111\\
0.707707506500943	0.0725845980788873\\
0.75	0.0680503411196409\\
0.790028454785599	0.0633749097252847\\
0.815276333542261	0.0601781873821301\\
0.839254705778566	0.0569151524570852\\
0.861867019052535	0.0535824203769617\\
0.883022221559489	0.0501752758014522\\
0.902635128765529	0.0466873052906744\\
0.92062676641559	0.0431099866754812\\
0.936924688534893	0.0394321648822538\\
0.951463269143311	0.0356391124142923\\
0.958054228716035	0.0336933954490253\\
0.964183966508036	0.0317108165680282\\
0.969846310392954	0.029687575935567\\
0.975035558870473	0.0276190732505519\\
0.979746486807249	0.0254996549628819\\
0.983974350698178	0.0233222334754064\\
0.987714893442704	0.021077688949523\\
0.990964348631353	0.0187538869207149\\
0.993719444338197	0.0163339700544789\\
0.995977406415398	0.0137931365060677\\
0.997735961286542	0.0110917920959595\\
0.998993338235942	0.00815802222630535\\
0.999748271191593	0.00482494072188033\\
1	0\\
};
\addplot [color=red!50!mycolor1, line width=1.0pt, forget plot]
  table[row sep=crcr]{%
-0.827430366972642	0.510327331238305\\
-0.815276333542261	0.491009688143185\\
-0.802804843568833	0.472990934422544\\
-0.790028454785599	0.456141650330005\\
-0.776960031933055	0.440349123759947\\
-0.763612733805251	0.425514738849828\\
-0.75	0.411551839321717\\
-0.736135537386341	0.398383968824148\\
-0.722033306302887	0.385943412874842\\
-0.707707506500943	0.374169983756004\\
-0.693172562846564	0.363010002388324\\
-0.678443110795936	0.352415440882061\\
-0.663533981658711	0.342343196905572\\
-0.648460187664137	0.332754476782117\\
-0.633236906845017	0.323614268732416\\
-0.617879467754714	0.314890891223109\\
-0.602403334032595	0.306555604182016\\
-0.586824088833465	0.298582273066726\\
-0.571157419136643	0.290947077550076\\
-0.555419099950506	0.283628258012031\\
-0.539624978428394	0.276605894178414\\
-0.523790957911871	0.269861711181862\\
-0.507932981917404	0.263378909085545\\
-0.492067018082596	0.257142012540297\\
-0.476209042088129	0.251136737767312\\
-0.460375021571606	0.245349874490657\\
-0.444580900049494	0.239769180801268\\
-0.428842580863357	0.234383289228469\\
-0.397596665967405	0.224154317962469\\
-0.366763093154983	0.2145865121641\\
-0.336466018341289	0.205612810013017\\
-0.306827437153436	0.197173787980884\\
-0.277966693697113	0.189216173956564\\
-0.236387266194749	0.178077506293598\\
-0.138132980947465	0.152035474539948\\
-0.116977778440511	0.146133068596356\\
-0.0973648712344708	0.140384879324039\\
-0.0793732335844095	0.134737319777987\\
-0.0630753114651075	0.129126149832626\\
-0.0555822756725383	0.12630975860211\\
-0.0485367308566894	0.123470226644348\\
-0.0419457712839653	0.120593058412174\\
-0.0358160334919637	0.117660974848949\\
-0.0301536896070458	0.114652995440812\\
-0.0249644411295273	0.111543111861158\\
-0.0202535131927513	0.108298307480132\\
-0.0160256493018218	0.104875483151599\\
-0.0122851065572964	0.101216456518283\\
-0.0090356513686467	0.0972393420894159\\
-0.00628055566180286	0.0928225510570609\\
-0.00402259358460233	0.0877720209141923\\
-0.00226403871345771	0.0817441301459724\\
-0.00100666176405784	0.0740212678370945\\
-0.000251728808407403	0.0625446954109524\\
0	0\\
};
\addplot [color=red, line width=1.0pt, forget plot]
  table[row sep=crcr]{%
0	0\\
0.000251728808407403	0.00136392667769925\\
0.00100666176405784	0.00244721130945624\\
0.00226403871345782	0.00344308494524026\\
0.00402259358460233	0.0043839569931361\\
0.00628055566180286	0.00528378017104569\\
0.0090356513686467	0.00615014808845027\\
0.0122851065572964	0.00698774608538066\\
0.0202535131927513	0.00858814906127048\\
0.0301536896070458	0.0101002728476045\\
0.0419457712839653	0.0115317198991229\\
0.0555822756725384	0.0128863249012374\\
0.0710082933825116	0.0141663452245224\\
0.0881617092850837	0.0153735611841874\\
0.116977778440511	0.017051274302271\\
0.149262556146839	0.0185732192531864\\
0.184723666457739	0.0199440402228983\\
0.223039968066945	0.0211695588939196\\
0.277966693697113	0.0225887401467657\\
0.336466018341289	0.0237771645176983\\
0.397596665967405	0.0247519924012329\\
0.476209042088129	0.02569507746746\\
0.555419099950506	0.0263595066951361\\
0.648460187664137	0.0268227339284803\\
0.736135537386341	0.0269483042880159\\
0.815276333542261	0.0267457033838672\\
0.872632224837877	0.026313548457237\\
0.911838290714917	0.0257708177859366\\
0.944417724327462	0.0250107244398075\\
0.964183966508036	0.0242511962010301\\
0.975035558870473	0.0236204765048318\\
0.983974350698178	0.0228501650491046\\
0.990964348631353	0.0218740066605119\\
0.993719444338197	0.0212700388112834\\
0.995977406415398	0.0205490569971936\\
0.997735961286542	0.01965116002773\\
0.998993338235942	0.0184475895648299\\
0.999748271191593	0.016553943895103\\
1	0\\
};
\addplot [color=red, line width=1.0pt, forget plot]
  table[row sep=crcr]{%
-0.872632224837877	0.504672890540153\\
-0.861867019052535	0.481782311941854\\
-0.850737443853161	0.460645996126868\\
-0.839254705778566	0.441061364313975\\
-0.827430366972642	0.422855832146377\\
-0.815276333542261	0.405881452384672\\
-0.802804843568833	0.390010665967422\\
-0.790028454785599	0.37513290280395\\
-0.776960031933055	0.361151840414641\\
-0.763612733805251	0.347983176514487\\
-0.75	0.335552806529665\\
-0.736135537386341	0.323795322689978\\
-0.722033306302887	0.312652770390576\\
-0.707707506500943	0.302073611799587\\
-0.693172562846564	0.292011857495404\\
-0.678443110795936	0.282426335166858\\
-0.663533981658711	0.273280070759657\\
-0.648460187664137	0.264539762378832\\
-0.633236906845017	0.256175331105036\\
-0.617879467754714	0.248159535906229\\
-0.602403334032595	0.240467642214165\\
-0.586824088833465	0.233077135630145\\
-0.571157419136643	0.225967473736461\\
-0.555419099950506	0.219119870204373\\
-0.539624978428394	0.212517106372469\\
-0.523790957911871	0.206143366271362\\
-0.507932981917404	0.199984091729415\\
-0.492067018082596	0.194025854736934\\
-0.476209042088129	0.188256244693177\\
-0.460375021571606	0.182663768527331\\
-0.428842580863357	0.171968310619388\\
-0.397596665967405	0.161862748662298\\
-0.366763093154983	0.152280258704218\\
-0.336466018341289	0.143162186960389\\
-0.292292493499057	0.130243884753931\\
-0.172569633027358	0.0956398990117382\\
-0.149262556146839	0.0885708387696932\\
-0.127367775162123	0.0816607710498186\\
-0.106973452628606	0.0748808670119571\\
-0.0881617092850836	0.0682022856805239\\
-0.0710082933825116	0.0615954117473764\\
-0.0630753114651075	0.0583091660501637\\
-0.0555822756725383	0.0550288825278477\\
-0.0485367308566894	0.0517501435606559\\
-0.0419457712839653	0.0484682132438061\\
-0.0358160334919637	0.0451779439539889\\
-0.0301536896070458	0.0418736524869929\\
-0.0249644411295273	0.0385489516511917\\
-0.0202535131927513	0.0351965142639156\\
-0.0160256493018218	0.0318077301426091\\
-0.0122851065572964	0.0283721852483564\\
-0.0090356513686467	0.0248768276629454\\
-0.00628055566180286	0.0213045410669584\\
-0.00402259358460233	0.017631486491863\\
-0.00226403871345771	0.0138215245677122\\
-0.00100666176405784	0.00981219975297687\\
-0.000251728808407403	0.0054660968663961\\
0	0\\
};
\addplot [color=mycolor2, line width=1.0pt, forget plot]
  table[row sep=crcr]{%
0	0\\
0.000251728808407403	0.000200914759776127\\
0.00226403871345782	0.000591891319563054\\
0.00628055566180286	0.000976112964592879\\
0.0160256493018218	0.00154002938743902\\
0.0358160334919637	0.00226511887564729\\
0.0630753114651075	0.00295314182554307\\
0.106973452628606	0.0037513083436449\\
0.172569633027358	0.00460855033818874\\
0.263864462613659	0.00546438511717251\\
0.382120532245287	0.00625933943220547\\
0.539624978428394	0.00700904167741645\\
0.736135537386341	0.00765164489651493\\
0.944417724327462	0.00806521596438325\\
0.995977406415398	0.00799881941316083\\
0.999748271191593	0.00783063074968315\\
1	0\\
};
\addplot [color=mycolor2, line width=1.0pt, forget plot]
  table[row sep=crcr]{%
-0.883022221559489	0.507531004960461\\
-0.872632224837877	0.483473007756267\\
-0.861867019052535	0.461333773806907\\
-0.850737443853161	0.440882850704251\\
-0.839254705778566	0.421925223319501\\
-0.827430366972642	0.404294750454823\\
-0.815276333542261	0.387849007935146\\
-0.802804843568833	0.372465198633311\\
-0.790028454785599	0.358036880456588\\
-0.776960031933055	0.344471327582308\\
-0.763612733805251	0.331687386422231\\
-0.75	0.319613721388591\\
-0.736135537386341	0.308187370229656\\
-0.722033306302887	0.297352547038926\\
-0.707707506500943	0.287059644787064\\
-0.693172562846564	0.277264399624245\\
-0.678443110795936	0.267927187138649\\
-0.663533981658711	0.259012426869273\\
-0.648460187664137	0.250488076115816\\
-0.633236906845017	0.242325197796847\\
-0.617879467754714	0.234497590022239\\
-0.602403334032595	0.226981467347172\\
-0.586824088833465	0.219755185499312\\
-0.571157419136643	0.21279900282393\\
-0.555419099950506	0.206094872857164\\
-0.539624978428394	0.199626263380472\\
-0.523790957911871	0.193377998079744\\
-0.507932981917404	0.18733611756739\\
-0.492067018082596	0.181487757051018\\
-0.476209042088129	0.175821038366525\\
-0.444580900049494	0.164989384614972\\
-0.413175911166535	0.154762492639674\\
-0.382120532245286	0.145072379510035\\
-0.351539812335863	0.13585991207559\\
-0.306827437153436	0.122825297395819\\
-0.25	0.106683674275144\\
-0.184723666457739	0.0880998295894372\\
-0.149262556146839	0.0776258074404911\\
-0.127367775162123	0.0708739910478952\\
-0.106973452628606	0.064281753354131\\
-0.0881617092850836	0.0578292615834481\\
-0.0710082933825116	0.0514978864542727\\
-0.0555822756725383	0.0452699477586238\\
-0.0485367308566894	0.0421894372466127\\
-0.0419457712839653	0.0391284495164589\\
-0.0358160334919637	0.0360849146698883\\
-0.0301536896070458	0.0330567601113937\\
-0.0249644411295273	0.0300418934521914\\
-0.0202535131927513	0.027038181271005\\
-0.0160256493018218	0.0240434213140687\\
-0.0122851065572964	0.0210553034777154\\
-0.0090356513686467	0.0180713504350579\\
-0.00628055566180286	0.0150888191326272\\
-0.00402259358460233	0.0121045211710764\\
-0.00226403871345771	0.00911445457004667\\
-0.00100666176405784	0.00611290746691473\\
-0.000251728808407403	0.00308949383892354\\
0	0\\
};
\addplot [color=mycolor3, line width=1.0pt, forget plot]
  table[row sep=crcr]{%
0	0\\
0.000251728808407403	0.000175511233273529\\
0.00226403871345782	0.000524478822045982\\
0.00628055566180286	0.000870818239696591\\
0.0160256493018218	0.00138324357785335\\
0.0358160334919637	0.00204680279764613\\
0.0710082933825116	0.00283253485193891\\
0.116977778440511	0.00355829489519044\\
0.184723666457739	0.00433951819197764\\
0.277966693697113	0.00512246034314656\\
0.413175911166535	0.00593463159479102\\
0.586824088833465	0.00666815952928257\\
0.827430366972643	0.00736610534830717\\
0.999748271191593	0.00769179919012419\\
1	0\\
};
\addplot [color=mycolor3, line width=1.0pt, forget plot]
  table[row sep=crcr]{%
-0.883022221559489	0.50531140357185\\
-0.872632224837877	0.481337330251329\\
-0.861867019052535	0.459274458455872\\
-0.850737443853161	0.43889324005219\\
-0.839254705778566	0.419999425154496\\
-0.827430366972642	0.402427524566384\\
-0.815276333542261	0.386035673136392\\
-0.802804843568833	0.370701555852409\\
-0.790028454785599	0.356319148681451\\
-0.776960031933055	0.342796090170395\\
-0.763612733805251	0.330051545834524\\
-0.75	0.318014460821717\\
-0.736135537386341	0.30662212093678\\
-0.722033306302887	0.295818960372949\\
-0.707707506500943	0.285555568187029\\
-0.693172562846564	0.275787855911553\\
-0.678443110795936	0.266476356604333\\
-0.663533981658711	0.257585631725202\\
-0.648460187664137	0.249083766957203\\
-0.633236906845017	0.240941941784739\\
-0.617879467754714	0.233134060545455\\
-0.602403334032595	0.225636434964906\\
-0.586824088833465	0.21842750999943\\
-0.571157419136643	0.211487626258929\\
-0.555419099950506	0.204798813440819\\
-0.539624978428394	0.198344610144847\\
-0.523790957911871	0.192109906205847\\
-0.507932981917404	0.186080804314509\\
-0.492067018082596	0.18024449822068\\
-0.476209042088129	0.174589165247288\\
-0.444580900049494	0.163778486040199\\
-0.413175911166535	0.153570501569891\\
-0.382120532245286	0.143897603359941\\
-0.351539812335863	0.134701018915969\\
-0.306827437153436	0.121688511473968\\
-0.25	0.105575087627394\\
-0.184723666457739	0.0870290125449802\\
-0.149262556146839	0.0765818866704919\\
-0.127367775162123	0.0698509958241907\\
-0.106973452628606	0.0632829271675931\\
-0.0881617092850836	0.0568587174622022\\
-0.0710082933825116	0.0505607761651019\\
-0.0555822756725383	0.0443726873692459\\
-0.0485367308566894	0.0413149847946229\\
-0.0419457712839653	0.0382790307744379\\
-0.0358160334919637	0.0352630236312189\\
-0.0301536896070458	0.0322652006775261\\
-0.0249644411295273	0.0292838322275757\\
-0.0202535131927513	0.0263172159053208\\
-0.0160256493018218	0.0233636713299503\\
-0.0122851065572964	0.0204215349355049\\
-0.0090356513686467	0.0174891540698461\\
-0.00628055566180286	0.0145648783935453\\
-0.00402259358460233	0.0116470442709377\\
-0.00226403871345771	0.00873394181679044\\
-0.00100666176405784	0.00582373337625064\\
-0.000251728808407403	0.00291418339879657\\
0	0\\
};
\end{axis}

\begin{axis}[%
width=0.307\fwidth,
height=0.307\fheight,
at={(0.577\fwidth,0.601\fheight)},
scale only axis,
xmin=-1,
xmax=1,
xlabel style={font=\color{white!15!black}},
xlabel={$x$},
ymode=log,
ymin=0.005,
ymax=120,
yminorticks=true,
ylabel style={font=\color{white!15!black}},
ylabel={$1/\Phi$},
axis background/.style={fill=white},
xminorgrids,
yminorgrids,
axis on top=false
]
\addplot [color=mycolor1, line width=2.0pt, forget plot]
  table[row sep=crcr]{%
0.000251728808407403	0.0100000000000002\\
1	0.0100000000000002\\
};
\addplot [color=red!50!mycolor1, line width=2.0pt, forget plot]
  table[row sep=crcr]{%
0.000251728808407403	0.1\\
1	0.1\\
};
\addplot [color=red, line width=2.0pt, forget plot]
  table[row sep=crcr]{%
0.000251728808407403	1\\
1	1\\
};
\addplot [color=mycolor2, line width=2.0pt, forget plot]
  table[row sep=crcr]{%
0.000251728808407403	10\\
1	10\\
};
\addplot [color=mycolor3, line width=2.0pt, forget plot]
  table[row sep=crcr]{%
0.000251728808407403	100\\
1	100\\
};
\end{axis}
\end{tikzpicture}%

%% file: sections/asymp-results.tex
\section{Scaling laws for unsteady porous aerofoils}    \label{sec:scaling}
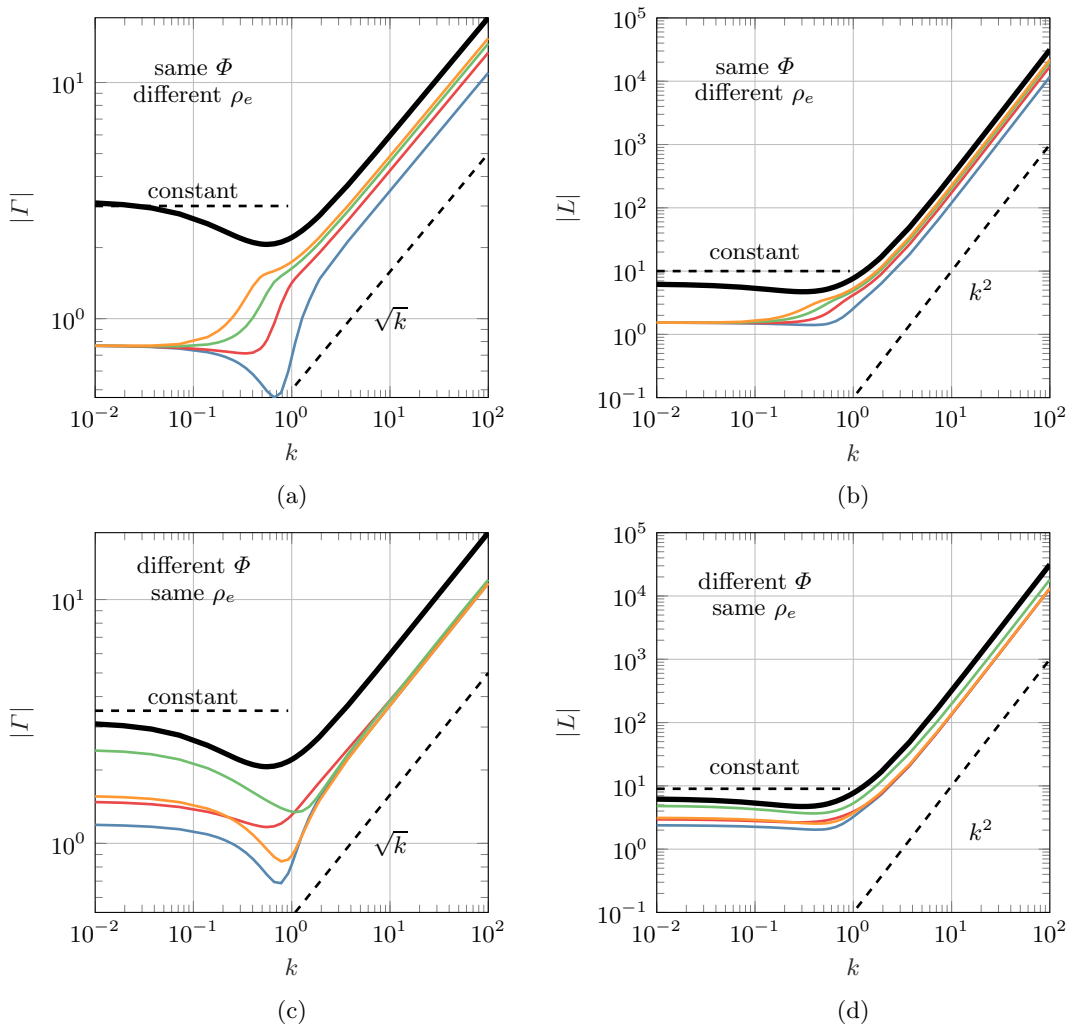
\begin{figure}
	\begin{subfigure}[t]{.45\linewidth}
		\setlength{\fheight}{5cm}
		\setlength{\fwidth}{.9\linewidth}
		\centering
		\input{images/asympCirc1}
		\caption{}
		\label{Fig:circ1}
	\end{subfigure}
	\hfill
	\begin{subfigure}[t]{.45\linewidth}
		\setlength{\fheight}{5cm}
		\setlength{\fwidth}{.9\linewidth}
		\centering
		\input{images/asympLift1}
		\caption{}
		\label{Fig:lift1}
	\end{subfigure}

	\begin{subfigure}[t]{.45\linewidth}
		\setlength{\fheight}{5cm}
		\setlength{\fwidth}{.9\linewidth}
		\centering
		\input{images/asympCirc2}
		\caption{}
		\label{Fig:circ2}
	\end{subfigure}
	\hfill
	\begin{subfigure}[t]{.45\linewidth}
		\setlength{\fheight}{5cm}
		\setlength{\fwidth}{.9\linewidth}
		\centering
		\input{images/asympLift2}
		\caption{}
		\label{Fig:lift2}
	\end{subfigure}
	\caption{Magnitude of the bound circulation $\Gamma$ and lift $L$ as a function of reduced frequency $k$ for a range of different dimensionless flow resistance distributions $\Phi(x)$ and effective densities $\rho_{\rm e}$ of the porous aerofoil. The aerofoil pitches harmonically about the leading edge.
In (a) and (b), $\Phi = (1+x)^{-1}$ and $\rho_{\rm e} = $ 1 (blue), 2 (red), 3 (green), 4 (orange).
In (c) and (d), $\rho_{\rm e} = 1.2$ and $\Phi =  2(1+x)^{-1}$ (blue), $2 \sec(\pi x/2)$ (red), 
$2/\max(0,x)$ (green) and $\exp(1/(x+1))$ (orange).
The thick black line denotes the impermeable case.
}
	\label{Fig:asympScales}
\end{figure}

The asymptotic scaling laws derived in appendix \ref{Sec:asymp} for porous aerofoils undergoing low- or high-frequency motions are now {verified} by results from the numerical scheme for pitching motions about the leading edge. Figure~\ref{Fig:asympScales} plots the magnitudes of the aerofoil circulation and the lift as a function of reduced frequency for various specifications of nondimensional flow resistance $\Phi(x)$ and effective fluid density $\rho_{\rm e}(x)$ of the porous material. In all cases examined, the lift coefficient and circulation tend to constant values in the low-frequency limit ($k \ll 1)$. However, in the high-frequency limit ($k \gg 1$), figures \ref{Fig:circ1} and \ref{Fig:circ2} indicate that the aerofoil circulation scales as $\sqrt{k}$, whilst figures \ref{Fig:lift1} and \ref{Fig:lift2} show that the lift coefficient scales as $k^2$.
Moreover, the curves in figures \ref{Fig:circ1} and \ref{Fig:lift1} collapse at low frequencies,  indicating that 
the value of the  effective density {of the porous medium} is irrelevant in this regime.
Conversely, the curves in figures \ref{Fig:circ2} and \ref{Fig:lift2} collapse at high frequencies, confirming that the high-frequency
circulation is only a function of the effective density of the porous material and the {aerofoil camber line}. {These results confirm that the asymptotic scaling laws with respect to reduced frequency are identical for porous and impermeable unsteady aerofoils, where the details of the porosity distribution are reflected only in the scaling coefficients.}

%% file: images/asympCirc1.tex
%
%
\definecolor{mycolor1}{rgb}{0.34667,0.53600,0.69067}%
\definecolor{mycolor2}{rgb}{0.91529,0.28157,0.28784}%
\definecolor{mycolor3}{rgb}{0.44157,0.74902,0.43216}%
\definecolor{mycolor4}{rgb}{1.00000,0.59843,0.20000}%
\begin{tikzpicture}[%
trim axis left, trim axis right
]

\begin{axis}[%
width=0.951\fwidth,
height=\fheight,
at={(0\fwidth,0\fheight)},
scale only axis,
xmode=log,
xmin=0.01,
xmax=100,
xminorticks=true,
xlabel style={font=\color{white!15!black}},
xlabel={$k$},
ymode=log,
ymin=0.46348000684272,
ymax=18.8083227369961,
yminorticks=true,
ylabel style={font=\color{white!15!black}},
ylabel={$|\Gamma| $},
axis background/.style={fill=white},
xmajorgrids,
ymajorgrids
]
\addplot [color=black, line width=2.0pt, forget plot]
  table[row sep=crcr]{%
0.01	3.0892395564342\\
0.0193069772888325	3.03917673095194\\
0.0372759372031494	2.94430215122822\\
0.0719685673001152	2.77704081472304\\
0.138949549437314	2.52327157952198\\
0.177827941003892	2.41283918636018\\
0.268269579527972	2.23399044723178\\
0.291263265490874	2.2021016021294\\
0.343332001828199	2.14542354511194\\
0.404708995075976	2.10107340572526\\
0.477058269614393	2.07241173401628\\
0.517947467923121	2.06492660080544\\
0.562341325190349	2.06248715332603\\
0.662870316182644	2.07387696389642\\
0.781370737651809	2.10850186328503\\
0.921055317689482	2.16758268409199\\
1	2.20650770567572\\
1.0857111194022	2.25170760702308\\
1.27980221399795	2.36099193090471\\
1.50859070860018	2.49528740263838\\
1.77827941003892	2.65438327692633\\
1.93069772888325	2.74317247834635\\
3.72759372031494	3.67989508682818\\
7.19685673001152	5.06313708014435\\
13.8949549437314	7.01670690510505\\
100	18.8083227369961\\
};
\addplot [color=mycolor1, line width=1.0pt, forget plot]
  table[row sep=crcr]{%
0.01	0.76754372346729\\
0.0193069772888325	0.76466784148557\\
0.0372759372031494	0.758943175268385\\
0.0719685673001152	0.747132930055105\\
0.138949549437314	0.721232044459298\\
0.177827941003892	0.704196508558568\\
0.209617999245313	0.689177106567067\\
0.24709112279856	0.670254208759101\\
0.268269579527972	0.659013966207146\\
0.291263265490874	0.646400499865612\\
0.343332001828199	0.61649839760432\\
0.404708995075976	0.5797266669087\\
0.477058269614393	0.536667729303314\\
0.562341325190349	0.492385998881851\\
0.662870316182644	0.46348000684272\\
0.781370737651809	0.485726359734236\\
0.921055317689482	0.596700377258035\\
1	0.685999460347617\\
1.0857111194022	0.790627260933391\\
1.27980221399795	1.00767385336323\\
1.50859070860018	1.19370910021193\\
1.77827941003892	1.38158126853332\\
1.93069772888325	1.47605851954122\\
3.72759372031494	2.12262108325345\\
100	10.9952434231809\\
};
\addplot [color=mycolor2, line width=1.0pt, forget plot]
  table[row sep=crcr]{%
0.01	0.767695011405766\\
0.0193069772888325	0.765178483503929\\
0.0372759372031494	0.760640021907944\\
0.0719685673001152	0.752657934948367\\
0.138949549437314	0.738894188439315\\
0.209617999245313	0.725935953520679\\
0.291263265490874	0.714722136904017\\
0.343332001828199	0.712238921775351\\
0.404708995075976	0.718093447248868\\
0.477058269614393	0.744419829109418\\
0.517947467923121	0.771894581998883\\
0.562341325190349	0.813723447967712\\
0.662870316182644	0.952507914103588\\
0.781370737651809	1.15130956576321\\
0.921055317689482	1.33838798309645\\
1	1.41544683345657\\
1.0857111194022	1.48016793076776\\
1.27980221399795	1.59365942068423\\
1.50859070860018	1.71027932086014\\
1.93069772888325	1.90755814982631\\
3.72759372031494	2.6020138530922\\
7.19685673001152	3.59660354957269\\
26.8269579527973	6.93183524945031\\
100	13.3824820842165\\
};
\addplot [color=mycolor3, line width=1.0pt, forget plot]
  table[row sep=crcr]{%
0.01	0.767952924377731\\
0.0372759372031494	0.763809392562072\\
0.0719685673001152	0.763658626396821\\
0.138949549437314	0.777019279829152\\
0.177827941003892	0.792464803406949\\
0.209617999245313	0.809353814815547\\
0.24709112279856	0.834444922556406\\
0.268269579527972	0.851258693529788\\
0.291263265490874	0.871797497453658\\
0.343332001828199	0.927714143338293\\
0.404708995075976	1.01150609192214\\
0.477058269614393	1.13344054067766\\
0.562341325190349	1.28862527139557\\
0.662870316182644	1.42888021853709\\
0.781370737651809	1.51553807249911\\
0.921055317689482	1.59004926130794\\
1	1.63089895983668\\
1.0857111194022	1.67517541087098\\
1.27980221399795	1.77565019088592\\
1.50859070860018	1.89236613561877\\
1.77827941003892	2.02622310303126\\
1.93069772888325	2.09937962878764\\
3.72759372031494	2.84890750163421\\
7.19685673001152	3.93281040354572\\
13.8949549437314	5.45509901408112\\
100	14.6269566296665\\
};
\addplot [color=mycolor4, line width=1.0pt, forget plot]
  table[row sep=crcr]{%
0.01	0.768317469848946\\
0.0193069772888325	0.767388939116681\\
0.0372759372031494	0.768450305966975\\
0.0719685673001152	0.780105480310064\\
0.138949549437314	0.835028916683651\\
0.177827941003892	0.885315195087232\\
0.209617999245313	0.936138688607459\\
0.24709112279856	1.00686401799859\\
0.268269579527972	1.05169694608908\\
0.291263265490874	1.1038911420307\\
0.343332001828199	1.23126536214069\\
0.404708995075976	1.37627758908903\\
0.477058269614393	1.49474884564716\\
0.517947467923121	1.53200103862702\\
0.562341325190349	1.55517298042698\\
0.662870316182644	1.59148501768086\\
0.781370737651809	1.63786758102764\\
0.921055317689482	1.7013865430507\\
1	1.74006184790208\\
1.0857111194022	1.78322850726647\\
1.27980221399795	1.88421839156015\\
1.50859070860018	2.00359434925456\\
1.77827941003892	2.1420379050645\\
1.93069772888325	2.21801653985605\\
3.72759372031494	3.00247508055801\\
7.19685673001152	4.14199850065387\\
13.8949549437314	5.74420850534383\\
100	15.4012819627895\\
};
\addplot [color=black, dashed, line width=1.0pt, forget plot]
  table[row sep=crcr]{%
1.0857111194022	0.520987312561977\\
100	5\\
};
\addplot [color=black, dashed, line width=1.0pt, forget plot]
  table[row sep=crcr]{%
0.01	3\\
0.921055317689482	3\\
};
\node[align=center]
at (axis cs:0.1,10) {same $\Phi$\\different $\rho_e$};
\node[align=center]
at (axis cs:10,1) {$\sqrt{k}$};
\node[align=center]
at (axis cs:0.1,3.5) {constant};
\end{axis}
\end{tikzpicture}%

%% file: images/asympLift1.tex
%
%
\definecolor{mycolor1}{rgb}{0.34667,0.53600,0.69067}%
\definecolor{mycolor2}{rgb}{0.91529,0.28157,0.28784}%
\definecolor{mycolor3}{rgb}{0.44157,0.74902,0.43216}%
\definecolor{mycolor4}{rgb}{1.00000,0.59843,0.20000}%
\begin{tikzpicture}[%
trim axis left, trim axis right
]

\begin{axis}[%
width=0.951\fwidth,
height=\fheight,
at={(0\fwidth,0\fheight)},
scale only axis,
xmode=log,
xmin=0.01,
xmax=100,
xminorticks=true,
xlabel style={font=\color{white!15!black}},
xlabel={$k$},
ymode=log,
ymin=0.1,
ymax=100000,
yminorticks=true,
ylabel style={font=\color{white!15!black}},
ylabel={$|L| $},
axis background/.style={fill=white},
xmajorgrids,
ymajorgrids
]
\addplot [color=black, line width=2.0pt, forget plot]
  table[row sep=crcr]{%
0.01	6.17887017206095\\
0.0193069772888325	6.07980571063691\\
0.0372759372031494	5.89397942784674\\
0.0719685673001151	5.57389661023921\\
0.138949549437314	5.11948843636013\\
0.177827941003892	4.94469025677098\\
0.209617999245313	4.8428155993479\\
0.24709112279856	4.76397554709473\\
0.268269579527973	4.73700281288639\\
0.291263265490874	4.72069433073663\\
0.343332001828199	4.72823870461335\\
0.404708995075976	4.80486318072216\\
0.477058269614393	4.97206093166264\\
0.517947467923121	5.09735316210764\\
0.562341325190349	5.25516268283595\\
0.662870316182644	5.68481549098304\\
0.78137073765181	6.29983151396808\\
0.921055317689482	7.1520890630535\\
1	7.68873036104095\\
1.0857111194022	8.31392615062244\\
1.27980221399795	9.8884763748692\\
1.50859070860018	12.0235740455121\\
1.77827941003892	14.9302085015396\\
1.93069772888325	16.7619668158415\\
3.72759372031494	49.0007370228617\\
7.19685673001151	168.167439357108\\
13.8949549437314	612.015985343652\\
26.8269579527973	2266.41497196502\\
100	31421.2925389719\\
};
\addplot [color=mycolor1, line width=1.0pt, forget plot]
  table[row sep=crcr]{%
0.01	1.53523400149355\\
0.0193069772888325	1.52988118477817\\
0.0372759372031494	1.51991433536564\\
0.0719685673001151	1.50179292099757\\
0.138949549437314	1.4704052005836\\
0.209617999245313	1.44209732784333\\
0.291263265490874	1.41735619195607\\
0.343332001828199	1.40821134583062\\
0.404708995075976	1.40727777980977\\
0.477058269614393	1.42529380413223\\
0.517947467923121	1.44729932341261\\
0.562341325190349	1.48278157467377\\
0.662870316182644	1.61445111610714\\
0.78137073765181	1.86785949467379\\
0.921055317689482	2.28806197747999\\
1.27980221399795	3.62620981639702\\
1.50859070860018	4.5174025143134\\
1.93069772888325	6.39380587110162\\
3.72759372031494	18.328603661187\\
7.19685673001151	62.8884500749943\\
13.8949549437314	229.110789048554\\
26.8269579527973	848.783702900972\\
100	11769.2545066835\\
};
\addplot [color=mycolor2, line width=1.0pt, forget plot]
  table[row sep=crcr]{%
0.01	1.53560674362076\\
0.0372759372031494	1.52427380715444\\
0.0719685673001151	1.51640409680731\\
0.138949549437314	1.51876760496673\\
0.177827941003892	1.53001434629161\\
0.209617999245313	1.54475760752811\\
0.24709112279856	1.56906388717849\\
0.268269579527973	1.58643850324587\\
0.291263265490874	1.60857344829855\\
0.343332001828199	1.67291872710901\\
0.404708995075976	1.77881267986142\\
0.477058269614393	1.95438521656083\\
0.517947467923121	2.08103770833641\\
0.562341325190349	2.24167447389415\\
0.662870316182644	2.68226831996155\\
0.78137073765181	3.25660738154205\\
0.921055317689482	3.87400963786623\\
1.0857111194022	4.53798190095731\\
1.27980221399795	5.36317135142223\\
1.50859070860018	6.48424857169816\\
1.77827941003892	8.02513744262113\\
1.93069772888325	9.00169079820271\\
3.72759372031494	26.4045176056748\\
7.19685673001151	91.0882401043521\\
13.8949549437314	332.186123547378\\
51.7947467923121	4581.22714743398\\
100	17069.8501688184\\
};
\addplot [color=mycolor3, line width=1.0pt, forget plot]
  table[row sep=crcr]{%
0.01	1.53619270398316\\
0.0372759372031494	1.53157133289514\\
0.0719685673001151	1.54186518924713\\
0.138949549437314	1.60663737932544\\
0.177827941003892	1.66927372348653\\
0.209617999245313	1.73368428828328\\
0.24709112279856	1.82476808909449\\
0.268269579527973	1.88354275753148\\
0.291263265490874	1.95337527521402\\
0.343332001828199	2.13487381591576\\
0.404708995075976	2.38979599450368\\
0.477058269614393	2.73808669609076\\
0.562341325190349	3.16839819421489\\
0.662870316182644	3.60167904925813\\
0.78137073765181	4.0197847724632\\
0.921055317689482	4.54913030247204\\
1	4.88098015226884\\
1.0857111194022	5.26797699821179\\
1.27980221399795	6.24838927916403\\
1.50859070860018	7.58586376215354\\
1.77827941003892	9.41858697947838\\
1.93069772888325	10.5772762032435\\
3.72759372031494	31.1332948248132\\
7.19685673001151	107.379638699687\\
13.8949549437314	391.497791868415\\
26.8269579527973	1450.58930128906\\
100	20114.9473111898\\
};
\addplot [color=mycolor4, line width=1.0pt, forget plot]
  table[row sep=crcr]{%
0.01	1.536991857959\\
0.0193069772888325	1.53610347464838\\
0.0372759372031494	1.54179757382187\\
0.0719685673001151	1.57802343190006\\
0.138949549437314	1.73191485525956\\
0.177827941003892	1.866746759772\\
0.209617999245313	1.99956646913613\\
0.24709112279856	2.18001830142213\\
0.268269579527973	2.29226721932997\\
0.291263265490874	2.42123846831061\\
0.404708995075976	3.08311314071491\\
0.477058269614393	3.40579927032441\\
0.517947467923121	3.54069639699314\\
0.562341325190349	3.66702532941752\\
0.662870316182644	3.96690982468481\\
0.78137073765181	4.38572275398002\\
0.921055317689482	4.96782453766962\\
1	5.3353464223329\\
1.0857111194022	5.76392651675525\\
1.27980221399795	6.84952202409789\\
1.50859070860018	8.32756983557693\\
1.77827941003892	10.3495688947192\\
1.93069772888325	11.6265958771725\\
3.72759372031494	34.230767615042\\
7.19685673001151	117.989493811569\\
13.8949549437314	430.057181543665\\
26.8269579527973	1593.31242343497\\
100	22093.2647790922\\
};
\addplot [color=black, dashed, line width=1.0pt, forget plot]
  table[row sep=crcr]{%
1.0857111194022	0.117876863479359\\
100	1000\\
};
\addplot [color=black, dashed, line width=1.0pt, forget plot]
  table[row sep=crcr]{%
0.01	10\\
0.921055317689482	10\\
};
\node[align=center]
at (axis cs:0.1,10000) {same $\Phi$\\different $\rho_e$};
\node[align=center]
at (axis cs:20,5) {$k^2$};
\node[align=center]
at (axis cs:0.1,20) {constant};
\end{axis}
\end{tikzpicture}%

%% file: images/asympCirc2.tex
%
%
\definecolor{mycolor1}{rgb}{0.34667,0.53600,0.69067}%
\definecolor{mycolor2}{rgb}{0.91529,0.28157,0.28784}%
\definecolor{mycolor3}{rgb}{0.44157,0.74902,0.43216}%
\definecolor{mycolor4}{rgb}{1.00000,0.59843,0.20000}%
\begin{tikzpicture}[%
trim axis left, trim axis right
]

\begin{axis}[%
width=0.951\fwidth,
height=\fheight,
at={(0\fwidth,0\fheight)},
scale only axis,
xmode=log,
xmin=0.01,
xmax=100,
xminorticks=true,
xlabel style={font=\color{white!15!black}},
xlabel={$k$},
ymode=log,
ymin=0.520987312561977,
ymax=18.8083227369961,
yminorticks=true,
ylabel style={font=\color{white!15!black}},
ylabel={$|\Gamma| $},
axis background/.style={fill=white},
xmajorgrids,
ymajorgrids
]
\addplot [color=black, line width=2.0pt, forget plot]
  table[row sep=crcr]{%
0.01	3.0892395564342\\
0.0193069772888325	3.03917673095194\\
0.0372759372031494	2.94430215122822\\
0.0719685673001152	2.77704081472304\\
0.138949549437314	2.52327157952198\\
0.177827941003892	2.41283918636018\\
0.268269579527972	2.23399044723178\\
0.291263265490874	2.2021016021294\\
0.343332001828199	2.14542354511194\\
0.404708995075976	2.10107340572526\\
0.477058269614393	2.07241173401628\\
0.517947467923121	2.06492660080544\\
0.562341325190349	2.06248715332603\\
0.662870316182644	2.07387696389642\\
0.781370737651809	2.10850186328503\\
0.921055317689482	2.16758268409199\\
1	2.20650770567572\\
1.0857111194022	2.25170760702308\\
1.27980221399795	2.36099193090471\\
1.50859070860018	2.49528740263838\\
1.77827941003892	2.65438327692633\\
1.93069772888325	2.74317247834635\\
3.72759372031494	3.67989508682818\\
7.19685673001152	5.06313708014435\\
13.8949549437314	7.01670690510505\\
51.7947467923121	13.5362693857883\\
100	18.8083227369961\\
};
\addplot [color=mycolor1, line width=1.0pt, forget plot]
  table[row sep=crcr]{%
0.01	1.19093115480189\\
0.0193069772888325	1.1839335410746\\
0.0372759372031494	1.17018582515715\\
0.0719685673001152	1.14305373223223\\
0.138949549437314	1.08907089203545\\
0.177827941003892	1.05683936117137\\
0.209617999245313	1.03004192094421\\
0.24709112279856	0.998008063076861\\
0.268269579527972	0.979731652429996\\
0.291263265490874	0.959790432258707\\
0.343332001828199	0.914506541076547\\
0.404708995075976	0.861760352190295\\
0.477058269614393	0.802666410817187\\
0.562341325190349	0.742282394713134\\
0.662870316182644	0.694400758791516\\
0.781370737651809	0.686814656463727\\
0.921055317689482	0.753205390179439\\
1	0.818023272714789\\
1.27980221399795	1.0890752209779\\
1.50859070860018	1.27990736670743\\
1.77827941003892	1.45997907288671\\
1.93069772888325	1.54507778289314\\
3.72759372031494	2.23077546436678\\
7.19685673001152	3.11684925933765\\
26.8269579527973	6.02754708564538\\
100	11.639249961766\\
};
\addplot [color=mycolor2, line width=1.0pt, forget plot]
  table[row sep=crcr]{%
0.01	1.47709370387487\\
0.0193069772888325	1.46731146593006\\
0.0372759372031494	1.44718430243739\\
0.0719685673001152	1.41020355433504\\
0.138949549437314	1.34737170040307\\
0.177827941003892	1.31537074593003\\
0.209617999245313	1.29180368037819\\
0.268269579527972	1.25409932872785\\
0.343332001828199	1.21566342401678\\
0.404708995075976	1.19266634388134\\
0.477058269614393	1.17492318793574\\
0.517947467923121	1.16928041447781\\
0.562341325190349	1.16690590335842\\
0.662870316182644	1.17502311348059\\
0.781370737651809	1.20604761860711\\
0.921055317689482	1.26566067073288\\
1	1.30648254146171\\
1.0857111194022	1.353832966548\\
1.27980221399795	1.46399214837681\\
1.77827941003892	1.71833155812027\\
1.93069772888325	1.787060176588\\
3.72759372031494	2.43304903323795\\
7.19685673001152	3.30614559080377\\
13.8949549437314	4.50564369872844\\
26.8269579527973	6.17061895541931\\
51.7947467923121	8.49152209828962\\
100	11.728852764149\\
};
\addplot [color=mycolor3, line width=1.0pt, forget plot]
  table[row sep=crcr]{%
0.01	2.40181931944767\\
0.0193069772888325	2.37797756503495\\
0.0372759372031494	2.31999932400494\\
0.0719685673001152	2.20894671953555\\
0.138949549437314	2.03489431279436\\
0.177827941003892	1.94643716276504\\
0.209617999245313	1.88417268506488\\
0.24709112279856	1.82098054191368\\
0.291263265490874	1.75383741689728\\
0.517947467923121	1.5242013399721\\
0.562341325190349	1.49504884069607\\
0.662870316182644	1.44236866557738\\
0.781370737651809	1.39530322599857\\
0.921055317689482	1.36116488048356\\
1	1.34956581003665\\
1.0857111194022	1.34271394276285\\
1.27980221399795	1.35589638555152\\
1.50859070860018	1.41416891449859\\
1.77827941003892	1.52623423739573\\
1.93069772888325	1.59699210040776\\
3.72759372031494	2.31960168011749\\
7.19685673001152	3.26503534106678\\
13.8949549437314	4.51579165380351\\
26.8269579527973	6.29159374030029\\
51.7947467923121	8.74577786843093\\
100	12.0590445366308\\
};
\addplot [color=mycolor4, line width=1.0pt, forget plot]
  table[row sep=crcr]{%
0.01	1.55880283350591\\
0.0193069772888325	1.54659922473584\\
0.0372759372031494	1.52262424110826\\
0.0719685673001152	1.4762105629032\\
0.138949549437314	1.38877708396021\\
0.177827941003892	1.33983983572289\\
0.209617999245313	1.30084579080815\\
0.24709112279856	1.256064510972\\
0.268269579527972	1.23133368480879\\
0.291263265490874	1.20496181604232\\
0.343332001828199	1.14717373780975\\
0.404708995075976	1.08282782301012\\
0.477058269614393	1.0132773013826\\
0.562341325190349	0.942518112834708\\
0.662870316182644	0.879868005510327\\
0.781370737651809	0.843270413827984\\
0.921055317689482	0.857961226379168\\
1	0.890849437701486\\
1.0857111194022	0.940914107745692\\
1.27980221399795	1.08082819292643\\
1.50859070860018	1.24706770792207\\
1.77827941003892	1.41871623148152\\
1.93069772888325	1.50441961741708\\
3.72759372031494	2.21651517987062\\
7.19685673001152	3.11429865987505\\
13.8949549437314	4.33829781894085\\
100	11.6446605141622\\
};
\addplot [color=black, dashed, line width=1.0pt, forget plot]
  table[row sep=crcr]{%
1.0857111194022	0.520987312561977\\
100	5\\
};
\addplot [color=black, dashed, line width=1.0pt, forget plot]
  table[row sep=crcr]{%
0.01	3.5\\
0.921055317689482	3.5\\
};
\node[align=center]
at (axis cs:0.1,12) {different $\Phi$\\same $\rho_e$};
\node[align=center]
at (axis cs:10,1) {$\sqrt{k}$};
\node[align=center]
at (axis cs:0.1,4) {constant};
\end{axis}
\end{tikzpicture}%

%% file: images/asympLift2.tex
%
%
\definecolor{mycolor1}{rgb}{0.34667,0.53600,0.69067}%
\definecolor{mycolor2}{rgb}{0.91529,0.28157,0.28784}%
\definecolor{mycolor3}{rgb}{0.44157,0.74902,0.43216}%
\definecolor{mycolor4}{rgb}{1.00000,0.59843,0.20000}%
\begin{tikzpicture}[%
trim axis left, trim axis right
]

\begin{axis}[%
width=0.951\fwidth,
height=\fheight,
at={(0\fwidth,0\fheight)},
scale only axis,
xmode=log,
xmin=0.01,
xmax=100,
xminorticks=true,
xlabel style={font=\color{white!15!black}},
xlabel={$k$},
ymode=log,
ymin=0.1,
ymax=100000,
yminorticks=true,
ylabel style={font=\color{white!15!black}},
ylabel={$|L| $},
axis background/.style={fill=white},
xmajorgrids,
ymajorgrids
]
\addplot [color=black, line width=2.0pt, forget plot]
  table[row sep=crcr]{%
0.01	6.17887017206095\\
0.0193069772888325	6.07980571063691\\
0.0372759372031494	5.89397942784674\\
0.0719685673001151	5.57389661023921\\
0.138949549437314	5.11948843636013\\
0.177827941003892	4.94469025677098\\
0.209617999245313	4.8428155993479\\
0.24709112279856	4.76397554709473\\
0.268269579527973	4.73700281288639\\
0.291263265490874	4.72069433073663\\
0.343332001828199	4.72823870461335\\
0.404708995075976	4.80486318072216\\
0.477058269614393	4.97206093166264\\
0.517947467923121	5.09735316210764\\
0.562341325190349	5.25516268283595\\
0.662870316182644	5.68481549098304\\
0.78137073765181	6.29983151396808\\
0.921055317689482	7.1520890630535\\
1	7.68873036104095\\
1.0857111194022	8.31392615062244\\
1.27980221399795	9.8884763748692\\
1.50859070860018	12.0235740455121\\
1.77827941003892	14.9302085015396\\
1.93069772888325	16.7619668158415\\
3.72759372031494	49.0007370228617\\
7.19685673001151	168.167439357108\\
13.8949549437314	612.015985343652\\
26.8269579527973	2266.41497196502\\
100	31421.2925389719\\
};
\addplot [color=mycolor1, line width=1.0pt, forget plot]
  table[row sep=crcr]{%
0.01	2.38204814467826\\
0.0193069772888325	2.36855871148883\\
0.0372759372031494	2.34294207884441\\
0.0719685673001151	2.29564016678826\\
0.138949549437314	2.21345912665844\\
0.209617999245313	2.14044004087952\\
0.343332001828199	2.04820727690946\\
0.404708995075976	2.0320801217394\\
0.477058269614393	2.04100009381603\\
0.517947467923121	2.06163804605782\\
0.562341325190349	2.0982432033451\\
0.662870316182644	2.23979388629908\\
0.78137073765181	2.51119964032707\\
0.921055317689482	2.9526276721047\\
1	3.24239287023554\\
1.0857111194022	3.57627573578086\\
1.50859070860018	5.34268928342514\\
1.77827941003892	6.58545897822213\\
1.93069772888325	7.34611045034414\\
3.72759372031494	20.6714228613037\\
7.19685673001151	70.3693665261793\\
13.8949549437314	255.743859315254\\
26.8269579527973	946.73846642567\\
100	13123.2378232403\\
};
\addplot [color=mycolor2, line width=1.0pt, forget plot]
  table[row sep=crcr]{%
0.01	2.95438051938816\\
0.0193069772888325	2.93534324887934\\
0.0372759372031494	2.89706628451857\\
0.0719685673001151	2.83054449962144\\
0.209617999245313	2.67138888294672\\
0.268269579527973	2.65175390034353\\
0.291263265490874	2.65112546717345\\
0.343332001828199	2.66351850375481\\
0.404708995075976	2.70180953842517\\
0.477058269614393	2.77558677443127\\
0.517947467923121	2.82931466142297\\
0.562341325190349	2.89675151378927\\
0.662870316182644	3.07918824524276\\
0.78137073765181	3.33742511203182\\
0.921055317689482	3.6901471832559\\
1	3.90961060607679\\
1.0857111194022	4.1635245106116\\
1.27980221399795	4.80060615249001\\
1.50859070860018	5.66710863500231\\
1.77827941003892	6.85626902831092\\
1.93069772888325	7.60979285443539\\
3.72759372031494	21.023750164289\\
7.19685673001151	70.8111311694298\\
13.8949549437314	256.23191220988\\
26.8269579527973	947.237792400885\\
100	13123.6880655157\\
};
\addplot [color=mycolor3, line width=1.0pt, forget plot]
  table[row sep=crcr]{%
0.01	4.80391067834327\\
0.0193069772888325	4.75696587513427\\
0.0372759372031494	4.64374180814623\\
0.0719685673001151	4.43169107582525\\
0.138949549437314	4.12041048608563\\
0.209617999245313	3.88279794189403\\
0.268269579527973	3.76219992396832\\
0.291263265490874	3.72821379057267\\
0.343332001828199	3.67610172505805\\
0.404708995075976	3.66605809519228\\
0.477058269614393	3.70639050028871\\
0.517947467923121	3.75088494200191\\
0.562341325190349	3.81928757581214\\
0.662870316182644	4.03910317990859\\
0.78137073765181	4.39537149452723\\
0.921055317689482	4.9453745317741\\
1	5.31133008237403\\
1.0857111194022	5.75031532234938\\
1.27980221399795	6.89876024236284\\
1.50859070860018	8.47844306213978\\
1.77827941003892	10.586615666816\\
1.93069772888325	11.8778394795646\\
3.72759372031494	32.7627042329877\\
7.19685673001151	105.426148441002\\
13.8949549437314	369.447074728922\\
26.8269579527973	1336.83443510553\\
51.7947467923121	4916.06982743117\\
100	18205.98082683\\
};
\addplot [color=mycolor4, line width=1.0pt, forget plot]
  table[row sep=crcr]{%
0.01	3.11783148137429\\
0.0193069772888325	3.09403880489787\\
0.0372759372031494	3.04837055714613\\
0.0719685673001151	2.96399116572635\\
0.138949549437314	2.82035005947568\\
0.209617999245313	2.69883068134203\\
0.291263265490874	2.59802511870288\\
0.343332001828199	2.55684016446295\\
0.404708995075976	2.53263341676754\\
0.477058269614393	2.53985039415077\\
0.517947467923121	2.56177206098619\\
0.562341325190349	2.60050128998272\\
0.662870316182644	2.74552471988423\\
0.78137073765181	3.01265356066728\\
0.921055317689482	3.43834307628651\\
1	3.71869861906844\\
1.0857111194022	4.04607011241076\\
1.27980221399795	4.84552680862757\\
1.50859070860018	5.86224124586057\\
1.77827941003892	7.17370456687165\\
1.93069772888325	7.9789628946252\\
3.72759372031494	21.7908521430098\\
7.19685673001151	72.5823527592978\\
13.8949549437314	260.953180383904\\
26.8269579527973	960.392148670959\\
100	13232.1620560349\\
};
\addplot [color=black, dashed, line width=1.0pt, forget plot]
  table[row sep=crcr]{%
1.0857111194022	0.117876863479359\\
100	1000\\
};
\addplot [color=black, dashed, line width=1.0pt, forget plot]
  table[row sep=crcr]{%
0.01	9\\
0.921055317689482	9\\
};
\node[align=center]
at (axis cs:0.1,10000) {different $\Phi$\\same $\rho_e$};
\node[align=center]
at (axis cs:20,2) {$k^2$};
\node[align=center]
at (axis cs:0.1,20) {constant};
\end{axis}
\end{tikzpicture}%

%% file: sections/results-theodorsen.tex
\section{Porous extensions of standard unsteady aerodynamic functions}  \label{sec:standard}
{The generalised unsteady solution procedure developed in \S\ref{Sec:numUnsteady} permits the numerical construction of porous analogues of classical unsteady aerodynamic functions. The effects of the aerofoil porosity parameters on the Theodorsen and Sears functions are presented, and the corresponding Wagner and K\"ussner functions are then determined.}

\subsection{Theodorsen function}    \label{sec:theo}
The Theodorsen function \citep{Theodorsen1935} may be interpreted as the ratio of the wake-induced (circulatory) lift $L^C$ to the quasi-steady lift $L^Q$ \citep[cf.][p.~279]{Taha2019,A}},
\begin{align}
C(k) &= \frac{L^C}{L^{Q}}.
\label{eq:C_ratio}
\end{align}
This ratio can be expressed in closed form for an impermeable aerofoil as
\begin{align}
C(k) &= \frac{K_1(\i k)}{K_0(\i k) + K_1(\i k)} \qquad \mbox{for}\  \psi(x,k) \equiv 0,
\label{eq:C_orig}
\end{align}
where $K_0$  and $K_1$ are modified Bessel functions of the second kind. The lack of an analytic solution to SF--VIE \eqref{Eq:SIEfin} precludes the derivation of a similar expression for the permeable case in terms of standard functions. However, we may use the numerical solution presented in \S\ref{Sec:numSol} to compute the relevant circulatory and quasi-steady lift quantities to construct numerically  a porous extension to the Theodorsen function via \eqref{eq:C_ratio}. Specifically, the wake-induced lift is computed by the numerical scheme in \S\ref{Sec:nc}, whereas the quasi-steady lift may be determined analytically using the solution presented in \S\ref{Sec:qs}. {Results in this section are constructed using harmonically heaving motions and a particular family of porosity distributions, although other motions and porosity profiles can be examined using our numerical method.}
\begin{figure}
\centering
\vspace{.5cm}

{\large \uline{Porous Theodorsen function}}\\

\vspace{.5cm}
%
%
	\begin{subfigure}[t]{.45\linewidth}
	\setlength{\fheight}{3cm}
		\setlength{\fwidth}{\linewidth}
		\centering
		{\large Varying flow resistance, $\Phi$}
		
	\input{images/1theo.tex}
	\end{subfigure}
	\hfill
	\begin{subfigure}[t]{.45\linewidth}
	\setlength{\fheight}{3cm}
		\setlength{\fwidth}{\linewidth}
		\centering
		{\large Varying effective density, $\rho_e$}
		
	\input{images/2theo.tex}
	\end{subfigure}
\vspace{.2cm}

%
%
	\begin{subfigure}[t]{.45\linewidth}
	\setlength{\fheight}{2cm}
		\setlength{\fwidth}{\linewidth}
		\centering
	\input{images/1absTheo.tex}

	\label{Fig:theoA}
	\end{subfigure}
	\hfill
	\begin{subfigure}[t]{.45\linewidth}
	\setlength{\fheight}{2cm}
		\setlength{\fwidth}{\linewidth}
		\centering
	\input{images/2absTheo.tex}
	\end{subfigure}
\vspace{.2cm}

%
%
	\begin{subfigure}[t]{.45\linewidth}
	\setlength{\fheight}{2cm}
		\setlength{\fwidth}{\linewidth}
		\centering
	\input{images/1phaseTheo.tex}
	\end{subfigure}
	\hfill
	\begin{subfigure}[t]{.45\linewidth}
	\setlength{\fheight}{2cm}
		\setlength{\fwidth}{\linewidth}
		\centering
	\input{images/2phaseTheo.tex}
	\end{subfigure}
\caption{
Complex, magnitude and phase plots of the porous Theodorsen function for a range of flow resistance distributions (left) and effective densities (right). The dimensionless flow resistance is given by the reciprocal linear profile $\Phi=1/(\mu(1+x))$ so that the leading edge is impermeable in every case.
On the left, $\mu$ ranges from 0, 0.05, 0.10, 0.15, 0.2 (black to orange) and $\rho_{\rm e}=1.5$;
on the right, the effective density takes values $\infty$, 4, 3, 2, 1 (black to green) and $\mu=0.05$.
Thus, the black curve is the classical Theodorsen function for impermeable aerofoils.
In the complex plots, the points $k=0,10^{-3},10^{-2},10^{-1},1,10$ are indicated by $\square$ with $k=0$ representing the rightmost part of the curve.
}
\label{Fig:theoPlot}
\end{figure}
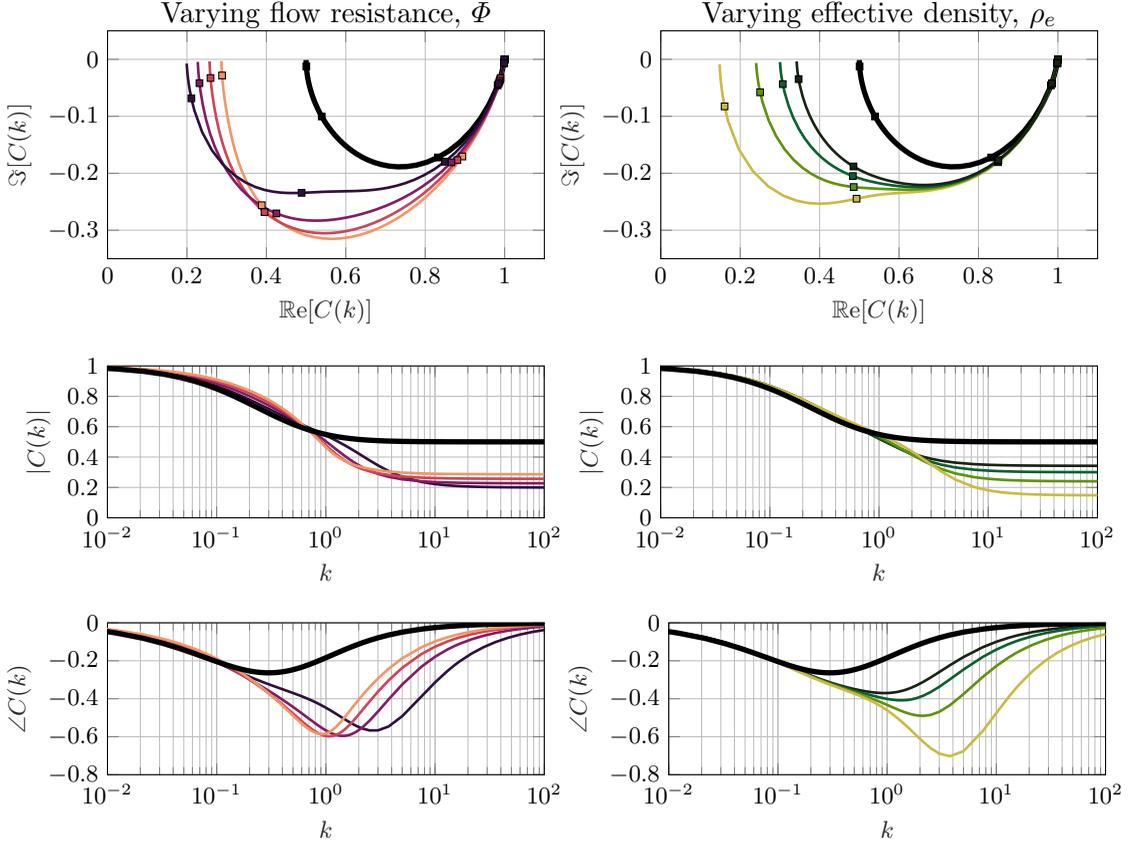

{The porosity profile \eqref{Eq:psidef} is set by two dimensionless properties of the porous medium, flow resistance $\Phi$ and effective density $\rho_{\rm e}$,} each of which has different effects on the relationship between the wake and quasi-steady lift.
 {Figure~\ref{Fig:theoPlot} illustrates these separate effects of $\Phi$ and $\rho_{\rm e}$ on the magnitude and phase of the porous extension of the Theodorsen function, whose classical result for impermeable aerofoils \eqref{eq:C_orig} is indicated by the thick black curves.}
The left {column} of figure \ref{Fig:theoPlot} examines the effect of reducing the dimensionless flow resistance of the porous medium.  
For these {porosity} configurations, the magnitude of the Theodorsen function is larger than the traditional, impermeable Theodorsen function at low reduced frequency; however, this trend reverses at high frequency, as illustrated in figure \ref{Fig:theoPlot}. Given that {aerofoil} porosity cannot increase the magnitude of the quasi-steady lift (cf.~\S\ref{Sec:qs}), figure \ref{Fig:theoPlot} implies that an effect of {the dimensionless flow resistance} is to modify the wake-induced lift. {This observation is consistent with the weakening of the trailing-edge zero due to porosity at that position, as discussed in \S\ref{Sec:steady}.}

{The right column of figure \ref{Fig:theoPlot} fixes the chordwise flow resistance distribution $\Phi(x)$ and steps through a set of constant effective density values representative of the full range of physically-relevant values. The influence of effective density on the magnitude of the Theodorsen function is marginal at low reduced frequencies, say, below $k \approx 1$, as would be expected of the term arising from the unsteady contribution of the aerofoil boundary condition \eqref{Eq:morseBC}. The role of effective density becomes pronounced at larger reduced frequencies , and the magnitude of the porous Theodorsen function decreases with decreasing effective density. }

{Although the magnitude of the Theodorsen function is relatively insensitive to changes in dimensionless flow resistance and effective density below $k \approx 1$, the bottom row of figure~\ref{Fig:theoPlot} demonstrates that the phase depends strongly on these parameters at lower reduced frequencies and in a complicated way. The reduction of effective density increases the maximum value of phase lag and the reduced frequency at which it occurs. Smaller values of the dimensionless flow resistance of the porous medium lead to larger maximum phase lag but lower reduced frequencies at these extrema. Viewed together, the phase lag of the porous Theodorsen function is greater and has peaks at larger reduced frequencies than the classical impermeable scenario. The phase angle of the Theodorsen function plays a critical role in the energy transfer between the fluid flow and aerofoil motions, and the above observations suggest that the porosity distribution of the aerofoil may be tuned to shift the flutter boundary of an aeroelastic aerofoil in a desired way \mbox{\citep[pp.~279-280]{A}}.}  

 {%
 Lastly, we note in the first row of figure~\mbox{\ref{Fig:theoPlot}} that the ratio between the wake-induced lift and quasi-steady lift tends to a finite, real value with a large reduced frequency that will depend on both the nondimensional flow resistance $\Phi$ and effective density $\rho_{\rm e}$ of the porous material. 
 However, caution should be exercised in the large frequency limit, which is beyond the realm of physical validity of the mathematical modelling employed in this paper: as discussed by \mbox{\cite{howe1979theory}} and noted more recently by \mbox{\cite{Weidenfeld2016}}, the present modelling assumptions for porosity are only valid when there is Stokes flow in the pores passing through the wing, which is rendered invalid at large frequencies. We expect that a higher-order (e.g., quadratic) porosity law would yield more physically meaningful results at high frequencies; an exploration of such porosity models is beyond the scope of the present work and is not pursued here.%
}

%% file: images/1theo.tex
%
%
\definecolor{mycolor1}{rgb}{0.94901,0.58547,0.40375}%
\definecolor{mycolor2}{rgb}{0.81038,0.26571,0.33825}%
\definecolor{mycolor3}{rgb}{0.51514,0.10993,0.38770}%
\definecolor{mycolor4}{rgb}{0.18517,0.05913,0.24304}%
\begin{tikzpicture}[%
trim axis left, trim axis right
]

\begin{axis}[%
width=0.951\fwidth,
height=\fheight,
at={(0\fwidth,0\fheight)},
scale only axis,
xmin=0,
xmax=1.1,
xlabel style={font=\color{white!15!black}},
xlabel={$\Re[C(k)]$},
ymin=-0.35,
ymax=0.05,
ylabel style={font=\color{white!15!black}},
ylabel={$\Im[C(k)]$},
axis background/.style={fill=white},
xmajorgrids,
ymajorgrids
]
\addplot [color=black, line width=2.0pt, forget plot]
  table[row sep=crcr]{%
1	-7.99360577730113e-15\\
0.998382581346416	-0.00700130186594028\\
0.997704828554474	-0.00936686937947895\\
0.990545455816093	-0.028861858990112\\
0.982421502833096	-0.0456520927493174\\
0.978288240159941	-0.0530407209015981\\
0.961139194654815	-0.078668282491596\\
0.939741156241609	-0.103475023965488\\
0.915045393818586	-0.125920777440791\\
0.88814419742074	-0.145084239141885\\
0.860116725139841	-0.160553820606936\\
0.83191743983797	-0.172304597403703\\
0.804315257153654	-0.180574366675261\\
0.777876570667823	-0.185754322590537\\
0.752978528411413	-0.18830350484444\\
0.729838625415966	-0.188688933442845\\
0.708549791556282	-0.187348427290084\\
0.689114121062283	-0.184670937325692\\
0.671471774168853	-0.180989035075001\\
0.65552390989786	-0.17657903167312\\
0.641149852440241	-0.171665356666385\\
0.628219320410753	-0.166426910843116\\
0.616600733099336	-0.161003967982198\\
0.606166559303723	-0.155504815121849\\
0.596796526366295	-0.150011726872571\\
0.588379335014207	-0.14458611935127\\
0.580813366186546	-0.139272872318397\\
0.5740067332996	-0.134103882133296\\
0.567876929654923	-0.129100940207709\\
0.562350242947747	-0.124278039676286\\
0.55736105232033	-0.119643208339959\\
0.552851083294492	-0.115199955390465\\
0.548768668017829	-0.110948406893224\\
0.545068039211433	-0.106886192581771\\
0.538654691054873	-0.0993117839106327\\
0.535596460840255	-0.0954267976975405\\
0.531817921943106	-0.0903431156276531\\
0.5281151636435	-0.0850115246561891\\
0.52450653577312	-0.0794218024629193\\
0.52101322924872	-0.0735641024948883\\
0.517659739536189	-0.0674293407075309\\
0.514474392444483	-0.0610097557479758\\
0.511489918390405	-0.0542997190546581\\
0.50874403260703	-0.0472969044488504\\
0.506279919221585	-0.0400039694602287\\
0.504146397018089	-0.0324309439727134\\
0.502397311392114	-0.0245985259426313\\
0.501089279876031	-0.0165423156849837\\
0.500372078870497	-0.00965328567445567\\
0.500080598765295	-0.00448973700755373\\
0.500006249258149	-0.00124994532645506\\
};
\addplot[only marks, mark=square*, mark options={}, mark size=1.2500pt, draw=black, fill=black] table[row sep=crcr]{%
x	y\\
1	-8.02904074443401e-15\\
0.998382581346416	-0.00700130186594032\\
0.982421502833096	-0.0456520927493173\\
0.831924104965276	-0.172302228734195\\
0.539434871077794	-0.100272902864108\\
0.500617885388891	-0.0124466215539119\\
};
\addplot [color=mycolor1, line width=1.0pt, forget plot]
  table[row sep=crcr]{%
1	-4.9960036108132e-15\\
0.998623699054577	-0.006410714869312\\
0.994424272403864	-0.0204500036208511\\
0.989719341255274	-0.0331543780451194\\
0.987332530992353	-0.0389355889426489\\
0.977398187626897	-0.060048496552785\\
0.964792549673804	-0.0825054474011196\\
0.949784543633551	-0.105364504692971\\
0.932700214156614	-0.127943168220367\\
0.913884093258024	-0.149765041395806\\
0.893672223989151	-0.170518440111803\\
0.872365990447344	-0.190015725636223\\
0.850219719427003	-0.208156503168994\\
0.827439969262591	-0.224902102530912\\
0.80418480697861	-0.240245237818514\\
0.780571647730553	-0.25419051683932\\
0.756689372616511	-0.266746941961224\\
0.732606644479822	-0.277913571267288\\
0.708383427257976	-0.287677109452997\\
0.684080828267905	-0.296013355740003\\
0.659768691571591	-0.302892127473875\\
0.635530166420023	-0.308277734313495\\
0.611466184998574	-0.312134498942277\\
0.58769505876419	-0.314441539243274\\
0.564351967456944	-0.315190156898085\\
0.541585172060057	-0.314392774265803\\
0.519550428690457	-0.312088960506727\\
0.498404691220528	-0.308350019408988\\
0.478298344215282	-0.303275550774189\\
0.459368548522129	-0.296992792867085\\
0.441730391181004	-0.289659460387922\\
0.425470357510948	-0.281452761126657\\
0.410643281342628	-0.272562059896068\\
0.397267250421292	-0.263183662632017\\
0.388347457870194	-0.256089327516487\\
0.385324869327422	-0.253512227181732\\
0.373729819625344	-0.242701013005863\\
0.360158702671085	-0.227791590164587\\
0.347950705324349	-0.211612895415015\\
0.337314408877901	-0.194543060120355\\
0.328283823094351	-0.177010498102645\\
0.320698076361201	-0.159399900558249\\
0.314272609252436	-0.141939475214794\\
0.308710462520504	-0.124674060900533\\
0.303795911061495	-0.107520038835106\\
0.299418129335474	-0.0903460214597083\\
0.295558631228872	-0.0730170793231784\\
0.292251902293633	-0.0554234334197379\\
0.289571465894715	-0.0374763762093031\\
0.288499647262798	-0.0283431608484367\\
0.287482200296264	-0.017197265395838\\
0.286912959689416	-0.00808265157675492\\
0.286696497329637	-0.0029087679097779\\
};
\addplot[only marks, mark=square*, mark options={}, mark size=1.2500pt, draw=black, fill=mycolor1] table[row sep=crcr]{%
x	y\\
1	-4.99444416819013e-15\\
};
\addplot[only marks, mark=square*, mark options={}, mark size=1.2500pt, draw=black, fill=mycolor1] table[row sep=crcr]{%
x	y\\
0.999027343348101	-0.00476311868975967\\
};
\addplot[only marks, mark=square*, mark options={}, mark size=1.2500pt, draw=black, fill=mycolor1] table[row sep=crcr]{%
x	y\\
0.989719341255274	-0.0331543780451195\\
};
\addplot[only marks, mark=square*, mark options={}, mark size=1.2500pt, draw=black, fill=mycolor1] table[row sep=crcr]{%
x	y\\
0.893677173903375	-0.170513647485101\\
};
\addplot[only marks, mark=square*, mark options={}, mark size=1.2500pt, draw=black, fill=mycolor1] table[row sep=crcr]{%
x	y\\
0.388347457870194	-0.256089327516487\\
};
\addplot[only marks, mark=square*, mark options={}, mark size=1.2500pt, draw=black, fill=mycolor1] table[row sep=crcr]{%
x	y\\
0.288499647262798	-0.0283431608484366\\
};
\addplot [color=mycolor2, line width=1.0pt, forget plot]
  table[row sep=crcr]{%
1	-5.55111512312578e-15\\
0.998458709416066	-0.00703392526485325\\
0.993727693131948	-0.0223261811699107\\
0.988403116405662	-0.0360651907037587\\
0.985696990215265	-0.0422863444979412\\
0.974420823599227	-0.0648326778338064\\
0.960123615426436	-0.0884873024909946\\
0.943169873651772	-0.112167258270792\\
0.924007035357824	-0.135104012398205\\
0.903109683608044	-0.156787699300214\\
0.88094205685534	-0.176915256026212\\
0.857918221924672	-0.195343828028373\\
0.834382989103326	-0.212044681430988\\
0.81061220890718	-0.22706578056345\\
0.786808473951891	-0.240495675119055\\
0.763110417929922	-0.252438444379281\\
0.739609986013792	-0.263000219692231\\
0.716359747144563	-0.272273045318415\\
0.693387836116587	-0.280329418251458\\
0.670709558000274	-0.287222472405453\\
0.64833744688535	-0.292989130395443\\
0.626286374682401	-0.297648468244905\\
0.604579829321286	-0.301206882908808\\
0.583253504426686	-0.303670138910749\\
0.562354494451532	-0.305041060455085\\
0.541942897342155	-0.305325657559586\\
0.522087232526799	-0.304540931327521\\
0.502864274390191	-0.302717287301164\\
0.484352519437576	-0.299897935752737\\
0.466630665355612	-0.296139395242802\\
0.449771040437037	-0.291518478021188\\
0.433836762263607	-0.286122612805837\\
0.418879623407802	-0.280049802628321\\
0.404934789477564	-0.27340765061266\\
0.392021453146277	-0.26631026607664\\
0.378949679654239	-0.258072642878433\\
0.362776017211614	-0.246158554487357\\
0.347152351693095	-0.232443344469827\\
0.332446403415649	-0.216999682787036\\
0.319006517258829	-0.200028068616601\\
0.307083967563933	-0.181859679644129\\
0.296774011042123	-0.162887261597128\\
0.287997268188653	-0.143465832866961\\
0.280557698227419	-0.123827564405033\\
0.274236868087936	-0.104056102560886\\
0.268874760479498	-0.0841130605844749\\
0.264392993108075	-0.063901512058585\\
0.260795434163043	-0.0432957821510896\\
0.259351383780191	-0.0327970179326126\\
0.257964252525336	-0.0199537568603748\\
0.257174314404167	-0.0094112478838041\\
0.256888997026476	-0.00341368614084736\\
};
\addplot[only marks, mark=square*, mark options={}, mark size=1.2500pt, draw=black, fill=mycolor2] table[row sep=crcr]{%
x	y\\
1	-5.55595724021972e-15\\
};
\addplot[only marks, mark=square*, mark options={}, mark size=1.2500pt, draw=black, fill=mycolor2] table[row sep=crcr]{%
x	y\\
0.998911506258969	-0.00523060951249984\\
};
\addplot[only marks, mark=square*, mark options={}, mark size=1.2500pt, draw=black, fill=mycolor2] table[row sep=crcr]{%
x	y\\
0.988403116405662	-0.0360651907037587\\
};
\addplot[only marks, mark=square*, mark options={}, mark size=1.2500pt, draw=black, fill=mycolor2] table[row sep=crcr]{%
x	y\\
0.880947452152502	-0.176910657379468\\
};
\addplot[only marks, mark=square*, mark options={}, mark size=1.2500pt, draw=black, fill=mycolor2] table[row sep=crcr]{%
x	y\\
0.39533739796832	-0.268225990783993\\
};
\addplot[only marks, mark=square*, mark options={}, mark size=1.2500pt, draw=black, fill=mycolor2] table[row sep=crcr]{%
x	y\\
0.259351383780191	-0.0327970179326126\\
};
\addplot [color=mycolor3, line width=1.0pt, forget plot]
  table[row sep=crcr]{%
0.999999999999999	-6.21724893790088e-15\\
0.998251710259919	-0.00775616758311926\\
0.992852471998004	-0.0244439303066957\\
0.986750884795209	-0.0392862213079809\\
0.983645816598243	-0.0459607416620952\\
0.970706531096841	-0.069894857452761\\
0.954346554618995	-0.0945326014493639\\
0.935074207080992	-0.118628709879011\\
0.913509057066359	-0.141330898079634\\
0.890299284438226	-0.162118589475819\\
0.866072411629749	-0.18072980413452\\
0.841375658418083	-0.197106369512367\\
0.81665034796427	-0.211332401768563\\
0.792240479667831	-0.223577449025985\\
0.768383248698875	-0.234055944045439\\
0.74522336443245	-0.242995060553719\\
0.722841742564278	-0.250614372334391\\
0.701260614224105	-0.257109158885089\\
0.680461729747045	-0.262646153249948\\
0.660404262020536	-0.267359122256995\\
0.641035772827496	-0.271354645418861\\
0.622296326779055	-0.274708917646676\\
0.604127131585631	-0.277472620939049\\
0.586478561782789	-0.279682395507619\\
0.569308498727359	-0.28135602703077\\
0.552585234757179	-0.28250155042478\\
0.536288759336389	-0.283120060021334\\
0.520411109317953	-0.283213361626938\\
0.504952340494677	-0.282780133977052\\
0.48992151756256	-0.281820681750031\\
0.475335236172674	-0.280343539446793\\
0.461212972381825	-0.27835981412437\\
0.447578436521543	-0.275886631094092\\
0.434454833479553	-0.272946986367497\\
0.421865225773816	-0.269573469404591\\
0.408595751177551	-0.265381271642949\\
0.391255057315445	-0.258819118511426\\
0.373245012033351	-0.250539747173625\\
0.354781371047971	-0.240257163858477\\
0.336191603135215	-0.227706188307596\\
0.317918200508012	-0.212704520280862\\
0.300495681609099	-0.195213505919862\\
0.284472283711019	-0.175393043719307\\
0.270295856603821	-0.153604409699048\\
0.258205767629558	-0.130325266308876\\
0.248211362303829	-0.105990535033495\\
0.240176246116008	-0.0808722051535442\\
0.233955525090305	-0.0550301251308137\\
0.231507104028784	-0.0417992107992792\\
0.229513651435987	-0.0283027160053109\\
0.228543507278533	-0.0199391401209359\\
0.227635269485793	-0.0094232643018668\\
0.227360574345255	-0.00442665096996775\\
};
\addplot[only marks, mark=square*, mark options={}, mark size=1.2500pt, draw=black, fill=mycolor3] table[row sep=crcr]{%
x	y\\
0.999999999999999	-6.24934880421672e-15\\
};
\addplot[only marks, mark=square*, mark options={}, mark size=1.2500pt, draw=black, fill=mycolor3] table[row sep=crcr]{%
x	y\\
0.998766247487827	-0.00577476521562612\\
};
\addplot[only marks, mark=square*, mark options={}, mark size=1.2500pt, draw=black, fill=mycolor3] table[row sep=crcr]{%
x	y\\
0.986750884795209	-0.0392862213079808\\
};
\addplot[only marks, mark=square*, mark options={}, mark size=1.2500pt, draw=black, fill=mycolor3] table[row sep=crcr]{%
x	y\\
0.866078255145068	-0.180725634821642\\
};
\addplot[only marks, mark=square*, mark options={}, mark size=1.2500pt, draw=black, fill=mycolor3] table[row sep=crcr]{%
x	y\\
0.425143132314303	-0.270506368479212\\
};
\addplot[only marks, mark=square*, mark options={}, mark size=1.2500pt, draw=black, fill=mycolor3] table[row sep=crcr]{%
x	y\\
0.231507104028784	-0.0417992107992792\\
};
\addplot [color=mycolor4, line width=1.0pt, forget plot]
  table[row sep=crcr]{%
0.999999999999999	-7.105427357601e-15\\
0.997996745092171	-0.00855984506534968\\
0.991774938099978	-0.0267122510162389\\
0.984722397460611	-0.0426341437458664\\
0.981131941981768	-0.0497265424539458\\
0.966191849426031	-0.0747912588256247\\
0.947403098224663	-0.0999229737072221\\
0.925477362840599	-0.123707799779075\\
0.901260450054296	-0.145231707660628\\
0.875610429119569	-0.16400742912532\\
0.849338702049374	-0.179860647196504\\
0.823126327079069	-0.19286356785714\\
0.797494917208019	-0.203253187244425\\
0.772838018440293	-0.211343201965533\\
0.749404905186951	-0.217485086273635\\
0.72732379632443	-0.222032274338574\\
0.706653464857698	-0.225308411265954\\
0.687379521411525	-0.227599459383346\\
0.66943845343697	-0.229152864156653\\
0.652742963357651	-0.230169337821553\\
0.637194424262424	-0.230814088209251\\
0.622681187177132	-0.231213838321565\\
0.596306438038171	-0.231646511267094\\
0.561867901564088	-0.232162956580503\\
0.541332667560853	-0.232667068446317\\
0.503901181179613	-0.233931592797432\\
0.486394129666223	-0.234455166801884\\
0.476972297957855	-0.234642191418098\\
0.464091006808394	-0.2347293551744\\
0.449803073912232	-0.234519084058899\\
0.433845032750844	-0.233797310380258\\
0.415919955851161	-0.232228789886146\\
0.395727701139732	-0.229312844143851\\
0.373025699341515	-0.224318198696134\\
0.34774815255843	-0.216222127111328\\
0.320219095101294	-0.203715554717568\\
0.29143138077827	-0.185362739193187\\
0.26325363920886	-0.16003948009001\\
0.238205485593487	-0.127618625064112\\
0.218546628824258	-0.0893011304851996\\
0.211116590915118	-0.0684929305640662\\
0.207014983895997	-0.0538889033782192\\
0.204356315829492	-0.0423090031325306\\
0.20260936189365	-0.0331632507884783\\
0.201451035590617	-0.0259616028379978\\
0.200153417756263	-0.0158544587969266\\
0.19939754174316	-0.00750287002851546\\
};
\addplot[only marks, mark=square*, mark options={}, mark size=1.2500pt, draw=black, fill=mycolor4] table[row sep=crcr]{%
x	y\\
0.999999999999999	-7.08819333362731e-15\\
};
\addplot[only marks, mark=square*, mark options={}, mark size=1.2500pt, draw=black, fill=mycolor4] table[row sep=crcr]{%
x	y\\
0.998587382919152	-0.00638401770555096\\
};
\addplot[only marks, mark=square*, mark options={}, mark size=1.2500pt, draw=black, fill=mycolor4] table[row sep=crcr]{%
x	y\\
0.984722397460611	-0.0426341437458664\\
};
\addplot[only marks, mark=square*, mark options={}, mark size=1.2500pt, draw=black, fill=mycolor4] table[row sep=crcr]{%
x	y\\
0.849344975600682	-0.179857208155207\\
};
\addplot[only marks, mark=square*, mark options={}, mark size=1.2500pt, draw=black, fill=mycolor4] table[row sep=crcr]{%
x	y\\
0.488674753967596	-0.234397620114281\\
};
\addplot[only marks, mark=square*, mark options={}, mark size=1.2500pt, draw=black, fill=mycolor4] table[row sep=crcr]{%
x	y\\
0.211116590915118	-0.0684929305640662\\
};
\end{axis}
\end{tikzpicture}%

%% file: images/2theo.tex
%
%
\definecolor{mycolor1}{rgb}{0.79072,0.73098,0.27064}%
\definecolor{mycolor2}{rgb}{0.37697,0.57257,0.04618}%
\definecolor{mycolor3}{rgb}{0.04342,0.37461,0.17524}%
\definecolor{mycolor4}{rgb}{0.09053,0.13734,0.07326}%
\begin{tikzpicture}[%
trim axis left, trim axis right
]

\begin{axis}[%
width=0.951\fwidth,
height=\fheight,
at={(0\fwidth,0\fheight)},
scale only axis,
xmin=0,
xmax=1.1,
xlabel style={font=\color{white!15!black}},
xlabel={$\Re[C(k)]$},
ymin=-0.35,
ymax=0.05,
ylabel style={font=\color{white!15!black}},
ylabel={$\Im[C(k)]$},
axis background/.style={fill=white},
xmajorgrids,
ymajorgrids
]
\addplot [color=black, line width=2.0pt, forget plot]
  table[row sep=crcr]{%
1	-7.99360577730113e-15\\
0.998382581346416	-0.00700130186594028\\
0.997704828554474	-0.00936686937947895\\
0.990545455816093	-0.028861858990112\\
0.982421502833096	-0.0456520927493174\\
0.978288240159941	-0.0530407209015981\\
0.961139194654815	-0.078668282491596\\
0.939741156241609	-0.103475023965488\\
0.915045393818586	-0.125920777440791\\
0.88814419742074	-0.145084239141885\\
0.860116725139841	-0.160553820606936\\
0.83191743983797	-0.172304597403703\\
0.804315257153654	-0.180574366675261\\
0.777876570667823	-0.185754322590537\\
0.752978528411413	-0.18830350484444\\
0.729838625415966	-0.188688933442845\\
0.708549791556282	-0.187348427290084\\
0.689114121062283	-0.184670937325692\\
0.671471774168853	-0.180989035075001\\
0.65552390989786	-0.17657903167312\\
0.641149852440241	-0.171665356666385\\
0.628219320410753	-0.166426910843116\\
0.616600733099336	-0.161003967982198\\
0.606166559303723	-0.155504815121849\\
0.596796526366295	-0.150011726872571\\
0.588379335014207	-0.14458611935127\\
0.580813366186546	-0.139272872318397\\
0.5740067332996	-0.134103882133296\\
0.567876929654923	-0.129100940207709\\
0.562350242947747	-0.124278039676286\\
0.55736105232033	-0.119643208339959\\
0.552851083294492	-0.115199955390465\\
0.548768668017829	-0.110948406893224\\
0.545068039211433	-0.106886192581771\\
0.538654691054873	-0.0993117839106327\\
0.535596460840255	-0.0954267976975405\\
0.531817921943106	-0.0903431156276531\\
0.5281151636435	-0.0850115246561891\\
0.52450653577312	-0.0794218024629193\\
0.52101322924872	-0.0735641024948883\\
0.517659739536189	-0.0674293407075309\\
0.514474392444483	-0.0610097557479758\\
0.511489918390405	-0.0542997190546581\\
0.50874403260703	-0.0472969044488504\\
0.506279919221585	-0.0400039694602287\\
0.504146397018089	-0.0324309439727134\\
0.502397311392114	-0.0245985259426313\\
0.501089279876031	-0.0165423156849837\\
0.500372078870497	-0.00965328567445567\\
0.500080598765295	-0.00448973700755373\\
0.500006249258149	-0.00124994532645506\\
};
\addplot[only marks, mark=square*, mark options={}, mark size=1.2500pt, draw=black, fill=black] table[row sep=crcr]{%
x	y\\
1	-8.02904074443401e-15\\
0.998382581346416	-0.00700130186594032\\
0.982421502833096	-0.0456520927493173\\
0.831924104965276	-0.172302228734195\\
0.539434871077794	-0.100272902864108\\
0.500617885388891	-0.0124466215539119\\
};
\addplot [color=mycolor1, line width=1.0pt, forget plot]
  table[row sep=crcr]{%
0.999999999999999	-7.105427357601e-15\\
0.997996565680895	-0.00857226338845984\\
0.991772859972809	-0.0267604059089468\\
0.984716643791312	-0.0427199273365169\\
0.981123884547719	-0.0498306585550577\\
0.966172114989806	-0.0749673265335998\\
0.947365827722821	-0.100182635470634\\
0.925418274692878	-0.124058332877578\\
0.901178321515529	-0.145676934765886\\
0.87550790104251	-0.164549384296146\\
0.849222846068704	-0.180500475503967\\
0.82300854710521	-0.193602918635354\\
0.797390316862671	-0.204095618052265\\
0.77276497284851	-0.212294565785257\\
0.749384282776246	-0.218554297322594\\
0.727378332107414	-0.223231542593297\\
0.706807340388883	-0.226653206675102\\
0.687657607473396	-0.229108954059241\\
0.669865577065397	-0.230850226749083\\
0.653343780231153	-0.232081316017959\\
0.6379928095569	-0.232971226154878\\
0.610345313102433	-0.234221114355499\\
0.586031596031587	-0.235321134489333\\
0.564239636614842	-0.23664077759047\\
0.554073598001255	-0.237433780490775\\
0.544295584139047	-0.238320569254088\\
0.525653233024908	-0.240351057287695\\
0.507903954287246	-0.242637193009717\\
0.468612462183296	-0.248104742810699\\
0.454207883267423	-0.24990753128749\\
0.437852774708581	-0.25160950336883\\
0.419078259699903	-0.252940419508393\\
0.397325286945219	-0.253433378456805\\
0.371972718508652	-0.252304798169828\\
0.34244554487874	-0.248283241365998\\
0.308499622367029	-0.239426433571162\\
0.270793921039154	-0.223060955818974\\
0.231777020051968	-0.196255627061838\\
0.196243153377542	-0.157471975593398\\
0.1697001405619	-0.108902385497443\\
0.160652376839854	-0.0826012024080833\\
0.156022860250102	-0.0644544539193516\\
0.154199780719603	-0.0557417629488274\\
0.153177302021107	-0.0502807918559456\\
0.151376110048064	-0.0392320307320968\\
0.150205003765584	-0.0306116196993472\\
0.148903683417719	-0.0186251750367331\\
0.14813254260213	-0.00880693294450208\\
};
\addplot[only marks, mark=square*, mark options={}, mark size=1.2500pt, draw=black, fill=mycolor1] table[row sep=crcr]{%
x	y\\
0.999999999999999	-7.09016063014689e-15\\
};
\addplot[only marks, mark=square*, mark options={}, mark size=1.2500pt, draw=black, fill=mycolor1] table[row sep=crcr]{%
x	y\\
0.998587286981112	-0.00639284709801578\\
};
\addplot[only marks, mark=square*, mark options={}, mark size=1.2500pt, draw=black, fill=mycolor1] table[row sep=crcr]{%
x	y\\
0.984716643791312	-0.0427199273365169\\
};
\addplot[only marks, mark=square*, mark options={}, mark size=1.2500pt, draw=black, fill=mycolor1] table[row sep=crcr]{%
x	y\\
0.849229129053837	-0.180496994304226\\
};
\addplot[only marks, mark=square*, mark options={}, mark size=1.2500pt, draw=black, fill=mycolor1] table[row sep=crcr]{%
x	y\\
0.492988992440812	-0.244716268362176\\
};
\addplot[only marks, mark=square*, mark options={}, mark size=1.2500pt, draw=black, fill=mycolor1] table[row sep=crcr]{%
x	y\\
0.160652376839854	-0.0826012024080833\\
};
\addplot [color=mycolor2, line width=1.0pt, forget plot]
  table[row sep=crcr]{%
0.999999999999999	-7.105427357601e-15\\
0.997996926870559	-0.0085474260532461\\
0.991777052218506	-0.0266640957699503\\
0.984728266304823	-0.0425483725162866\\
0.981140171259393	-0.0496224381110556\\
0.966212086976808	-0.0746152813486701\\
0.947441529352524	-0.0996634566373968\\
0.925538665256658	-0.123357656527558\\
0.901346338604625	-0.14478727963151\\
0.875718858337078	-0.163466776787602\\
0.849463243619304	-0.179222773213036\\
0.823256360315898	-0.192126784135536\\
0.797616364886382	-0.202413608210528\\
0.772933414894294	-0.210394942581145\\
0.749454552150452	-0.216418916162362\\
0.727306589053588	-0.220835244708483\\
0.706547548339314	-0.223963719990056\\
0.687162462099944	-0.22608684859632\\
0.669088517821465	-0.227447927266219\\
0.652239401112404	-0.228243777836669\\
0.636518000836809	-0.228635953159336\\
0.621814858242322	-0.228747488493264\\
0.608016059162895	-0.228672702046174\\
0.582733689514203	-0.228222458170814\\
0.549246085255114	-0.227274558322469\\
0.519451408616503	-0.226165010480237\\
0.501031988504011	-0.225227459710855\\
0.483516836061945	-0.224004134108145\\
0.474183598091046	-0.223168377879889\\
0.461550159240355	-0.221778401789399\\
0.447731670114285	-0.219847530561979\\
0.432569158905991	-0.217142511472712\\
0.415910422319983	-0.213332316864116\\
0.397648687279647	-0.207971040802375\\
0.377776886573336	-0.20047820099686\\
0.356471003631773	-0.190138044228248\\
0.334198225166205	-0.176168086718751\\
0.311800739525572	-0.157860721273016\\
0.290474897262565	-0.134803741835816\\
0.271565273084419	-0.10707469722353\\
0.256231658944349	-0.0751927275313434\\
0.250166105343397	-0.0578837522821554\\
0.246720651214593	-0.0456668698700829\\
0.244444097094931	-0.0359243236648977\\
0.242931836681212	-0.0281964743210739\\
0.24192245736751	-0.0220895317889525\\
0.240783425989985	-0.0134808949368419\\
0.240145346080921	-0.00637875787326547\\
};
\addplot[only marks, mark=square*, mark options={}, mark size=1.2500pt, draw=black, fill=mycolor2] table[row sep=crcr]{%
x	y\\
0.999999999999999	-7.08622603710773e-15\\
};
\addplot[only marks, mark=square*, mark options={}, mark size=1.2500pt, draw=black, fill=mycolor2] table[row sep=crcr]{%
x	y\\
0.998587480033127	-0.00637518924975628\\
};
\addplot[only marks, mark=square*, mark options={}, mark size=1.2500pt, draw=black, fill=mycolor2] table[row sep=crcr]{%
x	y\\
0.984728266304823	-0.0425483725162866\\
};
\addplot[only marks, mark=square*, mark options={}, mark size=1.2500pt, draw=black, fill=mycolor2] table[row sep=crcr]{%
x	y\\
0.849469486838363	-0.17921941631486\\
};
\addplot[only marks, mark=square*, mark options={}, mark size=1.2500pt, draw=black, fill=mycolor2] table[row sep=crcr]{%
x	y\\
0.485786548260047	-0.224186046142\\
};
\addplot[only marks, mark=square*, mark options={}, mark size=1.2500pt, draw=black, fill=mycolor2] table[row sep=crcr]{%
x	y\\
0.250166105343397	-0.0578837522821554\\
};
\addplot [color=mycolor3, line width=1.0pt, forget plot]
  table[row sep=crcr]{%
0.999999999999999	-7.105427357601e-15\\
0.997997297479734	-0.00852258857358135\\
0.991781388004693	-0.0265677914892184\\
0.984740352799234	-0.0423768321540652\\
0.981157143416434	-0.049414284278911\\
0.966254101210987	-0.0742634216073428\\
0.94752185084664	-0.0991449699639878\\
0.925667884839055	-0.122658671257662\\
0.901529413402309	-0.143900834718321\\
0.875953391263441	-0.162389617570528\\
0.84973832980566	-0.177953341140835\\
0.823552988597508	-0.190661932841797\\
0.797908850380475	-0.200745706720055\\
0.773190229105787	-0.208511301725199\\
0.749640120765134	-0.214300197312863\\
0.727383866901405	-0.218453577202101\\
0.706478363928075	-0.221284312702975\\
0.686909733517902	-0.223066793200579\\
0.668618289649477	-0.224035915632014\\
0.651521703650535	-0.224381030023017\\
0.635528272751925	-0.224256886484058\\
0.620535382612818	-0.22378006774579\\
0.606437685847028	-0.223039163757435\\
0.593144816237559	-0.222100721505392\\
0.580567999976568	-0.221008157399707\\
0.568627408349822	-0.219791565123297\\
0.557254112313526	-0.218466831199576\\
0.546392425275388	-0.217046424546187\\
0.535992288280451	-0.215531813578952\\
0.526011793547713	-0.21392136016825\\
0.516421180918554	-0.212215341408291\\
0.507193535746533	-0.210408935506808\\
0.498309235992458	-0.208499490487573\\
0.483662286352329	-0.204934103389016\\
0.481512592714058	-0.20436274907381\\
0.472764151453753	-0.201897080827897\\
0.461172719471545	-0.198252298050514\\
0.448847062604775	-0.193842326084384\\
0.43576984821091	-0.188476652122383\\
0.421957252514038	-0.181922071876946\\
0.407479342577958	-0.173912498435804\\
0.39248398290556	-0.164157796071501\\
0.377210206764043	-0.152362001112831\\
0.362003005404841	-0.138265928334142\\
0.347301773671006	-0.121683614988633\\
0.333608265554041	-0.102526976241537\\
0.321444311986204	-0.0808099674002957\\
0.311325299076604	-0.056590186168097\\
0.30720311808389	-0.0435450455132831\\
0.304825439692871	-0.0343428354631247\\
0.303243057867353	-0.0270023950560111\\
0.302186590285113	-0.0211741605288296\\
0.301011756356493	-0.0129391960576868\\
0.300372817211765	-0.00613202728100593\\
0.300257347465491	-0.00474407315692627\\
};
\addplot[only marks, mark=square*, mark options={}, mark size=1.2500pt, draw=black, fill=mycolor3] table[row sep=crcr]{%
x	y\\
0.999999999999999	-7.08229144406858e-15\\
};
\addplot[only marks, mark=square*, mark options={}, mark size=1.2500pt, draw=black, fill=mycolor3] table[row sep=crcr]{%
x	y\\
0.998587677835948	-0.00635753156405206\\
};
\addplot[only marks, mark=square*, mark options={}, mark size=1.2500pt, draw=black, fill=mycolor3] table[row sep=crcr]{%
x	y\\
0.984740352799234	-0.0423768321540652\\
};
\addplot[only marks, mark=square*, mark options={}, mark size=1.2500pt, draw=black, fill=mycolor3] table[row sep=crcr]{%
x	y\\
0.849744575729806	-0.177950008289032\\
};
\addplot[only marks, mark=square*, mark options={}, mark size=1.2500pt, draw=black, fill=mycolor3] table[row sep=crcr]{%
x	y\\
0.483662286352329	-0.204934103389016\\
};
\addplot[only marks, mark=square*, mark options={}, mark size=1.2500pt, draw=black, fill=mycolor3] table[row sep=crcr]{%
x	y\\
0.30720311808389	-0.0435450455132831\\
};
\addplot [color=mycolor4, line width=1.0pt, forget plot]
  table[row sep=crcr]{%
0.999999999999999	-7.105427357601e-15\\
0.997997677492629	-0.00849775184965551\\
0.991785867488849	-0.0264714905673678\\
0.984752899597404	-0.0422053477178201\\
0.981174800760773	-0.0492062007817842\\
0.96629813767695	-0.073911872939871\\
0.947606768238304	-0.0986273000658747\\
0.92580591843626	-0.12196147737303\\
0.901727476813747	-0.143017913778376\\
0.87621134172537	-0.161318550418601\\
0.850047787425885	-0.176693395530539\\
0.823897764813828	-0.189210652692021\\
0.798266581591986	-0.199095840481738\\
0.773533514844489	-0.20664989028765\\
0.749938479996669	-0.212207194477109\\
0.727606357910351	-0.216100527226068\\
0.70659478640668	-0.218635166520943\\
0.686893116427521	-0.220077262300061\\
0.668446727603118	-0.220654152454793\\
0.651179871704603	-0.220548616719783\\
0.635008641723408	-0.219910285983194\\
0.619839790033331	-0.218851750436012\\
0.60557890439812	-0.217458670709991\\
0.592146211380366	-0.215797876350373\\
0.579464569760419	-0.213913792680215\\
0.567465634754122	-0.211840686733718\\
0.556091124634415	-0.209599896661179\\
0.545295198123461	-0.207211868298682\\
0.535035979309665	-0.204687615329995\\
0.525278521703104	-0.202036409631475\\
0.515997921444177	-0.199270391894648\\
0.507169151997556	-0.196397592903269\\
0.498773690723364	-0.193428056840903\\
0.490793190771302	-0.190370033953734\\
0.483214636198335	-0.187238413085054\\
0.475287457968025	-0.18370328783792\\
0.464984105538636	-0.178673352867965\\
0.454286873132223	-0.172869815251967\\
0.443234586598677	-0.166169217313792\\
0.431891842303558	-0.158434399890726\\
0.420354829783609	-0.149524732368825\\
0.408760281332272	-0.139305617802962\\
0.397280426756395	-0.127649687827451\\
0.386126603632363	-0.114456004556951\\
0.375541559310212	-0.0996539548113728\\
0.365792665378209	-0.0832022581973546\\
0.357168867960755	-0.0650829744713446\\
0.349990724832463	-0.0452757973621206\\
0.347067504465186	-0.0347221121158828\\
0.34467645894242	-0.0237060083227688\\
0.343529666632349	-0.0167691579469915\\
0.342718197018761	-0.0102159614101309\\
0.342173920367125	-0.00369755208977551\\
};
\addplot[only marks, mark=square*, mark options={}, mark size=1.2500pt, draw=black, fill=mycolor4] table[row sep=crcr]{%
x	y\\
0.999999999999999	-7.07835685102942e-15\\
};
\addplot[only marks, mark=square*, mark options={}, mark size=1.2500pt, draw=black, fill=mycolor4] table[row sep=crcr]{%
x	y\\
0.998587880395397	-0.00633987359741893\\
};
\addplot[only marks, mark=square*, mark options={}, mark size=1.2500pt, draw=black, fill=mycolor4] table[row sep=crcr]{%
x	y\\
0.984752899597404	-0.0422053477178202\\
};
\addplot[only marks, mark=square*, mark options={}, mark size=1.2500pt, draw=black, fill=mycolor4] table[row sep=crcr]{%
x	y\\
0.850054034966155	-0.176690082994421\\
};
\addplot[only marks, mark=square*, mark options={}, mark size=1.2500pt, draw=black, fill=mycolor4] table[row sep=crcr]{%
x	y\\
0.485181671791429	-0.188073538969102\\
};
\addplot[only marks, mark=square*, mark options={}, mark size=1.2500pt, draw=black, fill=mycolor4] table[row sep=crcr]{%
x	y\\
0.347067504465186	-0.0347221121158828\\
};
\end{axis}
\end{tikzpicture}%

%% file: images/1absTheo.tex
%
%
\definecolor{mycolor1}{rgb}{0.18517,0.05913,0.24304}%
\definecolor{mycolor2}{rgb}{0.51514,0.10993,0.38770}%
\definecolor{mycolor3}{rgb}{0.81038,0.26571,0.33825}%
\definecolor{mycolor4}{rgb}{0.94901,0.58547,0.40375}%
\begin{tikzpicture}[%
trim axis left, trim axis right
]

\begin{axis}[%
width=0.951\fwidth,
height=\fheight,
at={(0\fwidth,0\fheight)},
scale only axis,
xmode=log,
xmin=0.01,
xmax=100,
xminorticks=true,
xlabel style={font=\color{white!15!black}},
xlabel={$k$},
ymin=0,
ymax=1,
ylabel style={font=\color{white!15!black}},
ylabel={$|C(k)|$},
axis background/.style={fill=white},
xmajorgrids,
xminorgrids,
ymajorgrids
]
\addplot [color=mycolor1, line width=1.0pt, forget plot]
  table[row sep=crcr]{%
0.00553220719809771	0.992134603870649\\
0.01	0.985644900698784\\
0.012230757294587	0.982391274697276\\
0.0213755451908948	0.96908225775422\\
0.0328492202200639	0.952657982279152\\
0.0465441846922757	0.933708717350636\\
0.0623614859398746	0.912886985197014\\
0.0802098649172976	0.890837841804699\\
0.100004935222997	0.868173993626197\\
0.121668471397888	0.84541901216927\\
0.145127789278423	0.82298845745104\\
0.17031520429155	0.801214423087704\\
0.197167556059509	0.78032523646856\\
0.225625789674533	0.760459226749036\\
0.287142031567055	0.724080189183165\\
0.39018234102294	0.677710762513766\\
0.465546310209241	0.651584791865883\\
0.545889678899884	0.628544253764147\\
0.720456888900558	0.589215598663131\\
0.862619486763344	0.563609651156781\\
1	0.541982711422172\\
1.07142857142857	0.531563289752673\\
1.15384615384615	0.520075314527635\\
1.25	0.507269165324207\\
1.36363636363636	0.492831304589738\\
1.5	0.476360809184819\\
1.66666666666667	0.457367241873179\\
1.875	0.43527787289903\\
2.5	0.379526410281786\\
3	0.345386153143884\\
3.75	0.308082965683255\\
5	0.270237611797624\\
7.5	0.236087527998373\\
10	0.221949310647443\\
12.9154966501488	0.213914042233713\\
15	0.210589078056173\\
16.6810053720006	0.208690094555232\\
21.5443469003188	0.205305515585459\\
27.8255940220713	0.203117021843198\\
35.9381366380463	0.201697577191917\\
46.4158883361278	0.200780363838839\\
77.4263682681127	0.199786285924646\\
100	0.199538649669381\\
};
\addplot [color=mycolor2, line width=1.0pt, forget plot]
  table[row sep=crcr]{%
0.00553220719809771	0.993153329995619\\
0.01	0.987532640386528\\
0.012230757294587	0.984718986455199\\
0.0213755451908948	0.973219636368073\\
0.0328492202200639	0.959017079644534\\
0.0465441846922757	0.942569118715283\\
0.0623614859398746	0.92437720660684\\
0.0802098649172976	0.904939364225497\\
0.100004935222997	0.884728593574702\\
0.121668471397888	0.864155032086725\\
0.145127789278423	0.843551524725925\\
0.17031520429155	0.823183972959436\\
0.197167556059509	0.803240313868794\\
0.225625789674533	0.783839564164457\\
0.255634585613067	0.765054082018993\\
0.287142031567055	0.746908005476288\\
0.320099330538618	0.729390956529717\\
0.35446054042221	0.712470834174228\\
0.39018234102294	0.696103588293396\\
0.42722382506056	0.680233568531657\\
0.465546310209241	0.664801208248502\\
0.50511316964193	0.649753296096056\\
0.545889678899884	0.63503809387291\\
0.587842877204593	0.620610640952966\\
0.630941441580349	0.60643433426633\\
0.675155572369048	0.592484203085258\\
0.720456888900558	0.578741281008303\\
0.766818334237483	0.565195532566257\\
0.862619486763344	0.538703621681377\\
0.962365914765952	0.513079973981489\\
1.07142857142857	0.487214231338921\\
1.36363636363636	0.428477917783399\\
1.5	0.406047906301954\\
1.66666666666667	0.382511692843715\\
1.875	0.358337784163548\\
2.14285714285714	0.334196349448772\\
2.5	0.310892529302629\\
3	0.289231556846152\\
3.75	0.26989419018091\\
5	0.253426405026734\\
7.5	0.240340388599572\\
10	0.235250320380731\\
12.9154966501488	0.23241633563189\\
16.6810053720006	0.230590638621617\\
21.5443469003188	0.229411647542001\\
35.9381366380463	0.228157309044101\\
59.9484250318941	0.227625121727558\\
100	0.227403663131037\\
};
\addplot [color=mycolor3, line width=1.0pt, forget plot]
  table[row sep=crcr]{%
0.00553220719809771	0.993978463782276\\
0.01	0.989060877044948\\
0.012230757294587	0.986603614148271\\
0.0213755451908948	0.976575249317173\\
0.0328492202200639	0.964192594662323\\
0.0465441846922757	0.94981624769862\\
0.0623614859398746	0.93383194181654\\
0.0802098649172976	0.916618504765464\\
0.100004935222997	0.898530976289264\\
0.121668471397888	0.879876631499649\\
0.145127789278423	0.860905290626193\\
0.17031520429155	0.841814125524315\\
0.197167556059509	0.822742817916677\\
0.225625789674533	0.803780242450517\\
0.255634585613067	0.784979010527977\\
0.287142031567055	0.766357552671066\\
0.320099330538618	0.747911274157332\\
0.35446054042221	0.72962186086193\\
0.39018234102294	0.711466145058001\\
0.42722382506056	0.69341851270453\\
0.465546310209241	0.675457146186045\\
0.50511316964193	0.657571443793055\\
0.545889678899884	0.63975981898944\\
0.587842877204593	0.622033649526939\\
0.630941441580349	0.604417287328321\\
0.675155572369048	0.586949942064014\\
0.720456888900558	0.569680730720532\\
0.814214088046517	0.53598228687486\\
1.07142857142857	0.458480478005672\\
1.15384615384615	0.438406743345978\\
1.25	0.417785427790828\\
1.36363636363636	0.397000596314004\\
1.5	0.376532052139118\\
1.66666666666667	0.356894250745332\\
1.875	0.338536665105334\\
2.14285714285714	0.321752811462988\\
2.5	0.306669019858789\\
3	0.293314732495389\\
3.75	0.281724411054127\\
5	0.272005621353714\\
7.5	0.264364867621181\\
10	0.261416879053308\\
12.9154966501488	0.259782164719835\\
16.6810053720006	0.258734821765062\\
27.8255940220713	0.257626178383472\\
59.9484250318941	0.257044054268889\\
100	0.256911677520383\\
};
\addplot [color=mycolor4, line width=1.0pt, forget plot]
  table[row sep=crcr]{%
0.00553220719809771	0.994634523930296\\
0.01	0.990274500953312\\
0.012230757294587	0.988099947799856\\
0.0213755451908948	0.979241052608903\\
0.0328492202200639	0.968313901974426\\
0.0465441846922757	0.955610986842652\\
0.0623614859398746	0.941434620025235\\
0.0802098649172976	0.926074350975302\\
0.100004935222997	0.909794802330656\\
0.121668471397888	0.89282047315137\\
0.145127789278423	0.87533005267389\\
0.17031520429155	0.857460120621421\\
0.197167556059509	0.839303865158322\\
0.225625789674533	0.82091711889315\\
0.255634585613067	0.802329569239745\\
0.287142031567055	0.783548625568652\\
0.320099330538618	0.764568636105999\\
0.35446054042221	0.74537942444111\\
0.39018234102294	0.725974081674917\\
0.42722382506056	0.706352428963987\\
0.465546310209241	0.686526649757063\\
0.50511316964193	0.666527541589637\\
0.545889678899884	0.646403881623527\\
0.587842877204593	0.626224652266158\\
0.630941441580349	0.60607934069939\\
0.675155572369048	0.58607761491133\\
0.720456888900558	0.566343858249093\\
0.814214088046517	0.528231333304968\\
0.912010950317074	0.492867102783934\\
1	0.465183288287626\\
1.01366277027041	0.46124148149623\\
1.07142857142857	0.445620645607073\\
1.15384615384615	0.42614938655292\\
1.25	0.407246498869712\\
1.36363636363636	0.389394417881972\\
1.5	0.37296512564044\\
1.66666666666667	0.358127888441761\\
1.875	0.344839219855552\\
2.14285714285714	0.332935085461196\\
2.5	0.322261561978444\\
3	0.312751690272575\\
3.75	0.304444409320932\\
5	0.297460806437217\\
7.5	0.291986493924283\\
10	0.28988856693157\\
15	0.288262484787515\\
21.5443469003188	0.287524784719825\\
35.9381366380463	0.287026785673832\\
100	0.286711252851777\\
};
\addplot [color=black, line width=2.0pt, forget plot]
  table[row sep=crcr]{%
0.00553220719809771	0.990965845497349\\
0.01	0.983481633179406\\
0.012230757294587	0.979725062917651\\
0.0213755451908948	0.964353280790751\\
0.0328492202200639	0.945420817053959\\
0.0465441846922757	0.923668834020021\\
0.0623614859398746	0.899916413818256\\
0.0802098649172976	0.874973319694246\\
0.100004935222997	0.849573717221181\\
0.121668471397888	0.824336178261224\\
0.145127789278423	0.799747852485396\\
0.17031520429155	0.776166782454208\\
0.197167556059509	0.753835481224428\\
0.320099330538618	0.678890087476901\\
0.35446054042221	0.663733476603026\\
0.39018234102294	0.64989032243151\\
0.42722382506056	0.637274463449346\\
0.465546310209241	0.62579521022791\\
0.50511316964193	0.615361367069925\\
0.545889678899884	0.605883972209878\\
0.587842877204593	0.597278075359178\\
0.630941441580349	0.589463808114204\\
0.675155572369048	0.582366946174675\\
0.720456888900558	0.575919114884348\\
0.766818334237483	0.570057751412525\\
0.862619486763344	0.559871949636884\\
0.962365914765952	0.551415604448068\\
1.07142857142857	0.544031104426405\\
1.15384615384615	0.539436910714492\\
1.25	0.534913624237194\\
1.36363636363636	0.530485559440762\\
1.5	0.526181016597955\\
1.66666666666667	0.522032874371841\\
1.875	0.518079232142679\\
2.14285714285714	0.514364069608714\\
2.5	0.510937851292804\\
3	0.507857927160332\\
3.75	0.505188435886336\\
5	0.502999151065461\\
7.5	0.501362258865684\\
12.9154966501488	0.5004651868387\\
27.8255940220713	0.500100752848717\\
100	0.5000078115995\\
};
\end{axis}
\end{tikzpicture}%

%% file: images/2absTheo.tex
%
%
\definecolor{mycolor1}{rgb}{0.09053,0.13734,0.07326}%
\definecolor{mycolor2}{rgb}{0.04342,0.37461,0.17524}%
\definecolor{mycolor3}{rgb}{0.37697,0.57257,0.04618}%
\definecolor{mycolor4}{rgb}{0.79072,0.73098,0.27064}%
\begin{tikzpicture}[%
trim axis left, trim axis right
]

\begin{axis}[%
width=0.951\fwidth,
height=\fheight,
at={(0\fwidth,0\fheight)},
scale only axis,
xmode=log,
xmin=0.01,
xmax=100,
xminorticks=true,
xlabel style={font=\color{white!15!black}},
xlabel={$k$},
ymin=0,
ymax=1,
ylabel style={font=\color{white!15!black}},
ylabel={$|C(k)|$},
axis background/.style={fill=white},
xmajorgrids,
xminorgrids,
ymajorgrids
]
\addplot [color=mycolor1, line width=1.0pt, forget plot]
  table[row sep=crcr]{%
0.00553220719809771	0.992139076321192\\
0.01	0.985656920353871\\
0.012230757294587	0.98240787855316\\
0.0213755451908948	0.969120764321672\\
0.0328492202200639	0.952725527908918\\
0.0465441846922757	0.933804690807783\\
0.0623614859398746	0.912998667086872\\
0.0802098649172976	0.890937702691569\\
0.100004935222997	0.868217597686061\\
0.121668471397888	0.845345017112754\\
0.145127789278423	0.822720298147363\\
0.17031520429155	0.800661149140863\\
0.197167556059509	0.779384126838339\\
0.225625789674533	0.759019400238854\\
0.255634585613067	0.739646894279117\\
0.320099330538618	0.703924202339187\\
0.50511316964193	0.630242698561001\\
0.912010950317074	0.534966922090566\\
1.07142857142857	0.50955379074654\\
1.25	0.486066390656155\\
1.36363636363636	0.473359385182103\\
1.5	0.460034805767016\\
1.66666666666667	0.446156730883191\\
1.875	0.431846063715189\\
2.14285714285714	0.417284291923611\\
2.5	0.402733076629924\\
3	0.388538767021635\\
3.75	0.375135828485675\\
5	0.363050125749055\\
7.5	0.352907077423404\\
10	0.348800054078952\\
12.9154966501488	0.346465082754949\\
16.6810053720006	0.344934559739713\\
21.5443469003188	0.343938710404607\\
35.9381366380463	0.342870425139474\\
59.9484250318941	0.342417323239831\\
100	0.342193897769765\\
};
\addplot [color=mycolor2, line width=1.0pt, forget plot]
  table[row sep=crcr]{%
0.00553220719809771	0.992137172540738\\
0.01	0.985651742926767\\
0.012230757294587	0.982400687890585\\
0.0213755451908948	0.969103732268059\\
0.0328492202200639	0.952694800500665\\
0.0465441846922757	0.933759167375135\\
0.0623614859398746	0.912941801793598\\
0.0802098649172976	0.890878628972895\\
0.100004935222997	0.8681719995278\\
0.121668471397888	0.845335139257122\\
0.145127789278423	0.822774192766172\\
0.17031520429155	0.800812146094074\\
0.197167556059509	0.77966972830109\\
0.225625789674533	0.759479594999652\\
0.287142031567055	0.722221417731726\\
0.814214088046517	0.558324446205208\\
1.01366277027041	0.523085566758651\\
1.5	0.459503605192817\\
1.66666666666667	0.44304059829766\\
1.875	0.425430909606324\\
2.14285714285714	0.406819025452446\\
2.5	0.387509539057092\\
3	0.368001934983674\\
3.75	0.349007529579532\\
5	0.331446370533033\\
7.5	0.3164267545826\\
10	0.310273954351986\\
12.9154966501488	0.306753938901523\\
16.6810053720006	0.304442903486851\\
21.5443469003188	0.302927516779581\\
27.8255940220713	0.301937474823351\\
46.4158883361278	0.300875205239688\\
100	0.300294823360528\\
};
\addplot [color=mycolor3, line width=1.0pt, forget plot]
  table[row sep=crcr]{%
0.00553220719809771	0.992135421860574\\
0.01	0.985647057756217\\
0.012230757294587	0.982394229432867\\
0.0213755451908948	0.969088869624876\\
0.0328492202200639	0.952669017093955\\
0.0465441846922757	0.933723156138396\\
0.0623614859398746	0.912901187675345\\
0.0802098649172976	0.890845051600178\\
0.100004935222997	0.86816392731926\\
0.121668471397888	0.845377865798983\\
0.145127789278423	0.822899224889276\\
0.17031520429155	0.801056861729477\\
0.197167556059509	0.780076453312062\\
0.225625789674533	0.760094125609599\\
0.287142031567055	0.723399967118795\\
0.42722382506056	0.662555002564236\\
0.545889678899884	0.6258306826198\\
0.862619486763344	0.557828453893408\\
1	0.535021451677589\\
1.07142857142857	0.524074622152334\\
1.15384615384615	0.512068558881579\\
1.25	0.498795133414037\\
1.36363636363636	0.484011309294849\\
1.5	0.467431446110548\\
2.14285714285714	0.404009965586477\\
2.5	0.377787835275816\\
3	0.349484918829883\\
3.75	0.320230721125621\\
5	0.291912124330123\\
7.5	0.267036719046231\\
10	0.256775405833475\\
12.9154966501488	0.250911424091219\\
15	0.248469172079703\\
16.6810053720006	0.247069774831598\\
21.5443469003188	0.244562708599341\\
27.8255940220713	0.242928843066009\\
35.9381366380463	0.241863570435531\\
59.9484250318941	0.240708876363995\\
100	0.240230047655016\\
};
\addplot [color=mycolor4, line width=1.0pt, forget plot]
  table[row sep=crcr]{%
0.00553220719809771	0.992133824190596\\
0.01	0.985642866737879\\
0.012230757294587	0.982388503272528\\
0.0213755451908948	0.969076186804459\\
0.0328492202200639	0.952648188990452\\
0.0465441846922757	0.933696660105361\\
0.0623614859398746	0.912876846289974\\
0.0802098649172976	0.89083701352161\\
0.100004935222997	0.868193448456153\\
0.121668471397888	0.845473334122582\\
0.145127789278423	0.823095582988079\\
0.17031520429155	0.801395461631465\\
0.197167556059509	0.780604114868889\\
0.225625789674533	0.760862378901695\\
0.42722382506056	0.666028366709831\\
0.50511316964193	0.642266785033307\\
0.587842877204593	0.621401339910264\\
0.720456888900558	0.594183790289552\\
0.962365914765952	0.555624685834185\\
1.07142857142857	0.540806913559893\\
1.15384615384615	0.530239194249778\\
1.25	0.518419304633306\\
1.36363636363636	0.504997420295899\\
1.5	0.489495090449517\\
1.66666666666667	0.471270475377072\\
1.875	0.449467923765683\\
2.14285714285714	0.422981700727707\\
2.5	0.390508686322347\\
3	0.350835485210139\\
3.75	0.303705215920974\\
5	0.251612397041261\\
7.5	0.201637960894674\\
10	0.180643695775843\\
12.9154966501488	0.168812053925852\\
15	0.163965595508988\\
16.6810053720006	0.161218621393828\\
21.5443469003188	0.156377360665308\\
27.8255940220713	0.153292577826967\\
35.9381366380463	0.151323260377274\\
46.4158883361278	0.150063999948399\\
77.4263682681127	0.148732944802072\\
100	0.148394111222989\\
};
\addplot [color=black, line width=2.0pt, forget plot]
  table[row sep=crcr]{%
0.00553220719809771	0.990965845497349\\
0.01	0.983481633179406\\
0.012230757294587	0.979725062917651\\
0.0213755451908948	0.964353280790751\\
0.0328492202200639	0.945420817053959\\
0.0465441846922757	0.923668834020021\\
0.0623614859398746	0.899916413818256\\
0.0802098649172976	0.874973319694246\\
0.100004935222997	0.849573717221181\\
0.121668471397888	0.824336178261224\\
0.145127789278423	0.799747852485396\\
0.17031520429155	0.776166782454208\\
0.197167556059509	0.753835481224428\\
0.320099330538618	0.678890087476901\\
0.35446054042221	0.663733476603026\\
0.39018234102294	0.64989032243151\\
0.42722382506056	0.637274463449346\\
0.465546310209241	0.62579521022791\\
0.50511316964193	0.615361367069925\\
0.545889678899884	0.605883972209878\\
0.587842877204593	0.597278075359178\\
0.630941441580349	0.589463808114204\\
0.675155572369048	0.582366946174675\\
0.720456888900558	0.575919114884348\\
0.766818334237483	0.570057751412525\\
0.862619486763344	0.559871949636884\\
0.962365914765952	0.551415604448068\\
1.07142857142857	0.544031104426405\\
1.15384615384615	0.539436910714492\\
1.25	0.534913624237194\\
1.36363636363636	0.530485559440762\\
1.5	0.526181016597955\\
1.66666666666667	0.522032874371841\\
1.875	0.518079232142679\\
2.14285714285714	0.514364069608714\\
2.5	0.510937851292804\\
3	0.507857927160332\\
3.75	0.505188435886336\\
5	0.502999151065461\\
7.5	0.501362258865684\\
12.9154966501488	0.5004651868387\\
27.8255940220713	0.500100752848717\\
100	0.5000078115995\\
};
\end{axis}
\end{tikzpicture}%

%% file: images/1phaseTheo.tex
%
%
\definecolor{mycolor1}{rgb}{0.18517,0.05913,0.24304}%
\definecolor{mycolor2}{rgb}{0.51514,0.10993,0.38770}%
\definecolor{mycolor3}{rgb}{0.81038,0.26571,0.33825}%
\definecolor{mycolor4}{rgb}{0.94901,0.58547,0.40375}%
\begin{tikzpicture}[%
trim axis left, trim axis right
]

\begin{axis}[%
width=0.951\fwidth,
height=\fheight,
at={(0\fwidth,0\fheight)},
scale only axis,
xmode=log,
xmin=0.01,
xmax=100,
xminorticks=true,
xlabel style={font=\color{white!15!black}},
xlabel={$k$},
ymin=-0.8,
ymax=0,
ylabel style={font=\color{white!15!black}},
ylabel={$\angle C(k)$},
axis background/.style={fill=white},
xmajorgrids,
xminorgrids,
ymajorgrids
]
\addplot [color=mycolor1, line width=1.0pt, forget plot]
  table[row sep=crcr]{%
0.00553220719809771	-0.0269272730361814\\
0.01	-0.0432685743915799\\
0.012230757294587	-0.050639498617552\\
0.0213755451908948	-0.0772542321658554\\
0.0328492202200639	-0.105081894800071\\
0.0465441846922757	-0.132881497670144\\
0.0623614859398746	-0.159769419011913\\
0.0802098649172976	-0.18516090953515\\
0.100004935222997	-0.208682536215744\\
0.17031520429155	-0.266937434534078\\
0.197167556059509	-0.282451475527568\\
0.225625789674533	-0.296287307001841\\
0.255634585613067	-0.308649082697105\\
0.287142031567055	-0.31974982894176\\
0.320099330538618	-0.329804188067391\\
0.39018234102294	-0.34753269892273\\
0.50511316964193	-0.370526428583519\\
0.587842877204593	-0.384793397955128\\
0.630941441580349	-0.39183233426708\\
0.675155572369048	-0.398870173192105\\
0.720456888900558	-0.405932822910606\\
0.766818334237483	-0.413037466043869\\
0.814214088046517	-0.420192177852992\\
0.862619486763344	-0.427393265082853\\
0.912010950317074	-0.434633437052276\\
1	-0.447243418661619\\
1.07142857142857	-0.457179584962131\\
1.15384615384615	-0.468263322257297\\
1.25	-0.480606274277523\\
1.5	-0.509231132977879\\
1.66666666666667	-0.525188085265092\\
1.875	-0.541409970063\\
2.14285714285714	-0.556278966548553\\
2.5	-0.566595323481522\\
3	-0.566500598981586\\
3.75	-0.54622914347044\\
5	-0.491837043762917\\
7.5	-0.387909791485156\\
10	-0.313717874227537\\
12.9154966501488	-0.254662175690319\\
15	-0.224056221703454\\
16.6810053720006	-0.2041511787721\\
21.5443469003188	-0.162242054666644\\
27.8255940220713	-0.128166591800275\\
35.9381366380463	-0.100801585824338\\
46.4158883361278	-0.0790464825795469\\
59.9484250318941	-0.061789312461026\\
77.4263682681127	-0.0482244727601899\\
100	-0.0376099524837126\\
};
\addplot [color=mycolor2, line width=1.0pt, forget plot]
  table[row sep=crcr]{%
0.00553220719809771	-0.0246149291866433\\
0.01	-0.0397927011595511\\
0.012230757294587	-0.0466909300432392\\
0.0213755451908948	-0.0718800559092925\\
0.0328492202200639	-0.0987327177303068\\
0.0465441846922757	-0.126191426695972\\
0.0623614859398746	-0.153495135437134\\
0.0802098649172976	-0.180120948021263\\
0.100004935222997	-0.205725192925274\\
0.121668471397888	-0.230117003619995\\
0.145127789278423	-0.253224538637584\\
0.17031520429155	-0.275055995960985\\
0.197167556059509	-0.295679246489436\\
0.225625789674533	-0.315199467635774\\
0.255634585613067	-0.333738281440295\\
0.287142031567055	-0.351419989284004\\
0.320099330538618	-0.36836401594159\\
0.35446054042221	-0.384673194925157\\
0.39018234102294	-0.40043541139055\\
0.42722382506056	-0.415715932460176\\
0.465546310209241	-0.43055677818766\\
0.50511316964193	-0.44498460117357\\
0.587842877204593	-0.472596074098127\\
0.814214088046517	-0.532871407315388\\
0.862619486763344	-0.543017322697425\\
0.912010950317074	-0.552389968278781\\
0.962365914765952	-0.560934197718177\\
1.01366277027041	-0.568606124305453\\
1.07142857142857	-0.576020774041227\\
1.15384615384615	-0.584424076256225\\
1.25	-0.591167185450982\\
1.36363636363636	-0.595257995047208\\
1.5	-0.595335431552793\\
1.66666666666667	-0.589653718533315\\
1.875	-0.576120912974118\\
2.14285714285714	-0.552504042802346\\
2.5	-0.516771259141225\\
3	-0.467427745429549\\
3.75	-0.403578040572851\\
5	-0.324795674097869\\
7.5	-0.231016817726257\\
10	-0.178628149489299\\
12.9154966501488	-0.141130652759322\\
15	-0.122696605371398\\
16.6810053720006	-0.111008726618874\\
21.5443469003188	-0.0870240448534023\\
27.8255940220713	-0.0680593480596747\\
35.9381366380463	-0.0531231804571766\\
46.4158883361278	-0.041372706024299\\
59.9484250318941	-0.0322145717106466\\
77.4263682681127	-0.025090587373902\\
100	-0.0194672791092585\\
};
\addplot [color=mycolor3, line width=1.0pt, forget plot]
  table[row sep=crcr]{%
0.00553220719809771	-0.0224633226262667\\
0.01	-0.0364721611644474\\
0.012230757294587	-0.0428736540648926\\
0.0213755451908948	-0.0664366579513929\\
0.0328492202200639	-0.0919027881615526\\
0.0465441846922757	-0.118369868444072\\
0.0623614859398746	-0.145186539525817\\
0.0802098649172976	-0.17189539198951\\
0.100004935222997	-0.198188776685596\\
0.121668471397888	-0.223878250362083\\
0.145127789278423	-0.248865230051519\\
0.17031520429155	-0.273116645297607\\
0.197167556059509	-0.296641096046931\\
0.225625789674533	-0.319470617408219\\
0.255634585613067	-0.341648835401819\\
0.287142031567055	-0.363215733766721\\
0.320099330538618	-0.384198739502455\\
0.35446054042221	-0.40460899511522\\
0.39018234102294	-0.424439809143204\\
0.42722382506056	-0.443659938274591\\
0.50511316964193	-0.480029729100718\\
0.587842877204593	-0.513065914642087\\
0.630941441580349	-0.528060222616114\\
0.675155572369048	-0.541878564944128\\
0.720456888900558	-0.554398035734451\\
0.766818334237483	-0.565496715102806\\
0.814214088046517	-0.575072620193078\\
0.862619486763344	-0.583036315500497\\
0.912010950317074	-0.589318235776743\\
0.962365914765952	-0.593879725526349\\
1.01366277027041	-0.596715453574384\\
1.07142857142857	-0.597874409508384\\
1.15384615384615	-0.596178594840063\\
1.25	-0.590010993709251\\
1.36363636363636	-0.578296079625682\\
1.5	-0.560060936291455\\
1.66666666666667	-0.534675444788829\\
1.875	-0.501967233515066\\
2.14285714285714	-0.462166496598954\\
2.5	-0.415647559087161\\
3	-0.362656512816251\\
3.75	-0.303188764241413\\
5	-0.237143645008623\\
7.5	-0.164513915157262\\
10	-0.125790155865886\\
12.9154966501488	-0.0986720664591463\\
15	-0.0855016613884931\\
16.6810053720006	-0.0771971443404662\\
21.5443469003188	-0.0602725750427306\\
27.8255940220713	-0.0469809093969746\\
35.9381366380463	-0.0365784979612074\\
46.4158883361278	-0.0284604088425189\\
59.9484250318941	-0.0220941794170155\\
77.4263682681127	-0.0171985850048091\\
100	-0.0132877829151781\\
};
\addplot [color=mycolor4, line width=1.0pt, forget plot]
  table[row sep=crcr]{%
0.00553220719809771	-0.0205617683650017\\
0.01	-0.0334862454630493\\
0.012230757294587	-0.0394147090634984\\
0.0213755451908948	-0.0613599622201173\\
0.0328492202200639	-0.085308705712543\\
0.0465441846922757	-0.110483417008192\\
0.0623614859398746	-0.136324196227697\\
0.0802098649172976	-0.162433673271528\\
0.100004935222997	-0.188540199606871\\
0.121668471397888	-0.214466676491837\\
0.145127789278423	-0.240103796153687\\
0.17031520429155	-0.26539317957579\\
0.197167556059509	-0.290303933800573\\
0.225625789674533	-0.314816631335298\\
0.255634585613067	-0.338916636402634\\
0.287142031567055	-0.362577998098983\\
0.320099330538618	-0.385757061900927\\
0.35446054042221	-0.408388756997861\\
0.39018234102294	-0.430386054903099\\
0.465546310209241	-0.47198764701815\\
0.50511316964193	-0.491286658637038\\
0.545889678899884	-0.509345128500049\\
0.587842877204593	-0.525961417537669\\
0.630941441580349	-0.540927000738356\\
0.675155572369048	-0.554037336270096\\
0.720456888900558	-0.5650963631996\\
0.766818334237483	-0.573927754998506\\
0.814214088046517	-0.580398484197653\\
0.862619486763344	-0.584423881124467\\
0.912010950317074	-0.585976565991007\\
0.962365914765952	-0.585102132901445\\
1	-0.58297830314935\\
1.01366277027041	-0.581921462146414\\
1.07142857142857	-0.575954946196732\\
1.15384615384615	-0.56395697271082\\
1.25	-0.546404608501724\\
1.36363636363636	-0.523141726799488\\
1.5	-0.494513340733788\\
1.875	-0.424220598228393\\
2.14285714285714	-0.383824663789763\\
2.5	-0.340164472813113\\
3	-0.293051119884361\\
3.75	-0.242198101103109\\
5	-0.187417051871253\\
7.5	-0.128704726902276\\
10	-0.0979290565965036\\
12.9154966501488	-0.0765688047602247\\
15	-0.0662441670164911\\
16.6810053720006	-0.0597490767579418\\
21.5443469003188	-0.0465554945031688\\
27.8255940220713	-0.0362304897781338\\
35.9381366380463	-0.0281636439291297\\
46.4158883361278	-0.0219002692967432\\
59.9484250318941	-0.017001527498727\\
77.4263682681127	-0.0132031745464465\\
100	-0.0101454609177036\\
};
\addplot [color=black, line width=2.0pt, forget plot]
  table[row sep=crcr]{%
0.00553220719809771	-0.0291290977323722\\
0.01	-0.0464355424958702\\
0.012230757294587	-0.0541648542151782\\
0.0213755451908948	-0.0816669558691929\\
0.0328492202200639	-0.109668342073344\\
0.0465441846922757	-0.136752600739155\\
0.0623614859398746	-0.16192637845729\\
0.100004935222997	-0.204229776318171\\
0.121668471397888	-0.220845102456179\\
0.145127789278423	-0.234406862529273\\
0.17031520429155	-0.245052249049447\\
0.197167556059509	-0.252995466969203\\
0.225625789674533	-0.258495388874061\\
0.255634585613067	-0.261831022247178\\
0.287142031567055	-0.263283780432803\\
0.320099330538618	-0.26312533871952\\
0.35446054042221	-0.261609867309062\\
0.39018234102294	-0.258969563070517\\
0.42722382506056	-0.255412579530849\\
0.465546310209241	-0.251122636293842\\
0.50511316964193	-0.24625975335341\\
0.545889678899884	-0.240961694335202\\
0.587842877204593	-0.235345814608767\\
0.630941441580349	-0.229511097908425\\
0.675155572369048	-0.223540232254205\\
0.720456888900558	-0.217501626374775\\
0.814214088046517	-0.205434647798771\\
0.912010950317074	-0.1936398479718\\
1.07142857142857	-0.176319038781258\\
1.36363636363636	-0.150280316883378\\
1.5	-0.140267098687494\\
1.66666666666667	-0.129528742803772\\
1.875	-0.118035332576786\\
2.14285714285714	-0.105763771401587\\
2.5	-0.0927015171423289\\
3	-0.0788516868843159\\
3.75	-0.0642399117482273\\
5	-0.0489232264504231\\
7.5	-0.0330007261942566\\
10	-0.0248573977536779\\
12.9154966501488	-0.0192898219785729\\
16.6810053720006	-0.0149559470367082\\
21.5443469003188	-0.0115894616345149\\
27.8255940220713	-0.00897778556651563\\
35.9381366380463	-0.00695326231416171\\
46.4158883361278	-0.00538462997889377\\
59.9484250318941	-0.00416957488679159\\
100	-0.00249985420094134\\
};
\end{axis}
\end{tikzpicture}%

%% file: images/2phaseTheo.tex
%
%
\definecolor{mycolor1}{rgb}{0.09053,0.13734,0.07326}%
\definecolor{mycolor2}{rgb}{0.04342,0.37461,0.17524}%
\definecolor{mycolor3}{rgb}{0.37697,0.57257,0.04618}%
\definecolor{mycolor4}{rgb}{0.79072,0.73098,0.27064}%
\begin{tikzpicture}[%
trim axis left, trim axis right
]

\begin{axis}[%
width=0.951\fwidth,
height=\fheight,
at={(0\fwidth,0\fheight)},
scale only axis,
xmode=log,
xmin=0.01,
xmax=100,
xminorticks=true,
xlabel style={font=\color{white!15!black}},
xlabel={$k$},
ymin=-0.8,
ymax=0,
ylabel style={font=\color{white!15!black}},
ylabel={$\angle C(k)$},
axis background/.style={fill=white},
xmajorgrids,
xminorgrids,
ymajorgrids
]
\addplot [color=mycolor1, line width=1.0pt, forget plot]
  table[row sep=crcr]{%
0.00553220719809771	-0.026684396366687\\
0.01	-0.0428326071922651\\
0.012230757294587	-0.050108309791725\\
0.0213755451908948	-0.076341068176899\\
0.0328492202200639	-0.103707007999864\\
0.0465441846922757	-0.130981251381145\\
0.0623614859398746	-0.157294168030164\\
0.0802098649172976	-0.18207028721665\\
0.100004935222997	-0.204944479260087\\
0.145127789278423	-0.244423495248995\\
0.17031520429155	-0.261054079132567\\
0.197167556059509	-0.27575707724881\\
0.225625789674533	-0.288704050765404\\
0.255634585613067	-0.300077219385738\\
0.287142031567055	-0.310061400420766\\
0.320099330538618	-0.318837585663752\\
0.35446054042221	-0.326564479358667\\
0.39018234102294	-0.33338436285614\\
0.42722382506056	-0.339414215310899\\
0.465546310209241	-0.344751714997128\\
0.50511316964193	-0.349474803821767\\
0.545889678899884	-0.353638818835285\\
0.587842877204593	-0.357288344003066\\
0.630941441580349	-0.360449792516641\\
0.675155572369048	-0.363146518006005\\
0.720456888900558	-0.365389017393529\\
0.766818334237483	-0.367184130887223\\
0.814214088046517	-0.368540012029224\\
0.862619486763344	-0.369460631186305\\
0.912010950317074	-0.369951404802032\\
0.962365914765952	-0.370016710500827\\
1.01366277027041	-0.369671228630029\\
1.07142857142857	-0.368823125564117\\
1.15384615384615	-0.366861511776555\\
1.25	-0.363610113617897\\
1.36363636363636	-0.35868412526026\\
1.5	-0.351595936986446\\
1.66666666666667	-0.341753261908451\\
1.875	-0.328455684470498\\
2.14285714285714	-0.310889699630929\\
2.5	-0.288170012465796\\
3	-0.259382691895068\\
3.75	-0.223652221419349\\
5	-0.180241520011034\\
7.5	-0.128648391198654\\
10	-0.0997124753981464\\
12.9154966501488	-0.0789258950223886\\
15	-0.0686694265691083\\
16.6810053720006	-0.0621578128043976\\
21.5443469003188	-0.0487755766232687\\
27.8255940220713	-0.0381771300303271\\
35.9381366380463	-0.0297998101579307\\
46.4158883361278	-0.0231970754572202\\
59.9484250318941	-0.0180691436722662\\
77.4263682681127	-0.0140803453700489\\
100	-0.0108056399321006\\
};
\addplot [color=mycolor2, line width=1.0pt, forget plot]
  table[row sep=crcr]{%
0.00553220719809771	-0.0267815463969212\\
0.01	-0.0430069734476155\\
0.012230757294587	-0.0503207554189808\\
0.0213755451908948	-0.0767062337229829\\
0.0328492202200639	-0.104256690188225\\
0.0465441846922757	-0.131740815650203\\
0.0623614859398746	-0.158283329796629\\
0.0802098649172976	-0.18330510162348\\
0.100004935222997	-0.206437912471231\\
0.17031520429155	-0.263410371341394\\
0.197167556059509	-0.278444325654394\\
0.225625789674533	-0.291757476945854\\
0.255634585613067	-0.303542171588571\\
0.287142031567055	-0.313996359616009\\
0.320099330538618	-0.32331540879328\\
0.35446054042221	-0.3316730928278\\
0.39018234102294	-0.339226557017025\\
0.465546310209241	-0.352430932151899\\
0.545889678899884	-0.36373735937628\\
0.630941441580349	-0.373626974728067\\
0.720456888900558	-0.382330369177969\\
0.814214088046517	-0.389897087794189\\
0.912010950317074	-0.396278927773321\\
1	-0.400780226162092\\
1.07142857142857	-0.403611368859219\\
1.15384615384615	-0.406002946119613\\
1.25	-0.407672822752839\\
1.36363636363636	-0.408217997088959\\
1.5	-0.407058611623202\\
1.66666666666667	-0.403394929182794\\
1.875	-0.396142444716718\\
2.14285714285714	-0.383879413544561\\
2.5	-0.364847156377774\\
3	-0.337003101358886\\
3.75	-0.298165591392708\\
5	-0.246292561399566\\
7.5	-0.179808683195538\\
10	-0.140808707018569\\
12.9154966501488	-0.112190851887013\\
15	-0.097917681950447\\
16.6810053720006	-0.0888111497338411\\
21.5443469003188	-0.0699554826288096\\
27.8255940220713	-0.0549018982518557\\
35.9381366380463	-0.0429592368409129\\
46.4158883361278	-0.0335646062762902\\
59.9484250318941	-0.0261442498414355\\
77.4263682681127	-0.0204118857622668\\
100	-0.0157987089567375\\
};
\addplot [color=mycolor3, line width=1.0pt, forget plot]
  table[row sep=crcr]{%
0.00553220719809771	-0.0268786962907974\\
0.01	-0.0431813780572621\\
0.012230757294587	-0.0505332404205414\\
0.0213755451908948	-0.0770715714310795\\
0.0328492202200639	-0.104806755634445\\
0.0465441846922757	-0.132501108169174\\
0.0623614859398746	-0.159273821160431\\
0.0802098649172976	-0.184541976832493\\
0.100004935222997	-0.207934017894113\\
0.17031520429155	-0.265764185882389\\
0.197167556059509	-0.281121143551764\\
0.225625789674533	-0.294787692794759\\
0.255634585613067	-0.306964000192442\\
0.287142031567055	-0.317859152211216\\
0.320099330538618	-0.327682012116561\\
0.39018234102294	-0.344845358099993\\
0.545889678899884	-0.373279716889223\\
0.630941441580349	-0.386094930085639\\
0.720456888900558	-0.398508414131534\\
0.814214088046517	-0.410639771929172\\
0.912010950317074	-0.422460597346178\\
1.07142857142857	-0.439882560523674\\
1.25	-0.456442021651041\\
1.36363636363636	-0.465233022467789\\
1.5	-0.473936849172536\\
1.66666666666667	-0.481879387362202\\
1.875	-0.487888403922087\\
2.14285714285714	-0.49000138957706\\
2.5	-0.485120403207636\\
3	-0.468664704034266\\
3.75	-0.434501397031327\\
5	-0.37557182450148\\
7.5	-0.285442399849968\\
10	-0.227379867739726\\
12.9154966501488	-0.183024055632364\\
15	-0.160463819859774\\
16.6810053720006	-0.14591880500292\\
21.5443469003188	-0.115550397767559\\
27.8255940220713	-0.0910558240593016\\
35.9381366380463	-0.0714877097273359\\
46.4158883361278	-0.0559292464059635\\
59.9484250318941	-0.0436909420740941\\
77.4263682681127	-0.0341053352476122\\
100	-0.0265558273272748\\
};
\addplot [color=mycolor4, line width=1.0pt, forget plot]
  table[row sep=crcr]{%
0.00553220719809771	-0.0269758485268206\\
0.01	-0.0433557785715823\\
0.012230757294587	-0.0507457605333017\\
0.0213755451908948	-0.0774369480557082\\
0.0328492202200639	-0.105357063241592\\
0.0465441846922757	-0.133261997612613\\
0.0623614859398746	-0.160265238996637\\
0.0802098649172976	-0.185780065747509\\
0.100004935222997	-0.209431158688509\\
0.17031520429155	-0.268106567294758\\
0.197167556059509	-0.283774286423173\\
0.225625789674533	-0.297773896688717\\
0.255634585613067	-0.310312254863764\\
0.287142031567055	-0.32160625531588\\
0.35446054042221	-0.341318295183387\\
0.465546310209241	-0.366421320899886\\
0.50511316964193	-0.37418988957382\\
0.545889678899884	-0.381842085577807\\
0.587842877204593	-0.389463137744412\\
0.630941441580349	-0.397115853144629\\
0.675155572369048	-0.404851634262642\\
0.720456888900558	-0.412705327100682\\
0.766818334237483	-0.420703106855314\\
0.814214088046517	-0.428860535968094\\
0.862619486763344	-0.437181399108511\\
0.912010950317074	-0.445667381210429\\
0.962365914765952	-0.454308670287151\\
1.01366277027041	-0.463096206897962\\
1.07142857142857	-0.472956647089403\\
1.15384615384615	-0.486925553882047\\
1.25	-0.503000726998191\\
1.36363636363636	-0.521566752724024\\
1.5	-0.543035738989972\\
1.66666666666667	-0.567785452188953\\
1.875	-0.596005354882519\\
2.14285714285714	-0.627327756466408\\
2.5	-0.659996684416062\\
3	-0.689043636548858\\
3.75	-0.702600311435944\\
5	-0.676222698335968\\
7.5	-0.570542507129185\\
10	-0.474912200308631\\
12.9154966501488	-0.391755944330581\\
15	-0.346874487775812\\
16.6810053720006	-0.317170617221464\\
21.5443469003188	-0.25358974052291\\
27.8255940220713	-0.201045698442969\\
35.9381366380463	-0.158475245679987\\
46.4158883361278	-0.124435761998082\\
59.9484250318941	-0.0974592582054155\\
77.4263682681127	-0.0761262434909975\\
100	-0.0593831592838425\\
};
\addplot [color=black, line width=2.0pt, forget plot]
  table[row sep=crcr]{%
0.00553220719809771	-0.0291290977323722\\
0.01	-0.0464355424958702\\
0.012230757294587	-0.0541648542151782\\
0.0213755451908948	-0.0816669558691929\\
0.0328492202200639	-0.109668342073344\\
0.0465441846922757	-0.136752600739155\\
0.0623614859398746	-0.16192637845729\\
0.100004935222997	-0.204229776318171\\
0.121668471397888	-0.220845102456179\\
0.145127789278423	-0.234406862529273\\
0.17031520429155	-0.245052249049447\\
0.197167556059509	-0.252995466969203\\
0.225625789674533	-0.258495388874061\\
0.255634585613067	-0.261831022247178\\
0.287142031567055	-0.263283780432803\\
0.320099330538618	-0.26312533871952\\
0.35446054042221	-0.261609867309062\\
0.39018234102294	-0.258969563070517\\
0.42722382506056	-0.255412579530849\\
0.465546310209241	-0.251122636293842\\
0.50511316964193	-0.24625975335341\\
0.545889678899884	-0.240961694335202\\
0.587842877204593	-0.235345814608767\\
0.630941441580349	-0.229511097908425\\
0.675155572369048	-0.223540232254205\\
0.720456888900558	-0.217501626374775\\
0.814214088046517	-0.205434647798771\\
0.912010950317074	-0.1936398479718\\
1.07142857142857	-0.176319038781258\\
1.36363636363636	-0.150280316883378\\
1.5	-0.140267098687494\\
1.66666666666667	-0.129528742803772\\
1.875	-0.118035332576786\\
2.14285714285714	-0.105763771401587\\
2.5	-0.0927015171423289\\
3	-0.0788516868843159\\
3.75	-0.0642399117482273\\
5	-0.0489232264504231\\
7.5	-0.0330007261942566\\
10	-0.0248573977536779\\
12.9154966501488	-0.0192898219785729\\
16.6810053720006	-0.0149559470367082\\
21.5443469003188	-0.0115894616345149\\
27.8255940220713	-0.00897778556651563\\
35.9381366380463	-0.00695326231416171\\
46.4158883361278	-0.00538462997889377\\
59.9484250318941	-0.00416957488679159\\
100	-0.00249985420094134\\
};
\end{axis}
\end{tikzpicture}%

%% file: sections/results-gust.tex
\subsection{Sears function}
\label{Sec:Gust}
\begin{figure}
	\centering
\includegraphics[width = \linewidth]{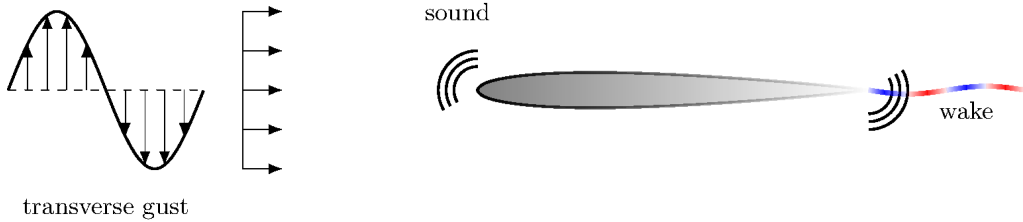}
\caption{The harmonic gust response problem considered by \cite{Sears1941}. A symmetric, stationary, porous aerofoil is subjected to an unsteady vertical sinusoidal disturbance convected at the velocity of the freestream, resulting in unsteady wake generation. Acoustic waves are also generated that can be modelled using the Sears response function.}
\label{Fig:searsSchematic}
\end{figure}

We now consider a uniform flow in the horizontal direction with a transverse sinusoidal gust, as illustrated in figure \ref{Fig:searsSchematic}. The gust convects at the free-stream velocity and has unit amplitude and reduced frequency $k$.  The interaction between the {harmonic gust generates unsteady lift on the stationary aerofoil}, as well as pressure perturbations that propagate to the acoustic far field as sound waves. {The unsteady lift response is described by the Sears function~\mbox{\citep{Sears1941}}, which may be written in closed form for an impermeable aerofoil as}
\begin{equation}
    S(k) = C(k) \left[ J_0(k)- \i J_1(k) \right] + \i J_1(k) \qquad \mbox{for } \psi(x,k) \equiv 0,
\end{equation}
where $J_n$ are Bessel functions of the first kind of order $n$. 

{To construct a porous extension of the Sears function, the harmonic gust of unitary amplitude enters the porous boundary condition \eqref{Eq:w01} as}
\begin{align}
w(x,t) &= w_s(x,t) -\e^{\i k (t-x) } .
\end{align}
Therefore, the forcing function $f_{\rm a}$ in the SF--VIE becomes
\begin{align}
f_{\rm a}(x) = \e^{-\i k x}, \label{Eq:searsForcing}
\end{align}
and we may apply the numerical scheme developed in \S\ref{Sec:numSol} {to find the resulting total unsteady lift: the Sears function}. Whilst the Volterra part of the SF--VIE renders it impossible to find analytic forms for the Sears function, our numerical scheme is sufficiently fast and robust that we may produce an accurate approximation for a range of porosity gradients.

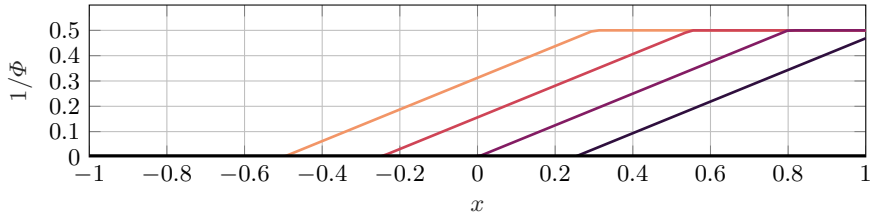
\begin{figure}
    \centering
    	\setlength{\fheight}{2cm}
		\setlength{\fwidth}{0.8\linewidth}
		\centering
		\input{images/searsPorosity.tex}
    \caption{The (reciprocal) piecewise linear flow resistance profiles used to compute the porous Sears function.
    The colours correspond to the curves in figure \ref{Fig:sears}.}
    \label{Fig:searsPorousProfiles}
\end{figure}

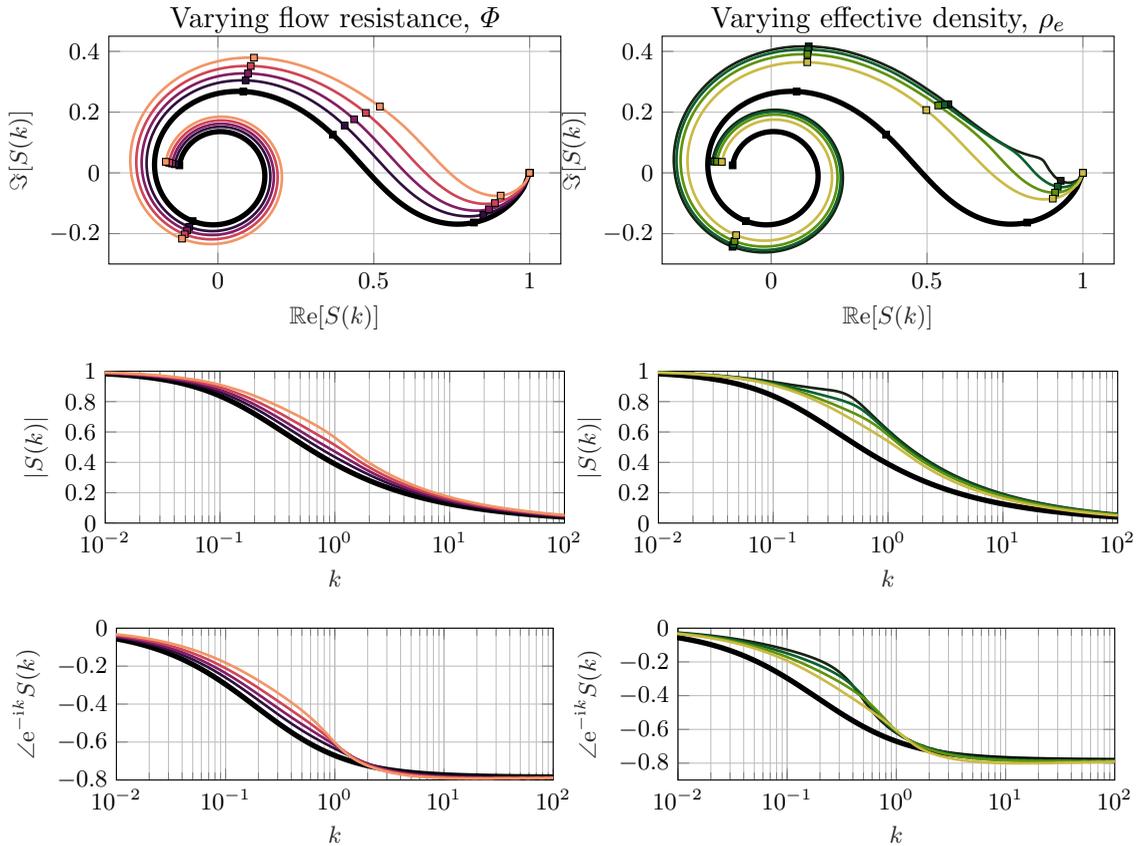
\begin{figure}
\centering
\vspace{.5cm}

{\large \uline{Porous Sears function}}\\

\vspace{.5cm}
%
%
	\begin{subfigure}[t]{.45\linewidth}
	\setlength{\fheight}{3cm}
		\setlength{\fwidth}{\linewidth}
		\centering
		{\large Varying flow resistance, $\Phi$}
		
	\input{images/1Sears.tex}
	\end{subfigure}
	\hfill
	\begin{subfigure}[t]{.45\linewidth}
	\setlength{\fheight}{3cm}
		\setlength{\fwidth}{\linewidth}
		\centering
		{\large Varying effective density, $\rho_e$}
		
	\input{images/2Sears.tex}
	\end{subfigure}
\vspace{.2cm}

%
%
	\begin{subfigure}[t]{.45\linewidth}
	\setlength{\fheight}{2cm}
		\setlength{\fwidth}{\linewidth}
		\centering
	\input{images/1absSears.tex}
	\end{subfigure}
	\hfill
	\begin{subfigure}[t]{.45\linewidth}
	\setlength{\fheight}{2cm}
		\setlength{\fwidth}{\linewidth}
		\centering
	\input{images/2absSears.tex}
	\end{subfigure}
\vspace{.2cm}

%
%
	\begin{subfigure}[t]{.45\linewidth}
	\setlength{\fheight}{2cm}
		\setlength{\fwidth}{\linewidth}
		\centering
	\input{images/1phaseSears.tex}
	\end{subfigure}
	\hfill
	\begin{subfigure}[t]{.45\linewidth}
	\setlength{\fheight}{2cm}
		\setlength{\fwidth}{\linewidth}
		\centering
	\input{images/2phaseSears.tex}
	\end{subfigure}
\caption{
Complex, magnitude and phase plots of the porous Sears function for a range of flow resistance distributions (left) and effective densities (right). 
The dimensionless flow resistance is given by a reciprocal piecewise linear profiles illustrated in figure \ref{Fig:searsPorousProfiles}.
The leading edge is impermeable in every case.
On the left, the colours correspond to the flow resistance profiles in figure \ref{Fig:searsPorousProfiles} and $\rho_e=1.5$;
on the right, the effective density takes values $\infty$, 4, 3, 2, 1 (black to green) and the flow resistance profile is given by the rightmost curve in figure \ref{Fig:searsPorousProfiles}.
The black curves represents the impermeable Sears function.
The points $k=0,0.1,1,2,5,10$ are indicated by $\square$ with $k=0$ representing the rightmost part of the curve.}
\label{Fig:sears}
\end{figure}

{We now explore the effects of aerofoil porosity distributions illustrated in figure~\ref{Fig:searsPorousProfiles} on the Sears gust response function. Figure~\ref{Fig:sears} presents the complex-valued information of the porous Sears function analogue as a function of the dimensionless flow resistance $\Phi$ and effective density $\rho_{\rm e}$ in the left and right columns, respectively. The porosity distribution has a fixed value of unity at the trailing edge for all cases considered, and the permeable length of the aerofoil is varied. We plot the unsteady lift normalised by the quasi-steady lift particular to the given porosity distribution, such that the Sears function asymptotes to unity for vanishing reduced frequency. This quasi-steady lift follows from \eqref{Eq:qsSol} with the forcing on the right-hand side of \eqref{Eq:qsSIE} replaced by \eqref{Eq:searsForcing}. Note that the phase comparison in the bottom row of figure~\ref{Fig:sears} involves a factor of $\e^{-\i k}$ that unwinds the Argand diagram representation of the Sears function to more easily distinguish differences due to $\Phi$ and $\rho_{\rm e}$. }

{The results show that the magnitude of the porous Sears function increases with decreasing flow resistance. However, the actual unsteady lift is  smaller at a given reduced frequency when compared to the impermeable case, as the Sears function must be multiplied by the quasi-steady lift introduced in the normalisation. Changes in the phase relative to the impermeable aerofoil limit are most pronounced in this porosity configuration for reduced frequencies between {\it O}($10^{-1}$) and {\it O}(1) that lie in the typical range of aerospace interest. Note the crossover in the porous Sears function phase curves with the impermeable limit at a particular value of reduced frequency, which occurs for variations in both the dimensionless flow resistance and effective density in the bottom row of figure~\ref{Fig:sears}. Effective density variations do not lead to significant changes in the porous Sears function for the cases considered here, where the deviation between the porous and impermeable results in the right column of figure~\ref{Fig:sears} depend on the dimensionless flow resistance. At the limits of very small or large reduced frequency, the curves for the porous Sears function collapse to the classical impermeable result in terms of both phase and amplitude.}

%% file: images/searsPorosity.tex
%
%
\definecolor{mycolor1}{rgb}{0.94901,0.58547,0.40375}%
\definecolor{mycolor2}{rgb}{0.81038,0.26571,0.33825}%
\definecolor{mycolor3}{rgb}{0.51514,0.10993,0.38770}%
\definecolor{mycolor4}{rgb}{0.18517,0.05913,0.24304}%
\begin{tikzpicture}[%
trim axis left, trim axis right
]

\begin{axis}[%
width=0.951\fwidth,
height=\fheight,
at={(0\fwidth,0\fheight)},
scale only axis,
xmin=-1,
xmax=1,
xlabel style={font=\color{white!15!black}},
xlabel={$x$},
ymin=0,
ymax=0.6,
ylabel style={font=\color{white!15!black}},
ylabel={$1/\Phi$},
ytick = {0,0.1,0.2,0.3,0.4,0.5},
axis background/.style={fill=white},
xmajorgrids,
ymajorgrids
]
\addplot [color=mycolor1, line width=1.0pt, forget plot]
  table[row sep=crcr]{%
-1	2.22044604925031e-16\\
-0.515151515151515	2.22044604925031e-16\\
-0.494949494949495	0.00315656565656597\\
0.292929292929293	0.495580808080808\\
0.313131313131313	0.5\\
1	0.5\\
};
\addplot [color=mycolor2, line width=1.0pt, forget plot]
  table[row sep=crcr]{%
-1	2.22044604925031e-16\\
-0.252525252525253	2.22044604925031e-16\\
-0.232323232323232	0.0110479797979801\\
0.535353535353535	0.49084595959596\\
0.555555555555556	0.5\\
1	0.5\\
};
\addplot [color=mycolor3, line width=1.0pt, forget plot]
  table[row sep=crcr]{%
-1	2.22044604925031e-16\\
-0.0101010101010102	2.22044604925031e-16\\
0.0101010101010102	0.00631313131313149\\
0.797979797979798	0.498737373737374\\
0.818181818181818	0.5\\
1	0.5\\
};
\addplot [color=mycolor4, line width=1.0pt, forget plot]
  table[row sep=crcr]{%
-1	2.22044604925031e-16\\
0.232323232323232	2.22044604925031e-16\\
0.252525252525253	0.0015782828282831\\
1	0.46875\\
};
\addplot [color=black, line width=2.0pt, forget plot]
  table[row sep=crcr]{%
-1	2.22044604925031e-16\\
1	2.22044604925031e-16\\
};
\end{axis}
\end{tikzpicture}%

%% file: images/1Sears.tex
%
%
\definecolor{mycolor1}{rgb}{0.18517,0.05913,0.24304}%
\definecolor{mycolor2}{rgb}{0.51514,0.10993,0.38770}%
\definecolor{mycolor3}{rgb}{0.81038,0.26571,0.33825}%
\definecolor{mycolor4}{rgb}{0.94901,0.58547,0.40375}%
\begin{tikzpicture}[%
trim axis left, trim axis right
]

\begin{axis}[%
width=0.984\fwidth,
height=\fheight,
at={(0\fwidth,0\fheight)},
scale only axis,
xmin=-0.35,
xmax=1.1,
xlabel style={font=\color{white!15!black}},
xlabel={$\Re[S(k)]$},
ymin=-0.3,
ymax=0.45,
ylabel style={font=\color{white!15!black}},
ylabel={$\Im[S(k)]$},
axis background/.style={fill=white},
xmajorgrids,
ymajorgrids
]
\addplot [color=black, line width=2.0pt, forget plot]
  table[row sep=crcr]{%
1	-7.99360577730113e-15\\
0.981169634186679	-0.0473526935450792\\
0.975268104274319	-0.0572020387558559\\
0.969250159220743	-0.0662045018490595\\
0.96179894026922	-0.0762190791908565\\
0.955089041374125	-0.0843709356309031\\
0.947260980824945	-0.0930243239134291\\
0.93816286261376	-0.102103922357411\\
0.927637109934235	-0.111497621664997\\
0.919748936191671	-0.117859809709497\\
0.911111402206164	-0.124235586663284\\
0.901682744789075	-0.130556691091301\\
0.891425823994952	-0.136744181202643\\
0.880309851433787	-0.142708347454063\\
0.868312225503855	-0.14834898478152\\
0.855420391150047	-0.153556089833553\\
0.841633615179317	-0.158211033668738\\
0.826964543805801	-0.162188236031879\\
0.819307060233782	-0.163882063696621\\
0.803370321202988	-0.16659743199815\\
0.786647890344584	-0.168305997546178\\
0.769204741155661	-0.168877599492619\\
0.760236971327944	-0.168697801104816\\
0.751119434179592	-0.168187652747411\\
0.741863704712643	-0.167332969139702\\
0.732481873852179	-0.166120293321165\\
0.722986426197111	-0.164536976834457\\
0.713390104090285	-0.162571251099379\\
0.703705758999715	-0.16021228878552\\
0.693946191406285	-0.1574502542351\\
0.684123980561909	-0.15427634227784\\
0.674251305608945	-0.150682805110534\\
0.664339759631732	-0.146662967278347\\
0.654400158241382	-0.142211229184615\\
0.644442344274809	-0.137323059962614\\
0.634474990120298	-0.1319949809579\\
0.624505399069119	-0.126224541485567\\
0.61453930694264	-0.120010288936335\\
0.60458068506634	-0.11335173570327\\
0.594631545467335	-0.106249325783741\\
0.584691748973709	-0.09870440427718\\
0.574758816706798	-0.0907191933488399\\
0.564827745297672	-0.0822967785653936\\
0.544937468169077	-0.0641570214983668\\
0.524923633055708	-0.0443276395846814\\
0.504637440174471	-0.0228674490332269\\
0.483873491400335	0.000144213700549067\\
0.415723976688649	0.0772064911581731\\
0.389729326888095	0.104877858410841\\
0.368649165757727	0.125943361459841\\
0.352634315254576	0.141007303114233\\
0.33661244517805	0.155217856313328\\
0.320553419412667	0.168576603460661\\
0.304438033489619	0.181084867080371\\
0.288255978143667	0.192743815058999\\
0.272004237050724	0.203554606299833\\
0.255411263358432	0.213678645094021\\
0.234055716454205	0.225377999325029\\
0.222884997833748	0.230912887496748\\
0.206430334724944	0.238348090272243\\
0.187710215152482	0.245784286475277\\
0.17350431403378	0.250711995151768\\
0.157076612099213	0.255650316936921\\
0.136478809423763	0.260700549379704\\
0.124413573265463	0.263071509148478\\
0.10823064299701	0.265570520460204\\
0.0921831414008394	0.26727460145429\\
0.0762990722262464	0.268194933630997\\
0.0606067746712325	0.268344002957401\\
0.0451347399143001	0.267735617387968\\
0.0299114430411733	0.266384915162501\\
0.0149651883007658	0.264308364490887\\
0.000323966045512769	0.261523755190835\\
-0.0139846799547296	0.258050182805319\\
-0.0279337758698565	0.25390802569132\\
-0.041497032669719	0.249118915540627\\
-0.0546489531490399	0.243705701766897\\
-0.0686288832486455	0.237050215781053\\
-0.0796213468434863	0.231104196272839\\
-0.0972980097826508	0.220063655865227\\
-0.102666435162272	0.216308957959196\\
-0.11341353241244	0.20815740601082\\
-0.124234464526802	0.198991188452407\\
-0.133262477932314	0.190491938823868\\
-0.142330678715705	0.181038668032532\\
-0.150808001703733	0.171213324701674\\
-0.158681296369406	0.16104790129127\\
-0.169735296616782	0.144580211551503\\
-0.172570962965696	0.139827344744887\\
-0.178568894971875	0.128838512723266\\
-0.186501662329346	0.111639879169801\\
-0.18863705633415	0.106271724853657\\
-0.192698680664992	0.0947614979007226\\
-0.196109042081408	0.0831452984049081\\
-0.198868010821759	0.0714570064448898\\
-0.200977125904808	0.0597303527414158\\
-0.202439587028727	0.0479988325827109\\
-0.203260241489297	0.0362956203842292\\
-0.203445566208174	0.0246534851506137\\
-0.20300364497148	0.0131047071018584\\
-0.201333266630901	-0.00296295333153773\\
-0.200278264890796	-0.00958659053960986\\
-0.198018738299201	-0.0206677226165993\\
-0.195179753101202	-0.0315328770836255\\
-0.19156384721065	-0.0427578267705442\\
-0.187827252585544	-0.0525016282125648\\
-0.183349048824465	-0.0625508577896976\\
-0.178361900595442	-0.0722755095613348\\
-0.172886601005078	-0.0816511414075646\\
-0.166945087947514	-0.0906545172207666\\
-0.160560378018716	-0.0992636623413177\\
-0.153756497577954	-0.107457915317088\\
-0.146558411171408	-0.115217975843798\\
-0.138991947539056	-0.122525948755346\\
-0.131083723431618	-0.129365383945554\\
-0.122861065469326	-0.135721312115347\\
-0.112649286092279	-0.142671687872974\\
-0.10558482314758	-0.146930358762174\\
-0.0965887154405601	-0.151761204186247\\
-0.0811661765059353	-0.158635640810862\\
-0.071602364869464	-0.162076607473955\\
-0.0685213335637237	-0.163058547021581\\
-0.058905319596345	-0.165740676073208\\
-0.0492092069196274	-0.167875703215548\\
-0.0394629908115571	-0.169462870575536\\
-0.0258420158591939	-0.170763511849407\\
-0.0199395160943607	-0.170998554728085\\
-0.0102212849083549	-0.170953471429388\\
-0.000570841107991771	-0.170373288187499\\
0.00898327478092154	-0.169265042993867\\
0.0217843553004768	-0.166923211546473\\
0.0276910962602491	-0.165499905585972\\
0.0367906834332662	-0.162864368201025\\
0.0456858521321468	-0.159743397391186\\
0.0543514628733444	-0.156151066861503\\
0.0675986825787007	-0.149506422539469\\
0.0708979700068796	-0.147614931217082\\
0.0787332934161302	-0.142705420654688\\
0.0862480312663607	-0.137393023713003\\
0.0934221048326952	-0.131697614306333\\
0.100236613169862	-0.125640042803275\\
0.107265210377755	-0.118617351246705\\
0.112717497409968	-0.11252629039562\\
0.118352364162714	-0.105516074471686\\
0.123564721998202	-0.0982354810420756\\
0.128342186362612	-0.0907091890456738\\
0.132673773575952	-0.0829624195171222\\
0.136549923681114	-0.0750208577948502\\
0.139962519024264	-0.0669105747938097\\
0.142904898529865	-0.058657947576618\\
0.145371867646469	-0.0502895794580825\\
0.147359703952259	-0.0418322198786559\\
0.148866158422125	-0.0333126842822457\\
0.149890452370863	-0.0247577742329811\\
0.150433270099795	-0.0161941980040229\\
0.150496747286654	-0.00764849186929972\\
0.150084455171048	0.000853057673820823\\
0.149201380600005	0.00928449052646751\\
0.147853902010134	0.0176202482345498\\
0.145167205785574	0.0292099677329655\\
0.14379803263494	0.0339049527548516\\
0.14110908546753	0.0418054445707958\\
0.137994546607896	0.0495134887461394\\
0.134467256763089	0.0570066008410373\\
0.130541224514939	0.06426310900226\\
0.126231576945126	0.0712622138957334\\
0.120681089701074	0.0791548800998992\\
0.116527219180574	0.0844097181843579\\
0.111167869484359	0.0905213791864155\\
0.105495506866547	0.0963022582512683\\
0.099530009328079	0.101736710408229\\
0.0932920190778062	0.106810256509216\\
0.0868028755548973	0.111509619845336\\
0.0780363497013806	0.117034728281297\\
0.0731595588390656	0.119738896681972\\
0.0660509249406488	0.123248544996804\\
0.0587820722241779	0.126343526220585\\
0.0513767685820889	0.129016989910211\\
0.0438590484940528	0.131263425790089\\
0.036253138505795	0.133078672570692\\
0.0285833825187629	0.134459922665789\\
0.0208741671149411	0.135405722812535\\
0.013149847140167	0.135915970610858\\
0.00543467176766843	0.135991907010718\\
-0.00224728873871449	0.135636104787802\\
-0.00987221534511362	0.134852453060116\\
-0.017416611439057	0.133646137909559\\
-0.0248573726829798	0.132023619183995\\
-0.0371081350613749	0.128366596684498\\
-0.0463341102314899	0.124741955641358\\
-0.0531394872149213	0.121543678264606\\
-0.0597339177791402	0.117979535572902\\
-0.0660980182668967	0.114062941224785\\
-0.0722132324286016	0.10980832192033\\
-0.0780618837542911	0.105231067635578\\
-0.0836272248725045	0.100347479177515\\
-0.0888934838797204	0.0951747132298284\\
-0.0938459074737688	0.0897307250665687\\
-0.098470800774755	0.0840342091172249\\
-0.102755563727472	0.0781045375724664\\
-0.10668872399	0.0719616972250166\\
-0.110259966224164	0.0656262247446991\\
-0.113460157714717	0.0591191405906693\\
-0.116281370255489	0.0524618817671815\\
-0.120271490489491	0.0405714411106214\\
-0.122410273134046	0.0318082385334016\\
-0.123660931160608	0.0247705812964559\\
};
\addplot[only marks, mark=square*, mark options={}, mark size=1.2500pt, draw=black, fill=black] table[row sep=crcr]{%
x	y\\
1	-8.02904074443401e-15\\
};
\addplot[only marks, mark=square*, mark options={}, mark size=1.2500pt, draw=black, fill=black] table[row sep=crcr]{%
x	y\\
0.821241247189739	-0.163478447925458\\
};
\addplot[only marks, mark=square*, mark options={}, mark size=1.2500pt, draw=black, fill=black] table[row sep=crcr]{%
x	y\\
0.368649165757727	0.125943361459841\\
};
\addplot[only marks, mark=square*, mark options={}, mark size=1.2500pt, draw=black, fill=black] table[row sep=crcr]{%
x	y\\
0.081573858278389	0.267974495775782\\
};
\addplot[only marks, mark=square*, mark options={}, mark size=1.2500pt, draw=black, fill=black] table[row sep=crcr]{%
x	y\\
-0.0811661765059352	-0.158635640810862\\
};
\addplot[only marks, mark=square*, mark options={}, mark size=1.2500pt, draw=black, fill=black] table[row sep=crcr]{%
x	y\\
-0.123660931160608	0.0247705812964559\\
};
\addplot [color=mycolor1, line width=1.0pt, forget plot]
  table[row sep=crcr]{%
1	-0\\
0.98515143062174	-0.0389892540267567\\
0.97921010005526	-0.0488964136712877\\
0.972561179560874	-0.0586056106031074\\
0.96763602176106	-0.0650757266700273\\
0.964840569364267	-0.0671997236924016\\
0.9565498685523	-0.0767219908519134\\
0.949125206824771	-0.0843265156480011\\
0.934739872594101	-0.0977737809694224\\
0.923801838555356	-0.105925871907415\\
0.915642712661842	-0.111316665402057\\
0.906744174540994	-0.116593378131691\\
0.880582733917748	-0.128608455810312\\
0.868876332327727	-0.132791724734174\\
0.856403977400867	-0.136444268681152\\
0.843016713384769	-0.139405413321628\\
0.828806323726891	-0.14156611234469\\
0.821567143030649	-0.142371720829265\\
0.806171400194203	-0.143082376743897\\
0.790045573401288	-0.142658621406831\\
0.773005184032694	-0.140869700713621\\
0.764381168410732	-0.139512354043183\\
0.755615567500048	-0.137792067722159\\
0.738160266648153	-0.133375319166668\\
0.729016627920293	-0.130476970641972\\
0.719770316377722	-0.127162220573907\\
0.710430898662248	-0.12341975074772\\
0.700602041675002	-0.119107601610415\\
0.691112864449567	-0.114488145659617\\
0.681555363179551	-0.109413653985191\\
0.671935735896002	-0.10387714805507\\
0.662406400449614	-0.0979091107026222\\
0.652672763044243	-0.0914270027722022\\
0.642887883498644	-0.0844675808790718\\
0.632990794847577	-0.077022406591662\\
0.62310358982609	-0.0691028926058868\\
0.61315848928153	-0.0607009445887301\\
0.603119970701073	-0.0518027111928574\\
0.582860640698022	-0.032601166612612\\
0.562303877261741	-0.0115192628876317\\
0.540872023447828	0.0113899943630835\\
0.518870295718411	0.0360650192148906\\
0.470671513249621	0.089940965038074\\
0.443864226270025	0.118756728769488\\
0.42955141241555	0.133473215783736\\
0.406662207601443	0.155726508716223\\
0.389234744662812	0.171673233901842\\
0.371797300513473	0.186817132143532\\
0.354209587982836	0.200960161309152\\
0.336484704443896	0.214084320941051\\
0.318682232579945	0.226403141035401\\
0.300757019914595	0.237867766199042\\
0.282419339676647	0.248535435522412\\
0.258783196334557	0.260814166690461\\
0.246395146465455	0.266621033064745\\
0.228132324754826	0.274377432570387\\
0.207326947693516	0.282126192481068\\
0.191523808738982	0.287234671233988\\
0.173210045091807	0.292387016675873\\
0.150257051732662	0.297526436986022\\
0.136842040770598	0.299831216866483\\
0.118798745949406	0.302273362729106\\
0.10083558569231	0.303909570759701\\
0.0831108535515295	0.304629552565223\\
0.0655970335116551	0.304507539595044\\
0.0483888361246076	0.30352673190352\\
0.0313970831061863	0.301768059294155\\
0.0147133540696462	0.299213925979118\\
-0.00162673648235634	0.295875696751876\\
-0.0175985854744696	0.291787715355443\\
-0.033194567198529	0.286981017828585\\
-0.0483358900245157	0.281445621803492\\
-0.0630175829895048	0.275223497626895\\
-0.0785908035729894	0.267594459385351\\
-0.090859905694505	0.26081385064544\\
-0.110706842299147	0.24826254190129\\
-0.116698590251451	0.243999218020004\\
-0.128692429623291	0.234755149795755\\
-0.140767565758312	0.22437625935576\\
-0.150724564944669	0.214796077125238\\
-0.160842078759078	0.20412311033548\\
-0.170299930824754	0.193039923542693\\
-0.179099039783376	0.18156253178615\\
-0.187192424647809	0.169763801578183\\
-0.194586514757285	0.157662634385815\\
-0.201268437696603	0.145307811480307\\
-0.210108677610275	0.125963931255854\\
-0.212487577714462	0.11992839996522\\
-0.217008937071095	0.10698525832668\\
-0.220803809742702	0.0939319842686301\\
-0.223870916952161	0.0808010977259903\\
-0.226212007502711	0.0676300769404921\\
-0.227883540697951	0.0543988114353164\\
-0.228782052121319	0.0412553860994804\\
-0.228970389777144	0.0281825269792819\\
-0.228410491625533	0.0153051664651065\\
-0.226523742220752	-0.00272854510612253\\
-0.225336484302356	-0.0101618691922802\\
-0.222796849904558	-0.0226552428981495\\
-0.21960654410541	-0.0348483443605894\\
-0.215546693220701	-0.0474456362979876\\
-0.211354072661489	-0.0583802206102331\\
-0.206342043622525	-0.069612351572754\\
-0.200753410614986	-0.0805266584128292\\
-0.194619116437328	-0.0910495462164456\\
-0.187963611530408	-0.101156339749867\\
-0.180803658682896	-0.110852952014063\\
-0.17318209108506	-0.120053170968025\\
-0.165119546899932	-0.128767068393807\\
-0.156652887259051	-0.136997633771803\\
-0.147794908109887	-0.144682440130859\\
-0.13858508017273	-0.15182577781464\\
-0.127147567343741	-0.159640684556575\\
-0.11925568321905	-0.164374502180992\\
-0.109182897782502	-0.169812603324904\\
-0.0988862959828503	-0.174659997417525\\
-0.0883990862229351	-0.178908881792726\\
-0.0776836519002368	-0.182600473014979\\
-0.0669120993770715	-0.185631499132608\\
-0.0560500768444701	-0.188050109943848\\
-0.0451317403155675	-0.189855894048774\\
-0.0299481093045486	-0.191328777908415\\
-0.0233359777365745	-0.191612464381351\\
-0.0124483962146464	-0.191596233523163\\
-0.00163567726876068	-0.190981400637962\\
0.00912618603263482	-0.189792374783963\\
0.0234740623310985	-0.187214308847847\\
0.0300953601512488	-0.18564147178878\\
0.040297064830388	-0.182723193868432\\
0.0502713151716576	-0.179261264860967\\
0.0599548865533439	-0.17528654147874\\
0.074816320462199	-0.167901792338643\\
0.0785186794948993	-0.165798154155916\\
0.087312338015568	-0.160335371987173\\
0.0957236555803773	-0.154404517782966\\
0.103779192478488	-0.148062075699917\\
0.11143376534052	-0.141313646727576\\
0.11933204941532	-0.133487074044184\\
0.125502961686256	-0.126676541350789\\
0.13184078764294	-0.118856965576079\\
0.137706872041914	-0.110733629539133\\
0.143087526000736	-0.102334367717015\\
0.147969996372784	-0.0936870504565539\\
0.152314001041502	-0.084857023833353\\
0.156172453401505	-0.0758022590805942\\
0.159506019397858	-0.0665867784660761\\
0.16230807794613	-0.0572397424371793\\
0.164623783542283	-0.0477276617970577\\
0.166349683708615	-0.0382054561575582\\
0.167536429017831	-0.0286417451112055\\
0.168185298294583	-0.0190671143470511\\
0.168298042876517	-0.00951030393380359\\
0.167862122522282	-7.811251601475e-05\\
0.166920816497158	0.00935336870990633\\
0.165460409671465	0.0186796444117026\\
0.16252133341295	0.0316503001859803\\
0.161017161247794	0.03690588011628\\
0.158031334851612	0.0457684838972643\\
0.154592674589062	0.0543999424021364\\
0.150692011096057	0.0627929711915276\\
0.146344776296713	0.0709238046553238\\
0.141577583941191	0.0787795006588359\\
0.135419271881569	0.0876302392568975\\
0.130807289893569	0.0935252761609151\\
0.124853602215437	0.100384505141013\\
0.118548663473915	0.10687529633799\\
0.111922884022345	0.112965320798062\\
0.104983907214131	0.118669407948227\\
0.0977627262919485	0.123956686078305\\
0.0880031579418517	0.130179703305096\\
0.0825719961028191	0.133228804304291\\
0.0746097994794075	0.13725114202863\\
0.0665067584295556	0.140748654700737\\
0.058249442248085	0.143775995603312\\
0.0498646677708952	0.146327001368708\\
0.0413791810641617	0.148396543684035\\
0.0328957165637724	0.149935015882765\\
0.0242936314391109	0.151038447055033\\
0.0156721168792815	0.151655459626678\\
0.00705962201319998	0.151788919601508\\
-0.00151872059099301	0.151439685472052\\
-0.0100357281766157	0.150612547849486\\
-0.0185155215652595	0.149300082965763\\
-0.0268310250446515	0.147532853459479\\
-0.0405265983959735	0.14352381439877\\
-0.0508452148784466	0.139533409768227\\
-0.0584323426514848	0.136027021223992\\
-0.0658144244182222	0.132090627408291\\
-0.0729410828212751	0.12775982496637\\
-0.0797919784676442	0.123050591759772\\
-0.0863472690630003	0.117980015597676\\
-0.0925941688774186	0.112558274977626\\
-0.0985025968944691	0.106819689997866\\
-0.104062001556291	0.10077676883779\\
-0.109257588243233	0.0944505336203183\\
-0.114074749866118	0.0878623349402718\\
-0.118500727493021	0.0810343018492357\\
-0.122554960680428	0.0739846262107493\\
-0.126165879202783	0.0667451081523356\\
-0.129353898024666	0.0593354307530043\\
-0.13387848670033	0.0460958274314494\\
-0.13443583182716	0.0441053418517612\\
-0.136276795133075	0.0363710668834818\\
-0.137717450216776	0.0285312511647846\\
};
\addplot[only marks, mark=square*, mark options={}, mark size=1.2500pt, draw=black, fill=mycolor1] table[row sep=crcr]{%
x	y\\
1	0\\
};
\addplot[only marks, mark=square*, mark options={}, mark size=1.2500pt, draw=black, fill=mycolor1] table[row sep=crcr]{%
x	y\\
0.851481434057751	-0.137640226204982\\
};
\addplot[only marks, mark=square*, mark options={}, mark size=1.2500pt, draw=black, fill=mycolor1] table[row sep=crcr]{%
x	y\\
0.406662207601443	0.155726508716223\\
};
\addplot[only marks, mark=square*, mark options={}, mark size=1.2500pt, draw=black, fill=mycolor1] table[row sep=crcr]{%
x	y\\
0.0889972923747828	0.304483787336953\\
};
\addplot[only marks, mark=square*, mark options={}, mark size=1.2500pt, draw=black, fill=mycolor1] table[row sep=crcr]{%
x	y\\
-0.0919139954781958	-0.177559550603946\\
};
\addplot[only marks, mark=square*, mark options={}, mark size=1.2500pt, draw=black, fill=mycolor1] table[row sep=crcr]{%
x	y\\
-0.137717450216776	0.0285312511647845\\
};
\addplot [color=mycolor2, line width=1.0pt, forget plot]
  table[row sep=crcr]{%
1	-0\\
0.986303896072296	-0.0339544077498461\\
0.981106445879612	-0.0425436396744101\\
0.9752994134147	-0.0509577959294532\\
0.971001430947157	-0.0565646828067456\\
0.970023517436316	-0.0585990627476232\\
0.962765391959474	-0.0668792500163948\\
0.9562641464944	-0.0734949370768729\\
0.948716500931758	-0.0803826319946843\\
0.942591466601585	-0.0850003711656375\\
0.933011418996288	-0.0920930133059015\\
0.925852465170984	-0.0967875605268194\\
0.914260054936447	-0.103718023402143\\
0.90536879435846	-0.108040424894324\\
0.895731473089831	-0.112072528997617\\
0.885324041708472	-0.115719287263732\\
0.86787661133002	-0.12013263765998\\
0.855500708451643	-0.122337574461168\\
0.842612102801336	-0.123824172076866\\
0.828667456686518	-0.124319861361331\\
0.813980371251931	-0.123762344387975\\
0.798368329504466	-0.121964821867204\\
0.79042880292244	-0.120607687863255\\
0.782332397340916	-0.118904479515196\\
0.774086267586436	-0.116839443282577\\
0.765924624899627	-0.114466799958998\\
0.757396187348592	-0.111629853424933\\
0.748740736978315	-0.108386594901769\\
0.739965721505897	-0.104723663344913\\
0.730894709754652	-0.100581737401873\\
0.721906401070577	-0.0960461139990894\\
0.712818464436278	-0.0910558242801749\\
0.703636012139526	-0.0856011171533642\\
0.685151112242402	-0.0732777023733222\\
0.666012655939314	-0.0589599562058314\\
0.656388971043363	-0.0510739217875231\\
0.646670451540333	-0.0426885219012068\\
0.627038052944993	-0.0244275693155933\\
0.606674076981766	-0.00417103704017374\\
0.585751057755104	0.0180448248021692\\
0.540996763369722	0.067989184932993\\
0.503582851507743	0.109660522915506\\
0.475842987014739	0.138849932911557\\
0.460972316558752	0.153822551521287\\
0.437056946876361	0.176365261708169\\
0.418722281806502	0.192512896434661\\
0.400237118342238	0.207786635707232\\
0.381559286017496	0.222093515410511\\
0.362694387235108	0.235448342827874\\
0.343642291093101	0.24793613319051\\
0.324395750682989	0.259561810707804\\
0.304683682283185	0.27041232121034\\
0.279249679851447	0.282916375711277\\
0.265907450499745	0.288847182878412\\
0.246217326849622	0.296825303113699\\
0.223802053026003	0.304755845705068\\
0.206780100165345	0.309985092826298\\
0.18710580939677	0.315171850135808\\
0.16241014780079	0.320417418073724\\
0.147900124464601	0.322846385567569\\
0.128494846381906	0.325301154024464\\
0.109284655037522	0.326844152488296\\
0.090242913173729	0.327499655923954\\
0.0714343789192715	0.327269325680528\\
0.0528944521584707	0.326160531877609\\
0.0346563881008526	0.324197518078571\\
0.0167535519003128	0.32139354998589\\
-0.000767678480404932	0.317771502457537\\
-0.0179013581397838	0.313345161767508\\
-0.0346478435893369	0.308133649728573\\
-0.050884421095053	0.30217075123752\\
-0.0666263785288543	0.295478413615807\\
-0.0833371130339624	0.287313194155106\\
-0.0964922361864922	0.280034917040945\\
-0.110581902573012	0.271325995222996\\
-0.124049790628624	0.261994037453723\\
-0.136907132000074	0.252094958424477\\
-0.14985234882265	0.240984335017027\\
-0.16065041391382	0.230708755447784\\
-0.171499552804745	0.219281839396114\\
-0.181642000587495	0.207417980169259\\
-0.19109466546954	0.195127158373894\\
-0.199777494687363	0.182501129930082\\
-0.207712532682809	0.169552844735115\\
-0.214888597093607	0.156346622375853\\
-0.221301424912704	0.142872516811672\\
-0.226942917983903	0.129195384523205\\
-0.231778602280219	0.11535386517887\\
-0.235865256162829	0.101391747046564\\
-0.239173658609018	0.0873464690676227\\
-0.241706123131171	0.073258527953971\\
-0.243474580959959	0.0591747988912512\\
-0.24446878579838	0.0451191097402448\\
-0.244704087182861	0.0311382717737378\\
-0.244197324296733	0.0172350867495585\\
-0.242219016126611	-0.00205795018766564\\
-0.24096593988709	-0.010010793264942\\
-0.238295916482757	-0.0232984898100246\\
-0.23491796208722	-0.0363453152175095\\
-0.230612188450397	-0.0498258431971867\\
-0.226160598659125	-0.0615280319715217\\
-0.220795343548331	-0.0735794112279156\\
-0.214850232011383	-0.0852609489772249\\
-0.208321531477602	-0.0965249806410806\\
-0.201235366167797	-0.107344369261736\\
-0.193631128075265	-0.117696699942598\\
-0.185513361768266	-0.127550587334816\\
-0.176923804270241	-0.13688619733847\\
-0.167876437681977	-0.145708532042081\\
-0.158435162367595	-0.153944585843247\\
-0.148616927744647	-0.161602120963348\\
-0.13642169564776	-0.169983465978342\\
-0.128009064094721	-0.175127651446223\\
-0.117262730797493	-0.180966070703386\\
-0.106276469967447	-0.186174305500709\\
-0.0950853533046714	-0.190743739143721\\
-0.0837165888796578	-0.194632608274883\\
-0.0722245221063305	-0.19790508744249\\
-0.0606343435813437	-0.200523803954725\\
-0.048982482713962	-0.202488422303531\\
-0.0326988070065641	-0.204152060968905\\
-0.0256390768869408	-0.20447494481493\\
-0.0140131980790221	-0.204490253207126\\
-0.00246566287116146	-0.203866122704776\\
0.00899487342071192	-0.202613698122643\\
0.0243198928916337	-0.199907081702511\\
0.0313935470080542	-0.198249053199281\\
0.0422926269700445	-0.19516622088465\\
0.0529501066537228	-0.191502827114255\\
0.0633246640270215	-0.187299569170705\\
0.079210741961494	-0.179462973287905\\
0.0831690149480877	-0.177229044110462\\
0.0925721310927088	-0.171425599598238\\
0.101572854985435	-0.165109284862111\\
0.110189081956094	-0.158364821517759\\
0.118378772813798	-0.151186019045593\\
0.126831590167056	-0.142857804813957\\
0.133403996078566	-0.135642499138316\\
0.140192865668467	-0.127319534391388\\
0.146478976187762	-0.118671297517249\\
0.152247878014106	-0.109727078209362\\
0.157486362376177	-0.100517192337328\\
0.162194231568919	-0.0910538510687677\\
0.166338016139923	-0.0814037160945156\\
0.169921425726813	-0.0715804548498113\\
0.172938832476143	-0.0616162346832079\\
0.175398663428757	-0.0515653682478179\\
0.177273065133656	-0.0414136740192579\\
0.178572453730508	-0.0312161150277996\\
0.179297852752142	-0.0210051195085894\\
0.17945179077544	-0.0108119816440948\\
0.178999961464366	-0.000667726775918531\\
0.178024018221344	0.00939440164861161\\
0.176494022230294	0.0193458098490289\\
0.173398010826763	0.0331875672276389\\
0.171809380213708	0.0387970057476634\\
0.168695449204937	0.048244378298044\\
0.165055083038123	0.0574615298877144\\
0.160921155069215	0.0664264582478127\\
0.156310393381747	0.0751117136306769\\
0.151231029452098	0.0835095887822341\\
0.144690499533578	0.0929665956210282\\
0.139790154930309	0.0992667385934121\\
0.133461942020533	0.106599296509781\\
0.126758363425134	0.113539547068602\\
0.119730017529571	0.120071631155151\\
0.112347351082287	0.126176041831453\\
0.104662595243229	0.131836816426528\\
0.0942737429285765	0.138502654415487\\
0.0884912750437736	0.141770647717043\\
0.0800377331741924	0.145987811178665\\
0.0714129207678511	0.149743080976818\\
0.0626222866170887	0.152997787549039\\
0.0536941254006973	0.155745003062145\\
0.0446575493385799	0.157979949546596\\
0.0355469898494649	0.159719750033852\\
0.0263789410355946	0.160921065685012\\
0.0171895903095587	0.161604308730576\\
0.00800750618842438	0.161770715401585\\
-0.00114150233412125	0.161420973545778\\
-0.0102209050058739	0.16056576170168\\
-0.0192254362375188	0.159199387743804\\
-0.0280949558945029	0.157341527609078\\
-0.0427068064092404	0.153110477013639\\
-0.0453672832414824	0.152175251329055\\
-0.0537182500862163	0.148888895821561\\
-0.0618370042222613	0.145177619482403\\
-0.0697173972275345	0.141003573676999\\
-0.0773272627225354	0.136408340753531\\
-0.0846440708756884	0.131408796033691\\
-0.0916468527796488	0.126023222035689\\
-0.0982942016678947	0.120237219434109\\
-0.104607683360558	0.11413921037939\\
-0.110550241247094	0.107716050207157\\
-0.116105620809437	0.100989815527586\\
-0.121258395593276	0.0939833318591063\\
-0.125995022591572	0.0867204760227782\\
-0.130322969838061	0.0792388534340065\\
-0.134192785143568	0.071535806676462\\
-0.137613499066249	0.0636509349635088\\
-0.142475955136758	0.0495582026602097\\
-0.145102353069788	0.0391499516220746\\
-0.146658335977389	0.0307989275281779\\
};
\addplot[only marks, mark=square*, mark options={}, mark size=1.2500pt, draw=black, fill=mycolor2] table[row sep=crcr]{%
x	y\\
1	-6.6982597067695e-31\\
};
\addplot[only marks, mark=square*, mark options={}, mark size=1.2500pt, draw=black, fill=mycolor2] table[row sep=crcr]{%
x	y\\
0.869367723843047	-0.119807866753323\\
};
\addplot[only marks, mark=square*, mark options={}, mark size=1.2500pt, draw=black, fill=mycolor2] table[row sep=crcr]{%
x	y\\
0.437056946876361	0.176365261708169\\
};
\addplot[only marks, mark=square*, mark options={}, mark size=1.2500pt, draw=black, fill=mycolor2] table[row sep=crcr]{%
x	y\\
0.0965658601626538	0.32738002448608\\
};
\addplot[only marks, mark=square*, mark options={}, mark size=1.2500pt, draw=black, fill=mycolor2] table[row sep=crcr]{%
x	y\\
-0.0988363529012426	-0.189292126057132\\
};
\addplot[only marks, mark=square*, mark options={}, mark size=1.2500pt, draw=black, fill=mycolor2] table[row sep=crcr]{%
x	y\\
-0.146658335977389	0.0307989275281779\\
};
\addplot [color=mycolor3, line width=1.0pt, forget plot]
  table[row sep=crcr]{%
1	-0\\
0.988135465265402	-0.0287004378742668\\
0.983666724560261	-0.035876947300151\\
0.978687584751696	-0.0428853533862554\\
0.975008860244698	-0.047544288611497\\
0.974344428594665	-0.0492537967179929\\
0.968140522312062	-0.0561173149782592\\
0.960544723672237	-0.0634610268590474\\
0.953618971217462	-0.0691617177147958\\
0.945806207114456	-0.0749786812128745\\
0.936822843683167	-0.0807519127091676\\
0.930123459119115	-0.0844942871018319\\
0.910279913592323	-0.0929274971263647\\
0.897455547379252	-0.097300558821503\\
0.887412301916684	-0.099476421369143\\
0.876641258490956	-0.101052558424117\\
0.870980901201253	-0.101579953177861\\
0.852182410310765	-0.101737457684907\\
0.83925662650563	-0.100798884965332\\
0.832971010905552	-0.100035548798942\\
0.818995067008036	-0.0973986872609074\\
0.804367568897086	-0.0934667830905618\\
0.796531110541119	-0.0909064911419333\\
0.788841647191137	-0.088045839718476\\
0.781008445136835	-0.0847982271744274\\
0.773036401636269	-0.0811488861587104\\
0.76500856589569	-0.0771202966957749\\
0.756770540789186	-0.0726250949113341\\
0.748406379931653	-0.0676878278940471\\
0.739919016804743	-0.0622968533843578\\
0.722646444053037	-0.050083140809714\\
0.704657890881759	-0.0359690686911829\\
0.686327299982546	-0.0198815727374324\\
0.658063381476984	0.00801156229493794\\
0.647819146368067	0.0183155192095661\\
0.627360805484774	0.0403609239357632\\
0.58284352623869	0.0899264787286838\\
0.557917286188852	0.11708680735996\\
0.530600156411607	0.14559844274935\\
0.515762521995296	0.160185426475692\\
0.499976684323771	0.174831627377394\\
0.474556351268842	0.197395061309676\\
0.454803985193481	0.213480237309926\\
0.446469404987567	0.21994623859795\\
0.42611807058162	0.234788434278424\\
0.404434640044489	0.249665056396591\\
0.393845291495209	0.256592627495709\\
0.373040575410913	0.269338097467129\\
0.352029599048464	0.281230435525304\\
0.330534308570485	0.292439058108854\\
0.302849168321189	0.305424490243547\\
0.288346781495487	0.311608246829389\\
0.266939048207837	0.320043618697451\\
0.242637519479371	0.328352128435033\\
0.224207548033358	0.333838419457158\\
0.202978321019353	0.339172629373641\\
0.176307321579243	0.344681377981923\\
0.160584225785499	0.347338915736536\\
0.139649527537839	0.349905866695748\\
0.119026359748944	0.351406523206544\\
0.0985175813474317	0.352064503815908\\
0.0782692623980363	0.351780788298751\\
0.0582930705153013	0.35059868290965\\
0.0386686465755446	0.348471170717027\\
0.019410080466755	0.345446988717526\\
0.000490020333719698	0.341542215809278\\
-0.0179336468193689	0.336782334671349\\
-0.035799732251266	0.331187137602605\\
-0.0532526614398718	0.324794285465337\\
-0.0701728639350594	0.317622237416335\\
-0.0882896044301176	0.308865969298303\\
-0.102427948328785	0.301059964153687\\
-0.117570424809064	0.291721737211177\\
-0.131926907974502	0.281733652458767\\
-0.145747397008292	0.271131833786841\\
-0.159663864249647	0.259233237743629\\
-0.17140646517697	0.248198301548773\\
-0.183072267824423	0.235954215242639\\
-0.193980133955489	0.223242401551327\\
-0.203996572150268	0.210139124080563\\
-0.213341418932753	0.196619178095592\\
-0.221885072294693	0.182753908192374\\
-0.229628281869125	0.168591063573478\\
-0.236540278740935	0.154161038016101\\
-0.242625954836571	0.139513146344052\\
-0.247950612966768	0.124606285214457\\
-0.252367472060708	0.109646763750383\\
-0.255950432233481	0.0945976319360977\\
-0.258701966978036	0.0795022083738797\\
-0.260568146623483	0.0645186256794195\\
-0.261677184655773	0.049460691282003\\
-0.261973342412713	0.0344818543041825\\
-0.261506180072719	0.0194871591660759\\
-0.259440511742549	-0.00119116347606374\\
-0.258120453379856	-0.00971568930160016\\
-0.25525552021126	-0.0238577171637133\\
-0.251680078512772	-0.0378412565980899\\
-0.247113124644835	-0.0522912875243824\\
-0.242384843253145	-0.0648367103088536\\
-0.23673434375128	-0.077851687749791\\
-0.230403884819417	-0.0903817537140357\\
-0.223447688965299	-0.102465960841984\\
-0.215893679427449	-0.114075138385547\\
-0.207750276453842	-0.125119263243647\\
-0.199092615318333	-0.135694528511114\\
-0.189928872452889	-0.145716047893065\\
-0.180314601069649	-0.155147076337254\\
-0.170239512366173	-0.163996219516506\\
-0.159759496845614	-0.17222728904288\\
-0.14673909744401	-0.181241325002437\\
-0.137644204034192	-0.186888996016563\\
-0.126159633376489	-0.193169690026213\\
-0.114416986953298	-0.198776331811033\\
-0.102453799184455	-0.203699699902656\\
-0.0904306351694466	-0.207831663354823\\
-0.0781470104341691	-0.211375223852225\\
-0.0657566555970674	-0.214219722217455\\
-0.0532984307794304	-0.216364628426761\\
-0.0357763470957917	-0.218226114840376\\
-0.0282243552049677	-0.218593867927973\\
-0.015786678715936	-0.218647059463405\\
-0.00343130924168822	-0.218015943822235\\
0.00873553133022598	-0.216695433357833\\
0.0251354869583298	-0.213856385857448\\
0.0327064478899926	-0.212108906992466\\
0.044373371086486	-0.208851756176848\\
0.0557836360114541	-0.20497304994598\\
0.0669269869591131	-0.200518388099948\\
0.0839426425944605	-0.192193769395408\\
0.0881831700530604	-0.189818789117762\\
0.0982583400249986	-0.183646083984196\\
0.107922361963868	-0.17691189090557\\
0.117158218096264	-0.169729563201985\\
0.125939059478063	-0.162081607830462\\
0.135004820150851	-0.153205901857461\\
0.142010805450457	-0.145544313642766\\
0.149298543553685	-0.136671870500147\\
0.156049834298791	-0.127450190393711\\
0.162249072484362	-0.117910531715837\\
0.16788215565188	-0.108085224492226\\
0.173063197296454	-0.0978926977307839\\
0.177522296957765	-0.0875868402624977\\
0.181382594958813	-0.077094508435831\\
0.184638002181181	-0.0664500636629846\\
0.187176629005824	-0.0558342964540743\\
0.189216898407596	-0.0449934907646705\\
0.19064287516582	-0.0341018675988436\\
0.191455575149367	-0.0231941292336855\\
0.191657514952028	-0.0123036690884881\\
0.19127326816636	-0.00138722818677373\\
0.190263699482084	0.00937035804840214\\
0.188661582167673	0.020011218986687\\
0.185398660975254	0.0348147476964451\\
0.183719471853934	0.040814903372264\\
0.180401302719834	0.050890089393105\\
0.176542827952032	0.0607506425834101\\
0.17215626384858	0.0703430377980365\\
0.167259337214195	0.0796380639027354\\
0.161902105804161	0.0885958038479018\\
0.154948791204743	0.0987241815579714\\
0.149736358081811	0.105473448030426\\
0.143002331910509	0.113330765726202\\
0.135865938245613	0.120770295409357\\
0.12828147483392	0.12780334769086\\
0.120414508110795	0.134346433000168\\
0.112223634003989	0.140416379407726\\
0.101147814823778	0.14756787810901\\
0.0949817654743259	0.151075986058844\\
0.0860644228832337	0.155603477358462\\
0.0768651168719372	0.15964610330067\\
0.0674870950810484	0.163154539165772\\
0.0579604885057299	0.166121378582834\\
0.0483163944847615	0.168541303061893\\
0.0384734232517079	0.170487429407326\\
0.0286822947547258	0.171796748340005\\
0.0188671063447081	0.172552555341826\\
0.00905839338738268	0.172756123079298\\
-0.00071631687309881	0.172408176456998\\
-0.0104177575635989	0.171519941554049\\
-0.0198923809341747	0.170045055529833\\
-0.0293698049368261	0.168093527064813\\
-0.0386922049363427	0.165621540875198\\
-0.0478311131782552	0.162639977446317\\
-0.0567594925620389	0.159161367457241\\
-0.0655012785516071	0.155206389433532\\
-0.0739292014353508	0.150775049419994\\
-0.0820694104643815	0.14589310756644\\
-0.0898978322971751	0.140578580423737\\
-0.0973920096350063	0.134850926506953\\
-0.10453751424719	0.128717001905414\\
-0.111298722003337	0.122225066714689\\
-0.117664774928408	0.115384830288834\\
-0.123618167268043	0.108219837481636\\
-0.129142444880702	0.100754425438986\\
-0.134223044010154	0.0930140068144727\\
-0.138820976672853	0.0850807766158737\\
-0.142979298138144	0.0768704371892019\\
-0.146658710260246	0.0684645489993825\\
-0.149849990319895	0.0598902838372315\\
-0.152546510212184	0.0511761054125297\\
-0.154761148984682	0.0422342614112585\\
-0.156445825491861	0.0333231287320395\\
};
\addplot[only marks, mark=square*, mark options={}, mark size=1.2500pt, draw=black, fill=mycolor3] table[row sep=crcr]{%
x	y\\
1	0\\
};
\addplot[only marks, mark=square*, mark options={}, mark size=1.2500pt, draw=black, fill=mycolor3] table[row sep=crcr]{%
x	y\\
0.888707332622025	-0.0992349192088727\\
};
\addplot[only marks, mark=square*, mark options={}, mark size=1.2500pt, draw=black, fill=mycolor3] table[row sep=crcr]{%
x	y\\
0.474556351268842	0.197395061309676\\
};
\addplot[only marks, mark=square*, mark options={}, mark size=1.2500pt, draw=black, fill=mycolor3] table[row sep=crcr]{%
x	y\\
0.105326528726876	0.351950228411229\\
};
\addplot[only marks, mark=square*, mark options={}, mark size=1.2500pt, draw=black, fill=mycolor3] table[row sep=crcr]{%
x	y\\
-0.106463752084128	-0.202135084808875\\
};
\addplot[only marks, mark=square*, mark options={}, mark size=1.2500pt, draw=black, fill=mycolor3] table[row sep=crcr]{%
x	y\\
-0.156445825491861	0.0333231287320396\\
};
\addplot [color=mycolor4, line width=1.0pt, forget plot]
  table[row sep=crcr]{%
1	-0\\
0.991052620252041	-0.0232143864475804\\
0.986541732259653	-0.0299165075206136\\
0.981123678717728	-0.036744224936726\\
0.980046298415156	-0.0379768795790407\\
0.977653304820334	-0.0391079755407819\\
0.972512080292793	-0.0443587912859962\\
0.96623464470111	-0.0499159797458051\\
0.958744168934729	-0.0556784998968494\\
0.951995953030559	-0.0599562504607374\\
0.944255064544939	-0.0640826467398559\\
0.941436365346471	-0.065398082947123\\
0.939268242493877	-0.0667966816864616\\
0.932976898847605	-0.0691969890272439\\
0.926140066422813	-0.0713387322016856\\
0.918731876691732	-0.0731490137129633\\
0.901499834953786	-0.0753044859667498\\
0.892272149037112	-0.075596462896552\\
0.882354576933567	-0.0751885150083034\\
0.871853355112963	-0.0739575932639667\\
0.861141392304377	-0.0718419852934973\\
0.849361277158155	-0.0686084810548313\\
0.83696427331022	-0.064191046306302\\
0.830262096406523	-0.0614883645216791\\
0.816978815898537	-0.0550738242836328\\
0.802990961405275	-0.0471326394466167\\
0.788569577121456	-0.0376434751945283\\
0.781146658935107	-0.0322706331974485\\
0.766241182224281	-0.0202659456954093\\
0.750091866149728	-0.00637922044282102\\
0.733540756586034	0.00933916747196051\\
0.71612202374844	0.0269841756932279\\
0.707133850373904	0.0364754554227018\\
0.678745494086651	0.0676673649733981\\
0.634741821947963	0.115066889897053\\
0.609410323189773	0.141076318571172\\
0.595467703286705	0.154506813259214\\
0.5804846562443	0.168140274352549\\
0.564559439637839	0.182126702187728\\
0.547479873326443	0.196326354324132\\
0.519499288505718	0.218384477175497\\
0.497603314700963	0.23432985695202\\
0.488341314571791	0.240803491290388\\
0.452707188449941	0.264158985056297\\
0.429886150572162	0.278129243460842\\
0.406879313057308	0.291398651065974\\
0.383703341303485	0.304091021974626\\
0.360084315800928	0.315928929644988\\
0.329807790361944	0.329561735715311\\
0.3139778685544	0.336141395262924\\
0.290709339889615	0.34499895883963\\
0.26429470950369	0.35387820504445\\
0.244292538517517	0.359749792793352\\
0.221071168386709	0.365771657718371\\
0.192163357294035	0.371668604207626\\
0.175410974677664	0.374220446802985\\
0.152779412985506	0.376977719678656\\
0.130371396008042	0.378711165143082\\
0.108212166548528	0.37942211265403\\
0.0863428572160487	0.379123677501053\\
0.0645979157316401	0.377933261344462\\
0.0434071701931322	0.375648979806082\\
0.0226162817687057	0.372402168534859\\
0.00248015485373698	0.368137058509095\\
-0.0173934047108792	0.363040704489955\\
-0.0367579726654588	0.357057563989698\\
-0.0555839425431368	0.350203159907217\\
-0.0738348362057735	0.34251210473338\\
-0.093465590096792	0.333117388892956\\
-0.1087190367077	0.324741608267056\\
-0.125056682500835	0.314720686033966\\
-0.140458429691608	0.304021563415257\\
-0.155370963927136	0.292645704761741\\
-0.170389581553637	0.279877203538292\\
-0.182926537538072	0.268062000872404\\
-0.195522492371127	0.254928103211365\\
-0.207304170152945	0.241291072237416\\
-0.218466018676277	0.227089703299913\\
-0.228563810718144	0.212563805547304\\
-0.237799895960488	0.197666000681614\\
-0.245964981980434	0.182574258990673\\
-0.253451530189172	0.167080126139602\\
-0.260050279823333	0.151350616513183\\
-0.265762880821819	0.135434903176916\\
-0.270573918569823	0.119371669517068\\
-0.274489302795631	0.103210393529518\\
-0.277511637766355	0.0869974565812142\\
-0.279767310326043	0.0705848856272921\\
-0.281008217560808	0.0543966789623667\\
-0.281375067401328	0.0382922561130992\\
-0.280781262459462	0.0225264039849316\\
-0.278656858792634	0.000303717107693169\\
-0.277278235052641	-0.0088589425753236\\
-0.274302026731231	-0.0241949982405516\\
-0.270519511546133	-0.0392330010085462\\
-0.265674553912251	-0.0547752305205667\\
-0.260648899232559	-0.0682712638571559\\
-0.254611755848406	-0.0824141926214399\\
-0.24785838510823	-0.0958978792376035\\
-0.240431823988142	-0.108904182703519\\
-0.232362012161963	-0.121401718264668\\
-0.223717852098607	-0.133141106899119\\
-0.214469369402927	-0.144534014867462\\
-0.204675870691359	-0.155334111783582\\
-0.194377564780758	-0.165526306913417\\
-0.183600413983162	-0.175069868173795\\
-0.172386619058301	-0.18395133641057\\
-0.158450314824846	-0.193683921186226\\
-0.148686650203217	-0.199831681612231\\
-0.136385744938761	-0.206621789684087\\
-0.123805659928577	-0.212689229345823\\
-0.11098671295126	-0.218023999466713\\
-0.0981191275639057	-0.222468231475909\\
-0.0849540074798922	-0.226323999558063\\
-0.0716720775733721	-0.22943005025046\\
-0.0583149732857715	-0.231785682584226\\
-0.0396342217801455	-0.233832219062956\\
-0.0315355313432124	-0.234263613775726\\
-0.018195334274226	-0.234381620927357\\
-0.00494068898117783	-0.233765355528917\\
0.00837678674154274	-0.232480884161241\\
0.025984729593975	-0.229496760793443\\
0.0341146617029711	-0.227650483970463\\
0.0466446015499549	-0.224199981414045\\
0.058900992745619	-0.220081777607198\\
0.0706465821608178	-0.21530306912853\\
0.0822521278798596	-0.209919349048308\\
0.0934850770399316	-0.20393001208446\\
0.104314577979717	-0.197359146895498\\
0.1147215056515	-0.190241125389627\\
0.124658405215305	-0.182584528538783\\
0.134109254385012	-0.174426867754089\\
0.143871130256064	-0.164954653594971\\
0.151623889012521	-0.1566471372475\\
0.159471743906853	-0.147158571568161\\
0.16674524603589	-0.137293850118369\\
0.173427515537025	-0.127086410702295\\
0.179503465120675	-0.116570947590556\\
0.184806996370075	-0.105903432567859\\
0.189638761946592	-0.0948856899537602\\
0.193829836447899	-0.0836654572813109\\
0.197373601126804	-0.0722795320859135\\
0.200278543285605	-0.0607629532571607\\
0.202512962165866	-0.0491535123490654\\
0.204089185581777	-0.0374872946313487\\
0.20500816438765	-0.0258015276862673\\
0.205272457779789	-0.0141319111030846\\
0.204942925805659	-0.00234596263950437\\
0.20390316666323	0.00918743828515822\\
0.202227906364751	0.020597715444999\\
0.198789037302985	0.0364751452062695\\
0.197012977123323	0.0429117229657039\\
0.193478984889997	0.0535643289134236\\
0.189391565399909	0.0641449523219297\\
0.184737329312919	0.0744403995347083\\
0.179535170857886	0.0844194213422678\\
0.173816008361662	0.0940679509207452\\
0.166413384741875	0.104947168477298\\
0.160860536841952	0.112199152989453\\
0.153683009254914	0.120644311035298\\
0.146072915359738	0.128643648694624\\
0.137979168966639	0.136326792313565\\
0.129576765998653	0.143368437646028\\
0.120825620409998	0.149904301570671\\
0.108988240244985	0.157610109048326\\
0.102396496535122	0.161392887785995\\
0.0928934397717198	0.166179566493646\\
0.0830599251410933	0.170548406893561\\
0.0730329654202413	0.174345819094024\\
0.0628448448027066	0.177563818469343\\
0.0525287815783901	0.180196524054568\\
0.0421163163780089	0.18225837835258\\
0.0316422831493952	0.183709296067568\\
0.0211399733335031	0.184567832859211\\
0.0106420942420746	0.184835215837311\\
0.000178026966191736	0.184512154503902\\
-0.010209975015093	0.183610325914021\\
-0.0206588740373275	0.182157986061968\\
-0.0308165912518341	0.180100083781565\\
-0.0408098006790691	0.177484201063732\\
-0.0506080194387586	0.174321818444688\\
-0.0601821456332272	0.170626395795263\\
-0.0693314148179056	0.166416107688729\\
-0.0783717284469572	0.161711145889193\\
-0.0871059056870989	0.156522689133665\\
-0.09550807776073	0.150869954460679\\
-0.103554176668923	0.144773662407306\\
-0.111236362800499	0.138267279517156\\
-0.118502939982025	0.131349887115669\\
-0.125348004442802	0.124058013928431\\
-0.131752726813877	0.116416781590922\\
-0.137699388080652	0.108452200142467\\
-0.143172334650668	0.100191403430403\\
-0.148271158282108	0.0915569841038855\\
-0.152749448360792	0.0827809906869603\\
-0.15671555729163	0.0737942038799535\\
-0.160159444895418	0.0646257119461622\\
-0.163074026938142	0.0553060391547751\\
-0.165352367385916	0.0459966643914398\\
-0.167198424906582	0.0364704893940515\\
};
\addplot[only marks, mark=square*, mark options={}, mark size=1.2500pt, draw=black, fill=mycolor4] table[row sep=crcr]{%
x	y\\
1	0\\
};
\addplot[only marks, mark=square*, mark options={}, mark size=1.2500pt, draw=black, fill=mycolor4] table[row sep=crcr]{%
x	y\\
0.906948155140847	-0.0748107623457157\\
};
\addplot[only marks, mark=square*, mark options={}, mark size=1.2500pt, draw=black, fill=mycolor4] table[row sep=crcr]{%
x	y\\
0.519499288505718	0.218384477175497\\
};
\addplot[only marks, mark=square*, mark options={}, mark size=1.2500pt, draw=black, fill=mycolor4] table[row sep=crcr]{%
x	y\\
0.115568020014227	0.379297735014196\\
};
\addplot[only marks, mark=square*, mark options={}, mark size=1.2500pt, draw=black, fill=mycolor4] table[row sep=crcr]{%
x	y\\
-0.115283798206811	-0.216327775901827\\
};
\addplot[only marks, mark=square*, mark options={}, mark size=1.2500pt, draw=black, fill=mycolor4] table[row sep=crcr]{%
x	y\\
-0.167198424906582	0.0364704893940515\\
};
\end{axis}
\end{tikzpicture}%

%% file: images/2Sears.tex
%
%
\definecolor{mycolor1}{rgb}{0.09053,0.13734,0.07326}%
\definecolor{mycolor2}{rgb}{0.04342,0.37461,0.17524}%
\definecolor{mycolor3}{rgb}{0.37697,0.57257,0.04618}%
\definecolor{mycolor4}{rgb}{0.79072,0.73098,0.27064}%
\begin{tikzpicture}[%
trim axis left, trim axis right
]

\begin{axis}[%
width=0.984\fwidth,
height=\fheight,
at={(0\fwidth,0\fheight)},
scale only axis,
xmin=-0.35,
xmax=1.1,
xlabel style={font=\color{white!15!black}},
xlabel={$\Re[S(k)]$},
ymin=-0.3,
ymax=0.45,
ylabel style={font=\color{white!15!black}},
ylabel={$\Im[S(k)]$},
axis background/.style={fill=white},
xmajorgrids,
ymajorgrids
]
\addplot [color=black, line width=2.0pt, forget plot]
  table[row sep=crcr]{%
1	-7.99360577730113e-15\\
0.981169634186679	-0.0473526935450792\\
0.975268104274319	-0.0572020387558559\\
0.969250159220743	-0.0662045018490595\\
0.96179894026922	-0.0762190791908565\\
0.955089041374125	-0.0843709356309031\\
0.947260980824945	-0.0930243239134291\\
0.93816286261376	-0.102103922357411\\
0.927637109934235	-0.111497621664997\\
0.919748936191671	-0.117859809709497\\
0.911111402206164	-0.124235586663284\\
0.901682744789075	-0.130556691091301\\
0.891425823994952	-0.136744181202643\\
0.880309851433787	-0.142708347454063\\
0.868312225503855	-0.14834898478152\\
0.855420391150047	-0.153556089833553\\
0.841633615179317	-0.158211033668738\\
0.826964543805801	-0.162188236031879\\
0.819307060233782	-0.163882063696621\\
0.803370321202988	-0.16659743199815\\
0.786647890344584	-0.168305997546178\\
0.769204741155661	-0.168877599492619\\
0.760236971327944	-0.168697801104816\\
0.751119434179592	-0.168187652747411\\
0.741863704712643	-0.167332969139702\\
0.732481873852179	-0.166120293321165\\
0.722986426197111	-0.164536976834457\\
0.713390104090285	-0.162571251099379\\
0.703705758999715	-0.16021228878552\\
0.693946191406285	-0.1574502542351\\
0.684123980561909	-0.15427634227784\\
0.674251305608945	-0.150682805110534\\
0.664339759631732	-0.146662967278347\\
0.654400158241382	-0.142211229184615\\
0.644442344274809	-0.137323059962614\\
0.634474990120298	-0.1319949809579\\
0.624505399069119	-0.126224541485567\\
0.61453930694264	-0.120010288936335\\
0.60458068506634	-0.11335173570327\\
0.594631545467335	-0.106249325783741\\
0.584691748973709	-0.09870440427718\\
0.574758816706798	-0.0907191933488399\\
0.564827745297672	-0.0822967785653936\\
0.544937468169077	-0.0641570214983668\\
0.524923633055708	-0.0443276395846814\\
0.504637440174471	-0.0228674490332269\\
0.483873491400335	0.000144213700549067\\
0.415723976688649	0.0772064911581731\\
0.389729326888095	0.104877858410841\\
0.368649165757727	0.125943361459841\\
0.352634315254576	0.141007303114233\\
0.33661244517805	0.155217856313328\\
0.320553419412667	0.168576603460661\\
0.304438033489619	0.181084867080371\\
0.288255978143667	0.192743815058999\\
0.272004237050724	0.203554606299833\\
0.255411263358432	0.213678645094021\\
0.234055716454205	0.225377999325029\\
0.222884997833748	0.230912887496748\\
0.206430334724944	0.238348090272243\\
0.187710215152482	0.245784286475277\\
0.17350431403378	0.250711995151768\\
0.157076612099213	0.255650316936921\\
0.136478809423763	0.260700549379704\\
0.124413573265463	0.263071509148478\\
0.10823064299701	0.265570520460204\\
0.0921831414008394	0.26727460145429\\
0.0762990722262464	0.268194933630997\\
0.0606067746712325	0.268344002957401\\
0.0451347399143001	0.267735617387968\\
0.0299114430411733	0.266384915162501\\
0.0149651883007658	0.264308364490887\\
0.000323966045512769	0.261523755190835\\
-0.0139846799547296	0.258050182805319\\
-0.0279337758698565	0.25390802569132\\
-0.041497032669719	0.249118915540627\\
-0.0546489531490399	0.243705701766897\\
-0.0686288832486455	0.237050215781053\\
-0.0796213468434863	0.231104196272839\\
-0.0972980097826508	0.220063655865227\\
-0.102666435162272	0.216308957959196\\
-0.11341353241244	0.20815740601082\\
-0.124234464526802	0.198991188452407\\
-0.133262477932314	0.190491938823868\\
-0.142330678715705	0.181038668032532\\
-0.150808001703733	0.171213324701674\\
-0.158681296369406	0.16104790129127\\
-0.169735296616782	0.144580211551503\\
-0.172570962965696	0.139827344744887\\
-0.178568894971875	0.128838512723266\\
-0.186501662329346	0.111639879169801\\
-0.18863705633415	0.106271724853657\\
-0.192698680664992	0.0947614979007226\\
-0.196109042081408	0.0831452984049081\\
-0.198868010821759	0.0714570064448898\\
-0.200977125904808	0.0597303527414158\\
-0.202439587028727	0.0479988325827109\\
-0.203260241489297	0.0362956203842292\\
-0.203445566208174	0.0246534851506137\\
-0.20300364497148	0.0131047071018584\\
-0.201333266630901	-0.00296295333153773\\
-0.200278264890796	-0.00958659053960986\\
-0.198018738299201	-0.0206677226165993\\
-0.195179753101202	-0.0315328770836255\\
-0.19156384721065	-0.0427578267705442\\
-0.187827252585544	-0.0525016282125648\\
-0.183349048824465	-0.0625508577896976\\
-0.178361900595442	-0.0722755095613348\\
-0.172886601005078	-0.0816511414075646\\
-0.166945087947514	-0.0906545172207666\\
-0.160560378018716	-0.0992636623413177\\
-0.153756497577954	-0.107457915317088\\
-0.146558411171408	-0.115217975843798\\
-0.138991947539056	-0.122525948755346\\
-0.131083723431618	-0.129365383945554\\
-0.122861065469326	-0.135721312115347\\
-0.112649286092279	-0.142671687872974\\
-0.10558482314758	-0.146930358762174\\
-0.0965887154405601	-0.151761204186247\\
-0.0811661765059353	-0.158635640810862\\
-0.071602364869464	-0.162076607473955\\
-0.0685213335637237	-0.163058547021581\\
-0.058905319596345	-0.165740676073208\\
-0.0492092069196274	-0.167875703215548\\
-0.0394629908115571	-0.169462870575536\\
-0.0258420158591939	-0.170763511849407\\
-0.0199395160943607	-0.170998554728085\\
-0.0102212849083549	-0.170953471429388\\
-0.000570841107991771	-0.170373288187499\\
0.00898327478092154	-0.169265042993867\\
0.0217843553004768	-0.166923211546473\\
0.0276910962602491	-0.165499905585972\\
0.0367906834332662	-0.162864368201025\\
0.0456858521321468	-0.159743397391186\\
0.0543514628733444	-0.156151066861503\\
0.0675986825787007	-0.149506422539469\\
0.0708979700068796	-0.147614931217082\\
0.0787332934161302	-0.142705420654688\\
0.0862480312663607	-0.137393023713003\\
0.0934221048326952	-0.131697614306333\\
0.100236613169862	-0.125640042803275\\
0.107265210377755	-0.118617351246705\\
0.112717497409968	-0.11252629039562\\
0.118352364162714	-0.105516074471686\\
0.123564721998202	-0.0982354810420756\\
0.128342186362612	-0.0907091890456738\\
0.132673773575952	-0.0829624195171222\\
0.136549923681114	-0.0750208577948502\\
0.139962519024264	-0.0669105747938097\\
0.142904898529865	-0.058657947576618\\
0.145371867646469	-0.0502895794580825\\
0.147359703952259	-0.0418322198786559\\
0.148866158422125	-0.0333126842822457\\
0.149890452370863	-0.0247577742329811\\
0.150433270099795	-0.0161941980040229\\
0.150496747286654	-0.00764849186929972\\
0.150084455171048	0.000853057673820823\\
0.149201380600005	0.00928449052646751\\
0.147853902010134	0.0176202482345498\\
0.145167205785574	0.0292099677329655\\
0.14379803263494	0.0339049527548516\\
0.14110908546753	0.0418054445707958\\
0.137994546607896	0.0495134887461394\\
0.134467256763089	0.0570066008410373\\
0.130541224514939	0.06426310900226\\
0.126231576945126	0.0712622138957334\\
0.120681089701074	0.0791548800998992\\
0.116527219180574	0.0844097181843579\\
0.111167869484359	0.0905213791864155\\
0.105495506866547	0.0963022582512683\\
0.099530009328079	0.101736710408229\\
0.0932920190778062	0.106810256509216\\
0.0868028755548973	0.111509619845336\\
0.0780363497013806	0.117034728281297\\
0.0731595588390656	0.119738896681972\\
0.0660509249406488	0.123248544996804\\
0.0587820722241779	0.126343526220585\\
0.0513767685820889	0.129016989910211\\
0.0438590484940528	0.131263425790089\\
0.036253138505795	0.133078672570692\\
0.0285833825187629	0.134459922665789\\
0.0208741671149411	0.135405722812535\\
0.013149847140167	0.135915970610858\\
0.00543467176766843	0.135991907010718\\
-0.00224728873871449	0.135636104787802\\
-0.00987221534511362	0.134852453060116\\
-0.017416611439057	0.133646137909559\\
-0.0248573726829798	0.132023619183995\\
-0.0371081350613749	0.128366596684498\\
-0.0463341102314899	0.124741955641358\\
-0.0531394872149213	0.121543678264606\\
-0.0597339177791402	0.117979535572902\\
-0.0660980182668967	0.114062941224785\\
-0.0722132324286016	0.10980832192033\\
-0.0780618837542911	0.105231067635578\\
-0.0836272248725045	0.100347479177515\\
-0.0888934838797204	0.0951747132298284\\
-0.0938459074737688	0.0897307250665687\\
-0.098470800774755	0.0840342091172249\\
-0.102755563727472	0.0781045375724664\\
-0.10668872399	0.0719616972250166\\
-0.110259966224164	0.0656262247446991\\
-0.113460157714717	0.0591191405906693\\
-0.116281370255489	0.0524618817671815\\
-0.120271490489491	0.0405714411106214\\
-0.122410273134046	0.0318082385334016\\
-0.123660931160608	0.0247705812964559\\
};
\addplot[only marks, mark=square*, mark options={}, mark size=1.2500pt, draw=black, fill=black] table[row sep=crcr]{%
x	y\\
1	-8.02904074443401e-15\\
};
\addplot[only marks, mark=square*, mark options={}, mark size=1.2500pt, draw=black, fill=black] table[row sep=crcr]{%
x	y\\
0.821241247189739	-0.163478447925458\\
};
\addplot[only marks, mark=square*, mark options={}, mark size=1.2500pt, draw=black, fill=black] table[row sep=crcr]{%
x	y\\
0.368649165757727	0.125943361459841\\
};
\addplot[only marks, mark=square*, mark options={}, mark size=1.2500pt, draw=black, fill=black] table[row sep=crcr]{%
x	y\\
0.081573858278389	0.267974495775782\\
};
\addplot[only marks, mark=square*, mark options={}, mark size=1.2500pt, draw=black, fill=black] table[row sep=crcr]{%
x	y\\
-0.0811661765059352	-0.158635640810862\\
};
\addplot[only marks, mark=square*, mark options={}, mark size=1.2500pt, draw=black, fill=black] table[row sep=crcr]{%
x	y\\
-0.123660931160608	0.0247705812964559\\
};
\addplot [color=mycolor1, line width=1.0pt, forget plot]
  table[row sep=crcr]{%
1	-0\\
0.991529432508639	-0.0166646611615908\\
0.987452701789317	-0.0206259272013085\\
0.982699233533697	-0.0242592772126375\\
0.98177120352889	-0.0248679246370771\\
0.979540605081045	-0.0253365653997697\\
0.975214103252257	-0.0276612325878769\\
0.970085851276922	-0.0297399422832187\\
0.96420338871014	-0.0314210341263321\\
0.959062785926366	-0.0321878495350942\\
0.953369505131259	-0.032410234610486\\
0.947902088608217	-0.0320523840388263\\
0.941119520845239	-0.0306955694946671\\
0.933231494249329	-0.0281235644352962\\
0.928200413118671	-0.0258083415319292\\
0.923044135065404	-0.0228774057585426\\
0.917755382326056	-0.0193009873853831\\
0.912501451555282	-0.015010902147147\\
0.907289114120443	-0.0100034948601657\\
0.892646274568513	0.00903525899464674\\
0.885983049074467	0.0201659910720318\\
0.877091182723682	0.0348335213697784\\
0.874924411802166	0.0380590444809208\\
0.872109110522234	0.0411163102282495\\
0.868851532176175	0.0438966565105881\\
0.864755729884	0.0463758320258156\\
0.86008825342226	0.0486010672010091\\
0.854403596965476	0.0506523527300073\\
0.847492336254544	0.0527502084968932\\
0.829371547206306	0.0575876798914334\\
0.818139780975714	0.0610271241240083\\
0.805264899789814	0.0654606810982998\\
0.79102904444971	0.0711354570018894\\
0.77548338979816	0.0780332836222195\\
0.759068910881948	0.0865184906612031\\
0.741806910264265	0.0964497338716974\\
0.723840171203801	0.107768079872339\\
0.705237508608928	0.120378918562078\\
0.666292452659782	0.149348804776415\\
0.645828646068082	0.164803823420525\\
0.54248847777886	0.245740798966793\\
0.531727400035566	0.253727802264163\\
0.478782916486708	0.290437667616374\\
0.450257707345393	0.308718506851045\\
0.420154793019003	0.326927639546384\\
0.414507368933093	0.330103664548019\\
0.388537531724056	0.343995494507658\\
0.355407614585742	0.35963269953934\\
0.338055648702145	0.367233704764842\\
0.312523220432684	0.377452298428613\\
0.283572374954546	0.387606964738869\\
0.244774201264182	0.399161177581902\\
0.236114938828222	0.401298426558195\\
0.204406801893818	0.407962095087391\\
0.18614367465582	0.410651303271612\\
0.161312860069607	0.413749897617028\\
0.136674748310413	0.415703862776155\\
0.112350975100598	0.416487158730988\\
0.0741039897190214	0.415661067359374\\
0.0643923232419459	0.414914611086224\\
0.0411258720511996	0.412330938306292\\
0.0182998924943396	0.408670710484991\\
-0.00366443937283911	0.403747932572276\\
-0.0254743121708749	0.398028847548522\\
-0.0467825147551564	0.391328066856647\\
-0.0674356616374132	0.383651040518834\\
-0.0874516690952509	0.37504445963864\\
-0.109053612793385	0.364627588078956\\
-0.125768669366099	0.355264947473463\\
-0.143663231130568	0.344070572002753\\
-0.160373561682085	0.332055645757677\\
-0.176684747145251	0.319366984868148\\
-0.193095711760832	0.30513330126291\\
-0.206852753340708	0.291957647598659\\
-0.220592152727036	0.277327586470971\\
-0.233427348278684	0.262143754123234\\
-0.245645360428976	0.246384459583295\\
-0.256610574376086	0.230220808532584\\
-0.266618614307102	0.213650248408343\\
-0.275297363221036	0.196860165646596\\
-0.283362945711835	0.179645764248608\\
-0.290443010911968	0.162176532096278\\
-0.296609448066181	0.144477362393807\\
-0.301707030941372	0.126649364266329\\
-0.305811389693423	0.108720088202097\\
-0.308925319156813	0.0907405578158085\\
-0.311246926708961	0.0725549306356461\\
-0.312384171198784	0.0546152780199325\\
-0.312553455882023	0.0367771976208158\\
-0.311541230870046	0.019368488755299\\
-0.308856515701241	-0.00521369149460083\\
-0.307191118855974	-0.0153447394625639\\
-0.303721593029392	-0.0323453331761454\\
-0.2992997462614	-0.0489577006358961\\
-0.293689703223622	-0.0661164177072293\\
-0.287908142784843	-0.0810065972383809\\
-0.2810291279705	-0.0966179192829355\\
-0.273323708955685	-0.11147441985588\\
-0.264875668441882	-0.125793786219957\\
-0.255719162759634	-0.13954059012641\\
-0.245860936207468	-0.15234811346815\\
-0.235411674485708	-0.164850681970912\\
-0.22436359857426	-0.17668805115672\\
-0.212800803750853	-0.187919519691748\\
-0.200672825943052	-0.19835020988923\\
-0.188067816771931	-0.208041258701925\\
-0.172419558212068	-0.218635797544786\\
-0.161472721526368	-0.225335727003559\\
-0.147681861407391	-0.232686908773123\\
-0.133589610789274	-0.239230889417259\\
-0.119240895581315	-0.244956414747803\\
-0.1093998009601	-0.248364611794131\\
-0.0901267469138896	-0.253657359829326\\
-0.0752965002751094	-0.256895646641291\\
-0.0603924322647154	-0.259297977053323\\
-0.0395636182230139	-0.2613571787856\\
-0.0305382629016753	-0.261704998766637\\
-0.0156792122639293	-0.261616920696207\\
-0.000925671077823598	-0.260712770773385\\
0.013904729544969	-0.259072743613211\\
0.0334776274122273	-0.255454870806353\\
0.0425079011700218	-0.253262109067768\\
0.0564175801889868	-0.249207817202695\\
0.0700131291799131	-0.24441272815583\\
0.0829782901913274	-0.238801714176071\\
0.0958266064381839	-0.232606919877396\\
0.108251331850977	-0.225742165499129\\
0.12021800620957	-0.218234596957971\\
0.131751254677179	-0.210202317272034\\
0.142711281739145	-0.201495413494059\\
0.153121787886529	-0.1922362957869\\
0.163859015036107	-0.181503641488088\\
0.172394622710846	-0.172088213045329\\
0.180997297004011	-0.161364337299992\\
0.18895337388554	-0.150228305377781\\
0.196245473706752	-0.138718473894492\\
0.202855846249296	-0.126873108890391\\
0.208511724182088	-0.114835248582687\\
0.213723585362467	-0.102450745337993\\
0.218218479973233	-0.0898502690133107\\
0.22198553640342	-0.0770723923892866\\
0.225121673670354	-0.0642022710237951\\
0.227426851520583	-0.0511912524163969\\
0.228996425466648	-0.0381269079688011\\
0.229830793311064	-0.0250493210119651\\
0.229933763296585	-0.0119992232362562\\
0.229371532428699	0.00120839768999925\\
0.228017319161699	0.0140884924645237\\
0.22595415342479	0.0268216736939841\\
0.221844078169176	0.0445246335593146\\
0.219750405669625	0.0516956713196355\\
0.215534762820826	0.0635218225600329\\
0.210787500941961	0.0752887341905963\\
0.205410626622161	0.086727928199507\\
0.199425462605767	0.0978056177329878\\
0.192961565321435	0.108519157892361\\
0.184498639036715	0.120574997480693\\
0.178164813353817	0.12860177348131\\
0.169992497898646	0.137938784463338\\
0.161342229178955	0.146771288593472\\
0.152124714169246	0.155275465682539\\
0.142599245192235	0.163022825374792\\
0.132690962592999	0.170199349011362\\
0.119305545781881	0.178636528007099\\
0.111859758462256	0.182766687525548\\
0.101096182258481	0.187894153009525\\
0.0900112587572195	0.192626569491747\\
0.0787177976812403	0.196715813124499\\
0.0672526300811722	0.200154190140316\\
0.0556528443937163	0.202935081004962\\
0.0440198140184347	0.205149056197593\\
0.0322560794278899	0.206605297565541\\
0.0204705574544324	0.207399505468426\\
0.0086986063056127	0.207531277833876\\
-0.00302331465174288	0.20700465232024\\
-0.0146589580000303	0.205825342375251\\
-0.0264040013407103	0.204029524830439\\
-0.0377622666632496	0.201557886625489\\
-0.0489274992659923	0.198462908731172\\
-0.0598663396209274	0.194758306013778\\
-0.0705461769701865	0.190459281207171\\
-0.0807084286258493	0.185510174696866\\
-0.0907711973246872	0.180088218148021\\
-0.100483460488876	0.174128587906053\\
-0.109816719726697	0.1676535154034\\
-0.118743761375389	0.16068617301122\\
-0.127260311891991	0.153361241948399\\
-0.135305358384817	0.145484752392752\\
-0.142872955606751	0.13719432245907\\
-0.149941127136624	0.128518316782757\\
-0.156490141836851	0.119485753505815\\
-0.1625051444717	0.110128070201329\\
-0.168134868274921	0.100310790535883\\
-0.173023317551057	0.0903862052401878\\
-0.177334403290235	0.0802326484390004\\
-0.181056785209569	0.0698828455491056\\
-0.184182966719897	0.0593708837824785\\
-0.186515891966093	0.0488533677013545\\
-0.188438230198684	0.0381290445282494\\
};
\addplot[only marks, mark=square*, mark options={}, mark size=1.2500pt, draw=black, fill=mycolor1] table[row sep=crcr]{%
x	y\\
1	0\\
};
\addplot[only marks, mark=square*, mark options={}, mark size=1.2500pt, draw=black, fill=mycolor1] table[row sep=crcr]{%
x	y\\
0.928837025744488	-0.0261303645720788\\
};
\addplot[only marks, mark=square*, mark options={}, mark size=1.2500pt, draw=black, fill=mycolor1] table[row sep=crcr]{%
x	y\\
0.568399349653536	0.226026796967463\\
};
\addplot[only marks, mark=square*, mark options={}, mark size=1.2500pt, draw=black, fill=mycolor1] table[row sep=crcr]{%
x	y\\
0.120425843980702	0.41635305898171\\
};
\addplot[only marks, mark=square*, mark options={}, mark size=1.2500pt, draw=black, fill=mycolor1] table[row sep=crcr]{%
x	y\\
-0.12404947175284	-0.243139283869475\\
};
\addplot[only marks, mark=square*, mark options={}, mark size=1.2500pt, draw=black, fill=mycolor1] table[row sep=crcr]{%
x	y\\
-0.188438230198684	0.0381290445282494\\
};
\addplot [color=mycolor2, line width=1.0pt, forget plot]
  table[row sep=crcr]{%
1	-0\\
0.991329850454347	-0.0192843137630594\\
0.987070134476456	-0.0243414927180843\\
0.982035337561272	-0.0292516312369064\\
0.981043922351956	-0.0301096267039906\\
0.978744340469279	-0.0308429727500013\\
0.974070915784344	-0.0343363097059279\\
0.968451219091321	-0.0378031567161459\\
0.961877533050262	-0.0411107180694699\\
0.956042336890473	-0.0432740938404623\\
0.949459687573443	-0.0450455644791909\\
0.940209900461709	-0.0466400286414501\\
0.934636888721915	-0.0468660028339432\\
0.928689802464703	-0.0466641820771896\\
0.918449605881063	-0.04526168976411\\
0.911603430653221	-0.0436700518279376\\
0.904361728284758	-0.0414021235800441\\
0.896859352646953	-0.0383581037484078\\
0.889069445078612	-0.0344737684527892\\
0.87733840921562	-0.0269539236751293\\
0.868972221611915	-0.0206864525291559\\
0.864441923224073	-0.0171786822144131\\
0.855849432606534	-0.00939425064796784\\
0.847006305308204	-0.000554796754825837\\
0.838128120374833	0.00916864443287846\\
0.829364278337117	0.0196475779797749\\
0.814526126542132	0.0366615665667467\\
0.803636008316602	0.0484425518084641\\
0.797546560704424	0.0544837917408807\\
0.790972311073226	0.0604694978829836\\
0.783713629107132	0.0665053660951256\\
0.775807582812931	0.0726214614564955\\
0.766706868357092	0.0789620773069155\\
0.75647510094905	0.0856132501747162\\
0.71799603575904	0.109011176600844\\
0.702654339625932	0.118624589795006\\
0.686144075087912	0.129347268497811\\
0.650103925105071	0.154441035880446\\
0.630608275522495	0.168310125616091\\
0.566977453513082	0.215663911504313\\
0.543871669065697	0.233015822940562\\
0.519720349598303	0.250449026177224\\
0.480103838731163	0.277414057190164\\
0.455133231062355	0.293520052008267\\
0.430191621624834	0.308673144393312\\
0.405143898189332	0.323147829556435\\
0.379779021273152	0.336461050273104\\
0.34743463700429	0.351549651490336\\
0.330498070138278	0.358850657233555\\
0.305598627695154	0.368663049883314\\
0.277357967114376	0.3784279420074\\
0.255974219120021	0.384859055914248\\
0.23106658764802	0.391534951015903\\
0.200156491466623	0.397941759853437\\
0.182343446786602	0.400589209989591\\
0.158140819970912	0.403558825982022\\
0.13415450098991	0.405414454223699\\
0.110452834239217	0.406145074554173\\
0.0731830592154215	0.405248831379424\\
0.0637217991265	0.404514716816412\\
0.0410557484592888	0.401984727077221\\
0.0188191592755245	0.398411289616333\\
-0.00260757670006462	0.393669783330102\\
-0.023855988409327	0.388099140450588\\
-0.0445818125049071	0.381570273336069\\
-0.0647028913747192	0.374099180833547\\
-0.0842020461587778	0.365726384642843\\
-0.105245375782772	0.355552183791459\\
-0.121530765051026	0.346447689582471\\
-0.138966234809155	0.335562530590776\\
-0.155291871059585	0.323922641311799\\
-0.171188361414257	0.311585544680965\\
-0.18718621389583	0.297745552947148\\
-0.200558700595062	0.284932783760184\\
-0.213954269596426	0.270708560102797\\
-0.226471068217276	0.255945822667541\\
-0.238380831542523	0.240593472456385\\
-0.249078845419798	0.224878552636524\\
-0.258847155786486	0.208767790906421\\
-0.26737231111396	0.192467119640013\\
-0.27525284815808	0.175727878868049\\
-0.282175206618797	0.158740326768954\\
-0.288167284492123	0.14153357170086\\
-0.293161544067857	0.124196273142454\\
-0.297190212698289	0.106759857742849\\
-0.300255600694959	0.0892735093297758\\
-0.30254252540424	0.0715679368691184\\
-0.303686606533915	0.0541195780735613\\
-0.303889153684201	0.0367683862371349\\
-0.302993505102303	0.0198360882248116\\
-0.300436794215437	-0.00408123931597482\\
-0.298839800038496	-0.0139382945844588\\
-0.295462707924092	-0.0304653117399054\\
-0.291200085876279	-0.0466310497325151\\
-0.285784090678492	-0.0633295519204153\\
-0.280196436273251	-0.077823040783398\\
-0.273537660861999	-0.0930274424399842\\
-0.266080936924959	-0.107490499945404\\
-0.257901921684794	-0.121432705902576\\
-0.249033084268322	-0.134819215436664\\
-0.239522920221029	-0.147313512829807\\
-0.22939491743665	-0.159494933408251\\
-0.218683821990327	-0.171030279330549\\
-0.207436840759116	-0.181947426395172\\
-0.195674021299201	-0.192116931566784\\
-0.183446164545194	-0.20156793538399\\
-0.168263602841223	-0.211904435408564\\
-0.157640180743986	-0.21844733980268\\
-0.144256687538088	-0.22562678523225\\
-0.130578504299867	-0.232021945814653\\
-0.116649839111468	-0.237622296559393\\
-0.102689116009897	-0.242190910910957\\
-0.0884043979489315	-0.246188622636704\\
-0.0740015179907678	-0.24937465416886\\
-0.0595247778312185	-0.251748309442919\\
-0.0392726712841793	-0.253762323973965\\
-0.0305043591288712	-0.254123199116639\\
-0.016067641146662	-0.254076302617079\\
-0.0017316482632066	-0.253236286129384\\
0.0126755974320349	-0.251686423500107\\
0.0316987340609434	-0.248223360390525\\
0.0404763662200109	-0.246117148903034\\
0.0539985685970781	-0.242215769971133\\
0.067217183194656	-0.237594139911083\\
0.079826404286004	-0.232218879903389\\
0.0923241530645651	-0.226236160407295\\
0.104411525892621	-0.219601623418392\\
0.116055792360534	-0.212341737483999\\
0.127277149340704	-0.204525014513191\\
0.13794415272644	-0.196097798000505\\
0.148079198563045	-0.18713328875135\\
0.158534847706371	-0.176738768423657\\
0.166842554610795	-0.167628221427982\\
0.17522471281576	-0.157237767582769\\
0.182980817965908	-0.146445915625812\\
0.190092089819101	-0.135289239150903\\
0.196542352125348	-0.123805376684665\\
0.202091216254046	-0.11216145552076\\
0.207185679182418	-0.10014971865602\\
0.211583776759724	-0.0879260230147902\\
0.215277922991396	-0.0755298630189669\\
0.218332293580806	-0.0630042409012597\\
0.22060467613478	-0.0503785863548361\\
0.222162813637494	-0.0376989470094165\\
0.223007658138205	-0.025005168104588\\
0.223142155849551	-0.0123365967816751\\
0.222630792248451	0.000469431124927588\\
0.22135098720543	0.0129756382481108\\
0.219382198237356	0.0253417108592622\\
0.215440333296187	0.0425358908969371\\
0.213427333136944	0.0495022333557051\\
0.209403279448735	0.0609868179633788\\
0.204827537154965	0.0724221427572842\\
0.199639662314846	0.0835404691631674\\
0.193860465279937	0.0943104212082067\\
0.187572775518309	0.10474324045618\\
0.179390486495164	0.11646661050391\\
0.173264231705393	0.124273727090345\\
0.165356857991187	0.133356652395881\\
0.156984794983161	0.1419514709349\\
0.148072962450665	0.150210799160143\\
0.138849924500136	0.15775483641649\\
0.129253733291798	0.16474463463608\\
0.116287137613274	0.172967917631712\\
0.10907256279611	0.176994689949691\\
0.0986712048911107	0.182019234698882\\
0.0879268058325295	0.18664230491687\\
0.0769784950388703	0.19064147067064\\
0.0658618305875607	0.194008956056244\\
0.0546132077285548	0.196739296549243\\
0.0432855926529623	0.198902204555043\\
0.031876842547681	0.20034794670023\\
0.020443826331294	0.201148962051378\\
0.00902176089952311	0.201306739485704\\
-0.00235315313186613	0.200825675241578\\
-0.013645304307349	0.199711381119375\\
-0.0250216177218234	0.197992152185607\\
-0.0360468182522802	0.195624105010061\\
-0.046887028033181	0.192650509531504\\
-0.0575088088270213	0.189085140516184\\
-0.0678809297076493	0.184942150769426\\
-0.0777474527470809	0.180201804424242\\
-0.087525092933803	0.174968373490302\\
-0.0969630723761217	0.169212684727547\\
-0.106035501853003	0.162955742040478\\
-0.114714668467758	0.156220412644957\\
-0.123018276312213	0.149098552647781\\
-0.130842587885372	0.141477964878235\\
-0.138204052992551	0.133454626489754\\
-0.145082849336241	0.125056276057487\\
-0.151459226352793	0.116311050009384\\
-0.157316978120291	0.10724874564649\\
-0.162776648625286	0.0977518932029982\\
-0.167543900332023	0.0881385604924438\\
-0.171750774265029	0.0783013383696205\\
-0.175387686997499	0.0682729327758351\\
-0.17844637446723	0.0580855144378736\\
-0.180758007380087	0.0479167194036776\\
-0.182650731301679	0.0375193784234813\\
};
\addplot[only marks, mark=square*, mark options={}, mark size=1.2500pt, draw=black, fill=mycolor2] table[row sep=crcr]{%
x	y\\
1	0\\
};
\addplot[only marks, mark=square*, mark options={}, mark size=1.2500pt, draw=black, fill=mycolor2] table[row sep=crcr]{%
x	y\\
0.919280532097948	-0.0454160959525164\\
};
\addplot[only marks, mark=square*, mark options={}, mark size=1.2500pt, draw=black, fill=mycolor2] table[row sep=crcr]{%
x	y\\
0.55554693148818	0.22442562544286\\
};
\addplot[only marks, mark=square*, mark options={}, mark size=1.2500pt, draw=black, fill=mycolor2] table[row sep=crcr]{%
x	y\\
0.118320939577774	0.406024526190401\\
};
\addplot[only marks, mark=square*, mark options={}, mark size=1.2500pt, draw=black, fill=mycolor2] table[row sep=crcr]{%
x	y\\
-0.121317761563927	-0.235844130609943\\
};
\addplot[only marks, mark=square*, mark options={}, mark size=1.2500pt, draw=black, fill=mycolor2] table[row sep=crcr]{%
x	y\\
-0.182650731301679	0.0375193784234813\\
};
\addplot [color=mycolor3, line width=1.0pt, forget plot]
  table[row sep=crcr]{%
1	-0\\
0.991142079534531	-0.0219042834379395\\
0.986711812629501	-0.0280579516942165\\
0.981416360924679	-0.0342461577628199\\
0.980366424800613	-0.0353538444900345\\
0.978003231991555	-0.0363522674961476\\
0.973010965047223	-0.0410166765641038\\
0.966942261113867	-0.0458760393869893\\
0.959741539736943	-0.0508182806560307\\
0.953280708394293	-0.054388791992082\\
0.945902854410048	-0.0577263236697898\\
0.935256750304422	-0.0616583738249969\\
0.928811467085718	-0.0631544382094509\\
0.921853924460319	-0.0642843305828054\\
0.905727336363487	-0.0649842092435775\\
0.897169783302499	-0.0645118333064285\\
0.888009999571365	-0.0633255244559778\\
0.878361945372019	-0.0613061375004085\\
0.86860661477528	-0.0584004086377083\\
0.857890494566012	-0.0543966977483268\\
0.846671834715429	-0.0492408244224218\\
0.840601289931157	-0.0461922509450927\\
0.828675531769956	-0.0391252608278168\\
0.81617036052402	-0.0306166661108271\\
0.803348008968735	-0.0206845910059295\\
0.796767164500222	-0.0151476957380106\\
0.776650748386703	0.00375493453401066\\
0.769306922749501	0.0108713612051881\\
0.754634054844888	0.0262341548553839\\
0.730965009893972	0.0521168086423325\\
0.713983687924476	0.0710115003387863\\
0.704731357022776	0.0809513093417471\\
0.684416970153151	0.101664442280334\\
0.67330260458293	0.112445153961784\\
0.661374398639272	0.123561641988167\\
0.634667181246314	0.147006131633631\\
0.619555368016244	0.15932613541722\\
0.585508775777865	0.185459437863704\\
0.535814231431844	0.221827922050165\\
0.502108279957468	0.245199619420789\\
0.464279922891095	0.269867284589616\\
0.440341597257998	0.284692154306297\\
0.416370885130329	0.298743584675667\\
0.39231055381057	0.312162642337217\\
0.367891777832348	0.324623669373198\\
0.336699269370647	0.338889991625085\\
0.320391772140911	0.345768986705617\\
0.29642911121551	0.355019684920289\\
0.269245709658801	0.364256552902206\\
0.248666568203384	0.370350035600362\\
0.224733517187948	0.376611768988196\\
0.19499661569996	0.382700617413051\\
0.177816575664243	0.385293262451961\\
0.154537637315058	0.388120637486556\\
0.131482786949374	0.389886710966908\\
0.108690283313435	0.39058734130872\\
0.0861960002739353	0.39024098973002\\
0.0728853093975299	0.389687320101231\\
0.0637879087903734	0.388990740003068\\
0.0419940409537172	0.386586767609397\\
0.0206124951007551	0.383186816191309\\
-4.49886901590091e-05	0.378717852720131\\
-0.020479485122348	0.373407844113496\\
-0.040396053921278	0.367183115909923\\
-0.0597485917353524	0.360057428170863\\
-0.0785068167231684	0.352068466007217\\
-0.0987158945535787	0.342323597165163\\
-0.114387091822821	0.333633006015906\\
-0.131168123817407	0.323240214016393\\
-0.146941907485534	0.312146263709845\\
-0.162250672275913	0.300359445781082\\
-0.177662791570267	0.287134051366954\\
-0.190530369476271	0.274897301941279\\
-0.20344549260026	0.261300898264484\\
-0.215519074404759	0.247187210098458\\
-0.226978976079195	0.232490053339028\\
-0.237312462930379	0.217462507388593\\
-0.246755757507945	0.202053896701133\\
-0.255062216009672	0.186458716683062\\
-0.262699178823994	0.170441549388746\\
-0.269418645050675	0.154183852368074\\
-0.275228276743493	0.137731226454284\\
-0.280101543249654	0.121134664860651\\
-0.284050382976466	0.104440431719429\\
-0.287076953060346	0.087695790171189\\
-0.289329638650657	0.0707360298553315\\
-0.290516616238494	0.0540229831845278\\
-0.290801595039321	0.0373997280242813\\
-0.290077595731696	0.0211504120934212\\
-0.287755060281056	-0.00177622507903052\\
-0.286278117180961	-0.0112272469829227\\
-0.283117112900601	-0.0270503786800549\\
-0.27912258021296	-0.0425551249501965\\
-0.274027414762571	-0.058574571218333\\
-0.268756705923695	-0.0724825180775832\\
-0.262445389375017	-0.087069280534382\\
-0.255388107287495	-0.100955146452973\\
-0.247637406931018	-0.114345892729034\\
-0.239224500372755	-0.127207838953986\\
-0.230218491104116	-0.139258793815569\\
-0.22059409005742	-0.150973955611996\\
-0.210409345849412	-0.162073549616452\\
-0.199703424559638	-0.172551731417303\\
-0.188507272886152	-0.182347819860642\\
-0.176863550342382	-0.191459190218582\\
-0.162399199477948	-0.201433023283943\\
-0.152267403837307	-0.207742273413154\\
-0.139509209232379	-0.214685861239574\\
-0.126465627177926	-0.220880693755458\\
-0.113178868883209	-0.22631668433263\\
-0.0998571622639994	-0.230810704554961\\
-0.0862206615516494	-0.234716125336475\\
-0.0724678333869535	-0.237846329221501\\
-0.0586404329994903	-0.240199766420346\\
-0.0393004323260973	-0.242208505728965\\
-0.0309216009138062	-0.24260459246178\\
-0.0171224357043003	-0.242644013633128\\
-0.00341583171311033	-0.241924308762485\\
0.0103608749127466	-0.24052104914136\\
0.0285590691796718	-0.23732377238441\\
0.0369584570094181	-0.235362847933317\\
0.0499016944292372	-0.231714478317983\\
0.062557773558866	-0.227376756425093\\
0.0746610793361706	-0.222349367079372\\
0.0866367876467056	-0.216706721128876\\
0.0982236765742921	-0.210439267650572\\
0.109389962288531	-0.20357221986988\\
0.120128437177212	-0.196143642730704\\
0.130366446986346	-0.188157222761257\\
0.140098104808969	-0.179654964994774\\
0.150144158886282	-0.169789340212091\\
0.158127201414731	-0.161140767810168\\
0.166191931792328	-0.151268575702569\\
0.173661360363467	-0.141010237647305\\
0.180516306029753	-0.130400220042187\\
0.186741531567995	-0.119474367346874\\
0.192145930076084	-0.10840164380127\\
0.197080557359032	-0.0969635807524705\\
0.201351400590507	-0.085319611669376\\
0.204950610014746	-0.0735069878119587\\
0.207901296265231	-0.0615579772967354\\
0.210143628532514	-0.0495205407302277\\
0.211704601609612	-0.0374282238813257\\
0.212584519836129	-0.0253186997710138\\
0.212786331280895	-0.0132295255838759\\
0.212375700665377	-0.0010161932027557\\
0.211227460104916	0.0109248369275052\\
0.209421476936587	0.0227348930082774\\
0.205760447785182	0.0391634709339734\\
0.203880701865563	0.0458209577148898\\
0.200142514966785	0.0568184247137953\\
0.195843322270718	0.0677552969457722\\
0.190958261033288	0.0783937100685386\\
0.185507142272553	0.0887023306722601\\
0.179530755749417	0.0986774701747011\\
0.171793443428935	0.109906413562117\\
0.165994768318832	0.117388294890847\\
0.158505093684628	0.126096282139858\\
0.150569582781237	0.134340983342967\\
0.142130375121337	0.142261323404512\\
0.133378416650079	0.149508343327148\\
0.124267948419126	0.156229622991384\\
0.111950860168595	0.164145118327283\\
0.105094892216365	0.168026383977212\\
0.0952189692174414	0.172910755603854\\
0.0849995613285954	0.177380613666985\\
0.0745823321587238	0.18125660471463\\
0.0640017508634541	0.184531473790674\\
0.0532913255197369	0.187198727430404\\
0.0424817310885706	0.189289991636576\\
0.0316142829831667	0.190732557173628\\
0.0207196767710425	0.191559957435813\\
0.00983236151385181	0.191774239933196\\
-0.00101289953430861	0.191379237798189\\
-0.0117821689096906	0.190380476143807\\
-0.0226162032986705	0.188810905725951\\
-0.0331378801475182	0.186616086618971\\
-0.0434856086480064	0.183843407099959\\
-0.0536285829509793	0.180505160165967\\
-0.0635358788273657	0.176615611018636\\
-0.0729821398447894	0.172185037563526\\
-0.0823305359516284	0.167252157779174\\
-0.0913583677553265	0.161819601019642\\
-0.100039400553664	0.155907437692927\\
-0.108348760135908	0.149537332639792\\
-0.11629244396597	0.142754781821832\\
-0.123789740174179	0.135536931751613\\
-0.130847670238225	0.127932622733134\\
-0.137447120987047	0.119968373500437\\
-0.14357000454327	0.111671232875668\\
-0.149199951202361	0.103069380199052\\
-0.154442046802978	0.0940727595816595\\
-0.159037131065085	0.084941444240588\\
-0.16309904959682	0.075594028192576\\
-0.166618487543661	0.0660611190538818\\
-0.16958832501528	0.0563743543756579\\
-0.171880032268313	0.0467086024959755\\
-0.173740510079624	0.0368142357225292\\
};
\addplot[only marks, mark=square*, mark options={}, mark size=1.2500pt, draw=black, fill=mycolor3] table[row sep=crcr]{%
x	y\\
1	0\\
};
\addplot[only marks, mark=square*, mark options={}, mark size=1.2500pt, draw=black, fill=mycolor3] table[row sep=crcr]{%
x	y\\
0.910798149597252	-0.0649567844847587\\
};
\addplot[only marks, mark=square*, mark options={}, mark size=1.2500pt, draw=black, fill=mycolor3] table[row sep=crcr]{%
x	y\\
0.535814231431844	0.221827922050165\\
};
\addplot[only marks, mark=square*, mark options={}, mark size=1.2500pt, draw=black, fill=mycolor3] table[row sep=crcr]{%
x	y\\
0.116256420947042	0.390470775813316\\
};
\addplot[only marks, mark=square*, mark options={}, mark size=1.2500pt, draw=black, fill=mycolor3] table[row sep=crcr]{%
x	y\\
-0.11763237831727	-0.22458956681994\\
};
\addplot[only marks, mark=square*, mark options={}, mark size=1.2500pt, draw=black, fill=mycolor3] table[row sep=crcr]{%
x	y\\
-0.173740510079624	0.0368142357225292\\
};
\addplot [color=mycolor4, line width=1.0pt, forget plot]
  table[row sep=crcr]{%
1	-0\\
0.990966110418654	-0.0245245669137546\\
0.986377699299556	-0.0317752879109958\\
0.980842183065299	-0.039242815197325\\
0.979738564652898	-0.0406005261698563\\
0.977317099107648	-0.041864408096522\\
0.972033864155203	-0.0477021765717194\\
0.965558141958495	-0.0539582226795403\\
0.957793621210321	-0.0605428585940393\\
0.950774772106959	-0.0655302057988891\\
0.942693549345041	-0.0704491329567212\\
0.939743426962415	-0.0720476986989149\\
0.937433040248527	-0.073739595044561\\
0.930826538677313	-0.076754041398452\\
0.92362717459937	-0.0795479145008287\\
0.915803750281939	-0.082047236920797\\
0.897561746125239	-0.0856857827944952\\
0.887727608552487	-0.0867633107316856\\
0.8771293190913	-0.0871626426170435\\
0.865867895977779	-0.0867587891198205\\
0.854310985408753	-0.0854849568044391\\
0.841600556702518	-0.0830911864897519\\
0.82818412755427	-0.0795054299397573\\
0.820939964490001	-0.0772070780696905\\
0.806502134690787	-0.0715891578633165\\
0.791267918655352	-0.0644077857092438\\
0.775517148612406	-0.0556174500992179\\
0.767399303407046	-0.050566441221064\\
0.751057324018505	-0.0391310242083887\\
0.733397027086713	-0.0257065771625575\\
0.715294732137494	-0.0103036796758413\\
0.696296997926933	0.00719684520420794\\
0.686530478467532	0.0167038841084577\\
0.644839710587044	0.0599129362529132\\
0.584032838466142	0.124968946997858\\
0.570166228216296	0.139228817596954\\
0.540084861053094	0.168624225073309\\
0.523837645418312	0.18368084076642\\
0.497620637533219	0.206823609745899\\
0.477322146720446	0.223325307432789\\
0.468769968612921	0.229953069957399\\
0.447907421575911	0.245163025220139\\
0.425681001706435	0.260342982088738\\
0.401965694081841	0.275205988943671\\
0.376637215368225	0.289870578472799\\
0.371840993910689	0.292455904750281\\
0.349668346237298	0.303776453003182\\
0.321046305332546	0.316761506870563\\
0.30601963132306	0.323002098169569\\
0.283861408058012	0.331386938791336\\
0.258621933143398	0.339811439165965\\
0.239464200785745	0.345394663076511\\
0.217226884527755	0.351131609891203\\
0.189459230835794	0.356787012097034\\
0.173300862134517	0.359269972688748\\
0.151528641289424	0.361954984800944\\
0.129968668234747	0.363665292556451\\
0.108636516216218	0.364419969909601\\
0.0875788119124701	0.364218162500525\\
0.0666901235356174	0.363169867418138\\
0.0462792639485439	0.361080524299107\\
0.026250210967993	0.358073354360746\\
0.00678513544330572	0.354092585955434\\
-0.0123687371383558	0.349315729966339\\
-0.0310232886138859	0.343686563125551\\
-0.0491763558635765	0.337226943138056\\
-0.0667797022969321	0.329965240173315\\
-0.0856751262532274	0.321093966823548\\
-0.100398672918369	0.313164034850808\\
-0.1161752063265	0.303666510352268\\
-0.131111736773198	0.29349978336365\\
-0.145526081538071	0.28269694943405\\
-0.160052216675395	0.270562471670881\\
-0.172175578817467	0.259328325146903\\
-0.184376901564227	0.246831617540028\\
-0.195800386605629	0.23384902616211\\
-0.206599275625546	0.220343636596267\\
-0.216414508343528	0.206501929362452\\
-0.225406303006036	0.192299718916029\\
-0.233409095614862	0.177877701124056\\
-0.240726605969644	0.16309322085062\\
-0.247194898949653	0.148078125031448\\
-0.25280989487241	0.132884763605635\\
-0.257568167592924	0.117539440743951\\
-0.261469923794247	0.10209528419929\\
-0.264516675354721	0.0865958250646237\\
-0.266813613287604	0.0709272684204409\\
-0.268156574187041	0.0554406818041304\\
-0.268662337996161	0.0400282360400233\\
-0.268262813875794	0.0249000636313452\\
-0.266442548409328	0.00361371377467479\\
-0.265211408090194	-0.0051665313489937\\
-0.262509007895084	-0.0198613986781122\\
-0.25903605135494	-0.034282863221462\\
-0.254552969177924	-0.0491939097226279\\
-0.249878129583591	-0.0621484024254004\\
-0.244240453072939	-0.075703521281522\\
-0.237916888627332	-0.088659563193151\\
-0.230946324796881	-0.101165057986376\\
-0.223357096364846	-0.113188783187686\\
-0.215210188536936	-0.124522898427539\\
-0.20648456650729	-0.13549995411257\\
-0.197233597573977	-0.145914306391332\\
-0.187496706665818	-0.155745702152148\\
-0.177296320165148	-0.164966470078118\\
-0.166674099574516	-0.17355969132863\\
-0.153461750098441	-0.182990733135941\\
-0.144209749306923	-0.188936948408174\\
-0.132534038400784	-0.195542170210753\\
-0.120585693339673	-0.201459092204746\\
-0.10840328667597	-0.206678117674654\\
-0.096149642260509	-0.211068379345119\\
-0.0836236186604811	-0.214877520406335\\
-0.0709804403877676	-0.21797235048994\\
-0.0582593039608086	-0.220351654403959\\
-0.0404623185590249	-0.222483362455922\\
-0.0327411093357373	-0.222974650722122\\
-0.0200177764262623	-0.223218646462399\\
-0.00737044058760938	-0.222761683958248\\
0.0053255230541196	-0.221660833774407\\
0.0221446906976857	-0.218990268842595\\
0.0299135756593996	-0.217310808722607\\
0.0418933765564296	-0.21414575312863\\
0.0536177660313919	-0.210341691536668\\
0.0648848027830626	-0.205907065180801\\
0.0759998861573783	-0.200889608453771\\
0.0867646468363419	-0.19529180216095\\
0.0971490712800636	-0.189136234704737\\
0.107129056949496	-0.182454864759354\\
0.116672354818582	-0.175257249107292\\
0.125756518411919	-0.167578172119079\\
0.135149241349549	-0.158650337074755\\
0.142601226824749	-0.150818919939046\\
0.150169527517925	-0.141858942708084\\
0.157194139419266	-0.132535654546673\\
0.163658856104518	-0.122881189800923\\
0.169548289120303	-0.112927758084652\\
0.174721899759542	-0.102810650240379\\
0.179430174209706	-0.0923673030919876\\
0.183529931310237	-0.0817256358313569\\
0.187014598755065	-0.0709206387287313\\
0.189885743714628	-0.0599897994396403\\
0.192125607907242	-0.048961151994223\\
0.193739786661993	-0.0378733800031383\\
0.194728368708793	-0.0267611185998131\\
0.195093421861117	-0.0156585232426412\\
0.194889505825791	-0.00445309500710644\\
0.194013109423126	0.00653077712286865\\
0.192530751264481	0.0174027066995859\\
0.189414216206484	0.0325412784645425\\
0.187787415093564	0.0386815054755878\\
0.184534159214213	0.0488694043056748\\
0.180748158654453	0.0589737790882789\\
0.176420564083465	0.0688106449253674\\
0.171568836897203	0.0783521623696986\\
0.166219960844448	0.0875747667478786\\
0.159285257306518	0.0979915767106225\\
0.154074925094634	0.104940579907229\\
0.147331478780161	0.113039145312364\\
0.14017279486813	0.120717861126215\\
0.132558318889983	0.128088567505161\\
0.124638760847265	0.134863813391184\\
0.116382889416671	0.141160234809984\\
0.105205520118983	0.148598578437571\\
0.0989768521848864	0.152257403844889\\
0.0899795710757085	0.156915082762923\\
0.0806749335572661	0.161160397530995\\
0.0711820367481608	0.164864945864817\\
0.0615305087636948	0.168019325390506\\
0.0517524857411422	0.170618590974713\\
0.0418814835721746	0.172667410466793\\
0.0319440590506548	0.174145277647447\\
0.0219741290814204	0.175060151395765\\
0.0120035280385933	0.175413128470527\\
0.00206307579641241	0.175206928528777\\
-0.00781642608887023	0.174445764299673\\
-0.0177423577396481	0.173164316360841\\
-0.0274100138184361	0.171305642009262\\
-0.0369256054260403	0.168915540783245\\
-0.0462612990420976	0.166004543307358\\
-0.0553880096891417	0.162585499817288\\
-0.0641357947701067	0.158675617377591\\
-0.0727652055350587	0.154291714819337\\
-0.0811073605645776	0.149446050386915\\
-0.0891381267394253	0.144156731997741\\
-0.0968341836335349	0.138443015281789\\
-0.104179531134273	0.132333391568302\\
-0.111142376604342	0.125833779495377\\
-0.117707736216911	0.118974793987114\\
-0.123858141884816	0.111780420699125\\
-0.129576712507369	0.104275300587225\\
-0.134847959931245	0.0964847767836148\\
-0.139760418118591	0.0883465028328496\\
-0.144092145671612	0.0800588112431477\\
-0.147939356880759	0.0715665828538257\\
-0.15129251413632	0.0628976470739422\\
-0.154144640991637	0.0540805833856535\\
-0.1564066913514	0.0452538149611323\\
-0.158245784472275	0.0362309613859633\\
};
\addplot[only marks, mark=square*, mark options={}, mark size=1.2500pt, draw=black, fill=mycolor4] table[row sep=crcr]{%
x	y\\
1	0\\
};
\addplot[only marks, mark=square*, mark options={}, mark size=1.2500pt, draw=black, fill=mycolor4] table[row sep=crcr]{%
x	y\\
0.903353602890586	-0.0847153751515627\\
};
\addplot[only marks, mark=square*, mark options={}, mark size=1.2500pt, draw=black, fill=mycolor4] table[row sep=crcr]{%
x	y\\
0.497620637533219	0.206823609745899\\
};
\addplot[only marks, mark=square*, mark options={}, mark size=1.2500pt, draw=black, fill=mycolor4] table[row sep=crcr]{%
x	y\\
0.115718606622367	0.364275163341364\\
};
\addplot[only marks, mark=square*, mark options={}, mark size=1.2500pt, draw=black, fill=mycolor4] table[row sep=crcr]{%
x	y\\
-0.112487715605355	-0.205016458115462\\
};
\addplot[only marks, mark=square*, mark options={}, mark size=1.2500pt, draw=black, fill=mycolor4] table[row sep=crcr]{%
x	y\\
-0.158245784472275	0.0362309613859632\\
};
\end{axis}
\end{tikzpicture}%

%% file: images/1absSears.tex
%
%
\definecolor{mycolor1}{rgb}{0.18517,0.05913,0.24304}%
\definecolor{mycolor2}{rgb}{0.51514,0.10993,0.38770}%
\definecolor{mycolor3}{rgb}{0.81038,0.26571,0.33825}%
\definecolor{mycolor4}{rgb}{0.94901,0.58547,0.40375}%
\begin{tikzpicture}[%
trim axis left, trim axis right
]

\begin{axis}[%
width=\fwidth,
height=1\fheight,
at={(0\fwidth,0\fheight)},
scale only axis,
xmode=log,
xmin=0.01,
xmax=100,
xminorticks=true,
xlabel style={font=\color{white!15!black}},
xlabel={$k$},
ymin=0,
ymax=1,
ylabel style={font=\color{white!15!black}},
ylabel={$|S(k)|$},
axis background/.style={fill=white},
xmajorgrids,
xminorgrids,
ymajorgrids
]
\addplot [color=black, line width=2.0pt, forget plot]
  table[row sep=crcr]{%
0.00398107170553497	0.983716406101629\\
0.01	0.983224957942028\\
0.0129339698301758	0.978133912262007\\
0.0158897592972195	0.972975934194222\\
0.0195210328954464	0.966619750373419\\
0.0227793799994691	0.960913222690428\\
0.0265815930918929	0.954265717760441\\
0.0310184513941747	0.946541658262916\\
0.036195886513146	0.937594559957146\\
0.042237511596681	0.927269669713438\\
0.0468156504137495	0.919542532036258\\
0.0518900153160213	0.911085518400478\\
0.0575143924243362	0.901853076048344\\
0.0637483977561905	0.891802168064446\\
0.0706581091302969	0.880893604384301\\
0.0783167665635036	0.869093503784415\\
0.0868055485839823	0.856374843963984\\
0.096214432689278	0.842719040142721\\
0.10664314906971	0.828117476750842\\
0.118202237706192	0.812572903865998\\
0.131014220047254	0.796100602379132\\
0.145214897684557	0.778729221662759\\
0.160954791792299	0.760501202452091\\
0.187820510554411	0.731679473743774\\
0.219170512366241	0.701303650758001\\
0.255753289931371	0.669674249775624\\
0.31420037852572	0.62614776948267\\
0.553371463857602	0.505155289243942\\
0.645737288779182	0.47338878664773\\
0.753520326496612	0.442654801374686\\
0.879293936264728	0.413138063031001\\
1.02606100879403	0.384972819980719\\
1.16981132075472	0.362177533719978\\
1.33962264150943	0.339736342947625\\
1.54861739809169	0.317013032188307\\
1.73584905660377	0.300050240157402\\
2.0188679245283	0.278836996913826\\
2.33730336219578	0.259574595601968\\
2.64150943396226	0.244435489256543\\
3.03773584905661	0.228160154131422\\
3.52765441849198	0.211894624480851\\
4.05660377358491	0.197710541847969\\
4.73584905660377	0.183071341204923\\
5.52830188679245	0.169503256622122\\
6.49056603773585	0.156477133303937\\
7.67924528301887	0.143887432856105\\
9.15094339622642	0.131830582869302\\
10.9422377115439	0.120571368928643\\
13.4428582788049	0.108790105177504\\
16.5149436036639	0.0981571331364948\\
20.2890900562591	0.0885616825317088\\
24.925739087578	0.0799032087314542\\
32.2388757353567	0.0702598262055933\\
41.6976646119888	0.0617797856115763\\
53.9316336080251	0.0543229833752559\\
73.4381601131354	0.0465528909277468\\
100	0.0398941033782756\\
};
\addplot [color=mycolor1, line width=1.0pt, forget plot]
  table[row sep=crcr]{%
0.00398107170553497	0.987070526737157\\
0.01	0.986680307853089\\
0.0129339698301758	0.982457332564831\\
0.0158897592972195	0.978178097964869\\
0.0195210328954464	0.972901267013274\\
0.0216369245535842	0.969821798482176\\
0.0227793799994691	0.967177919079782\\
0.0265815930918929	0.961657249464988\\
0.0310184513941747	0.955229983234784\\
0.036195886513146	0.947765851881483\\
0.0381070725305102	0.945023308467246\\
0.0401191714511029	0.942864943769433\\
0.0468156504137495	0.933340206229102\\
0.0546298700693171	0.922384397768614\\
0.0605512210031491	0.914209502186554\\
0.0824519853300611	0.884552814701241\\
0.0913889851748526	0.873235949417562\\
0.101294669592927	0.860949030326884\\
0.112274034647708	0.847752536994024\\
0.131014220047254	0.82640555033143\\
0.145214897684557	0.810908298027688\\
0.160954791792299	0.794253071332338\\
0.178400738587976	0.777008537631104\\
0.208178442106382	0.750113028163678\\
0.242926481699896	0.72107176941597\\
0.298442270379047	0.679917711978557\\
0.348256668654783	0.648413142193254\\
0.613351592496231	0.530453555881656\\
0.793307099263651	0.479187907377334\\
0.925721705771962	0.449810532602168\\
1.08023825493656	0.421410434868151\\
1.45283018867924	0.369521209511066\\
1.71647258545903	0.341675528054213\\
2.00297707054032	0.31699212168595\\
2.30188679245283	0.295880168649389\\
2.64150943396226	0.276187232057016\\
3.03773584905661	0.257422762241968\\
3.49056603773585	0.239953203907624\\
4	0.223945744228749\\
4.56603773584906	0.209392986203393\\
5.47169811320755	0.190988404961572\\
6.66037735849057	0.17284666410041\\
7.79245283018868	0.159612388550466\\
8.9811320754717	0.148508613614104\\
10.9422377115439	0.134381070313683\\
13.4428582788049	0.121072674176876\\
16.5149436036639	0.10914369674683\\
20.2890900562591	0.0983638560376656\\
24.925739087578	0.0886526556144465\\
32.2388757353567	0.0778796975588492\\
41.6976646119888	0.0684434745115228\\
53.9316336080251	0.0601236630743838\\
73.4381601131354	0.0515168434750524\\
100	0.0441514274005157\\
};
\addplot [color=mycolor2, line width=1.0pt, forget plot]
  table[row sep=crcr]{%
0.00398107170553497	0.987923351700439\\
0.01	0.987558871560612\\
0.0129339698301758	0.983821653898604\\
0.0167287575567897	0.978959756627618\\
0.0205517667358991	0.974044285926001\\
0.0216369245535842	0.972647594066039\\
0.0227793799994691	0.971791888489724\\
0.0279851329393847	0.965085506075827\\
0.0326562626566039	0.959084263058569\\
0.0381070725305102	0.952115731760898\\
0.0444677026307511	0.943598838111034\\
0.0518900153160213	0.934298605748284\\
0.0605512210031491	0.923612030394334\\
0.0637483977561905	0.920124380956642\\
0.0706581091302969	0.911792403570812\\
0.0783167665635036	0.902715416750817\\
0.0868055485839823	0.892854754297613\\
0.10664314906971	0.870286315128246\\
0.12444345696306	0.851661658980731\\
0.137931927266257	0.837941037126243\\
0.152882417635115	0.823335388934449\\
0.178400738587976	0.799577330132065\\
0.208178442106382	0.774430874462329\\
0.242926481699896	0.747339490906322\\
0.31420037852572	0.699073158396308\\
0.366645036497221	0.668617330247379\\
0.525618195660832	0.596344301842119\\
0.679832988486861	0.545252260191893\\
0.835194661128943	0.505488450960015\\
1.73584905660377	0.366526778050457\\
2.00297707054032	0.341068030418341\\
2.24528301886793	0.321829916996244\\
2.52830188679245	0.302896958104211\\
2.87144537350201	0.28377663077016\\
3.26415094339623	0.26574682592497\\
3.71698113207547	0.248597508622847\\
4.22641509433962	0.232732708249769\\
4.80357134908805	0.217957008947683\\
5.75471698113207	0.198688315175787\\
6.94339622641509	0.180524056699098\\
8.13207547169811	0.166488677526133\\
9.49056603773585	0.153881311483365\\
10.9422377115439	0.143152681728895\\
12.7686578989334	0.13234123155657\\
14.8999344410138	0.122353856069217\\
18.3050041805179	0.110237430419746\\
22.4882317016409	0.0993716401857827\\
27.6274487609748	0.0895371981807154\\
33.9411268599031	0.0807436398048278\\
41.6976646119888	0.0727580869714859\\
53.9316336080251	0.0639399417999407\\
69.7550121978289	0.0561727008507447\\
94.984694728691	0.0481016584888248\\
100	0.0469124523949929\\
};
\addplot [color=mycolor3, line width=1.0pt, forget plot]
  table[row sep=crcr]{%
0.00398107170553497	0.989454997552255\\
0.01	0.9891367433586\\
0.0136168988773609	0.985121468377355\\
0.0176120559260339	0.980665618068193\\
0.0239821584567228	0.974242165223342\\
0.0294627813663242	0.968115604336848\\
0.0343805523088551	0.962638804591099\\
0.0401191714511029	0.956123676865662\\
0.0468156504137495	0.948773515678349\\
0.0546298700693171	0.940296714794221\\
0.0605512210031491	0.933953389498937\\
0.0824519853300611	0.910795018883467\\
0.0868055485839823	0.906400053047536\\
0.0913889851748526	0.902714715881363\\
0.101294669592927	0.892970409364991\\
0.112274034647708	0.882446324516492\\
0.118202237706192	0.87688438071673\\
0.12444345696306	0.870355265672785\\
0.137931927266257	0.858233867159327\\
0.160954791792299	0.83895638505975\\
0.178400738587976	0.824766284509525\\
0.197737657725691	0.809779738839338\\
0.255753289931371	0.768885977278962\\
0.298442270379047	0.742536901016299\\
0.348256668654783	0.715104843616538\\
0.525618195660832	0.638374600540631\\
0.793307099263651	0.560151385728604\\
0.879293936264728	0.540065134914436\\
1.05660377358491	0.502414646153726\\
1.19732566135934	0.475289404945479\\
1.67924528301887	0.402140665564655\\
1.80710438704033	0.387155684408694\\
2.10873665095363	0.357447722210453\\
2.35849056603774	0.337259479681337\\
2.64150943396226	0.318007211577133\\
2.9811320754717	0.298646692493232\\
3.37735849056604	0.27987724445402\\
3.91001812828426	0.259443246209383\\
4.45283018867925	0.242518055825295\\
5.07547169811321	0.226653259561786\\
5.75471698113207	0.212428259068363\\
7.11320754716981	0.19049430162229\\
8.24528301886792	0.176561353792029\\
9.54716981132075	0.163796292129295\\
10.9422377115439	0.152737508561336\\
13.4428582788049	0.137509937428582\\
16.5149436036639	0.12385822925499\\
19.2715302531669	0.114501441016823\\
23.675637181206	0.103148884401584\\
29.0862110363026	0.0929488981760676\\
37.6200176017663	0.0816596336266562\\
46.2172892234977	0.0735899333829413\\
62.9336153742108	0.0630150177894064\\
77.315782635192	0.0568601925207295\\
94.984694728691	0.0512281909542676\\
100	0.049918183012557\\
};
\addplot [color=mycolor4, line width=1.0pt, forget plot]
  table[row sep=crcr]{%
0.00398107170553497	0.992063669838245\\
0.0105280119376742	0.991324469508634\\
0.0143358873934958	0.987713893345461\\
0.0185419935036271	0.983719560300386\\
0.0216369245535842	0.980781826105995\\
0.0227793799994691	0.978435188541898\\
0.0279851329393847	0.973523214248007\\
0.0343805523088551	0.967523123059425\\
0.042237511596681	0.960359556008637\\
0.0492875726425938	0.953882092585804\\
0.0575143924243362	0.946427077239421\\
0.0605512210031491	0.943705112442406\\
0.0637483977561905	0.941640392104044\\
0.0743889416417253	0.932282998351358\\
0.0868055485839823	0.921639321783034\\
0.096214432689278	0.913137948561982\\
0.10664314906971	0.904639551439367\\
0.118202237706192	0.89546882310317\\
0.137931927266257	0.880357299597807\\
0.160954791792299	0.864132957588607\\
0.187820510554411	0.845861442586871\\
0.208178442106382	0.832535865714448\\
0.242926481699896	0.811743452314575\\
0.31420037852572	0.774418191042582\\
0.348256668654783	0.758489372138536\\
0.499256838586894	0.699712746496845\\
0.58259013774611	0.672954978113871\\
0.679832988486861	0.645087257415925\\
0.753520326496612	0.625526553930253\\
0.835194661128943	0.604345586559887\\
0.925721705771962	0.581616926421285\\
1.02606100879403	0.55726830262306\\
1.13727612435045	0.531353853473067\\
1.45283018867924	0.469280280886192\\
1.62264150943396	0.442790300140667\\
1.73584905660377	0.427389011419234\\
1.9622641509434	0.400523217181122\\
2.13207547169811	0.383414189548762\\
2.35849056603774	0.363457127656067\\
2.64150943396226	0.342457210586263\\
2.9811320754717	0.321274622138073\\
3.35073178048779	0.302139678760384\\
3.71698113207547	0.286224766606311\\
4.16981132075472	0.269441671127163\\
5.18867924528302	0.240364378936672\\
5.9245283018868	0.224336598569939\\
6.77358490566038	0.20929317090898\\
7.79245283018868	0.1946445091729\\
10.9422377115439	0.163379998599376\\
12.7686578989334	0.150988697779268\\
14.8999344410138	0.139510358484006\\
17.3869523409391	0.128936468307783\\
20.2890900562591	0.11913267521266\\
24.925739087578	0.107286066811308\\
30.6219977011906	0.0967017517005879\\
37.6200176017663	0.0871274383596226\\
48.6576172671926	0.0765108685495739\\
62.9336153742108	0.0671952726061771\\
77.315782635192	0.060603989336423\\
94.984694728691	0.0546191744624913\\
100	0.0532622271616616\\
};
\end{axis}
\end{tikzpicture}%

%% file: images/2absSears.tex
%
%
\definecolor{mycolor1}{rgb}{0.09053,0.13734,0.07326}%
\definecolor{mycolor2}{rgb}{0.04342,0.37461,0.17524}%
\definecolor{mycolor3}{rgb}{0.37697,0.57257,0.04618}%
\definecolor{mycolor4}{rgb}{0.79072,0.73098,0.27064}%
\begin{tikzpicture}[%
trim axis left, trim axis right
]

\begin{axis}[%
width=\fwidth,
height=1\fheight,
at={(0\fwidth,0\fheight)},
scale only axis,
xmode=log,
xmin=0.01,
xmax=100,
xminorticks=true,
xlabel style={font=\color{white!15!black}},
xlabel={$k$},
ymin=0,
ymax=1,
ylabel style={font=\color{white!15!black}},
ylabel={$|S(k)|$},
axis background/.style={fill=white},
xmajorgrids,
xminorgrids,
ymajorgrids
]
\addplot [color=black, line width=2.0pt, forget plot]
  table[row sep=crcr]{%
0.00398107170553497	0.983716406101629\\
0.01	0.983224957942028\\
0.0129339698301758	0.978133912262007\\
0.0158897592972195	0.972975934194222\\
0.0195210328954464	0.966619750373419\\
0.0227793799994691	0.960913222690428\\
0.0265815930918929	0.954265717760441\\
0.0310184513941747	0.946541658262916\\
0.036195886513146	0.937594559957146\\
0.042237511596681	0.927269669713438\\
0.0468156504137495	0.919542532036258\\
0.0518900153160213	0.911085518400478\\
0.0575143924243362	0.901853076048344\\
0.0637483977561905	0.891802168064446\\
0.0706581091302969	0.880893604384301\\
0.0783167665635036	0.869093503784415\\
0.0868055485839823	0.856374843963984\\
0.096214432689278	0.842719040142721\\
0.10664314906971	0.828117476750842\\
0.118202237706192	0.812572903865998\\
0.131014220047254	0.796100602379132\\
0.145214897684557	0.778729221662759\\
0.160954791792299	0.760501202452091\\
0.187820510554411	0.731679473743774\\
0.219170512366241	0.701303650758001\\
0.255753289931371	0.669674249775624\\
0.31420037852572	0.62614776948267\\
0.553371463857602	0.505155289243942\\
0.645737288779182	0.47338878664773\\
0.753520326496612	0.442654801374686\\
0.879293936264728	0.413138063031001\\
1.02606100879403	0.384972819980719\\
1.16981132075472	0.362177533719978\\
1.33962264150943	0.339736342947625\\
1.54861739809169	0.317013032188307\\
1.73584905660377	0.300050240157402\\
2.0188679245283	0.278836996913826\\
2.33730336219578	0.259574595601968\\
2.64150943396226	0.244435489256543\\
3.03773584905661	0.228160154131422\\
3.52765441849198	0.211894624480851\\
4.05660377358491	0.197710541847969\\
4.73584905660377	0.183071341204923\\
5.52830188679245	0.169503256622122\\
6.49056603773585	0.156477133303937\\
7.67924528301887	0.143887432856105\\
9.15094339622642	0.131830582869302\\
10.9422377115439	0.120571368928643\\
13.4428582788049	0.108790105177504\\
16.5149436036639	0.0981571331364948\\
20.2890900562591	0.0885616825317088\\
24.925739087578	0.0799032087314542\\
32.2388757353567	0.0702598262055933\\
41.6976646119888	0.0617797856115763\\
53.9316336080251	0.0543229833752559\\
73.4381601131354	0.0465528909277468\\
100	0.0398941033782756\\
};
\addplot [color=mycolor1, line width=1.0pt, forget plot]
  table[row sep=crcr]{%
0.00398107170553497	0.992367950159234\\
0.0105280119376742	0.9916694643189\\
0.0143358873934958	0.98832572210605\\
0.0195210328954464	0.983870265402927\\
0.0216369245535842	0.982086100988258\\
0.0227793799994691	0.979868225093965\\
0.0294627813663242	0.974419305754631\\
0.036195886513146	0.969138535921083\\
0.0468156504137495	0.961369571541348\\
0.0575143924243362	0.953920246468083\\
0.0605512210031491	0.951898303705205\\
0.0637483977561905	0.95059736669214\\
0.0783167665635036	0.941619971380479\\
0.0868055485839823	0.936789554851343\\
0.0913889851748526	0.933655159432743\\
0.118202237706192	0.920676937392026\\
0.152882417635115	0.907344260196138\\
0.169453396941519	0.902622177204666\\
0.219170512366241	0.890464709809893\\
0.330790533594054	0.873077803876991\\
0.348256668654783	0.869959712525641\\
0.366645036497221	0.865998377690915\\
0.386004332113175	0.861460310988284\\
0.406385821648147	0.855903713828038\\
0.427843478161324	0.849132406934662\\
0.450434124553848	0.840987807027958\\
0.474217584043875	0.831368452727586\\
0.499256838586894	0.820412707784221\\
0.525618195660832	0.807921196406773\\
0.553371463857602	0.794221129412892\\
0.613351592496231	0.763983678290184\\
0.679832988486861	0.731818660931599\\
0.879293936264728	0.650516755463482\\
0.974600916933136	0.619206095684473\\
1.16981132075472	0.56643879584908\\
1.28301886792453	0.54117378889465\\
1.50943396226415	0.499141478475531\\
1.67924528301887	0.473208506626552\\
2.07547169811321	0.425408397447188\\
2.24528301886793	0.409080231340518\\
2.35849056603774	0.39884320736523\\
2.59064450009023	0.380586348220172\\
2.92452830188679	0.357809068577048\\
3.32075471698113	0.33551029733073\\
3.60377358490566	0.321976243920049\\
3.83018867924528	0.31214271878211\\
4.22641509433962	0.297174011472106\\
4.80357134908805	0.278442302860196\\
5.05720565066689	0.271391408931379\\
5.07547169811321	0.270707573698798\\
5.64150943396227	0.256804629147012\\
6.43396226415095	0.240322909748539\\
7.50943396226415	0.22211810821195\\
7.90566037735849	0.21658944185388\\
9.26415094339623	0.199800217874174\\
9.66037735849057	0.195784546447769\\
10.9422377115439	0.183727556799303\\
13.4428582788049	0.165721656407541\\
15.6866687665557	0.153334462139775\\
18.3050041805179	0.141880957787651\\
21.3603782316844	0.131379276810745\\
24.925739087578	0.121477533889844\\
29.0862110363026	0.11252386836581\\
35.7332588759175	0.101496796187011\\
43.8993510808154	0.0914616184182071\\
56.779288244356	0.0803897846863966\\
69.7550121978289	0.0725657596456015\\
85.6960676537351	0.0654613223641451\\
100	0.0606316313516069\\
};
\addplot [color=mycolor2, line width=1.0pt, forget plot]
  table[row sep=crcr]{%
0.00398107170553497	0.992233911047134\\
0.0105280119376742	0.991517401339558\\
0.0143358873934958	0.988055046896064\\
0.0185419935036271	0.984266206049007\\
0.0216369245535842	0.981505867126613\\
0.0227793799994691	0.979230194065063\\
0.0279851329393847	0.974675910824296\\
0.0343805523088551	0.969188754792982\\
0.042237511596681	0.962755669797403\\
0.0518900153160213	0.954943974637922\\
0.0671143892584777	0.943997266238697\\
0.0783167665635036	0.935811164702266\\
0.0868055485839823	0.929861438651406\\
0.0913889851748526	0.926093642546309\\
0.10664314906971	0.916153949765523\\
0.12444345696306	0.90530893700611\\
0.160954791792299	0.88609371024656\\
0.197737657725691	0.869218414008463\\
0.230742977057796	0.855900989155967\\
0.348256668654783	0.820378964345018\\
0.406385821648147	0.805094724047275\\
0.450434124553848	0.793280377331206\\
0.474217584043875	0.786530365699709\\
0.499256838586894	0.77919912873034\\
0.525618195660832	0.770762240667355\\
0.553371463857602	0.761304280141229\\
0.58259013774611	0.750605468812444\\
0.613351592496231	0.738946251628242\\
0.645737288779182	0.726224306939389\\
0.679832988486861	0.712597301636196\\
0.753520326496612	0.683449051509386\\
1.11320754716981	0.568313221933782\\
1.22641509433962	0.541572044097732\\
1.33962264150943	0.51823363262826\\
1.50943396226415	0.487855274197257\\
1.67924528301887	0.46221130857359\\
2.0188679245283	0.420896246331671\\
2.22008046346509	0.401155273792209\\
2.47169811320755	0.379653343528777\\
2.75471698113208	0.359223388397246\\
3.03773584905661	0.341756358946602\\
3.83018867924528	0.303642115870381\\
4.22641509433962	0.288923790915145\\
4.80357134908805	0.270584792231439\\
5.07547169811321	0.263061764372501\\
5.64150943396227	0.24942330926867\\
6.43396226415095	0.233319910941659\\
7.50943396226415	0.215583708909145\\
8.07547169811321	0.20790368978279\\
8.35849056603773	0.204178728728272\\
8.92452830188679	0.197637176302253\\
10.3934510867977	0.182935379018763\\
12.1282707262527	0.169122175414388\\
14.8999344410138	0.152538894870395\\
17.3869523409391	0.141124287220638\\
20.2890900562591	0.130469091690861\\
26.2418478669375	0.114627553650964\\
32.2388757353567	0.103448136160005\\
39.606399440691	0.0933131948502717\\
46.2172892234977	0.0862871566418617\\
59.7773024449226	0.0757984780005154\\
81.3981482553927	0.0649549594975576\\
100	0.0586611473138654\\
};
\addplot [color=mycolor3, line width=1.0pt, forget plot]
  table[row sep=crcr]{%
0.00398107170553497	0.992116308744977\\
0.0105280119376742	0.991384092800043\\
0.0143358873934958	0.987818966506808\\
0.0185419935036271	0.983887399039008\\
0.0216369245535842	0.981003680521415\\
0.0227793799994691	0.978678603596726\\
0.0279851329393847	0.973875097668327\\
0.0343805523088551	0.968029931003084\\
0.042237511596681	0.961086011106953\\
0.0492875726425938	0.9548310058283\\
0.0575143924243362	0.947662671220887\\
0.0605512210031491	0.945053338284822\\
0.0637483977561905	0.943111249826628\\
0.0743889416417253	0.934186552199032\\
0.0868055485839823	0.924092599906185\\
0.096214432689278	0.91603596104804\\
0.112274034647708	0.903845229120666\\
0.131014220047254	0.890265062431611\\
0.160954791792299	0.870567664779954\\
0.187820510554411	0.853925628531878\\
0.208178442106382	0.84186949860492\\
0.242926481699896	0.823288233302033\\
0.330790533594054	0.78361785997677\\
0.386004332113175	0.76227861975171\\
0.450434124553848	0.740334698066175\\
0.525618195660832	0.717506334329253\\
0.58259013774611	0.700748603067151\\
0.645737288779182	0.682627504564273\\
0.679832988486861	0.67281763840306\\
0.753520326496612	0.651470055865215\\
0.793307099263651	0.639713741813265\\
0.879293936264728	0.614178906838728\\
1	0.579917681749739\\
1.26054588560748	0.517094286158279\\
1.39717689987512	0.490637225365101\\
1.54861739809169	0.465185391743711\\
1.67924528301887	0.446087671888876\\
1.90566037735849	0.417755323831894\\
2.13207547169811	0.394186114818855\\
2.41509433962264	0.369398540578778\\
2.64150943396226	0.352697305743739\\
2.9811320754717	0.331161936057254\\
3.35073178048779	0.311692884549881\\
3.71698113207547	0.295496847737536\\
4.16981132075472	0.278359268582568\\
5.32423214614946	0.24537620958128\\
6.03773584905661	0.230006891206131\\
6.88679245283019	0.214987697986641\\
7.84905660377358	0.201095319858358\\
9.15094339622642	0.185827701468521\\
10.3934510867977	0.174173013743789\\
12.1282707262527	0.160946680043944\\
14.8999344410138	0.145018830177851\\
17.3869523409391	0.134095238427043\\
20.2890900562591	0.123939557361473\\
26.2418478669375	0.108813969345789\\
32.2388757353567	0.0981372819054673\\
39.606399440691	0.0884880961338737\\
46.2172892234977	0.0818132117255996\\
59.7773024449226	0.0718688755835619\\
81.3981482553927	0.0615449694234051\\
100	0.0555729285986231\\
};
\addplot [color=mycolor4, line width=1.0pt, forget plot]
  table[row sep=crcr]{%
0.00398107170553497	0.992015138197467\\
0.0105280119376742	0.991269532660306\\
0.0143358873934958	0.987617461807754\\
0.0185419935036271	0.983566090117407\\
0.0216369245535842	0.980579450015851\\
0.0227793799994691	0.978213341185582\\
0.0279851329393847	0.97320364298238\\
0.0343805523088551	0.967064638634397\\
0.042237511596681	0.95970519356619\\
0.0492875726425938	0.953030364231426\\
0.0575143924243362	0.945322277485888\\
0.0605512210031491	0.942501235758805\\
0.0637483977561905	0.940328789746933\\
0.0743889416417253	0.930593022525064\\
0.0868055485839823	0.919471727741969\\
0.096214432689278	0.910586185339429\\
0.10664314906971	0.901642468764916\\
0.118202237706192	0.891957498469205\\
0.137931927266257	0.875926783641236\\
0.160954791792299	0.858577275281573\\
0.178400738587976	0.84569240407743\\
0.197737657725691	0.831991624070058\\
0.242926481699896	0.801938941126215\\
0.283474479460912	0.777508937921266\\
0.330790533594054	0.752076020770132\\
0.366645036497221	0.733847414282454\\
0.427843478161324	0.705841424665732\\
0.525618195660832	0.667579546164119\\
0.613351592496231	0.637646460747354\\
0.793307099263651	0.586919237585414\\
1	0.538889881559574\\
1.11320754716981	0.51538861712323\\
1.26054588560748	0.487149623390145\\
1.84905660377359	0.398883569594362\\
2.0188679245283	0.380268072712542\\
2.24528301886793	0.359034261149296\\
2.46071576991633	0.341613940104008\\
2.727433622322	0.323180281839446\\
3.03773584905661	0.304996981020929\\
3.54716981132076	0.280695507809981\\
4	0.263259291162747\\
4.50943396226415	0.246973913137019\\
5.13207547169811	0.230575927568011\\
5.81132075471698	0.215888297792093\\
6.66037735849057	0.200903746203947\\
7.67924528301887	0.186417831373994\\
8.81132075471698	0.173484673218507\\
10	0.162340416613353\\
11.5200009252003	0.150803793089761\\
13.4428582788049	0.139192427363838\\
15.6866687665557	0.128498506027435\\
18.3050041805179	0.118659813734772\\
21.3603782316844	0.109619980937754\\
26.2418478669375	0.0986097488733124\\
32.2388757353567	0.0887865882292895\\
39.606399440691	0.0799579335074472\\
48.6576172671926	0.0720122565725601\\
62.9336153742108	0.0631930440722837\\
77.315782635192	0.0569758542291838\\
94.984694728691	0.0513206717242132\\
100	0.0500355880065353\\
};
\end{axis}
\end{tikzpicture}%

%% file: images/1phaseSears.tex
%
%
\definecolor{mycolor1}{rgb}{0.18517,0.05913,0.24304}%
\definecolor{mycolor2}{rgb}{0.51514,0.10993,0.38770}%
\definecolor{mycolor3}{rgb}{0.81038,0.26571,0.33825}%
\definecolor{mycolor4}{rgb}{0.94901,0.58547,0.40375}%
\begin{tikzpicture}[%
trim axis left, trim axis right
]

\begin{axis}[%
width=0.951\fwidth,
height=\fheight,
at={(0\fwidth,0\fheight)},
scale only axis,
xmode=log,
xmin=0.01,
xmax=100,
xminorticks=true,
xlabel style={font=\color{white!15!black}},
xlabel={$k$},
ymin=-0.8,
ymax=0,
ylabel style={font=\color{white!15!black}},
ylabel={$\angle \e^{-\textrm{i} k}S(k)$},
axis background/.style={fill=white},
xmajorgrids,
xminorgrids,
ymajorgrids
]
\addplot [color=black, line width=2.0pt, forget plot]
  table[row sep=crcr]{%
0.00398107170553497	-0.0547059668494683\\
0.01	-0.0563570241587534\\
0.0110839035359811	-0.0612412491899068\\
0.0122852917594934	-0.0665145277757659\\
0.0136168988773609	-0.0722024092588618\\
0.0150928393615877	-0.0783311795848705\\
0.0167287575567897	-0.0849276928289515\\
0.0185419935036271	-0.0920191579552929\\
0.0205517667358991	-0.0996328749388264\\
0.0227793799994691	-0.107795914439621\\
0.0252484450523573	-0.116534735594231\\
0.0279851329393847	-0.125874737277362\\
0.0310184513941747	-0.135839739506717\\
0.0343805523088551	-0.146451393634458\\
0.0381070725305102	-0.157728522705619\\
0.042237511596681	-0.169686396960493\\
0.0468156504137495	-0.182335953963912\\
0.0518900153160213	-0.195682978236126\\
0.0575143924243362	-0.209727261409779\\
0.0637483977561905	-0.224461770579847\\
0.0706581091302969	-0.239871859218751\\
0.0783167665635036	-0.255934561191397\\
0.0868055485839823	-0.272618013255527\\
0.096214432689278	-0.289881054087357\\
0.10664314906971	-0.307673047422476\\
0.118202237706192	-0.325933972546733\\
0.131014220047254	-0.344594816581464\\
0.152882417635115	-0.373164845755481\\
0.187820510554411	-0.411879697698342\\
0.242926481699896	-0.460206353105147\\
0.283474479460912	-0.488586913940669\\
0.31420037852572	-0.507058170387629\\
0.348256668654783	-0.525071653706872\\
0.386004332113175	-0.54255154995853\\
0.427843478161324	-0.55943104750023\\
0.474217584043875	-0.575653213786447\\
0.525618195660832	-0.59117152502979\\
0.58259013774611	-0.605950056679363\\
0.645737288779182	-0.619963359871683\\
0.715728981841442	-0.63319606211309\\
0.793307099263651	-0.645642239049179\\
0.879293936264728	-0.657304608407671\\
0.974600916933136	-0.66819359761381\\
1.05660377358491	-0.676210889776464\\
1.13727612435045	-0.683116034910873\\
1.22641509433962	-0.689815462221577\\
1.33962264150943	-0.697177814343797\\
1.47094950809279	-0.70443695348675\\
1.62264150943396	-0.71148607004595\\
1.73584905660377	-0.716007596450516\\
1.90566037735849	-0.721853587521267\\
2.10873665095363	-0.72768512554776\\
2.33730336219578	-0.733101868407327\\
2.59064450009023	-0.73804094844624\\
2.87144537350201	-0.742537836315194\\
3.20754716981132	-0.74692030755463\\
3.60377358490566	-0.751066718583754\\
4.05660377358491	-0.754836984874939\\
4.62264150943396	-0.75853086839592\\
5.32423214614946	-0.762035731303162\\
6.15094339622642	-0.765151576909418\\
7.22641509433962	-0.768147517304894\\
8.52830188679245	-0.770769891100128\\
10	-0.772916126072203\\
12.1282707262527	-0.775101780312511\\
14.8999344410138	-0.77701433832253\\
19.2715302531669	-0.778914448051601\\
26.2418478669375	-0.780635785853017\\
35.7332588759175	-0.781900420832077\\
53.9316336080251	-0.783080530356404\\
85.6960676537351	-0.783939548997831\\
100	-0.784148181623823\\
};
\addplot [color=mycolor1, line width=1.0pt, forget plot]
  table[row sep=crcr]{%
0.00398107170553497	-0.0466229024476119\\
0.01	-0.048030008258904\\
0.0110839035359811	-0.0522203963154757\\
0.0122852917594934	-0.0567494571948912\\
0.0136168988773609	-0.0616403649407902\\
0.0150928393615877	-0.0669171582116994\\
0.0167287575567897	-0.0726046488784191\\
0.0185419935036271	-0.0787282631902042\\
0.0205517667358991	-0.0853138785103211\\
0.0227793799994691	-0.0923156131275893\\
0.0252484450523573	-0.0998984176530282\\
0.0279851329393847	-0.108020794261462\\
0.0310184513941747	-0.116707621425205\\
0.0343805523088551	-0.125982565704587\\
0.0381070725305102	-0.135867481688855\\
0.042237511596681	-0.146458504454452\\
0.0468156504137495	-0.157623149473989\\
0.0518900153160213	-0.169446702665825\\
0.0575143924243362	-0.181937405828872\\
0.0637483977561905	-0.195032035915257\\
0.0706581091302969	-0.208857698777793\\
0.0783167665635036	-0.223340687444654\\
0.0868055485839823	-0.23846359046627\\
0.096214432689278	-0.254208860840829\\
0.10664314906971	-0.270525020766173\\
0.118202237706192	-0.287376846368575\\
0.137931927266257	-0.313586582867512\\
0.160954791792299	-0.340551073909168\\
0.187820510554411	-0.368195895064539\\
0.298442270379047	-0.451821947764707\\
0.330790533594054	-0.469965841154289\\
0.366645036497221	-0.487729918277646\\
0.406385821648147	-0.505061457742046\\
0.450434124553848	-0.521876657745972\\
0.499256838586894	-0.538175761976524\\
0.553371463857602	-0.553880198173655\\
0.613351592496231	-0.569029398621149\\
0.679832988486861	-0.583549365829201\\
0.753520326496612	-0.597641899372343\\
0.835194661128943	-0.611202483960027\\
1	-0.63428800719126\\
1.11320754716981	-0.647585584964689\\
1.33962264150943	-0.670457128447297\\
1.50943396226415	-0.684630770919691\\
1.63038624539993	-0.693339307408467\\
1.71647258545903	-0.698695164135982\\
1.84905660377359	-0.706421124495474\\
1.90566037735849	-0.709336646648243\\
2.0188679245283	-0.714415510135076\\
2.24528301886793	-0.72362047533189\\
2.35849056603774	-0.727454236634862\\
2.69811320754717	-0.736877638695201\\
2.727433622322	-0.737176067896649\\
2.87144537350201	-0.740345255810318\\
2.92452830188679	-0.741859251279891\\
3.09433962264151	-0.744975507200023\\
3.37735849056604	-0.749598916197393\\
3.60377358490566	-0.752690365695868\\
3.66037735849056	-0.753112622894072\\
3.77358490566038	-0.754460003415066\\
3.83018867924528	-0.755503290988423\\
3.9433962264151	-0.756737684231378\\
4	-0.757070011801665\\
4.45283018867925	-0.761232364436185\\
5.05720565066689	-0.765629477905868\\
5.07547169811321	-0.765317210413698\\
5.24528301886793	-0.766278079654973\\
5.32423214614946	-0.767110242109311\\
5.47169811320755	-0.767873512765373\\
5.52830188679245	-0.767864811754258\\
6.32075471698114	-0.771223064939565\\
6.49056603773585	-0.77181852796524\\
6.54716981132076	-0.772276922514796\\
6.71698113207547	-0.772839349118265\\
6.77358490566038	-0.772582857771517\\
7	-0.773263361199223\\
7.05660377358491	-0.773883803726228\\
7.56603773584906	-0.775077952261801\\
8.07547169811321	-0.776323867563113\\
8.13207547169811	-0.776010588295442\\
8.35849056603773	-0.776443449717282\\
8.41509433962264	-0.777090473031633\\
8.69811320754717	-0.777597176935783\\
8.75471698113208	-0.777349842580154\\
9.88679245283019	-0.779027161816052\\
9.94339622641509	-0.779430200290802\\
10	-0.779504504814725\\
10.3934510867977	-0.779445189199951\\
11.5200009252003	-0.780981306081823\\
12.1282707262527	-0.78114818019526\\
12.7686578989334	-0.78160649522565\\
14.152657243572	-0.782889818810304\\
14.8999344410138	-0.783367623578286\\
17.3869523409391	-0.783699553221025\\
19.2715302531669	-0.784731541173218\\
20.2890900562591	-0.784538015586204\\
21.3603782316844	-0.785085992489932\\
22.4882317016409	-0.785380826485142\\
23.675637181206	-0.785206144910582\\
24.925739087578	-0.785691855976759\\
26.2418478669375	-0.785233993592662\\
27.6274487609748	-0.785897248969491\\
29.0862110363026	-0.785333544229817\\
30.6219977011906	-0.786314528252559\\
33.9411268599031	-0.786184482771216\\
35.7332588759175	-0.785776507253272\\
37.6200176017663	-0.786723596363877\\
41.6976646119888	-0.786486947612469\\
43.8993510808154	-0.78690936338682\\
46.2172892234977	-0.786535479248688\\
48.6576172671926	-0.786688696568753\\
51.2267975447787	-0.786208612415003\\
53.9316336080251	-0.786838056836827\\
56.779288244356	-0.78639104721978\\
62.9336153742108	-0.787182293380105\\
66.2565853940689	-0.786587515708939\\
69.7550121978289	-0.786574291205725\\
73.4381601131354	-0.785739517836814\\
77.315782635192	-0.786746627232879\\
81.3981482553927	-0.787257854887652\\
85.6960676537351	-0.785835687781773\\
90.2209223270261	-0.78736745021711\\
94.984694728691	-0.786195414551539\\
100	-0.787196221903066\\
};
\addplot [color=mycolor2, line width=1.0pt, forget plot]
  table[row sep=crcr]{%
0.00398107170553497	-0.0418304293927529\\
0.01	-0.0430928956313905\\
0.0116691468742838	-0.0488598218338594\\
0.0129339698301758	-0.0530963293048976\\
0.0143358873934958	-0.0576716583029948\\
0.0158897592972195	-0.0626084671634186\\
0.0176120559260339	-0.0679301913632853\\
0.0195210328954464	-0.0736608882331313\\
0.0216369245535842	-0.0798251274003468\\
0.0239821584567228	-0.0864875035175805\\
0.0265815930918929	-0.0935961622677\\
0.0294627813663242	-0.101212896467239\\
0.0326562626566039	-0.109361782937442\\
0.036195886513146	-0.118065881051646\\
0.0401191714511029	-0.127309505077436\\
0.0444677026307511	-0.137185008949415\\
0.0492875726425938	-0.14767401184171\\
0.0546298700693171	-0.158790393757416\\
0.0605512210031491	-0.170543949945943\\
0.0671143892584777	-0.182989548781673\\
0.0743889416417253	-0.196029456530507\\
0.0824519853300611	-0.209704923241438\\
0.0913889851748526	-0.22394435701285\\
0.101294669592927	-0.23884195696118\\
0.112274034647708	-0.254312201179739\\
0.12444345696306	-0.270351892397264\\
0.137931927266257	-0.286845175358094\\
0.152882417635115	-0.303772601908721\\
0.169453396941519	-0.321048948829824\\
0.197737657725691	-0.34754535641994\\
0.348256668654783	-0.446176790638326\\
0.406385821648147	-0.472302975679371\\
0.450434124553848	-0.489371511065376\\
0.525618195660832	-0.514388237813233\\
0.613351592496231	-0.538725149947957\\
0.879293936264728	-0.595378986282652\\
0.974600916933136	-0.61210204081353\\
1.08023825493656	-0.629378692649203\\
1.19732566135934	-0.646563899713828\\
1.26054588560748	-0.655138879003868\\
1.33962264150943	-0.66479672499917\\
1.47094950809279	-0.679028994531422\\
1.56603773584906	-0.687714593715893\\
1.67924528301887	-0.696727042145888\\
1.73584905660377	-0.700799508895781\\
1.80710438704033	-0.705436977590906\\
2.0188679245283	-0.716950154217377\\
2.13207547169811	-0.722052506065483\\
2.30188679245283	-0.728674651209388\\
2.41509433962264	-0.732324147413885\\
2.64150943396226	-0.738881939846217\\
2.87144537350201	-0.744329358907757\\
3.03773584905661	-0.747696622469488\\
3.18268235287355	-0.750130081686924\\
3.54716981132076	-0.755729407514171\\
3.77358490566038	-0.758560714073318\\
3.91001812828426	-0.75992944203654\\
4.39622641509434	-0.7646035467346\\
4.62264150943396	-0.76622514994518\\
5.30188679245283	-0.770547065056393\\
5.69811320754717	-0.772323223354418\\
6.49056603773585	-0.77545831305901\\
6.60377358490566	-0.775709755196662\\
7.56603773584906	-0.778322668389564\\
8.69811320754717	-0.780561773753171\\
8.86792452830189	-0.780726348699849\\
10.9422377115439	-0.783283457636366\\
15.6866687665557	-0.785822393316766\\
18.3050041805179	-0.786557173303851\\
20.2890900562591	-0.786995381410988\\
21.3603782316844	-0.78727020408833\\
29.0862110363026	-0.787496769526776\\
30.6219977011906	-0.787955632152941\\
32.2388757353567	-0.787546826015439\\
33.9411268599031	-0.788149700923312\\
35.7332588759175	-0.787561368842274\\
37.6200176017663	-0.788155196702854\\
39.606399440691	-0.787679659057628\\
41.6976646119888	-0.787529877723678\\
43.8993510808154	-0.788286144539821\\
48.6576172671926	-0.78748027451282\\
51.2267975447787	-0.787361645036312\\
53.9316336080251	-0.78802914611253\\
62.9336153742108	-0.788270495481291\\
66.2565853940689	-0.788121870013653\\
69.7550121978289	-0.787271146692847\\
73.4381601131354	-0.786884249309327\\
77.315782635192	-0.787291787771687\\
81.3981482553927	-0.78851985944034\\
85.6960676537351	-0.786615180865439\\
90.2209223270261	-0.787662315619452\\
94.984694728691	-0.787329345372977\\
100	-0.78864680731549\\
};
\addplot [color=mycolor3, line width=1.0pt, forget plot]
  table[row sep=crcr]{%
0.00398107170553497	-0.036825236773506\\
0.01	-0.0379366434416006\\
0.0116691468742838	-0.0430201417562941\\
0.0136168988773609	-0.0487354500754731\\
0.0150928393615877	-0.052929519161804\\
0.0167287575567897	-0.0574549104983659\\
0.0185419935036271	-0.0623332269878469\\
0.0205517667358991	-0.0675867237966217\\
0.0227793799994691	-0.0732870911636989\\
0.0252484450523573	-0.0793634264122338\\
0.0279851329393847	-0.0858843654516019\\
0.0310184513941747	-0.0928730833158578\\
0.0343805523088551	-0.100352423984785\\
0.0381070725305102	-0.10834446889977\\
0.042237511596681	-0.116847977105145\\
0.0468156504137495	-0.125925094001598\\
0.0518900153160213	-0.135572606486246\\
0.0575143924243362	-0.145805147552631\\
0.0637483977561905	-0.156582656345673\\
0.0706581091302969	-0.168013171440104\\
0.0783167665635036	-0.180051055780605\\
0.0868055485839823	-0.192694420202999\\
0.101294669592927	-0.212925834082808\\
0.112274034647708	-0.227039916336925\\
0.12444345696306	-0.241573990247628\\
0.137931927266257	-0.256754146891848\\
0.152882417635115	-0.272414814055076\\
0.178400738587976	-0.296769427974503\\
0.208178442106382	-0.32181475139894\\
0.242926481699896	-0.347517631260441\\
0.330790533594054	-0.399985068412535\\
0.427843478161324	-0.444245682177797\\
0.525618195660832	-0.480020803339145\\
0.58259013774611	-0.498312989924583\\
0.645737288779182	-0.517114886714384\\
0.715728981841442	-0.53658682639526\\
0.793307099263651	-0.556961894975732\\
0.835194661128943	-0.567383266538355\\
0.879293936264728	-0.578159376853291\\
0.974600916933136	-0.600470575221231\\
1.02606100879403	-0.611397143687609\\
1.16981132075472	-0.639530370415945\\
1.22641509433962	-0.648981801090724\\
1.28301886792453	-0.657674950257297\\
1.33962264150943	-0.665563214272322\\
1.39717689987512	-0.672853373381811\\
1.47094950809279	-0.681317548773876\\
1.54861739809169	-0.689071265768619\\
1.56603773584906	-0.690414774160104\\
1.63038624539993	-0.695984813974396\\
1.73584905660377	-0.704332745507615\\
1.80710438704033	-0.709119718348914\\
1.84905660377359	-0.711318079556803\\
1.90566037735849	-0.71459949239072\\
1.9622641509434	-0.718053265841499\\
2.07547169811321	-0.723603690539297\\
2.24528301886793	-0.730615976796723\\
2.35849056603774	-0.734494522036434\\
2.41509433962264	-0.73662092375996\\
2.52830188679245	-0.740066612170745\\
2.59064450009023	-0.741422247619364\\
2.69811320754717	-0.744207616038247\\
2.727433622322	-0.745326117031782\\
2.87144537350201	-0.748614420754392\\
2.92452830188679	-0.749340675270175\\
3.03773584905661	-0.751562571673335\\
3.09433962264151	-0.752976265342463\\
3.37735849056604	-0.757610097750259\\
3.43396226415094	-0.758050967259769\\
3.60377358490566	-0.76033237246574\\
3.66037735849056	-0.761510332713486\\
3.77358490566038	-0.762863465082645\\
3.83018867924528	-0.76297747118362\\
3.9433962264151	-0.764181196374022\\
4	-0.765212080029429\\
4.16981132075472	-0.766843184623318\\
4.22641509433962	-0.767106204246172\\
4.80357134908805	-0.771770018651432\\
4.84905660377358	-0.771474189479055\\
5.05720565066689	-0.772714288693706\\
5.07547169811321	-0.773489413283656\\
5.24528301886793	-0.774421415893988\\
5.32423214614946	-0.7743392996751\\
5.47169811320755	-0.7750466330288\\
5.52830188679245	-0.775622237718095\\
5.9811320754717	-0.777456554649126\\
6.15094339622642	-0.777989696613643\\
6.21292469062015	-0.778205787162112\\
6.26415094339623	-0.778651288597636\\
6.49056603773585	-0.779396411471393\\
6.54716981132076	-0.778761433764032\\
6.71698113207547	-0.779257345639557\\
6.77358490566038	-0.780293617472983\\
7	-0.780922844484102\\
7.05660377358491	-0.780670937584492\\
7.50943396226415	-0.781874476046148\\
7.63275925040289	-0.782291832664335\\
7.79245283018868	-0.78262282836513\\
7.84905660377358	-0.782340196549914\\
8.07547169811321	-0.782747020433788\\
8.13207547169811	-0.78331465788866\\
8.35849056603773	-0.783695566576556\\
8.41509433962264	-0.783062121267449\\
8.69811320754717	-0.783468218228406\\
8.75471698113208	-0.784281644969038\\
8.9811320754717	-0.784593823878407\\
9.0377358490566	-0.784404463650767\\
9.60377358490566	-0.784995237684615\\
9.66037735849057	-0.78543662061213\\
9.88679245283019	-0.785694634971093\\
9.94339622641509	-0.785030739327844\\
10	-0.785086807538899\\
10.3934510867977	-0.786176898025984\\
12.7686578989334	-0.787722675505894\\
13.4428582788049	-0.787846982457912\\
14.152657243572	-0.787669093942222\\
14.8999344410138	-0.78791890344102\\
15.6866687665557	-0.788656256343771\\
17.3869523409391	-0.789273335064431\\
19.2715302531669	-0.788782601524169\\
20.2890900562591	-0.789542252145395\\
21.3603782316844	-0.789292081728627\\
22.4882317016409	-0.788800771479626\\
23.675637181206	-0.789618575444295\\
24.925739087578	-0.789267653265104\\
26.2418478669375	-0.789813042256395\\
29.0862110363026	-0.789974392421819\\
30.6219977011906	-0.789856620365676\\
32.2388757353567	-0.788766398889279\\
33.9411268599031	-0.789667839421591\\
35.7332588759175	-0.789867843316527\\
37.6200176017663	-0.789670454699237\\
39.606399440691	-0.788757450164895\\
41.6976646119888	-0.789590023937063\\
43.8993510808154	-0.788603474829943\\
46.2172892234977	-0.789261062341761\\
48.6576172671926	-0.7885921095136\\
51.2267975447787	-0.789445973326233\\
53.9316336080251	-0.788931617162543\\
56.779288244356	-0.789589998214019\\
59.7773024449226	-0.788891921509451\\
62.9336153742108	-0.788860736767761\\
66.2565853940689	-0.78940405686058\\
69.7550121978289	-0.789164231830515\\
73.4381601131354	-0.788280660634408\\
77.315782635192	-0.787855305737633\\
81.3981482553927	-0.788528240974697\\
85.6960676537351	-0.787611707449455\\
90.2209223270261	-0.788673494171254\\
94.984694728691	-0.787072063558225\\
100	-0.789348002623073\\
};
\addplot [color=mycolor4, line width=1.0pt, forget plot]
  table[row sep=crcr]{%
0.00398107170553497	-0.0315980725003913\\
0.01	-0.0325517203667136\\
0.0116691468742838	-0.03691020203004\\
0.0136168988773609	-0.041811084394384\\
0.0158897592972195	-0.0473124170573458\\
0.0176120559260339	-0.0513439894605234\\
0.0195210328954464	-0.0556891107023323\\
0.0227793799994691	-0.0627599495929312\\
0.0252484450523573	-0.067970719194391\\
0.0279851329393847	-0.0735661263154195\\
0.0310184513941747	-0.0795672053716499\\
0.0343805523088551	-0.085994977377958\\
0.0381070725305102	-0.092870147185903\\
0.042237511596681	-0.100246763288775\\
0.0468156504137495	-0.108079362827086\\
0.0518900153160213	-0.116417091353839\\
0.0575143924243362	-0.125276310598267\\
0.0637483977561905	-0.134744533328943\\
0.0706581091302969	-0.144690515159874\\
0.0783167665635036	-0.155192979156483\\
0.0868055485839823	-0.166257486999091\\
0.096214432689278	-0.177795595565318\\
0.10664314906971	-0.189982116825495\\
0.118202237706192	-0.202723937592729\\
0.131014220047254	-0.216022351699673\\
0.145214897684557	-0.229840322712493\\
0.160954791792299	-0.244188515885444\\
0.178400738587976	-0.25900229683986\\
0.197737657725691	-0.274282894518238\\
0.219170512366241	-0.290024000093091\\
0.242926481699896	-0.306187000316707\\
0.269257368949693	-0.322725916635856\\
0.298442270379047	-0.33973066757466\\
0.348256668654783	-0.366129492516781\\
0.386004332113175	-0.384324329808412\\
0.427843478161324	-0.403075684401069\\
0.474217584043875	-0.422681150455022\\
0.525618195660832	-0.443300740536797\\
0.553371463857602	-0.454005044382446\\
0.58259013774611	-0.465049748505511\\
0.613351592496231	-0.476457042897799\\
0.645737288779182	-0.488291189333722\\
0.679832988486861	-0.500499136781926\\
0.715728981841442	-0.513127007857844\\
0.753520326496612	-0.52603115103974\\
0.793307099263651	-0.539434273593343\\
0.879293936264728	-0.567234561782059\\
0.974600916933136	-0.595410226013287\\
1.11320754716981	-0.629833439144477\\
1.16981132075472	-0.641605376279228\\
1.22641509433962	-0.652163364544928\\
1.28301886792453	-0.661516465946004\\
1.39717689987512	-0.677003423309967\\
1.47094950809279	-0.685924511221061\\
1.54861739809169	-0.693797318209988\\
1.56603773584906	-0.695446077235752\\
1.63038624539993	-0.701073377865307\\
1.67924528301887	-0.704984160426126\\
1.73584905660377	-0.708698798790186\\
1.80710438704033	-0.713485685615449\\
1.84905660377359	-0.716586189840296\\
1.9622641509434	-0.723011176412447\\
2.07547169811321	-0.728599336377914\\
2.10873665095363	-0.72955150257712\\
2.35849056603774	-0.739820503556314\\
2.52830188679245	-0.745186037943617\\
2.59064450009023	-0.746303072332354\\
2.69811320754717	-0.749090436779895\\
2.727433622322	-0.750473618914957\\
2.92452830188679	-0.754913393172628\\
3.03773584905661	-0.757159852688569\\
3.09433962264151	-0.757497612049539\\
3.20754716981132	-0.759449184439999\\
3.26415094339623	-0.761098616770781\\
3.54716981132076	-0.765231962486824\\
3.60377358490566	-0.765968560440509\\
3.66037735849056	-0.765925532741344\\
3.77358490566038	-0.767251047232012\\
3.83018867924528	-0.76865214989104\\
4.16981132075472	-0.772045470164569\\
4.22641509433962	-0.771779710784481\\
4.39622641509434	-0.773173728834407\\
4.45283018867925	-0.774406936832686\\
4.80357134908805	-0.776855131882503\\
4.84905660377358	-0.776355923597432\\
5.05720565066689	-0.777557567418648\\
5.07547169811321	-0.778467114849009\\
5.47169811320755	-0.780441235570008\\
5.52830188679245	-0.779896335177366\\
5.75471698113207	-0.780824299262264\\
5.81132075471698	-0.781874919988301\\
6.21292469062015	-0.783301736692783\\
6.26415094339623	-0.782657266337171\\
6.49056603773585	-0.783339518256639\\
6.54716981132076	-0.784351189618807\\
7	-0.785550888908533\\
7.05660377358491	-0.784864873609633\\
7.28301886792453	-0.785371528708817\\
7.33962264150944	-0.786353582799486\\
7.79245283018868	-0.787228981031635\\
7.84905660377358	-0.786496905554101\\
8.07547169811321	-0.786859482987174\\
8.13207547169811	-0.787816268191388\\
8.69811320754717	-0.788582023270931\\
8.75471698113208	-0.7878060233848\\
8.9811320754717	-0.788060755134457\\
9.0377358490566	-0.789007283129046\\
9.60377358490566	-0.789584551679674\\
9.66037735849057	-0.788785039310936\\
9.88679245283019	-0.788988035977983\\
9.94339622641509	-0.78993255275982\\
10.9422377115439	-0.790720794883149\\
12.1282707262527	-0.791317544312937\\
12.7686578989334	-0.790654037272156\\
13.4428582788049	-0.791802044181365\\
14.152657243572	-0.792029499364756\\
14.8999344410138	-0.791246891722281\\
15.6866687665557	-0.792291812041639\\
16.5149436036639	-0.79241214231722\\
17.3869523409391	-0.791556937980225\\
18.3050041805179	-0.792594388738169\\
20.2890900562591	-0.792619795669909\\
21.3603782316844	-0.791674862824057\\
22.4882317016409	-0.792674037152818\\
27.6274487609748	-0.792486263946929\\
29.0862110363026	-0.791302197928626\\
30.6219977011906	-0.791327838729699\\
32.2388757353567	-0.791115226770204\\
33.9411268599031	-0.79117907067602\\
35.7332588759175	-0.790876922004867\\
37.6200176017663	-0.790943300883075\\
39.606399440691	-0.790754148893342\\
41.6976646119888	-0.791511635137333\\
43.8993510808154	-0.791623141770203\\
48.6576172671926	-0.791485003574174\\
51.2267975447787	-0.789930083072551\\
53.9316336080251	-0.789953617330508\\
56.779288244356	-0.790911509665719\\
59.7773024449226	-0.791332421303008\\
62.9336153742108	-0.789821954717569\\
66.2565853940689	-0.790494581205413\\
69.7550121978289	-0.789297452361864\\
73.4381601131354	-0.789740355271658\\
77.315782635192	-0.790559998304223\\
81.3981482553927	-0.790377527687848\\
85.6960676537351	-0.788406296676815\\
90.2209223270261	-0.790047693983078\\
94.984694728691	-0.788261383752454\\
100	-0.789329040842808\\
};
\end{axis}
\end{tikzpicture}%

%% file: images/2phaseSears.tex
%
%
\definecolor{mycolor1}{rgb}{0.09053,0.13734,0.07326}%
\definecolor{mycolor2}{rgb}{0.04342,0.37461,0.17524}%
\definecolor{mycolor3}{rgb}{0.37697,0.57257,0.04618}%
\definecolor{mycolor4}{rgb}{0.79072,0.73098,0.27064}%
\begin{tikzpicture}[%
trim axis left, trim axis right
]

\begin{axis}[%
width=0.951\fwidth,
height=\fheight,
at={(0\fwidth,0\fheight)},
scale only axis,
xmode=log,
xmin=0.01,
xmax=100,
xminorticks=true,
xlabel style={font=\color{white!15!black}},
xlabel={$k$},
ymin=-0.9,
ymax=0,
ylabel style={font=\color{white!15!black}},
ylabel={$\angle \e^{-\textrm{i} k}S(k)$},
axis background/.style={fill=white},
xmajorgrids,
xminorgrids,
ymajorgrids
]
\addplot [color=black, line width=2.0pt, forget plot]
  table[row sep=crcr]{%
0.00398107170553497	-0.0547059668494683\\
0.01	-0.0563570241587534\\
0.0110839035359811	-0.0612412491899068\\
0.0122852917594934	-0.0665145277757659\\
0.0136168988773609	-0.0722024092588618\\
0.0150928393615877	-0.0783311795848705\\
0.0167287575567897	-0.0849276928289515\\
0.0185419935036271	-0.0920191579552929\\
0.0205517667358991	-0.0996328749388264\\
0.0227793799994691	-0.107795914439621\\
0.0252484450523573	-0.116534735594231\\
0.0279851329393847	-0.125874737277362\\
0.0310184513941747	-0.135839739506717\\
0.0343805523088551	-0.146451393634458\\
0.0381070725305102	-0.157728522705619\\
0.042237511596681	-0.169686396960493\\
0.0468156504137495	-0.182335953963912\\
0.0518900153160213	-0.195682978236126\\
0.0575143924243362	-0.209727261409779\\
0.0637483977561905	-0.224461770579847\\
0.0706581091302969	-0.239871859218751\\
0.0783167665635036	-0.255934561191397\\
0.0868055485839823	-0.272618013255527\\
0.096214432689278	-0.289881054087357\\
0.10664314906971	-0.307673047422476\\
0.118202237706192	-0.325933972546733\\
0.137931927266257	-0.35405135213614\\
0.160954791792299	-0.382799866996852\\
0.255753289931371	-0.469741488967319\\
0.298442270379047	-0.497874704742804\\
0.330790533594054	-0.516127070427278\\
0.366645036497221	-0.533882698897967\\
0.406385821648147	-0.55107014629644\\
0.450434124553848	-0.567627452585531\\
0.499256838586894	-0.583502844996219\\
0.553371463857602	-0.598655095941077\\
0.613351592496231	-0.613053552411701\\
0.679832988486861	-0.626677869119361\\
0.753520326496612	-0.63951748850207\\
0.835194661128943	-0.651570917088106\\
0.925721705771962	-0.66284484994052\\
1.02606100879403	-0.673353193697116\\
1.13727612435045	-0.683116034910873\\
1.26054588560748	-0.692158594847493\\
1.33962264150943	-0.697177814343797\\
1.47094950809279	-0.70443695348675\\
1.63038624539993	-0.711813746176951\\
1.80710438704033	-0.718584679472396\\
2.0188679245283	-0.72524002377473\\
2.24528301886793	-0.731046487273036\\
2.52830188679245	-0.73691310388924\\
2.81132075471698	-0.741647564661151\\
3.18268235287355	-0.74662675606137\\
3.60377358490566	-0.751066718583754\\
4.11647175310993	-0.755274723101484\\
4.73584905660377	-0.759165528639489\\
5.47169811320755	-0.762659652015036\\
6.43396226415095	-0.766036172488453\\
7.63275925040289	-0.769061382008082\\
9.20754716981132	-0.771845296755399\\
10.9422377115439	-0.773988283003849\\
13.4428582788049	-0.776106987735167\\
17.3869523409391	-0.778212315189114\\
23.675637181206	-0.780119844331228\\
33.9411268599031	-0.78171578132622\\
51.2267975447787	-0.782958169811337\\
85.6960676537351	-0.783939548997831\\
100	-0.784148181623823\\
};
\addplot [color=mycolor1, line width=1.0pt, forget plot]
  table[row sep=crcr]{%
0.00398107170553497	-0.0254958809433083\\
0.01	-0.0262653611849055\\
0.0122852917594934	-0.0307784620858422\\
0.0143358873934958	-0.034609433781998\\
0.0167287575567897	-0.0388589932397441\\
0.0195210328954464	-0.0435600817412345\\
0.0239821584567228	-0.0504764281168617\\
0.0279851329393847	-0.0563417963248884\\
0.0326562626566039	-0.0627607009280791\\
0.0381070725305102	-0.0697594398251944\\
0.0444677026307511	-0.0774183265697079\\
0.0518900153160213	-0.0856517594056183\\
0.0605512210031491	-0.0945282346983292\\
0.0743889416417253	-0.107508361121491\\
0.0868055485839823	-0.117988447190026\\
0.096214432689278	-0.125258219861824\\
0.10664314906971	-0.133022139978648\\
0.118202237706192	-0.141193771000099\\
0.131014220047254	-0.149856298708199\\
0.145214897684557	-0.159058740932999\\
0.160954791792299	-0.168972156191937\\
0.178400738587976	-0.179720002576814\\
0.187820510554411	-0.185493251449374\\
0.197737657725691	-0.191582961837897\\
0.208178442106382	-0.198056907172001\\
0.219170512366241	-0.20492961248204\\
0.230742977057796	-0.212277599032301\\
0.242926481699896	-0.220169263363468\\
0.255753289931371	-0.228611564991124\\
0.269257368949693	-0.237829373558777\\
0.283474479460912	-0.24784310599833\\
0.298442270379047	-0.258748307602853\\
0.31420037852572	-0.270727976047954\\
0.330790533594054	-0.28367958518604\\
0.348256668654783	-0.297776969040314\\
0.366645036497221	-0.313067537972418\\
0.386004332113175	-0.329557290363392\\
0.406385821648147	-0.347171257017425\\
0.427843478161324	-0.365680974319718\\
0.450434124553848	-0.385054115326426\\
0.525618195660832	-0.444505685216615\\
0.553371463857602	-0.463684966065486\\
0.58259013774611	-0.482302370224398\\
0.613351592496231	-0.499861607550086\\
0.645737288779182	-0.516442617960592\\
0.679832988486861	-0.532034851160847\\
0.715728981841442	-0.546665625398049\\
0.753520326496612	-0.559989781432051\\
0.835194661128943	-0.58534483171763\\
0.879293936264728	-0.596687704160968\\
0.925721705771962	-0.607323712895127\\
0.974600916933136	-0.617355528821086\\
1	-0.621516878358332\\
1.05660377358491	-0.631267761186931\\
1.08023825493656	-0.635015615116034\\
1.11320754716981	-0.640679857786691\\
1.16981132075472	-0.648524856025649\\
1.22641509433962	-0.655573554767128\\
1.28301886792453	-0.661993025293945\\
1.33962264150943	-0.667095447615625\\
1.39717689987512	-0.672509463773473\\
1.45283018867924	-0.678149965233943\\
1.54861739809169	-0.685628811457586\\
1.67924528301887	-0.694313141029771\\
1.71647258545903	-0.695760744306377\\
1.80710438704033	-0.700790210693261\\
1.84905660377359	-0.703845880367277\\
1.9622641509434	-0.709113997831397\\
2.07547169811321	-0.713863020781056\\
2.10873665095363	-0.71436659082557\\
2.24528301886793	-0.719235863684526\\
2.30188679245283	-0.722014657553374\\
2.41509433962264	-0.725314626132322\\
2.52830188679245	-0.728422077453656\\
2.59064450009023	-0.729233630940302\\
2.69811320754717	-0.731781838705934\\
2.727433622322	-0.733383092269276\\
2.87144537350201	-0.73644317462409\\
3.02306111706086	-0.739159737871154\\
3.03773584905661	-0.739422719331596\\
3.09433962264151	-0.739647274328273\\
3.20754716981132	-0.741507630700298\\
3.26415094339623	-0.743334791166672\\
3.43396226415094	-0.745641067488445\\
3.60377358490566	-0.74787532515407\\
3.66037735849056	-0.747805312013379\\
3.77358490566038	-0.749120571263628\\
3.83018867924528	-0.750686018413435\\
4.11647175310993	-0.753447228506133\\
4.16981132075472	-0.753946829197525\\
4.22641509433962	-0.753681314167285\\
4.39622641509434	-0.755114029379073\\
4.45283018867925	-0.756493534927098\\
4.80357134908805	-0.758942832158407\\
4.84905660377358	-0.758439477808333\\
5.05720565066689	-0.759726147069958\\
5.07547169811321	-0.760751685196214\\
5.47169811320755	-0.762859657737055\\
5.52830188679245	-0.762293212191133\\
5.75471698113207	-0.76334411393771\\
5.81132075471698	-0.764505683663671\\
6.21292469062015	-0.766183304397207\\
6.26415094339623	-0.765474322540327\\
6.49056603773585	-0.766292661531571\\
6.54716981132076	-0.76739384175654\\
7	-0.768952961689408\\
7.05660377358491	-0.768150216689627\\
7.28301886792453	-0.76878733450923\\
7.33962264150944	-0.76983407576323\\
7.79245283018868	-0.771124971724491\\
7.84905660377358	-0.770223818486165\\
8.07547169811321	-0.770712456774709\\
8.13207547169811	-0.771715014106721\\
8.69811320754717	-0.77303125177399\\
8.75471698113208	-0.772037975870737\\
8.9811320754717	-0.772418460870151\\
9.0377358490566	-0.773391921033735\\
9.26415094339623	-0.773758816408149\\
9.37706743211719	-0.774216405656488\\
9.60377358490566	-0.774590816302511\\
9.66037735849057	-0.773521850463852\\
9.88679245283019	-0.773845255629391\\
9.94339622641509	-0.774789553926864\\
10	-0.774868892814997\\
10.3934510867977	-0.775735779205379\\
10.9422377115439	-0.776036792850187\\
11.5200009252003	-0.77703811324071\\
12.1282707262527	-0.777160368075796\\
12.7686578989334	-0.776878205308797\\
13.4428582788049	-0.778714488381969\\
14.152657243572	-0.778684440018255\\
14.8999344410138	-0.778327762380068\\
15.6866687665557	-0.780088392707421\\
17.3869523409391	-0.779515997328177\\
18.3050041805179	-0.781260545302219\\
19.2715302531669	-0.780885031130426\\
20.2890900562591	-0.781173974019084\\
21.3603782316844	-0.780814026488128\\
22.4882317016409	-0.782434529417575\\
23.675637181206	-0.782679323159218\\
24.925739087578	-0.7822218326352\\
27.6274487609748	-0.782664396960452\\
29.0862110363026	-0.782141929707857\\
30.6219977011906	-0.782469213799352\\
32.2388757353567	-0.782485901436956\\
33.9411268599031	-0.782852243360115\\
35.7332588759175	-0.7827729284366\\
37.6200176017663	-0.783165509578046\\
39.606399440691	-0.783177496754501\\
43.8993510808154	-0.784042265960048\\
46.2172892234977	-0.78409606712142\\
48.6576172671926	-0.784592768122739\\
51.2267975447787	-0.783503511614629\\
53.9316336080251	-0.783776512802726\\
59.7773024449226	-0.785256820781449\\
62.9336153742108	-0.784338569923456\\
66.2565853940689	-0.784610363427585\\
69.7550121978289	-0.78404166701108\\
73.4381601131354	-0.78411474779428\\
77.315782635192	-0.785207123314856\\
81.3981482553927	-0.785348318898434\\
85.6960676537351	-0.783940507441328\\
90.2209223270261	-0.785278084029474\\
94.984694728691	-0.784065122490068\\
100	-0.785575471369369\\
};
\addplot [color=mycolor2, line width=1.0pt, forget plot]
  table[row sep=crcr]{%
0.00398107170553497	-0.027936243191284\\
0.01	-0.0287793749586442\\
0.0116691468742838	-0.0325165468056898\\
0.0136168988773609	-0.0366956522185218\\
0.0158897592972195	-0.0413592678004071\\
0.0185419935036271	-0.046551739009804\\
0.0216369245535842	-0.0523187088310224\\
0.0239821584567228	-0.0564064823918691\\
0.0279851329393847	-0.0632208614849259\\
0.0326562626566039	-0.0707303004414586\\
0.0381070725305102	-0.0789775859070843\\
0.0444677026307511	-0.088050476694435\\
0.0492875726425938	-0.0945204822961974\\
0.0546298700693171	-0.101360742825381\\
0.0605512210031491	-0.108579050519163\\
0.0706581091302969	-0.120223454567883\\
0.0783167665635036	-0.128418343964766\\
0.0913889851748526	-0.1413426637125\\
0.101294669592927	-0.150535370833844\\
0.112274034647708	-0.160142099622786\\
0.12444345696306	-0.1701920046348\\
0.137931927266257	-0.180675241277319\\
0.152882417635115	-0.191638115517786\\
0.169453396941519	-0.203150996977793\\
0.187820510554411	-0.215237857540311\\
0.208178442106382	-0.228048396226527\\
0.230742977057796	-0.241719058867941\\
0.255753289931371	-0.256408298839812\\
0.269257368949693	-0.264278185342463\\
0.283474479460912	-0.27253548517646\\
0.298442270379047	-0.281231073323272\\
0.31420037852572	-0.290514883212306\\
0.330790533594054	-0.300254237799842\\
0.348256668654783	-0.310618352605995\\
0.366645036497221	-0.321665706542247\\
0.386004332113175	-0.333504823802872\\
0.406385821648147	-0.346179452686268\\
0.427843478161324	-0.359635206301121\\
0.450434124553848	-0.374132965824909\\
0.474217584043875	-0.389561132154398\\
0.499256838586894	-0.40592124985651\\
0.525618195660832	-0.422991403887414\\
0.553371463857602	-0.440677057160161\\
0.645737288779182	-0.495061035009989\\
0.679832988486861	-0.512586472891084\\
0.715728981841442	-0.529402347388652\\
0.753520326496612	-0.545099692098094\\
0.793307099263651	-0.560067085088645\\
0.835194661128943	-0.574373289686754\\
0.879293936264728	-0.587485720242338\\
0.925721705771962	-0.59972726869966\\
0.974600916933136	-0.611126476837455\\
1.02606100879403	-0.621281124399898\\
1.11320754716981	-0.637204319021597\\
1.16981132075472	-0.64585960494091\\
1.22641509433962	-0.65362628338717\\
1.28301886792453	-0.660627885779283\\
1.33962264150943	-0.666339200354523\\
1.39717689987512	-0.67218483813421\\
1.47094950809279	-0.679664278254946\\
1.56603773584906	-0.687379809423131\\
1.67924528301887	-0.695374196323106\\
1.71647258545903	-0.697013927057488\\
1.80710438704033	-0.702336315177715\\
1.84905660377359	-0.705420255760939\\
2	-0.712764246812684\\
2.07547169811321	-0.716024301001082\\
2.10873665095363	-0.716602710615613\\
2.24528301886793	-0.721687115555214\\
2.30188679245283	-0.724466796084349\\
2.47169811320755	-0.729639566411478\\
2.52830188679245	-0.731216503953398\\
2.59064450009023	-0.732060142292204\\
2.69811320754717	-0.734689828156791\\
2.727433622322	-0.736255256458473\\
2.92452830188679	-0.740406228421802\\
3.03773584905661	-0.742563600696665\\
3.09433962264151	-0.742764639500437\\
3.20754716981132	-0.744664020635167\\
3.26415094339623	-0.746480481841331\\
3.52765441849198	-0.750193775043751\\
3.60377358490566	-0.751181833466412\\
3.66037735849056	-0.751069741454678\\
3.77358490566038	-0.752399726152105\\
3.83018867924528	-0.753969777321922\\
4.11647175310993	-0.756803848548706\\
4.16981132075472	-0.757302639211981\\
4.22641509433962	-0.75700319867498\\
4.39622641509434	-0.758439863502182\\
4.45283018867925	-0.759840617791867\\
4.80357134908805	-0.762287281381106\\
4.84905660377358	-0.761769446703005\\
5.05720565066689	-0.763049700583539\\
5.07547169811321	-0.764106645564993\\
5.47169811320755	-0.766147099451742\\
5.52830188679245	-0.765592762986554\\
5.75471698113207	-0.766625139832287\\
5.81132075471698	-0.767831213599919\\
6.21292469062015	-0.769380001005576\\
6.26415094339623	-0.768712780237938\\
6.49056603773585	-0.769508083585163\\
6.54716981132076	-0.770661521885326\\
7	-0.772044287199711\\
7.05660377358491	-0.771309906199045\\
7.28301886792453	-0.771923879994551\\
7.33962264150944	-0.773035575602666\\
7.79245283018868	-0.774116424781646\\
7.84905660377358	-0.773306131902768\\
8.07547169811321	-0.773770787746334\\
8.13207547169811	-0.774845524879361\\
8.69811320754717	-0.775912476846551\\
8.75471698113208	-0.775024949224613\\
8.9811320754717	-0.775377728902694\\
9.0377358490566	-0.77643594986702\\
9.60377358490566	-0.777358840292803\\
9.66037735849057	-0.776406853025022\\
9.88679245283019	-0.776704737080274\\
9.94339622641509	-0.777745913744007\\
11.5200009252003	-0.779616359387161\\
12.1282707262527	-0.779925448521782\\
12.7686578989334	-0.779452418388074\\
13.4428582788049	-0.781131325355953\\
14.152657243572	-0.781286909818396\\
14.8999344410138	-0.780716152882396\\
15.6866687665557	-0.782345002567283\\
16.5149436036639	-0.782322801057916\\
17.3869523409391	-0.781718612237365\\
18.3050041805179	-0.78336101930774\\
19.2715302531669	-0.783149792159847\\
20.2890900562591	-0.783393783363416\\
21.3603782316844	-0.782759740568874\\
22.4882317016409	-0.784364762529167\\
23.675637181206	-0.784569089410183\\
24.925739087578	-0.784199235180063\\
27.6274487609748	-0.78452578953047\\
29.0862110363026	-0.783734737236613\\
30.6219977011906	-0.784010236441838\\
32.2388757353567	-0.783990583008921\\
33.9411268599031	-0.784285793956057\\
35.7332588759175	-0.784145689235893\\
37.6200176017663	-0.784487929904916\\
39.606399440691	-0.784467336723611\\
43.8993510808154	-0.785427334107909\\
46.2172892234977	-0.785447804259818\\
48.6576172671926	-0.785968083045742\\
51.2267975447787	-0.784585509316571\\
53.9316336080251	-0.784806636905015\\
59.7773024449226	-0.786468062285836\\
62.9336153742108	-0.785240667937405\\
66.2565853940689	-0.78564349616542\\
69.7550121978289	-0.784878861510289\\
73.4381601131354	-0.785081383790968\\
77.315782635192	-0.786275087187943\\
81.3981482553927	-0.786259063681368\\
85.6960676537351	-0.784672812376947\\
90.2209223270261	-0.786067209073744\\
94.984694728691	-0.784777407506966\\
100	-0.786181419792261\\
};
\addplot [color=mycolor3, line width=1.0pt, forget plot]
  table[row sep=crcr]{%
0.00398107170553497	-0.0303773086929229\\
0.01	-0.0312941132106435\\
0.0116691468742838	-0.0354453989371453\\
0.0136168988773609	-0.0401055505998711\\
0.0158897592972195	-0.0453274233370391\\
0.0185419935036271	-0.0511670444243624\\
0.0216369245535842	-0.0576831723061049\\
0.0239821584567228	-0.0623458737738627\\
0.0265815930918929	-0.0674347713075818\\
0.0294627813663242	-0.0728876575956492\\
0.0326562626566039	-0.0787228383777689\\
0.036195886513146	-0.0849583204278059\\
0.0401191714511029	-0.0916479046191734\\
0.0444677026307511	-0.0987388947919103\\
0.0492875726425938	-0.10628011545562\\
0.0546298700693171	-0.114285666093311\\
0.0605512210031491	-0.122767988115178\\
0.0706581091302969	-0.136489525880997\\
0.0783167665635036	-0.146207164254003\\
0.0868055485839823	-0.156426593010741\\
0.096214432689278	-0.167049355668419\\
0.10664314906971	-0.178268506349943\\
0.118202237706192	-0.189984628063652\\
0.131014220047254	-0.202205430085654\\
0.145214897684557	-0.214897876617143\\
0.160954791792299	-0.228088343473448\\
0.178400738587976	-0.241723473354242\\
0.197737657725691	-0.255830318891242\\
0.219170512366241	-0.270434934539836\\
0.242926481699896	-0.285541430653771\\
0.269257368949693	-0.301153450104496\\
0.298442270379047	-0.317451426176378\\
0.330790533594054	-0.334571424523698\\
0.366645036497221	-0.352514606475125\\
0.386004332113175	-0.361914448410521\\
0.427843478161324	-0.381620797372182\\
0.450434124553848	-0.39210629171566\\
0.474217584043875	-0.403039395665433\\
0.499256838586894	-0.414539952953987\\
0.525618195660832	-0.426486052046541\\
0.553371463857602	-0.439004397060031\\
0.58259013774611	-0.452143177326757\\
0.613351592496231	-0.465888185061046\\
0.645737288779182	-0.480259075429732\\
0.679832988486861	-0.495136653087493\\
0.753520326496612	-0.525907118461234\\
0.835194661128943	-0.557308531624728\\
0.879293936264728	-0.57254260488901\\
0.925721705771962	-0.587269699297872\\
0.974600916933136	-0.601163352096331\\
1.05660377358491	-0.620874801669213\\
1.11320754716981	-0.632929237794994\\
1.16981132075472	-0.643285313346655\\
1.22641509433962	-0.65248615078551\\
1.28301886792453	-0.66064824156891\\
1.39717689987512	-0.674177255059109\\
1.47094950809279	-0.682308672405408\\
1.56603773584906	-0.690942029625554\\
1.67924528301887	-0.699746554451717\\
1.73584905660377	-0.703060386731037\\
1.80710438704033	-0.707548849329974\\
1.84905660377359	-0.710644336632515\\
2	-0.718580262438364\\
2.07547169811321	-0.722063845913929\\
2.10873665095363	-0.722839379594046\\
2.24528301886793	-0.728227184096591\\
2.30188679245283	-0.730971673555517\\
2.47169811320755	-0.736458429624546\\
2.52830188679245	-0.738107603923942\\
2.59064450009023	-0.739094852644858\\
2.69811320754717	-0.741827843192166\\
2.727433622322	-0.743302035417751\\
2.92452830188679	-0.747654245708439\\
3.03773584905661	-0.749875913143056\\
3.09433962264151	-0.750139005034328\\
3.20754716981132	-0.752079157439942\\
3.26415094339623	-0.753809141050253\\
3.54716981132076	-0.757917341347313\\
3.60377358490566	-0.758656113997928\\
3.66037735849056	-0.758563711788544\\
3.77358490566038	-0.759899217802023\\
3.83018867924528	-0.76138017333799\\
4.16981132075472	-0.764790512426191\\
4.22641509433962	-0.764487187398562\\
4.39622641509434	-0.765909257340051\\
4.45283018867925	-0.767224062298149\\
4.80357134908805	-0.76970410426451\\
4.84905660377358	-0.769175604338096\\
5.05720565066689	-0.770418328719948\\
5.07547169811321	-0.771403825886679\\
5.47169811320755	-0.77342761792456\\
5.52830188679245	-0.772862730572168\\
5.75471698113207	-0.773842542459693\\
5.81132075471698	-0.774978377731269\\
6.21292469062015	-0.776464499865353\\
6.26415094339623	-0.775802506886081\\
6.49056603773585	-0.776541140224291\\
6.54716981132076	-0.777636431353709\\
7	-0.778907586036266\\
7.05660377358491	-0.778203314410011\\
7.28301886792453	-0.778762827730527\\
7.33962264150944	-0.779825247202861\\
7.79245283018868	-0.780768205553171\\
7.84905660377358	-0.780012682787484\\
8.07547169811321	-0.780424846495483\\
8.13207547169811	-0.78146107653509\\
8.69811320754717	-0.782322918189926\\
8.75471698113208	-0.781521042833443\\
8.9811320754717	-0.781824152991287\\
9.0377358490566	-0.782850588764497\\
9.60377358490566	-0.783529687431702\\
9.66037735849057	-0.782693179760495\\
9.88679245283019	-0.7829417772936\\
9.94339622641509	-0.783961577726927\\
10.9422377115439	-0.784934509215887\\
12.1282707262527	-0.785733664352117\\
12.7686578989334	-0.78511957608952\\
13.4428582788049	-0.786473219108181\\
14.152657243572	-0.786760978802718\\
14.8999344410138	-0.78602965413188\\
15.6866687665557	-0.787314431594997\\
16.5149436036639	-0.787460691631685\\
17.3869523409391	-0.786663147846544\\
18.3050041805179	-0.787967861342739\\
20.2890900562591	-0.78808571318925\\
21.3603782316844	-0.787210706872338\\
22.4882317016409	-0.788500488210445\\
23.675637181206	-0.788607774696556\\
26.2418478669375	-0.788412871824648\\
27.6274487609748	-0.788560036501745\\
29.0862110363026	-0.787473794688357\\
30.6219977011906	-0.787613885795341\\
32.2388757353567	-0.787485458697663\\
33.9411268599031	-0.787669100327153\\
35.7332588759175	-0.787437662670376\\
37.6200176017663	-0.787642928132952\\
39.606399440691	-0.787529877192576\\
41.6976646119888	-0.788295290075617\\
43.8993510808154	-0.788512398803192\\
46.2172892234977	-0.788459142772624\\
48.6576172671926	-0.78872515901139\\
51.2267975447787	-0.787176483843115\\
53.9316336080251	-0.78728661409915\\
59.7773024449226	-0.788931636355442\\
62.9336153742108	-0.787428943840836\\
66.2565853940689	-0.788031252283351\\
69.7550121978289	-0.786953352073242\\
73.4381601131354	-0.787338805529098\\
77.315782635192	-0.78845720209625\\
81.3981482553927	-0.788238386467186\\
85.6960676537351	-0.78646717642203\\
90.2209223270261	-0.787978534708054\\
94.984694728691	-0.786434402647766\\
100	-0.787684818220191\\
};
\addplot [color=mycolor4, line width=1.0pt, forget plot]
  table[row sep=crcr]{%
0.00398107170553497	-0.0328189758712396\\
0.01	-0.0338094712982668\\
0.0116691468742838	-0.0383752221540004\\
0.0136168988773609	-0.0435169567078182\\
0.0158897592972195	-0.0492979414931982\\
0.0185419935036271	-0.0557860378052726\\
0.0205517667358991	-0.0605397451500957\\
0.0239821584567228	-0.0682930246707767\\
0.0265815930918929	-0.0740056343009301\\
0.0294627813663242	-0.0801447719761081\\
0.0326562626566039	-0.08673429081386\\
0.036195886513146	-0.0937980899625006\\
0.0401191714511029	-0.101388264610254\\
0.0444677026307511	-0.109473390390898\\
0.0492875726425938	-0.118101699974021\\
0.0546298700693171	-0.127293906869935\\
0.0605512210031491	-0.137068948635117\\
0.0671143892584777	-0.147512294335926\\
0.0743889416417253	-0.158502223100128\\
0.0824519853300611	-0.170115178914162\\
0.0913889851748526	-0.182277919628118\\
0.101294669592927	-0.195148549087607\\
0.112274034647708	-0.208647073538167\\
0.12444345696306	-0.222774269917601\\
0.137931927266257	-0.237497884574876\\
0.152882417635115	-0.252805502519481\\
0.169453396941519	-0.268683222327539\\
0.187820510554411	-0.285066970810122\\
0.208178442106382	-0.301949793217294\\
0.230742977057796	-0.319275934823796\\
0.269257368949693	-0.34597758033064\\
0.298442270379047	-0.364240409485404\\
0.348256668654783	-0.39222835129692\\
0.406385821648147	-0.420789628409398\\
0.450434124553848	-0.44009860829973\\
0.525618195660832	-0.469743797553324\\
0.58259013774611	-0.489944751845796\\
0.645737288779182	-0.510664295258843\\
0.715728981841442	-0.531946520419544\\
0.753520326496612	-0.542723233334533\\
0.835194661128943	-0.565161644238387\\
0.879293936264728	-0.576665982799268\\
1.08023825493656	-0.624182690035203\\
1.13727612435045	-0.636468196032373\\
1.26054588560748	-0.660196094946416\\
1.28301886792453	-0.664131265412097\\
1.45283018867924	-0.689908531473588\\
1.50943396226415	-0.697044136005431\\
1.56603773584906	-0.70354657360356\\
1.63038624539993	-0.710141322083674\\
1.67924528301887	-0.714668177209592\\
1.73584905660377	-0.719071303059545\\
1.84905660377359	-0.727704233950653\\
1.9622641509434	-0.734706859260251\\
2.07547169811321	-0.740652431171138\\
2.13207547169811	-0.742889228895975\\
2.35849056603774	-0.75230055147026\\
2.47169811320755	-0.756096760169361\\
2.52830188679245	-0.757818697047308\\
2.59064450009023	-0.759100208668597\\
2.69811320754717	-0.76192173495213\\
2.727433622322	-0.763137873268275\\
2.92452830188679	-0.767627119662483\\
3.03773584905661	-0.769868351617514\\
3.09433962264151	-0.770326521353651\\
3.20754716981132	-0.772259923011574\\
3.26415094339623	-0.773749393749839\\
3.54716981132076	-0.77783820716292\\
3.60377358490566	-0.778558336883961\\
3.66037735849056	-0.778606533451596\\
3.77358490566038	-0.779895131730773\\
3.83018867924528	-0.781150494596877\\
4.16981132075472	-0.784449918832955\\
4.22641509433962	-0.78425793178093\\
4.39622641509434	-0.785591573727889\\
4.45283018867925	-0.786694019889604\\
4.80357134908805	-0.789039157825029\\
4.84905660377358	-0.788607059103688\\
5.05720565066689	-0.789733866717504\\
5.07547169811321	-0.790532393658252\\
5.47169811320755	-0.79238372932108\\
5.52830188679245	-0.791891979301774\\
5.75471698113207	-0.792735466499145\\
5.81132075471698	-0.79366457595843\\
6.21292469062015	-0.794957824244035\\
6.26415094339623	-0.794363054776987\\
6.49056603773585	-0.79495678025216\\
6.54716981132076	-0.795849142495328\\
7	-0.796904676014822\\
7.05660377358491	-0.796263823090217\\
7.28301886792453	-0.7966892861272\\
7.33962264150944	-0.797554390679411\\
7.79245283018868	-0.79830280920393\\
7.84905660377358	-0.79762013760919\\
8.07547169811321	-0.797909329154194\\
8.13207547169811	-0.798746246197158\\
8.69811320754717	-0.799354324675531\\
8.75471698113208	-0.798631981124103\\
8.9811320754717	-0.798811442568216\\
9.0377358490566	-0.799637220270113\\
9.60377358490566	-0.800053908668041\\
9.66037735849057	-0.799308202538436\\
9.88679245283019	-0.799440203443463\\
9.94339622641509	-0.800261487575069\\
11.5200009252003	-0.800932629356561\\
12.1282707262527	-0.801031902405844\\
12.7686578989334	-0.800306311032205\\
13.4428582788049	-0.801193851105529\\
14.152657243572	-0.801241289873409\\
14.8999344410138	-0.800405137141313\\
15.6866687665557	-0.801158729338052\\
16.5149436036639	-0.801102003793019\\
17.3869523409391	-0.80021090004294\\
18.3050041805179	-0.800932345149866\\
20.2890900562591	-0.800640194034125\\
21.3603782316844	-0.799645721609232\\
22.4882317016409	-0.800284630222957\\
27.6274487609748	-0.799496266410949\\
29.0862110363026	-0.798271140972128\\
33.9411268599031	-0.797643023058434\\
35.7332588759175	-0.797221233172468\\
39.606399440691	-0.796763461471417\\
41.6976646119888	-0.797295798321258\\
46.2172892234977	-0.796965131380185\\
48.6576172671926	-0.796602549480995\\
51.2267975447787	-0.795199860885736\\
53.9316336080251	-0.795040320690249\\
56.779288244356	-0.795852905070615\\
59.7773024449226	-0.795831389188415\\
62.9336153742108	-0.794445407666524\\
66.2565853940689	-0.795036731084799\\
69.7550121978289	-0.793610511758742\\
77.315782635192	-0.794457000225793\\
81.3981482553927	-0.794373737319329\\
85.6960676537351	-0.792360246146789\\
90.2209223270261	-0.793783628880301\\
94.984694728691	-0.791921073309894\\
100	-0.792672762844374\\
};
\end{axis}
\end{tikzpicture}%

%% file: sections/indicial.tex
\subsection{Wagner function}
With numerical solutions in hand for the porous Theodorsen and Sears functions, we now turn our attention to time-domain aerodynamic responses. It is sufficient in this linearised physical system to determine indicial functions, which are the aerodynamic responses to impulsive system changes. Duhamel's integral may then be used to construct the aerodynamic response to an arbitrary aerofoil motion or gust field \citep[pp.~277-280]{fung}.

\citet{Wagner1925} first solved the canonical problem of the transient lift experienced by an aerofoil whose angle of attack changes instantaneously from zero to a fixed value; herein we develop the porous analogue to the so-called Wagner function. 
%
%
%
%
%
Recalling the nondimensional scalings chosen in \S\ref{MM}, the linearised normal wing velocity for an impulsive unit change in angle of attack modifies the boundary condition \eqref{Eq:w01} to
\begin{align}
    w(x,t) = w_{\rm s}(x,t) + H(t).
    \label{Eq:wagVel}
\end{align}
%
Note that the Heaviside function $H$ may be expressed in the frequency domain as
\begin{align}
    H(t) = \int_{-\infty}^{\infty} \frac{\e^{\i k t}}{\i k} \d k, 
    \label{Eq:heaviside}
\end{align}
where the path of integration passes above the pole at $k=0$. Carrying through the analysis of \S\ref{Sec:numUnsteady} with boundary condition \eqref{Eq:wagVel} furnishes the circulatory lift history due to an impulsive change in angle of attack:
\begin{align}
    L^C(t) = L^{Q} \int_{-\infty}^\infty \frac{C(k)}{\i k} \e^{\i k t} \d k.
    \label{Eq:cLift2}
\end{align}
The quasi-steady lift $L^Q$ corresponding to the asymptotic lift value at long times for a given porosity profile is
\begin{align}
L^{Q} = \int_{-1}^1 \frac{4}{1+(\psi(x,0))^2} \Bigg\{&\psi(x,0) 
	+\frac{Z(x)}{\pi} \dashint_{-1}^1\frac{\d \xi }{Z(\xi)(\xi - x)} \Bigg\}\d x,
	\label{Eq:qsLift2}
\end{align}
where $Z$ is defined in \eqref{Eq:Zdef}. The ratio $L^C(t)/L^{Q}$ from \eqref{Eq:cLift2} defines the Wagner function,
\begin{align}
    \phi(t)= \int_{-\infty}^\infty \frac{C(k)}{\i k} \e^{\i k t} \d k.
    \label{eq:wagner}
\end{align}
Therefore, the porous Theodorsen function analogue in the frequency domain from \S\ref{sec:theo} generates the associated time-domain Wagner function for impulsive aerofoil motions.

Figure \ref{Fig:wagner} plots the porous Wagner function against convective time for the same ranges of dimensionless flow resistance and effective density examined in figure~\ref{Fig:theoPlot}, i.e., varying the flow resistance in figure \ref{Fig:wagner1} and setting different constant effective density values in figure \ref{Fig:wagner2}. In each case, the Wagner function asymptotes to unity at large time, which is anticipated from \eqref{eq:wagner} by the fact that the Theodorsen function is unity at $k=0$ regardless of the porosity profile. Aerofoil porosity decreases the magnitude of the Wagner function at short times, whose precise value depends on the porosity parameters. Figure~\ref{Fig:wagner2} makes clear that effective density controls the short-time behaviour of the Wagner function and has a marginal influence at large times. However, figure~\ref{Fig:wagner1} indicates counterintuitively that a smaller flow resistance of the porous aerofoil leads to a larger initial value of the Wagner function, and a lesser flow resistance yields a faster approach to the quasi-steady lift value. 
Additionally figures \ref{Fig:wagner1} and \ref{Fig:wagner2} show that the late-time behaviour is controlled by the flow resistance, whereas the early-time behaviour is controlled by the effective density.
These can observations can be interpreted in light of the asymptotic solutions derived in appendix \ref{Sec:asymp}.
Therein, we show that, to leading order, the low-frequency (large-time) lift depends only on the flow resistance of the porous medium, whereas as the high-frequency (small-time) lift depends only on the effective density.

Figure~\ref{Fig:dimWagner} clarifies the roles of the porosity parameters by considering only the unsteady circulatory lift, which is also the metric of practical interest. We show here that the dimensionless flow resistance controls the rise time and asymptotic value of the circulatory lift at long times, and the effective density governs the short-time aerodynamic response only. Therefore, the effective density of the porous medium is the key parameter to reduce the transient effect of aerofoil motion on the resulting unsteady lift, where the flow resistance distribution may be set by steady aerodynamic constraints of a particular aerial system. 

In complement to the passive unsteady aerodynamic control that tailored porosity distributions could provide, we note the possible predictive limitations that are tied to the underlying linear assumptions of the flow field. \citet{beckwith2009impulsively} demonstrate experimentally that the impermeable Wagner function predicts well the lift history at large times for small angles of attack, where measured early-time responses overshoot the theoretical predictions due to delayed stall effects not present in the Wagner model. Larger impulsive changes to the angle of attack lead to persistent nonlinear flow structures that must be integrated into the aerodynamic lift model to improve its predictive capacity \citep{pitt2014impulsively,li2015unsteady,stevens2017experiments}. However, unsteady porous aerofoils may introduce important nonlinear flow features in real flows even at small angles, and an experimental campaign beyond the scope of the present work is needed to examine the relevant fluid mechanics of porous aerofoils and the predictive capability of the present model.

\begin{figure}
	\begin{subfigure}[t]{.45\linewidth}
	\setlength{\fheight}{3cm}
		\setlength{\fwidth}{\linewidth}
		\centering
	\input{images/1wagner.tex}
	\caption{}
	\label{Fig:wagner1}
	\end{subfigure}
	\hfill
	\begin{subfigure}[t]{.45\linewidth}
	\setlength{\fheight}{3cm}
		\setlength{\fwidth}{\linewidth}
		\centering
	\input{images/2wagner.tex}
	\caption{}
	\label{Fig:wagner2}
	\end{subfigure}
\caption{%
{Porous extension of the Wagner function $\phi(t)$ for an impulsive change in angle of attack as a function of convective time. The ranges of the dimensionless flow resistance $\Phi$ and effective density $\rho_{\rm e}$ of the porous aerofoil are the same as in figure \mbox{\ref{Fig:theoPlot}}. The black curve indicates the classical Wagner function for an impermeable aerofoil.
In (a) the flow resistance ranges from high (black) to low (orange), and in (b) the effective density varies from high (black) to low (yellow).
}}
	\label{Fig:wagner}
\begin{subfigure}[t]{.45\linewidth}
	\setlength{\fheight}{3cm}
		\setlength{\fwidth}{\linewidth}
		\centering
	\input{images/1dimwagner.tex}
	\caption{}
	\label{Fig:dimWagner1}
	\end{subfigure}
	\hfill
	\begin{subfigure}[t]{.45\linewidth}
	\setlength{\fheight}{3cm}
		\setlength{\fwidth}{\linewidth}
		\centering
	\input{images/2dimwagner.tex}
	\caption{}
	\label{Fig:dimWagner2}
	\end{subfigure}
\caption{%
{%
Circulatory lift history due to an impulsive change in angle of attack of a porous aerofoil as a function of convective time. The ranges of the dimensionless flow resistance $\Phi$ and effective density $\rho_{\rm e}$ of the porous aerofoil are the same as in figure \mbox{\ref{Fig:theoPlot}}. The black curve corresponds to an impermeable aerofoil.
In (a) the flow resistance ranges from 0 (black) to high (orange), and in (b) the effective density varies from high (black) to low (yellow).
}}
\label{Fig:dimWagner}
\end{figure}



\subsection{K\"ussner function}
The K\"ussner function \citep{Kussner1936} is the indicial aerodynamic response to a stationary aerofoil encountering a sharp-edged gust. Like the relationship between the Theodorsen function and the Wagner function, the Sears harmonic gust response function is connected via the Fourier transform to the K\"{u}ssner sharp-edged gust function, which can be used to predict the aerodynamic response to arbitrary linear gusts by appeal to Duhamel's integral. The sharp-edged gust problem is carried out similarly to \S\ref{Sec:Gust} with boundary condition \eqref{Eq:w01} now given by
%
\begin{align}
    w(x,t) = w_s(x,t) + H(x-t+1).
\end{align}
The frequency domain representation of the Heaviside function \eqref{Eq:heaviside} may be used again to arrive at the transient total lift on the aerofoil, 
\begin{align}
    L(t) &= L^Q \int_{-\infty}^\infty \frac{S(k)}{\i k} \e^{-\i k (t-1)} \d k.
\end{align}
Note the dependence of the unsteady lift on the Sears function $S(k)$, into which all of the information about the porosity distribution is embedded. The K\"ussner function is the ratio $L(t)/L^Q$ of the unsteady lift to the quasi-steady lift \eqref{Eq:qsLift2},
\begin{align}
    \psi(t) = \int_{-\infty}^\infty \frac{S(k)}{\i k} \e^{-\i k (t-1)} \d k,
    \label{eq:kussner}
\end{align}
where the contour of integration passes above the origin.
\begin{figure}
	\begin{subfigure}[t]{.45\linewidth}
	\setlength{\fheight}{3cm}
		\setlength{\fwidth}{\linewidth}
		\centering
	\input{images/1kussner.tex}
	\caption{}
	\label{Fig:kussner1}
	\end{subfigure}
	\hfill
	\begin{subfigure}[t]{.45\linewidth}
	\setlength{\fheight}{3cm}
		\setlength{\fwidth}{\linewidth}
		\centering
	\input{images/2kussner.tex}
	\caption{}
	\label{Fig:kussner2}
	\end{subfigure}
\caption{{Porous extension of the K\"ussner function $\psi(t)$ for a sharp-edged gust as a function of convective time. The ranges of dimensionless flow resistance $\Phi$ and effective density $\rho_{\rm e}$ of the porous aerofoil are the same as in figure \ref{Fig:sears}. The black curve corresponds to the classical K\"ussner function for an impermeable aerofoil.
In (a) the flow resistance ranges from high (black) to low (orange) and in (b) the effective density varies from high (black) to low (yellow).}
}
\label{Fig:kussner}

	\begin{subfigure}[t]{.45\linewidth}
	\setlength{\fheight}{3cm}
		\setlength{\fwidth}{\linewidth}
		\centering
	\input{images/1dimkussner.tex}
	\caption{}
	\label{Fig:dimKussner1}
	\end{subfigure}
	\hfill
	\begin{subfigure}[t]{.45\linewidth}
	\setlength{\fheight}{3cm}
		\setlength{\fwidth}{\linewidth}
		\centering
	\input{images/2dimkussner.tex}
	\caption{}
	\label{Fig:dimKussner2}
	\end{subfigure}
\caption{{Circulatory lift history of a stationary porous aerofoil encountering a sharp-edged gust as a function of convective time. The ranges of dimensionless flow resistance $\Phi$ and effective density $\rho_{\rm e}$ of the porous aerofoil are the same as in figure \ref{Fig:sears}. The black curve indicates results for an impermeable aerofoil.
In (a) the flow resistance ranges from high (black) to low (orange) and in (b) the effective density varies from high (black) to low (yellow).} }
\label{Fig:dimKussner}
\end{figure}

Figures~\ref{Fig:kussner} and \ref{Fig:dimKussner} illustrate the effects of the dimensionless flow resistance $\Phi$ and effective density $\rho_{\rm e}$ of the porous aerofoil on the K\"ussner function and on the total unsteady lift, respectively. The descriptions of $\Phi$ and $\rho_{\rm e}$ are the same as in figure~\ref{Fig:sears}, where the porous Sears functions computed in \S\ref{Sec:Gust} generate the porous K\"ussner functions in figures \ref{Fig:kussner} and \ref{Fig:dimKussner}. The results in these figures differ by a factor of the quasi-steady lift that depends on the flow resistance distribution, where accelerated rise times in figure~\ref{Fig:kussner1} for aerofoils with less resistance have instead slower actual rise times in the lift coefficient and a smaller asymptotic lift value in figure~\ref{Fig:dimWagner1}, as discussed previously in the context of the Wagner function. Figures~\ref{Fig:kussner2} and \ref{Fig:dimKussner2} show that the effective density has a marginal influence on the sharp-edged gust response. Therefore, the dimensionless flow resistance distribution has the dominant influence on the porous K\"ussner function.

Comparisons between the impermeable K\"ussner function to both experimental and computational simulations \citep{biler2019experimental,sedky2020unsteady} are favourable for gust amplitudes on the order of or smaller than the freestream velocity, i.e., moderate to small gust ratios. Larger gust ratios can lead to signification deviations from the linear theory, especially at long times due to large-scale vortex structures generated by the gust encounter \citep{perrotta2017unsteady,andreu2020wing}. One might anticipate that the robustness of the K\"ussner function for gust encounters with significant nonlinear flow features might carry over to porous aerofoils, whose response in short times is similar to the impermeable case. Experimental unsteady lift measurements of porous aerofoils that are beyond the scope of the present work would be necessary to assess this claim and the effects of porosity on gust response in real flows.

%% file: images/1wagner.tex
%
%
\definecolor{mycolor1}{rgb}{0.18517,0.05913,0.24304}%
\definecolor{mycolor2}{rgb}{0.51514,0.10993,0.38770}%
\definecolor{mycolor3}{rgb}{0.81038,0.26571,0.33825}%
\definecolor{mycolor4}{rgb}{0.94901,0.58547,0.40375}%
\begin{tikzpicture}[%
trim axis left, trim axis right
]

\begin{axis}[%
width=0.951\fwidth,
height=\fheight,
at={(0\fwidth,0\fheight)},
scale only axis,
xmin=0,
xmax=10,
xlabel style={font=\color{white!15!black}},
xlabel={convective time, $t$},
ymin=0,
ymax=1,
ylabel style={font=\color{white!15!black}},
ylabel={$\phi(t)$},
axis background/.style={fill=white},
xmajorgrids,
ymajorgrids
]
\addplot [color=black, line width=2.0pt, forget plot]
  table[row sep=crcr]{%
1.99999999992428e-05	0.500002499990936\\
0.105283157894737	0.512824274168592\\
0.210546315789474	0.525012422552404\\
0.31580947368421	0.536619017317854\\
0.421072631578948	0.547689829141911\\
0.526335789473684	0.558265336275237\\
0.63159894736842	0.568381535317096\\
0.736862105263159	0.578070599233259\\
0.842125263157895	0.587361416322425\\
0.947388421052631	0.596280035405069\\
1.05265157894737	0.604850036415135\\
1.15791473684211	0.613092841112536\\
1.26317789473684	0.621027975326282\\
1.47370421052632	0.636045159693241\\
1.68423052631579	0.650027571789252\\
1.89475684210526	0.663082183794989\\
2.00002	0.66929072565013\\
2.42107263157895	0.692230965429422\\
2.84212526315789	0.712535667772793\\
3.26317789473684	0.730633155693887\\
3.68423052631579	0.746860303744159\\
4.10528315789474	0.761486764172949\\
4.52633578947368	0.774731595381725\\
4.94738842105263	0.786774923409453\\
5.36844105263158	0.797766889135886\\
5.78949368421053	0.807833065739317\\
6.21054631578947	0.817080088245799\\
6.63159894736842	0.825598837674677\\
7.05265157894737	0.833467353534777\\
7.47370421052632	0.840753052077465\\
7.89475684210526	0.847514490762167\\
8.31580947368421	0.853802785909501\\
8.73686210526316	0.859662762161394\\
9.15791473684211	0.865133892259388\\
9.57896736842105	0.870251071239858\\
10.00002	0.875045258667948\\
};
\addplot [color=mycolor1, line width=1.0pt, forget plot]
  table[row sep=crcr]{%
1.99999999992428e-05	0.19936694248036\\
0.105283157894737	0.268866502871685\\
0.210546315789474	0.327010572411341\\
0.31580947368421	0.37684818876\\
0.421072631578948	0.419796971267594\\
0.526335789473684	0.456893907760907\\
0.63159894736842	0.488998894239726\\
0.736862105263159	0.516850949193966\\
0.842125263157895	0.541090263569359\\
0.947388421052631	0.562270430439895\\
1.05265157894737	0.580867548130124\\
1.15791473684211	0.597288342936807\\
1.26317789473684	0.611878042987978\\
1.36844105263158	0.624928035904311\\
1.47370421052632	0.636683010780711\\
1.57896736842105	0.647347392498624\\
1.68423052631579	0.657091161926514\\
1.78949368421053	0.666055199877684\\
1.89475684210526	0.674356074479572\\
2.00002	0.682090088364149\\
2.42107263157895	0.708748909222574\\
2.84212526315789	0.730687761932224\\
3.26317789473684	0.749502201934948\\
3.68423052631579	0.766009571236051\\
4.10528315789474	0.780682980961414\\
4.52633578947368	0.793834443905615\\
4.94738842105263	0.805692100717229\\
5.36844105263158	0.81643476786569\\
5.78949368421053	0.826207063764548\\
6.21054631578947	0.835126551089774\\
6.63159894736842	0.843291929409007\\
7.05265157894737	0.850787334825156\\
7.47370421052632	0.85768518709857\\
7.89475684210526	0.864048266308073\\
8.31580947368421	0.869931298276834\\
8.73686210526316	0.875382203671638\\
9.15791473684211	0.880443104947641\\
9.57896736842105	0.885151151996547\\
10.00002	0.889539207854611\\
};
\addplot [color=mycolor2, line width=1.0pt, forget plot]
  table[row sep=crcr]{%
1.99999999992428e-05	0.227231476883285\\
0.105283157894737	0.270519649051815\\
0.210546315789474	0.30959670290439\\
0.31580947368421	0.345688393572564\\
0.421072631578948	0.379207769346861\\
0.526335789473684	0.410400243309946\\
0.63159894736842	0.439445454377356\\
0.736862105263159	0.466490992137329\\
0.842125263157895	0.49166674802593\\
0.947388421052631	0.515092068190411\\
1.05265157894737	0.536879115517527\\
1.15791473684211	0.557134300760833\\
1.26317789473684	0.575959289061576\\
1.36844105263158	0.593451653867222\\
1.47370421052632	0.609704778608885\\
1.57896736842105	0.624807628838656\\
1.68423052631579	0.638844956739463\\
1.78949368421053	0.65189736638817\\
1.89475684210526	0.664040923955392\\
2.00002	0.675347077486355\\
2.42107263157895	0.713475747823956\\
2.84212526315789	0.742812983898435\\
3.26317789473684	0.765990580917592\\
3.68423052631579	0.784842823869795\\
4.10528315789474	0.800609898051984\\
4.52633578947368	0.814116050002735\\
4.94738842105263	0.825907785880373\\
5.36844105263158	0.836351828022551\\
5.78949368421053	0.845701144906597\\
6.21054631578947	0.854136598932529\\
6.63159894736842	0.861792845733071\\
7.05265157894737	0.868774143582232\\
7.47370421052632	0.875164019010416\\
7.89475684210526	0.881031255213371\\
8.31580947368421	0.886433695444071\\
8.73686210526316	0.891420743296967\\
9.15791473684211	0.896035075029907\\
9.57896736842105	0.900313864145089\\
10.00002	0.904289694243893\\
};
\addplot [color=mycolor3, line width=1.0pt, forget plot]
  table[row sep=crcr]{%
1.99999999992428e-05	0.256774854172164\\
0.105283157894737	0.290917365034776\\
0.210546315789474	0.322524272305571\\
0.31580947368421	0.352363969551718\\
0.421072631578948	0.380677071464147\\
0.526335789473684	0.4075999018386\\
0.63159894736842	0.433224116690875\\
0.736862105263159	0.45761790072288\\
0.842125263157895	0.480835938009339\\
0.947388421052631	0.502925075885974\\
1.05265157894737	0.523927581979382\\
1.15791473684211	0.543882860090591\\
1.26317789473684	0.562828495786471\\
1.36844105263158	0.580801020310153\\
1.47370421052632	0.597836372607347\\
1.57896736842105	0.613970230695188\\
1.68423052631579	0.629238347339662\\
1.78949368421053	0.643676602701978\\
1.89475684210526	0.657320678198142\\
2.00002	0.670205854464182\\
2.42107263157895	0.714856083250124\\
2.84212526315789	0.750098478749489\\
3.26317789473684	0.777916377668728\\
3.68423052631579	0.800017751353284\\
4.10528315789474	0.817801128996916\\
4.52633578947368	0.832361196510357\\
4.94738842105263	0.844522037034238\\
5.36844105263158	0.854885338262957\\
5.78949368421053	0.863876632626701\\
6.21054631578947	0.871795791306946\\
6.63159894736842	0.878855010077832\\
7.05265157894737	0.885206802583477\\
7.47370421052632	0.890963392128727\\
7.89475684210526	0.896209705305731\\
8.31580947368421	0.901011946708682\\
8.73686210526316	0.905423228954559\\
9.15791473684211	0.909487275524572\\
10.00002	0.91671552408584\\
};
\addplot [color=mycolor4, line width=1.0pt, forget plot]
  table[row sep=crcr]{%
1.99999999992428e-05	0.286676731143849\\
0.105283157894737	0.31615293160659\\
0.210546315789474	0.34383544188446\\
0.31580947368421	0.37021350253498\\
0.421072631578948	0.395462956537708\\
0.526335789473684	0.419685323786572\\
0.63159894736842	0.442948718684963\\
0.736862105263159	0.465302677120969\\
0.842125263157895	0.48678524159657\\
0.947388421052631	0.507427030205548\\
1.05265157894737	0.527253672531726\\
1.15791473684211	0.546287322127318\\
1.26317789473684	0.564547824520622\\
1.36844105263158	0.582053635962851\\
1.47370421052632	0.59882236228786\\
1.57896736842105	0.614871111952443\\
1.68423052631579	0.630216901427788\\
1.78949368421053	0.644876959333752\\
1.89475684210526	0.658868789183156\\
2.00002	0.672210219812987\\
2.42107263157895	0.719445364248637\\
2.84212526315789	0.757815600633945\\
3.26317789473684	0.788630565376286\\
3.68423052631579	0.813198888739842\\
4.10528315789474	0.83274992923457\\
4.52633578947368	0.848375461044524\\
4.94738842105263	0.860998339640648\\
5.36844105263158	0.871361709193945\\
5.78949368421053	0.880040234835011\\
6.21054631578947	0.887463511908372\\
6.63159894736842	0.893944238307554\\
7.05265157894737	0.899705561281964\\
7.47370421052632	0.904904684806226\\
7.89475684210526	0.90965175064947\\
8.31580947368421	0.914024142231174\\
9.15791473684211	0.921849986605098\\
10.00002	0.928670612284042\\
};
\node[fill=white, left, align=right]
at (axis cs:9.95,0.2) {Varying flow resistance, $\Phi$};
\end{axis}
\end{tikzpicture}%

%% file: images/2wagner.tex
%
%
\definecolor{mycolor1}{rgb}{0.09053,0.13734,0.07326}%
\definecolor{mycolor2}{rgb}{0.04342,0.37461,0.17524}%
\definecolor{mycolor3}{rgb}{0.37697,0.57257,0.04618}%
\definecolor{mycolor4}{rgb}{0.79072,0.73098,0.27064}%
\begin{tikzpicture}[%
trim axis left, trim axis right
]

\begin{axis}[%
width=0.951\fwidth,
height=\fheight,
at={(0\fwidth,0\fheight)},
scale only axis,
xmin=0,
xmax=5,
xlabel style={font=\color{white!15!black}},
xlabel={convective time, $t$},
ymin=0,
ymax=1,
ylabel style={font=\color{white!15!black}},
ylabel={$\phi(t)$},
axis background/.style={fill=white},
xmajorgrids,
ymajorgrids
]
\addplot [color=black, line width=2.0pt, forget plot]
  table[row sep=crcr]{%
9.99999999962142e-06	0.500001250000305\\
0.105273157894737	0.51282308707433\\
0.210536315789474	0.525011293070514\\
0.31579947368421	0.536617940818587\\
0.421062631578947	0.547688801542653\\
0.526325789473685	0.558264353954926\\
0.631588947368421	0.568380595047046\\
0.736852105263158	0.57806969812076\\
0.894746842105263	0.591864875729579\\
1.00001	0.600606412945772\\
1.21053631578947	0.617096960630157\\
1.42106263157895	0.632391736645461\\
1.63158894736842	0.646622189603234\\
1.8421152631579	0.659899657082068\\
2.05264157894737	0.672319336474665\\
2.26316789473684	0.683963303925336\\
2.47369421052632	0.694902846357015\\
2.68422052631579	0.705200288555893\\
2.89474684210526	0.714910442889305\\
3.10527315789474	0.72408177277751\\
3.31579947368421	0.732757336106278\\
3.52632578947368	0.740975557391444\\
3.73685210526316	0.748770865188305\\
3.94737842105263	0.756174222385422\\
4.1579047368421	0.763213570575586\\
4.36843105263158	0.76991420523446\\
4.57895736842105	0.776299092377298\\
5.00001	0.788203456928525\\
};
\addplot [color=mycolor1, line width=1.0pt, forget plot]
  table[row sep=crcr]{%
9.99999999962142e-06	0.341911537818826\\
0.0526415789473687	0.360754545812142\\
0.105273157894737	0.377833975955888\\
0.157904736842105	0.393776001409464\\
0.210536315789474	0.408869319190235\\
0.263167894736842	0.423075391704201\\
0.31579947368421	0.436578758467551\\
0.368431052631579	0.449479301205735\\
0.421062631578947	0.4616912961176\\
0.473694210526316	0.473403299775152\\
0.526325789473685	0.484617408267463\\
0.578957368421053	0.495273189460385\\
0.631588947368421	0.505558474943429\\
0.68422052631579	0.515400302654647\\
0.789483684210526	0.533899137735036\\
0.842115263157894	0.542596322596363\\
0.947378421052631	0.55905382042635\\
1.00001	0.566785580406988\\
1.21053631578947	0.595153853072901\\
1.42106263157895	0.619843673283517\\
1.63158894736842	0.641438637896671\\
1.8421152631579	0.660606470047644\\
2.05264157894737	0.677574953263467\\
2.26316789473684	0.692875983670877\\
2.47369421052632	0.706617591865617\\
2.68422052631579	0.719124160722473\\
2.89474684210526	0.730558114669551\\
3.10527315789474	0.741003231284095\\
3.31579947368421	0.750722764853272\\
3.52632578947368	0.759633097261544\\
3.73685210526316	0.768017414845069\\
3.94737842105263	0.775776643230653\\
4.1579047368421	0.783082495253824\\
4.36843105263158	0.789955401628323\\
4.78948368421053	0.802534595847615\\
5.00001	0.80825910020705\\
};
\addplot [color=mycolor2, line width=1.0pt, forget plot]
  table[row sep=crcr]{%
9.99999999962142e-06	0.299873850719028\\
0.0526415789473687	0.323662817010882\\
0.105273157894737	0.344859066909168\\
0.157904736842105	0.364578929359848\\
0.210536315789474	0.382966109187589\\
0.263167894736842	0.400169378337143\\
0.31579947368421	0.416490054844378\\
0.368431052631579	0.431730899491297\\
0.421062631578947	0.44631779677381\\
0.473694210526316	0.459992182389767\\
0.526325789473685	0.473046273190309\\
0.578957368421053	0.485438532865335\\
0.631588947368421	0.497145755548851\\
0.68422052631579	0.508424471869055\\
0.736852105263158	0.518996012603903\\
0.789483684210526	0.529242841623388\\
0.842115263157894	0.538889984766887\\
0.894746842105263	0.548164477397781\\
0.947378421052631	0.557045409120033\\
1.00001	0.56544436775769\\
1.21053631578947	0.59590830334241\\
1.42106263157895	0.621802461290427\\
1.63158894736842	0.643925890061446\\
1.8421152631579	0.663175349998033\\
2.05264157894737	0.680172785693435\\
2.26316789473684	0.695162004584128\\
2.47369421052632	0.708551644702117\\
2.68422052631579	0.720771709632045\\
2.89474684210526	0.731877429430052\\
3.10527315789474	0.741969537507036\\
3.31579947368421	0.751378470110891\\
3.52632578947368	0.760161541294699\\
3.73685210526316	0.768236195971066\\
3.94737842105263	0.775821327411594\\
4.1579047368421	0.783053590158663\\
4.36843105263158	0.789785091743464\\
4.78948368421053	0.802178567017134\\
5.00001	0.807938813046421\\
};
\addplot [color=mycolor3, line width=1.0pt, forget plot]
  table[row sep=crcr]{%
9.99999999962142e-06	0.239748047821628\\
0.0526415789473687	0.271198531979397\\
0.105273157894737	0.299202888360269\\
0.157904736842105	0.324915599527889\\
0.210536315789474	0.348722852443187\\
0.263167894736842	0.370862597873433\\
0.31579947368421	0.391508004225734\\
0.368431052631579	0.41079663475253\\
0.421062631578947	0.428843698727022\\
0.473694210526316	0.44574900161082\\
0.526325789473685	0.461600955848762\\
0.578957368421053	0.476479066157191\\
0.631588947368421	0.490455561377057\\
0.68422052631579	0.503596521538592\\
0.736852105263158	0.515962693048311\\
0.789483684210526	0.527610103880568\\
0.842115263157894	0.53859054591961\\
0.894746842105263	0.548951965927961\\
0.947378421052631	0.558738791641531\\
1.00001	0.567992210700845\\
1.21053631578947	0.600395163344454\\
1.42106263157895	0.626863480480018\\
1.63158894736842	0.648903683168318\\
1.8421152631579	0.667619829833015\\
2.05264157894737	0.683816485584074\\
2.26316789473684	0.698077371137008\\
2.47369421052632	0.710825811728086\\
2.68422052631579	0.722369676552735\\
2.89474684210526	0.732934036243236\\
3.10527315789474	0.742684399808497\\
3.31579947368421	0.751744209437927\\
3.52632578947368	0.760207411028805\\
3.73685210526316	0.768147057296765\\
3.94737842105263	0.775620630881359\\
4.1579047368421	0.782674349579607\\
4.36843105263158	0.789346533919383\\
4.57895736842105	0.795669917354356\\
4.78948368421053	0.8016724515543\\
5.00001	0.807378154944664\\
};
\addplot [color=mycolor4, line width=1.0pt, forget plot]
  table[row sep=crcr]{%
9.99999999962142e-06	0.14758792997098\\
0.0526415789473687	0.191339372382343\\
0.105273157894737	0.230786460576727\\
0.157904736842105	0.267187865389499\\
0.210536315789474	0.300836211175517\\
0.263167894736842	0.331936220818002\\
0.31579947368421	0.360620454987663\\
0.368431052631579	0.387055520349335\\
0.421062631578947	0.411385422217422\\
0.473694210526316	0.433741936197939\\
0.526325789473685	0.45429173313337\\
0.578957368421053	0.473152773380849\\
0.631588947368421	0.490470220818568\\
0.68422052631579	0.506380416418566\\
0.736852105263158	0.520985951245289\\
0.789483684210526	0.534426616326527\\
0.842115263157894	0.546794419166476\\
0.894746842105263	0.558190466833485\\
0.947378421052631	0.568722932692738\\
1.00001	0.578453286688426\\
1.21053631578947	0.610969845935218\\
1.42106263157895	0.635936128687797\\
1.63158894736842	0.655999336135008\\
1.8421152631579	0.672795828910887\\
2.05264157894737	0.687386639806829\\
2.26316789473684	0.700383967135803\\
2.47369421052632	0.712201716591195\\
2.68422052631579	0.72307744349093\\
2.89474684210526	0.733177080776213\\
3.10527315789474	0.742616620468036\\
3.31579947368421	0.751463092642697\\
3.52632578947368	0.759798958970833\\
3.73685210526316	0.767648468805818\\
3.94737842105263	0.775080274592322\\
4.1579047368421	0.782100620537489\\
4.36843105263158	0.788766530913212\\
4.57895736842105	0.795084186832926\\
4.78948368421053	0.801093072730666\\
5.00001	0.806813773936163\\
};
\node[fill=white, left, align=right]
at (axis cs:4.95,0.2) {Varying effective density, $\rho_e$};
\end{axis}
\end{tikzpicture}%

%% file: images/1dimwagner.tex
%
%
\definecolor{mycolor1}{rgb}{0.18517,0.05913,0.24304}%
\definecolor{mycolor2}{rgb}{0.51514,0.10993,0.38770}%
\definecolor{mycolor3}{rgb}{0.81038,0.26571,0.33825}%
\definecolor{mycolor4}{rgb}{0.94901,0.58547,0.40375}%
\begin{tikzpicture}[%
trim axis left, trim axis right
]

\begin{axis}[%
width=0.951\fwidth,
height=\fheight,
at={(0\fwidth,0\fheight)},
scale only axis,
xmin=0,
xmax=10,
xlabel style={font=\color{white!15!black}},
xlabel={convective time, $t$},
ymin=0,
ymax=1,
ylabel style={font=\color{white!15!black}},
ylabel={$L^C$},
axis background/.style={fill=white},
xmajorgrids,
ymajorgrids
]
\addplot [color=black, line width=2.0pt, forget plot]
  table[row sep=crcr]{%
1.99999999992428e-05	0.500002499990936\\
0.105283157894737	0.512824274168592\\
0.210546315789474	0.525012422552404\\
0.31580947368421	0.536619017317854\\
0.421072631578948	0.547689829141911\\
0.526335789473684	0.558265336275237\\
0.63159894736842	0.568381535317096\\
0.736862105263159	0.578070599233259\\
0.842125263157895	0.587361416322425\\
0.947388421052631	0.596280035405069\\
1.05265157894737	0.604850036415135\\
1.15791473684211	0.613092841112536\\
1.26317789473684	0.621027975326282\\
1.47370421052632	0.636045159693241\\
1.68423052631579	0.650027571789252\\
1.89475684210526	0.663082183794989\\
2.00002	0.66929072565013\\
2.42107263157895	0.692230965429422\\
2.84212526315789	0.712535667772793\\
3.26317789473684	0.730633155693887\\
3.68423052631579	0.746860303744159\\
4.10528315789474	0.761486764172949\\
4.52633578947368	0.774731595381725\\
4.94738842105263	0.786774923409453\\
5.36844105263158	0.797766889135886\\
5.78949368421053	0.807833065739317\\
6.21054631578947	0.817080088245799\\
6.63159894736842	0.825598837674677\\
7.05265157894737	0.833467353534777\\
7.47370421052632	0.840753052077465\\
7.89475684210526	0.847514490762167\\
8.31580947368421	0.853802785909501\\
8.73686210526316	0.859662762161394\\
9.15791473684211	0.865133892259388\\
9.57896736842105	0.870251071239858\\
10.00002	0.875045258667948\\
};
\addplot [color=mycolor1, line width=1.0pt, forget plot]
  table[row sep=crcr]{%
1.99999999992428e-05	0.174616646505569\\
0.105283157894737	0.235488223398715\\
0.210546315789474	0.286414030410082\\
0.31580947368421	0.330064584149659\\
0.421072631578948	0.367681514417384\\
0.526335789473684	0.400173072774558\\
0.63159894736842	0.428292403919894\\
0.736862105263159	0.45268678131211\\
0.842125263157895	0.473916919755158\\
0.947388421052631	0.492467686824908\\
1.05265157894737	0.508756075889496\\
1.15791473684211	0.523138320440301\\
1.26317789473684	0.535916790455252\\
1.36844105263158	0.547346699404155\\
1.47370421052632	0.557642359593036\\
1.57896736842105	0.566982817692406\\
1.68423052631579	0.575516952392242\\
1.78949368421053	0.583368154937197\\
1.89475684210526	0.590638522170661\\
2.00002	0.597412401288986\\
2.42107263157895	0.620761677955301\\
2.84212526315789	0.639976944241063\\
3.26317789473684	0.656455676262947\\
3.68423052631579	0.670913747566676\\
4.10528315789474	0.683765535165804\\
4.52633578947368	0.695284317203527\\
4.94738842105263	0.705669911937028\\
5.36844105263158	0.715078936766581\\
5.78949368421053	0.723638056534828\\
6.21054631578947	0.731450239166024\\
6.63159894736842	0.738601931225977\\
7.05265157894737	0.745166823788828\\
7.47370421052632	0.75120834610486\\
7.89475684210526	0.75678148445563\\
8.31580947368421	0.76193417075803\\
8.73686210526316	0.766708376594856\\
9.15791473684211	0.771140995152964\\
10.00002	0.779107867524699\\
};
\addplot [color=mycolor2, line width=1.0pt, forget plot]
  table[row sep=crcr]{%
1.99999999992428e-05	0.174348115486731\\
0.105283157894737	0.20756187329866\\
0.210546315789474	0.237544562279155\\
0.31580947368421	0.265236668755927\\
0.421072631578948	0.290955112691149\\
0.526335789473684	0.314888192418595\\
0.63159894736842	0.337173739663083\\
0.736862105263159	0.357924995631015\\
0.842125263157895	0.377241622250425\\
0.947388421052631	0.395215190355378\\
1.05265157894737	0.411931759272694\\
1.15791473684211	0.427472974884381\\
1.26317789473684	0.44191684190189\\
1.36844105263158	0.455338225598821\\
1.47370421052632	0.467808776370864\\
1.57896736842105	0.479396754903396\\
1.68423052631579	0.490167189085946\\
1.78949368421053	0.500181924086686\\
1.89475684210526	0.509499323270063\\
2.00002	0.518174206646979\\
2.42107263157895	0.547429228488825\\
2.84212526315789	0.5699388380435\\
3.26317789473684	0.587722335370662\\
3.68423052631579	0.602187114091011\\
4.10528315789474	0.614284732379238\\
4.52633578947368	0.624647610675806\\
4.94738842105263	0.633695067290452\\
5.36844105263158	0.641708477626592\\
5.78949368421053	0.648881937052984\\
6.21054631578947	0.655354216038575\\
6.63159894736842	0.661228631941182\\
7.05265157894737	0.666585179107742\\
7.47370421052632	0.671487945020191\\
7.89475684210526	0.675989705027789\\
8.31580947368421	0.680134840579305\\
8.73686210526316	0.683961257618531\\
9.57896736842105	0.690784691070323\\
10.00002	0.693835230083359\\
};
\addplot [color=mycolor3, line width=1.0pt, forget plot]
  table[row sep=crcr]{%
1.99999999992428e-05	0.174321738619909\\
0.105283157894737	0.197500728920996\\
0.210546315789474	0.218958324703133\\
0.31580947368421	0.239216180249814\\
0.421072631578948	0.258437646335384\\
0.526335789473684	0.276715271746074\\
0.63159894736842	0.294111280783717\\
0.736862105263159	0.310671963322852\\
0.842125263157895	0.326434443804697\\
0.947388421052631	0.341430526391072\\
1.05265157894737	0.355688906127599\\
1.15791473684211	0.369236333838948\\
1.26317789473684	0.38209832596981\\
1.36844105263158	0.394299683195609\\
1.47370421052632	0.405864804087308\\
1.57896736842105	0.416817910074212\\
1.68423052631579	0.427183273331828\\
1.78949368421053	0.436985252522945\\
1.89475684210526	0.446248071384318\\
2.00002	0.454995681567416\\
2.42107263157895	0.485308250673283\\
2.84212526315789	0.50923394104661\\
3.26317789473684	0.528119219046236\\
3.68423052631579	0.543123608393477\\
4.10528315789474	0.555196555798574\\
4.52633578947368	0.565081231973542\\
4.94738842105263	0.573337098265577\\
5.36844105263158	0.580372634100488\\
5.78949368421053	0.586476728954263\\
6.21054631578947	0.591852962207312\\
6.63159894736842	0.596645391331288\\
7.05265157894737	0.600957556229625\\
7.47370421052632	0.604865644119634\\
7.89475684210526	0.608427310768452\\
8.73686210526316	0.61468227473857\\
9.57896736842105	0.619989591574864\\
10.00002	0.622348494729778\\
};
\addplot [color=mycolor4, line width=1.0pt, forget plot]
  table[row sep=crcr]{%
1.99999999992428e-05	0.174363710560497\\
0.105283157894737	0.192291847474161\\
0.210546315789474	0.209129018703297\\
0.31580947368421	0.225172792169191\\
0.421072631578948	0.240530119818267\\
0.526335789473684	0.25526274850152\\
0.63159894736842	0.269412107043919\\
0.736862105263159	0.283008324368797\\
0.842125263157895	0.296074538844506\\
0.947388421052631	0.308629373135053\\
1.05265157894737	0.320688415772224\\
1.15791473684211	0.332265141081438\\
1.26317789473684	0.343371619592236\\
1.36844105263158	0.354019076842292\\
1.47370421052632	0.364218221124897\\
1.57896736842105	0.373979458216613\\
1.68423052631579	0.38331313794613\\
1.78949368421053	0.392229739176081\\
1.89475684210526	0.40073990796564\\
2.00002	0.408854488243994\\
2.42107263157895	0.437584043725524\\
2.84212526315789	0.460921748060755\\
3.26317789473684	0.479664153737794\\
3.68423052631579	0.494607201284163\\
4.10528315789474	0.506498616232202\\
4.52633578947368	0.516002442004854\\
4.94738842105263	0.52367998158469\\
5.36844105263158	0.529983233201982\\
5.78949368421053	0.535261722066192\\
6.21054631578947	0.539776738439741\\
6.63159894736842	0.543718472732509\\
7.05265157894737	0.547222648490155\\
7.89475684210526	0.55327216104437\\
8.73686210526316	0.558396532804416\\
9.57896736842105	0.562834470798661\\
10.00002	0.564839891958592\\
};
\node[fill=white, left, align=right]
at (axis cs:9.95,0.2) {Varying flow resistance, $\Phi$};
\end{axis}
\end{tikzpicture}%

%% file: images/2dimwagner.tex
%
%
\definecolor{mycolor1}{rgb}{0.09053,0.13734,0.07326}%
\definecolor{mycolor2}{rgb}{0.04342,0.37461,0.17524}%
\definecolor{mycolor3}{rgb}{0.37697,0.57257,0.04618}%
\definecolor{mycolor4}{rgb}{0.79072,0.73098,0.27064}%
\begin{tikzpicture}[%
trim axis left, trim axis right
]

\begin{axis}[%
width=0.951\fwidth,
height=\fheight,
at={(0\fwidth,0\fheight)},
scale only axis,
xmin=0,
xmax=5,
xlabel style={font=\color{white!15!black}},
xlabel={convective time, $t$},
ymin=0,
ymax=1,
ylabel style={font=\color{white!15!black}},
ylabel={$L^C$},
axis background/.style={fill=white},
xmajorgrids,
ymajorgrids
]
\addplot [color=black, line width=2.0pt, forget plot]
  table[row sep=crcr]{%
9.99999999962142e-06	0.500001250000305\\
0.105273157894737	0.51282308707433\\
0.210536315789474	0.525011293070514\\
0.31579947368421	0.536617940818587\\
0.421062631578947	0.547688801542653\\
0.526325789473685	0.558264353954926\\
0.631588947368421	0.568380595047046\\
0.736852105263158	0.57806969812076\\
0.894746842105263	0.591864875729579\\
1.00001	0.600606412945772\\
1.21053631578947	0.617096960630157\\
1.42106263157895	0.632391736645461\\
1.63158894736842	0.646622189603234\\
1.8421152631579	0.659899657082068\\
2.05264157894737	0.672319336474665\\
2.26316789473684	0.683963303925336\\
2.47369421052632	0.694902846357015\\
2.68422052631579	0.705200288555893\\
2.89474684210526	0.714910442889305\\
3.10527315789474	0.72408177277751\\
3.31579947368421	0.732757336106278\\
3.52632578947368	0.740975557391444\\
3.73685210526316	0.748770865188305\\
3.94737842105263	0.756174222385422\\
4.1579047368421	0.763213570575586\\
4.36843105263158	0.76991420523446\\
4.57895736842105	0.776299092377298\\
5.00001	0.788203456928525\\
};
\addplot [color=mycolor1, line width=1.0pt, forget plot]
  table[row sep=crcr]{%
9.99999999962142e-06	0.299465120215987\\
0.0526415789473687	0.315968873467331\\
0.105273157894737	0.33092798725988\\
0.157904736842105	0.344890898834602\\
0.210536315789474	0.358110465078289\\
0.263167894736842	0.370552932625102\\
0.31579947368421	0.382379931435683\\
0.368431052631579	0.393678943474256\\
0.421062631578947	0.404374887073259\\
0.473694210526316	0.414632910553992\\
0.526325789473685	0.424454849787714\\
0.578957368421053	0.433787774953943\\
0.631588947368421	0.442796199394032\\
0.68422052631579	0.451416218880612\\
0.789483684210526	0.467618526373786\\
0.842115263157894	0.475236004060128\\
0.947378421052631	0.489650394979932\\
1.00001	0.496422299920178\\
1.21053631578947	0.521268809161757\\
1.42106263157895	0.542893525381193\\
1.63158894736842	0.561807595129156\\
1.8421152631579	0.578595847423854\\
2.05264157894737	0.593457787733111\\
2.26316789473684	0.606859280234985\\
2.47369421052632	0.618894944127027\\
2.68422052631579	0.62984889195254\\
2.89474684210526	0.639863384049382\\
3.10527315789474	0.64901179747409\\
3.31579947368421	0.657524704956841\\
3.52632578947368	0.665328869106521\\
3.73685210526316	0.672672320249167\\
3.94737842105263	0.679468283544513\\
4.1579047368421	0.685867154633673\\
4.36843105263158	0.6918868278708\\
4.78948368421053	0.70290438502354\\
5.00001	0.707918223974724\\
};
\addplot [color=mycolor2, line width=1.0pt, forget plot]
  table[row sep=crcr]{%
9.99999999962142e-06	0.262646180042079\\
0.0526415789473687	0.283481878482353\\
0.105273157894737	0.302046731848705\\
0.157904736842105	0.319318483057447\\
0.210536315789474	0.335422996778544\\
0.263167894736842	0.350490575747274\\
0.31579947368421	0.364785131041274\\
0.368431052631579	0.378133909594423\\
0.421062631578947	0.390909924711218\\
0.473694210526316	0.402886711409498\\
0.526325789473685	0.414320209443634\\
0.578957368421053	0.425174039005402\\
0.631588947368421	0.435427875107997\\
0.68422052631579	0.445306401528947\\
0.789483684210526	0.463540286469459\\
0.842115263157894	0.471989790448829\\
0.894746842105263	0.480112906404061\\
0.947378421052631	0.487891319848486\\
1.00001	0.495247594486046\\
1.21053631578947	0.521929601907464\\
1.42106263157895	0.544609144168805\\
1.63158894736842	0.563986072307783\\
1.8421152631579	0.58084581885819\\
2.05264157894737	0.595733117451261\\
2.26316789473684	0.608861508186283\\
2.47369421052632	0.620588898956421\\
2.68422052631579	0.631291910792952\\
2.89474684210526	0.641018917247735\\
3.10527315789474	0.649858146238978\\
3.31579947368421	0.65809901219228\\
3.52632578947368	0.665791713939809\\
3.73685210526316	0.672863945149105\\
3.94737842105263	0.679507424709577\\
4.1579047368421	0.685841842262236\\
4.36843105263158	0.691737665365705\\
4.78948368421053	0.702592559616305\\
5.00001	0.707637703139381\\
};
\addplot [color=mycolor3, line width=1.0pt, forget plot]
  table[row sep=crcr]{%
9.99999999962142e-06	0.209984662101279\\
0.0526415789473687	0.237530743701512\\
0.105273157894737	0.2620585151075\\
0.157904736842105	0.284579136298366\\
0.210536315789474	0.305430851273321\\
0.263167894736842	0.324822070536293\\
0.31579947368421	0.342904464600486\\
0.368431052631579	0.359798519006213\\
0.421062631578947	0.375605140436654\\
0.473694210526316	0.39041174406087\\
0.526325789473685	0.404295766410764\\
0.578957368421053	0.417326842134666\\
0.631588947368421	0.429568233264937\\
0.68422052631579	0.441077816363855\\
0.736852105263158	0.451908796509674\\
0.789483684210526	0.462110246115578\\
0.842115263157894	0.471727527391654\\
0.894746842105263	0.480802634776729\\
0.947378421052631	0.489374480550576\\
1.00001	0.497479139137399\\
1.21053631578947	0.52585944556231\\
1.42106263157895	0.549041868445836\\
1.63158894736842	0.568345902644229\\
1.8421152631579	0.584738543872941\\
2.05264157894737	0.598924474961678\\
2.26316789473684	0.61141495094806\\
2.47369421052632	0.622580743596464\\
2.68422052631579	0.632691501855276\\
2.89474684210526	0.641944355090505\\
3.10527315789474	0.650484265288811\\
3.31579947368421	0.658419349978857\\
3.52632578947368	0.665831892197669\\
3.73685210526316	0.672785875572839\\
3.94737842105263	0.679331646594164\\
4.1579047368421	0.685509685376408\\
4.36843105263158	0.691353555167197\\
4.78948368421053	0.70214927873272\\
5.00001	0.707146650804617\\
};
\addplot [color=mycolor4, line width=1.0pt, forget plot]
  table[row sep=crcr]{%
9.99999999962142e-06	0.129265710275828\\
0.0526415789473687	0.16758565473205\\
0.105273157894737	0.202135606579487\\
0.157904736842105	0.234017979677926\\
0.210536315789474	0.26348907069798\\
0.263167894736842	0.290728187316876\\
0.31579947368421	0.315851433536187\\
0.368431052631579	0.339004732731023\\
0.421062631578947	0.360314212757864\\
0.473694210526316	0.379895290015003\\
0.526325789473685	0.3978939440879\\
0.578957368421053	0.414413491212184\\
0.631588947368421	0.429581073978887\\
0.68422052631579	0.44351610738755\\
0.736852105263158	0.456308446393226\\
0.789483684210526	0.468080527745999\\
0.842115263157894	0.478912936730743\\
0.894746842105263	0.488894228536264\\
0.947378421052631	0.498119147406797\\
1.00001	0.506641532135244\\
1.21053631578947	0.535121341612816\\
1.42106263157895	0.556988199380886\\
1.63158894736842	0.574560671341058\\
1.8421152631579	0.589271973066367\\
2.05264157894737	0.60205141603528\\
2.26316789473684	0.613435197549105\\
2.47369421052632	0.623785839214134\\
2.68422052631579	0.633311405178348\\
2.89474684210526	0.642157228732258\\
3.10527315789474	0.650424902133331\\
3.31579947368421	0.658173134047131\\
3.52632578947368	0.665474148987058\\
3.73685210526316	0.672349185884288\\
3.94737842105263	0.678858374364765\\
4.1579047368421	0.685007183452083\\
4.36843105263158	0.690845558172307\\
4.78948368421053	0.701641828460746\\
5.00001	0.70665233646604\\
};
\node[fill=white, left, align=right]
at (axis cs:4.95,0.2) {Varying effective density, $\rho_e$};
\end{axis}
\end{tikzpicture}%

%% file: images/1kussner.tex
%
%
\definecolor{mycolor1}{rgb}{0.18517,0.05913,0.24304}%
\definecolor{mycolor2}{rgb}{0.51514,0.10993,0.38770}%
\definecolor{mycolor3}{rgb}{0.81038,0.26571,0.33825}%
\definecolor{mycolor4}{rgb}{0.94901,0.58547,0.40375}%
\begin{tikzpicture}[%
trim axis left, trim axis right
]

\begin{axis}[%
width=0.988\fwidth,
height=\fheight,
at={(0\fwidth,0\fheight)},
scale only axis,
xmin=0,
xmax=10,
xlabel style={font=\color{white!15!black}},
xlabel={convective time, $t$},
ymin=0,
ymax=1,
ylabel style={font=\color{white!15!black}},
ylabel={$\psi(t)$},
axis background/.style={fill=white},
xmajorgrids,
ymajorgrids
]
\addplot [color=black, line width=2.0pt, forget plot]
  table[row sep=crcr]{%
9.99999999962142e-06	0.00314607853641213\\
0.03449275862069	0.0833651113167555\\
0.0689755172413786	0.117552133702517\\
0.103458275862069	0.143560602275095\\
0.137941034482759	0.165301146947687\\
0.172423793103448	0.18429389722662\\
0.206906551724138	0.201321132597412\\
0.241389310344827	0.21684914339829\\
0.275872068965517	0.231182780139571\\
0.310354827586208	0.244534518856108\\
0.344837586206896	0.257059595186618\\
0.379320344827587	0.268875604955293\\
0.448285862068966	0.290728615730282\\
0.517251379310345	0.310632502432119\\
0.586216896551724	0.328953351736068\\
0.655182413793103	0.34595375511257\\
0.724147931034484	0.361829951783418\\
0.793113448275863	0.376733364923703\\
0.862078965517242	0.39078387707684\\
0.965527241379311	0.410467684821565\\
1.00001	0.416696766544232\\
1.31035482758621	0.466680447312608\\
1.62069965517241	0.508215388432095\\
1.93104448275862	0.543622826831616\\
2.24138931034483	0.574344658232166\\
2.55173413793104	0.601351892635527\\
2.86207896551724	0.625336120283706\\
3.17242379310345	0.646810610125096\\
3.48276862068966	0.666168472448758\\
3.79311344827586	0.683718365627804\\
4.10345827586207	0.699707562999722\\
4.41380310344828	0.714337479027623\\
4.72414793103448	0.727774479127088\\
5.03449275862069	0.740157621816572\\
5.3448375862069	0.751604338571896\\
5.6551824137931	0.762214687622151\\
5.96552724137931	0.772074597325814\\
6.27587206896552	0.781258378249628\\
6.58621689655172	0.789830695980569\\
7.20690655172414	0.805360482726849\\
7.82759620689655	0.819040840286158\\
8.44828586206896	0.831168220466473\\
9.06897551724138	0.841979369564877\\
9.68966517241379	0.851665818217777\\
10.00001	0.856137386564933\\
};
\addplot [color=mycolor1, line width=1.0pt, forget plot]
  table[row sep=crcr]{%
9.99999999962142e-06	0.0153696893895834\\
0.03449275862069	0.0924272395772299\\
0.0689755172413786	0.130767542380431\\
0.103458275862069	0.15993707544142\\
0.137941034482759	0.18432967000906\\
0.172423793103448	0.205651477243345\\
0.206906551724138	0.2247667525886\\
0.241389310344827	0.242183351389286\\
0.275872068965517	0.258233809501263\\
0.310354827586208	0.273155774491993\\
0.344837586206896	0.287128985043021\\
0.379320344827587	0.300291363489993\\
0.413803103448275	0.312746923627875\\
0.448285862068966	0.324573069094374\\
0.517251379310345	0.346568962129259\\
0.586216896551724	0.366690196568808\\
0.655182413793103	0.385282511509148\\
0.724147931034484	0.402557725158635\\
0.793113448275863	0.418647667500577\\
0.862078965517242	0.433720991694287\\
0.931044482758621	0.44792583801768\\
1.00001	0.461321594232082\\
1.31035482758621	0.513561338681955\\
1.62069965517241	0.555895473236673\\
1.93104448275862	0.59125103355753\\
2.24138931034483	0.621382213748474\\
2.55173413793104	0.647489115963443\\
2.86207896551724	0.670485171312954\\
3.17242379310345	0.690900695912282\\
3.48276862068966	0.709186885632766\\
3.79311344827586	0.725698449175606\\
4.10345827586207	0.740677022946924\\
4.41380310344828	0.754347050211933\\
4.72414793103448	0.76680488732816\\
5.03449275862069	0.778249378467017\\
5.3448375862069	0.788811443642663\\
5.6551824137931	0.798506967552671\\
5.96552724137931	0.807477946265408\\
6.27587206896552	0.815794960384169\\
6.58621689655172	0.823526518700605\\
7.20690655172414	0.837377627950262\\
7.82759620689655	0.849523432434109\\
8.44828586206896	0.860165856578021\\
9.06897551724138	0.869617270040743\\
9.68966517241379	0.877994793835887\\
10.00001	0.881889596109822\\
};
\addplot [color=mycolor2, line width=1.0pt, forget plot]
  table[row sep=crcr]{%
9.99999999962142e-06	0.0138406704618159\\
0.03449275862069	0.0984011111609817\\
0.0689755172413786	0.139258147173322\\
0.103458275862069	0.170424754843035\\
0.137941034482759	0.19653047002007\\
0.172423793103448	0.219361083313377\\
0.206906551724138	0.239832121811792\\
0.241389310344827	0.258489810786283\\
0.275872068965517	0.275692489410266\\
0.310354827586208	0.291691003173298\\
0.344837586206896	0.306669411328459\\
0.379320344827587	0.320767679542719\\
0.413803103448275	0.334095298713034\\
0.448285862068966	0.346739948930921\\
0.517251379310345	0.370254789942578\\
0.586216896551724	0.391756028184162\\
0.655182413793103	0.411565342223936\\
0.724147931034484	0.429924291653293\\
0.793113448275863	0.447019476091965\\
0.862078965517242	0.462999995075069\\
0.931044482758621	0.477988796320506\\
1.00001	0.492088453824801\\
1.31035482758621	0.546564747741346\\
1.62069965517241	0.589946512460525\\
1.93104448275862	0.625443559007223\\
2.24138931034483	0.655108983490177\\
2.55173413793104	0.680354692365981\\
2.86207896551724	0.702192922485928\\
3.17242379310345	0.721361228325332\\
3.48276862068966	0.738385256153354\\
3.79311344827586	0.753660156954396\\
4.10345827586207	0.767477489833224\\
4.41380310344828	0.780052993026391\\
4.72414793103448	0.791551494672179\\
5.03449275862069	0.802111444080436\\
5.3448375862069	0.811829918246922\\
5.6551824137931	0.820792467502313\\
5.96552724137931	0.829067210463132\\
6.27587206896552	0.836722487567139\\
6.89656172413793	0.850395681130587\\
7.51725137931034	0.86219430743461\\
8.13794103448276	0.872461703280017\\
8.75863068965517	0.881452253011183\\
9.37932034482759	0.88940348137662\\
10.00001	0.896466318930592\\
};
\addplot [color=mycolor3, line width=1.0pt, forget plot]
  table[row sep=crcr]{%
9.99999999962142e-06	0.0194658909541801\\
0.03449275862069	0.104786189749841\\
0.0689755172413786	0.148544535995649\\
0.103458275862069	0.18198963014002\\
0.137941034482759	0.210003366336728\\
0.172423793103448	0.234512180928011\\
0.206906551724138	0.256502205936421\\
0.241389310344827	0.276556003583179\\
0.275872068965517	0.295051767371662\\
0.310354827586208	0.31225314019442\\
0.344837586206896	0.328354917108742\\
0.379320344827587	0.34350768975756\\
0.413803103448275	0.357831233709938\\
0.448285862068966	0.371421803423424\\
0.482768620689654	0.384356579036256\\
0.551734137931035	0.408494033021297\\
0.620699655172414	0.430619958391219\\
0.689665172413793	0.451028346559642\\
0.758630689655172	0.469969351452772\\
0.827596206896551	0.487628534410383\\
0.89656172413793	0.504128126837871\\
0.965527241379311	0.519575638824062\\
1.00001	0.526942449631749\\
1.31035482758621	0.584392644817497\\
1.62069965517241	0.629432169579349\\
1.93104448275862	0.665546311077605\\
2.24138931034483	0.695048034367792\\
2.55173413793104	0.719531255435493\\
2.86207896551724	0.740191474312715\\
3.17242379310345	0.757883579446785\\
3.48276862068966	0.773284118054477\\
3.79311344827586	0.786880231570651\\
4.10345827586207	0.799021786563218\\
4.41380310344828	0.810008738436935\\
4.72414793103448	0.820050824013476\\
5.03449275862069	0.829269553258831\\
5.3448375862069	0.837765812707136\\
5.6551824137931	0.845611474515959\\
6.27587206896552	0.859647334195518\\
6.89656172413793	0.87177444737091\\
7.51725137931034	0.882245153667785\\
8.13794103448276	0.891305640701489\\
8.75863068965517	0.89919931472382\\
9.37932034482759	0.906111206503292\\
10.00001	0.912200325433018\\
};
\addplot [color=mycolor4, line width=1.0pt, forget plot]
  table[row sep=crcr]{%
9.99999999962142e-06	0.0229016485910361\\
0.03449275862069	0.111906111657056\\
0.0689755172413786	0.158814191467401\\
0.103458275862069	0.194799279511614\\
0.137941034482759	0.224956752988316\\
0.172423793103448	0.251350325476382\\
0.206906551724138	0.275059018321333\\
0.241389310344827	0.296712516126755\\
0.275872068965517	0.316697984699276\\
0.310354827586208	0.335277939702134\\
0.344837586206896	0.352661833747973\\
0.379320344827587	0.369029793040736\\
0.413803103448275	0.384522433999985\\
0.448285862068966	0.399231212499204\\
0.482768620689654	0.413213927661742\\
0.517251379310345	0.426526343290968\\
0.586216896551724	0.451427152805984\\
0.655182413793103	0.474393274327239\\
0.724147931034484	0.495577344706879\\
0.793113448275863	0.515239250386161\\
0.862078965517242	0.533604268240037\\
0.931044482758621	0.550705693297132\\
1.00001	0.56675856064647\\
1.31035482758621	0.627893285757388\\
1.62069965517241	0.675120567108452\\
1.93104448275862	0.712200537290881\\
2.24138931034483	0.741754162314573\\
2.55173413793104	0.765489539793959\\
2.86207896551724	0.784842932163253\\
3.17242379310345	0.800841167287254\\
3.48276862068966	0.814221722456837\\
3.79311344827586	0.825710799206542\\
4.10345827586207	0.835643024555782\\
4.41380310344828	0.844438972030815\\
4.72414793103448	0.852389924601209\\
5.3448375862069	0.866283208399247\\
5.96552724137931	0.878261270339232\\
6.58621689655172	0.888786104617953\\
7.20690655172414	0.898077476077457\\
7.82759620689655	0.906201110714036\\
8.44828586206896	0.913249625444044\\
9.06897551724138	0.919410220802529\\
9.68966517241379	0.924829867808263\\
10.00001	0.927271141713979\\
};
\node[fill=white, left, align=right]
at (axis cs:9.95,0.2) {Varying flow resistance, $\Phi$};
\end{axis}
\end{tikzpicture}%

%% file: images/2kussner.tex
%
%
\definecolor{mycolor1}{rgb}{0.09053,0.13734,0.07326}%
\definecolor{mycolor2}{rgb}{0.04342,0.37461,0.17524}%
\definecolor{mycolor3}{rgb}{0.37697,0.57257,0.04618}%
\definecolor{mycolor4}{rgb}{0.79072,0.73098,0.27064}%
\begin{tikzpicture}[%
trim axis left, trim axis right
]

\begin{axis}[%
width=0.988\fwidth,
height=\fheight,
at={(0\fwidth,0\fheight)},
scale only axis,
xmin=0,
xmax=5,
xlabel style={font=\color{white!15!black}},
xlabel={convective time, $t$},
ymin=0,
ymax=1,
ylabel style={font=\color{white!15!black}},
ylabel={$\psi(t)$},
axis background/.style={fill=white},
xmajorgrids,
ymajorgrids
]
\addplot [color=black, line width=2.0pt, forget plot]
  table[row sep=crcr]{%
9.99999999962142e-06	0.00314607853641125\\
0.03449275862069	0.0833651113167546\\
0.0689755172413795	0.117552133702516\\
0.103458275862069	0.143560602275094\\
0.137941034482759	0.165301146947687\\
0.172423793103448	0.18429389722662\\
0.206906551724138	0.201321132597412\\
0.241389310344828	0.21684914339829\\
0.275872068965517	0.231182780139571\\
0.310354827586207	0.244534518856107\\
0.344837586206896	0.257059595186619\\
0.379320344827586	0.268875604955293\\
0.448285862068966	0.290728615730283\\
0.517251379310345	0.310632502432119\\
0.586216896551724	0.328953351736067\\
0.655182413793104	0.345953755112569\\
0.724147931034483	0.361829951783418\\
0.793113448275862	0.376733364923704\\
0.862078965517242	0.390783877076841\\
0.96552724137931	0.410467684821564\\
1.00001	0.416696766544232\\
1.31035482758621	0.466680447312608\\
1.62069965517241	0.508215388432095\\
1.93104448275862	0.543622826831617\\
2.24138931034483	0.574344658232167\\
2.55173413793104	0.601351892635527\\
2.86207896551724	0.625336120283706\\
3.17242379310345	0.646810610125096\\
3.48276862068966	0.666168472448758\\
3.79311344827586	0.683718365627804\\
4.10345827586207	0.699707562999722\\
4.41380310344828	0.714337479027624\\
4.72414793103448	0.727774479127087\\
5.03449275862069	0.740157621816572\\
};
\addplot [color=mycolor1, line width=1.0pt, forget plot]
  table[row sep=crcr]{%
9.99999999962142e-06	0.0290365407904307\\
0.03449275862069	0.12657696544971\\
0.0689755172413795	0.17866067181919\\
0.103458275862069	0.218442338931115\\
0.137941034482759	0.251610211337987\\
0.172423793103448	0.280521616181899\\
0.206906551724138	0.306407068338346\\
0.241389310344828	0.329968017350617\\
0.275872068965517	0.351642365673277\\
0.310354827586207	0.371757018891786\\
0.344837586206896	0.390582475686088\\
0.379320344827586	0.408318343517817\\
0.413803103448276	0.425084185215769\\
0.448285862068966	0.440959228160009\\
0.482768620689655	0.456034729397849\\
0.517251379310345	0.470418555902528\\
0.551734137931034	0.484193707427439\\
0.620699655172414	0.510033247584768\\
0.689665172413793	0.533823217732766\\
0.758630689655172	0.555970620406643\\
0.827596206896552	0.57654762943958\\
0.896561724137931	0.595808786598806\\
0.96552724137931	0.613886726569715\\
1.00001	0.622475822529202\\
1.31035482758621	0.689598785424094\\
1.62069965517241	0.742027046730098\\
1.93104448275862	0.783648883869424\\
2.24138931034483	0.816953436722026\\
2.55173413793104	0.843743560958525\\
2.86207896551724	0.865281737289269\\
3.17242379310345	0.882486359873792\\
3.48276862068966	0.896196929121015\\
3.79311344827586	0.907037786439195\\
4.10345827586207	0.915543074765468\\
4.41380310344828	0.922153569238137\\
4.72414793103448	0.927172406829108\\
5.03449275862069	0.930898199152684\\
};
\addplot [color=mycolor2, line width=1.0pt, forget plot]
  table[row sep=crcr]{%
9.99999999962142e-06	0.0266051147577508\\
0.03449275862069	0.122552479975768\\
0.0689755172413795	0.173260982047551\\
0.103458275862069	0.211965585257464\\
0.137941034482759	0.244243078602467\\
0.172423793103448	0.272385555334959\\
0.206906551724138	0.297602427669039\\
0.241389310344828	0.320593162882925\\
0.275872068965517	0.341775576323792\\
0.310354827586207	0.361429962791044\\
0.344837586206896	0.379790181868379\\
0.379320344827586	0.397066637684561\\
0.413803103448276	0.41342077366697\\
0.448285862068966	0.42894600920935\\
0.482768620689655	0.443692290077704\\
0.517251379310345	0.457715775226576\\
0.586216896551724	0.483946046529157\\
0.655182413793104	0.508171445856565\\
0.724147931034483	0.53049022278271\\
0.793113448275862	0.551226834118285\\
0.862078965517242	0.570630895963104\\
0.931044482758621	0.588682081030511\\
1.00001	0.605672712225502\\
1.31035482758621	0.670533956550456\\
1.62069965517241	0.720943197540867\\
1.93104448275862	0.760621121110082\\
2.24138931034483	0.79223787970805\\
2.55173413793104	0.817421050092152\\
2.86207896551724	0.837519604669965\\
3.17242379310345	0.85363143294174\\
3.48276862068966	0.866413278187664\\
3.79311344827586	0.876670521012599\\
4.10345827586207	0.884832954871226\\
4.41380310344828	0.891293069837574\\
4.72414793103448	0.896553158815517\\
5.03449275862069	0.900701941626902\\
};
\addplot [color=mycolor3, line width=1.0pt, forget plot]
  table[row sep=crcr]{%
9.99999999962142e-06	0.0271512525835682\\
0.03449275862069	0.116622247421255\\
0.0689755172413795	0.16459754823796\\
0.103458275862069	0.201957746133331\\
0.137941034482759	0.233290891976928\\
0.172423793103448	0.260148636237948\\
0.206906551724138	0.284198873978639\\
0.241389310344828	0.306643804203105\\
0.275872068965517	0.327337369930793\\
0.310354827586207	0.346039886381942\\
0.379320344827586	0.380578291697725\\
0.413803103448276	0.396657891570842\\
0.448285862068966	0.411362567149041\\
0.517251379310345	0.439257331695662\\
0.551734137931034	0.452502423817496\\
0.620699655172414	0.47629076813273\\
0.689665172413793	0.499293555769477\\
0.793113448275862	0.529716792171956\\
0.827596206896552	0.539477521038356\\
0.931044482758621	0.565878824457526\\
0.96552724137931	0.574425088566927\\
1.00001	0.582155047685591\\
1.31035482758621	0.644263831593683\\
1.62069965517241	0.692754294861433\\
1.93104448275862	0.730787310911081\\
2.24138931034483	0.760399969112073\\
2.55173413793104	0.784149696440211\\
2.86207896551724	0.803770771926744\\
3.17242379310345	0.819502454419737\\
3.48276862068966	0.831887511646464\\
3.79311344827586	0.842336818245081\\
4.10345827586207	0.851585006129896\\
4.41380310344828	0.859280848144026\\
5.03449275862069	0.871304361536456\\
};
\addplot [color=mycolor4, line width=1.0pt, forget plot]
  table[row sep=crcr]{%
9.99999999962142e-06	0.02237816970207\\
0.03449275862069	0.105734309387446\\
0.0689755172413795	0.150424534746165\\
0.103458275862069	0.18496930912556\\
0.137941034482759	0.214076750712573\\
0.172423793103448	0.23961399038373\\
0.206906551724138	0.262559962505342\\
0.241389310344828	0.283536337706653\\
0.275872068965517	0.30295094536322\\
0.310354827586207	0.321049721121835\\
0.344837586206896	0.33799636185138\\
0.379320344827586	0.353948318856172\\
0.413803103448276	0.369059244302664\\
0.448285862068966	0.383431509886349\\
0.482768620689655	0.397109252327522\\
0.517251379310345	0.410133464730401\\
0.586216896551724	0.434530394989784\\
0.655182413793104	0.457032235529992\\
0.724147931034483	0.477792729706914\\
0.793113448275862	0.497154833958086\\
0.862078965517242	0.515147478916871\\
0.931044482758621	0.532012983856663\\
1.00001	0.547805536481395\\
1.31035482758621	0.608070613733961\\
1.62069965517241	0.654768057154349\\
1.93104448275862	0.691699120953102\\
2.24138931034483	0.721361253016584\\
2.55173413793104	0.745600533451066\\
2.86207896551724	0.765769395151875\\
3.17242379310345	0.78283644787474\\
3.48276862068966	0.797577431511866\\
3.79311344827586	0.810524642468793\\
4.10345827586207	0.822050395331014\\
4.41380310344828	0.832460753141175\\
4.72414793103448	0.841927856760371\\
5.03449275862069	0.850574313990775\\
};
\node[fill=white, left, align=right]
at (axis cs:4.95,0.2) {Varying effective density, $\rho_e$};
\end{axis}

\end{tikzpicture}%

%% file: images/1dimkussner.tex
%
%
\definecolor{mycolor1}{rgb}{0.18517,0.05913,0.24304}%
\definecolor{mycolor2}{rgb}{0.51514,0.10993,0.38770}%
\definecolor{mycolor3}{rgb}{0.81038,0.26571,0.33825}%
\definecolor{mycolor4}{rgb}{0.94901,0.58547,0.40375}%
\begin{tikzpicture}[%
trim axis left, trim axis right
]

\begin{axis}[%
width=0.988\fwidth,
height=\fheight,
at={(0\fwidth,0\fheight)},
scale only axis,
xmin=0,
xmax=10,
xlabel style={font=\color{white!15!black}},
xlabel={convective time, $t$},
ymin=0,
ymax=1,
ylabel style={font=\color{white!15!black}},
ylabel={$L$},
axis background/.style={fill=white},
xmajorgrids,
ymajorgrids
]
\addplot [color=black, line width=2.0pt, forget plot]
  table[row sep=crcr]{%
9.99999999962142e-06	0.00314607853641213\\
0.03449275862069	0.0833651113167555\\
0.0689755172413786	0.117552133702517\\
0.103458275862069	0.143560602275095\\
0.137941034482759	0.165301146947687\\
0.172423793103448	0.18429389722662\\
0.206906551724138	0.201321132597412\\
0.241389310344827	0.21684914339829\\
0.275872068965517	0.231182780139571\\
0.310354827586208	0.244534518856108\\
0.344837586206896	0.257059595186618\\
0.379320344827587	0.268875604955293\\
0.448285862068966	0.290728615730282\\
0.517251379310345	0.310632502432119\\
0.586216896551724	0.328953351736068\\
0.655182413793103	0.34595375511257\\
0.724147931034484	0.361829951783418\\
0.793113448275863	0.376733364923703\\
0.862078965517242	0.39078387707684\\
0.965527241379311	0.410467684821565\\
1.00001	0.416696766544232\\
1.31035482758621	0.466680447312608\\
1.62069965517241	0.508215388432095\\
1.93104448275862	0.543622826831616\\
2.24138931034483	0.574344658232166\\
2.55173413793104	0.601351892635527\\
2.86207896551724	0.625336120283706\\
3.17242379310345	0.646810610125096\\
3.48276862068966	0.666168472448758\\
3.79311344827586	0.683718365627804\\
4.10345827586207	0.699707562999722\\
4.41380310344828	0.714337479027623\\
4.72414793103448	0.727774479127088\\
5.03449275862069	0.740157621816572\\
5.3448375862069	0.751604338571896\\
5.6551824137931	0.762214687622151\\
5.96552724137931	0.772074597325814\\
6.27587206896552	0.781258378249628\\
6.58621689655172	0.789830695980569\\
7.20690655172414	0.805360482726849\\
7.82759620689655	0.819040840286158\\
8.44828586206896	0.831168220466473\\
9.06897551724138	0.841979369564877\\
9.68966517241379	0.851665818217777\\
10.00001	0.856137386564933\\
};
\addplot [color=mycolor1, line width=1.0pt, forget plot]
  table[row sep=crcr]{%
9.99999999962142e-06	0.0128787914053632\\
0.03449275862069	0.0774479632291971\\
0.0689755172413786	0.109574621725985\\
0.103458275862069	0.134016776812016\\
0.137941034482759	0.154456170823785\\
0.172423793103448	0.17232244650414\\
0.206906551724138	0.18833979321737\\
0.241389310344827	0.202933760425122\\
0.275872068965517	0.216382991359142\\
0.310354827586208	0.228886619090487\\
0.344837586206896	0.240595252842828\\
0.379320344827587	0.251624462485276\\
0.448285862068966	0.271970938820521\\
0.517251379310345	0.290402054179493\\
0.586216896551724	0.307262328619455\\
0.655182413793103	0.32284146882132\\
0.724147931034484	0.337316964547709\\
0.793113448275863	0.35079928067614\\
0.862078965517242	0.363429737489909\\
0.965527241379311	0.381027444950057\\
1.00001	0.386557231724602\\
1.31035482758621	0.43033071047137\\
1.62069965517241	0.465803937967189\\
1.93104448275862	0.49542957807288\\
2.24138931034483	0.520677530366537\\
2.55173413793104	0.542553401722447\\
2.86207896551724	0.561822587487075\\
3.17242379310345	0.578929457774512\\
3.48276862068966	0.594252085124978\\
3.79311344827586	0.60808769216\\
4.10345827586207	0.620638754335753\\
4.41380310344828	0.63209334038425\\
4.72414793103448	0.642532190611803\\
5.03449275862069	0.652121923389211\\
5.3448375862069	0.660972241099547\\
5.6551824137931	0.669096454077287\\
5.96552724137931	0.676613545712307\\
6.58621689655172	0.690061196572893\\
7.20690655172414	0.701667517444941\\
7.82759620689655	0.711844904796944\\
8.44828586206896	0.72076255805087\\
9.06897551724138	0.728682222488249\\
9.68966517241379	0.735702037834956\\
10.00001	0.738965626628438\\
};
\addplot [color=mycolor2, line width=1.0pt, forget plot]
  table[row sep=crcr]{%
9.99999999962142e-06	0.010377092856908\\
0.03449275862069	0.0737765898377098\\
0.0689755172413786	0.104409300711632\\
0.103458275862069	0.127776577803855\\
0.137941034482759	0.147349432401862\\
0.172423793103448	0.164466767488944\\
0.206906551724138	0.179815002819121\\
0.241389310344827	0.193803672769592\\
0.275872068965517	0.206701443434754\\
0.310354827586208	0.218696387129823\\
0.344837586206896	0.229926502947084\\
0.379320344827587	0.240496730652785\\
0.448285862068966	0.259969533786817\\
0.517251379310345	0.277599871086341\\
0.586216896551724	0.293720502408858\\
0.655182413793103	0.308572607427143\\
0.724147931034484	0.322337296320596\\
0.793113448275863	0.335154472830611\\
0.862078965517242	0.347135924874189\\
0.965527241379311	0.363738520013019\\
1.00001	0.368945102279508\\
1.31035482758621	0.409788901142555\\
1.62069965517241	0.442314536517616\\
1.93104448275862	0.468928575857491\\
2.24138931034483	0.491170335413031\\
2.55173413793104	0.510098397168811\\
2.86207896551724	0.526471689373865\\
3.17242379310345	0.540843196170004\\
3.48276862068966	0.55360702275321\\
3.79311344827586	0.565059435006631\\
4.10345827586207	0.575419030426083\\
4.41380310344828	0.584847559536566\\
4.72414793103448	0.593468602832326\\
5.03449275862069	0.601385963185354\\
5.3448375862069	0.608672424425167\\
5.96552724137931	0.621596146756591\\
6.58621689655172	0.632654023969495\\
7.20690655172414	0.64216832803757\\
7.82759620689655	0.650412273002162\\
8.44828586206896	0.657609511414639\\
9.06897551724138	0.663940494474812\\
10.00001	0.672128872679743\\
};
\addplot [color=mycolor3, line width=1.0pt, forget plot]
  table[row sep=crcr]{%
9.99999999962142e-06	0.0128780333675067\\
0.03449275862069	0.0693233128259489\\
0.0689755172413786	0.0982724857349577\\
0.103458275862069	0.120398728987036\\
0.137941034482759	0.138931753256974\\
0.172423793103448	0.155146029441276\\
0.206906551724138	0.169693952085927\\
0.241389310344827	0.18296092639746\\
0.275872068965517	0.195197153538889\\
0.310354827586208	0.206577051520432\\
0.344837586206896	0.217229490746917\\
0.379320344827587	0.227254098006922\\
0.448285862068966	0.245721215081581\\
0.517251379310345	0.262442705235832\\
0.586216896551724	0.27771947088014\\
0.655182413793103	0.291766593944555\\
0.724147931034484	0.304764324530121\\
0.793113448275863	0.31685888556326\\
0.862078965517242	0.328149453134401\\
0.965527241379311	0.343735225347263\\
1.00001	0.348608880276124\\
1.31035482758621	0.386616158356196\\
1.62069965517241	0.416412919475688\\
1.93104448275862	0.440304922176628\\
2.24138931034483	0.459822352836456\\
2.55173413793104	0.476019697134545\\
2.86207896551724	0.489687861037607\\
3.17242379310345	0.501392412388189\\
3.48276862068966	0.51158093396854\\
3.79311344827586	0.520575703534538\\
4.10345827586207	0.528608182021948\\
4.41380310344828	0.535876810679634\\
4.72414793103448	0.542520344923121\\
5.3448375862069	0.554240035331279\\
5.96552724137931	0.564244643198183\\
6.58621689655172	0.572877638609866\\
7.20690655172414	0.580324680677137\\
7.82759620689655	0.5867696492208\\
8.44828586206896	0.592354922859993\\
9.37932034482759	0.599455240938569\\
10.00001	0.603483614309216\\
};
\addplot [color=mycolor4, line width=1.0pt, forget plot]
  table[row sep=crcr]{%
9.99999999962142e-06	0.0131381630403649\\
0.03449275862069	0.0641980307364882\\
0.0689755172413786	0.0911081458755323\\
0.103458275862069	0.1117519851986\\
0.137941034482759	0.129052652521628\\
0.172423793103448	0.144194054119311\\
0.206906551724138	0.157795200378832\\
0.241389310344827	0.170217327258953\\
0.275872068965517	0.181682543114491\\
0.310354827586208	0.192341447303846\\
0.344837586206896	0.202314198101361\\
0.413803103448275	0.220591911123236\\
0.482768620689654	0.237051578648975\\
0.551734137931035	0.251981909273807\\
0.620699655172414	0.265692436532838\\
0.689665172413793	0.278347186764371\\
0.758630689655172	0.29003805171207\\
0.827596206896551	0.300941516658705\\
0.89656172413793	0.311107504446136\\
1.00001	0.325136696805743\\
1.31035482758621	0.360208319826343\\
1.62069965517241	0.387301553742537\\
1.93104448275862	0.408573502434445\\
2.24138931034483	0.425527755419839\\
2.55173413793104	0.439144210056682\\
2.86207896551724	0.450246817946292\\
3.17242379310345	0.45942464724455\\
3.48276862068966	0.467100772161487\\
3.79311344827586	0.473691798258182\\
4.10345827586207	0.479389693563546\\
4.72414793103448	0.488997015164996\\
5.3448375862069	0.496967281016342\\
5.96552724137931	0.503838826968572\\
6.58621689655172	0.509876688748566\\
7.20690655172414	0.515206940525781\\
8.13794103448276	0.521951527431472\\
9.06897551724138	0.527445058545204\\
10.00001	0.531954692870066\\
};
\node[fill=white, left, align=right]
at (axis cs:9.95,0.2) {Varying flow resistance, $\Phi$};
\end{axis}

\end{tikzpicture}%

%% file: images/2dimkussner.tex
%
%
\definecolor{mycolor1}{rgb}{0.09053,0.13734,0.07326}%
\definecolor{mycolor2}{rgb}{0.04342,0.37461,0.17524}%
\definecolor{mycolor3}{rgb}{0.37697,0.57257,0.04618}%
\definecolor{mycolor4}{rgb}{0.79072,0.73098,0.27064}%
\begin{tikzpicture}[%
trim axis left, trim axis right
]

\begin{axis}[%
width=0.988\fwidth,
height=\fheight,
at={(0\fwidth,0\fheight)},
scale only axis,
xmin=0,
xmax=5,
xlabel style={font=\color{white!15!black}},
xlabel={convective time, $t$},
ymin=0,
ymax=1,
ylabel style={font=\color{white!15!black}},
ylabel={$L$},
axis background/.style={fill=white},
xmajorgrids,
ymajorgrids
]
\addplot [color=black, line width=2.0pt, forget plot]
  table[row sep=crcr]{%
9.99999999962142e-06	0.00314607853641125\\
0.03449275862069	0.0833651113167546\\
0.0689755172413795	0.117552133702516\\
0.103458275862069	0.143560602275094\\
0.137941034482759	0.165301146947687\\
0.172423793103448	0.18429389722662\\
0.206906551724138	0.201321132597412\\
0.241389310344828	0.21684914339829\\
0.275872068965517	0.231182780139571\\
0.310354827586207	0.244534518856107\\
0.344837586206896	0.257059595186619\\
0.379320344827586	0.268875604955293\\
0.448285862068966	0.290728615730283\\
0.517251379310345	0.310632502432119\\
0.586216896551724	0.328953351736067\\
0.655182413793104	0.345953755112569\\
0.724147931034483	0.361829951783418\\
0.793113448275862	0.376733364923704\\
0.862078965517242	0.390783877076841\\
0.96552724137931	0.410467684821564\\
1.00001	0.416696766544232\\
1.31035482758621	0.466680447312608\\
1.62069965517241	0.508215388432095\\
1.93104448275862	0.543622826831617\\
2.24138931034483	0.574344658232167\\
2.55173413793104	0.601351892635527\\
2.86207896551724	0.625336120283706\\
3.17242379310345	0.646810610125096\\
3.48276862068966	0.666168472448758\\
3.79311344827586	0.683718365627804\\
4.10345827586207	0.699707562999722\\
4.41380310344828	0.714337479027624\\
4.72414793103448	0.727774479127087\\
5.03449275862069	0.740157621816572\\
};
\addplot [color=mycolor1, line width=1.0pt, forget plot]
  table[row sep=crcr]{%
9.99999999962142e-06	0.0166576133380278\\
0.03449275862069	0.072614371084347\\
0.0689755172413795	0.102493627300711\\
0.103458275862069	0.125315479031443\\
0.137941034482759	0.1443431448194\\
0.172423793103448	0.16092897046664\\
0.206906551724138	0.175778874806629\\
0.241389310344828	0.189295263737254\\
0.275872068965517	0.201729352092281\\
0.310354827586207	0.213268678286841\\
0.344837586206896	0.224068421357289\\
0.379320344827586	0.234243091635254\\
0.448285862068966	0.252968436341603\\
0.517251379310345	0.269868593088058\\
0.586216896551724	0.285343671614515\\
0.655182413793104	0.299547812268942\\
0.724147931034483	0.312707760495403\\
0.793113448275862	0.324958459387862\\
0.862078965517242	0.336359941863996\\
0.96552724137931	0.352173070420142\\
1.00001	0.35710044246659\\
1.31035482758621	0.395607383430246\\
1.62069965517241	0.425684303099282\\
1.93104448275862	0.449561819174259\\
2.24138931034483	0.468667895473726\\
2.55173413793104	0.484036789930892\\
2.86207896551724	0.49639275946298\\
3.17242379310345	0.506262666236934\\
3.48276862068966	0.514128112841364\\
3.79311344827586	0.520347269963499\\
4.10345827586207	0.525226563447184\\
4.41380310344828	0.529018856120531\\
4.72414793103448	0.531898050877238\\
5.03449275862069	0.534035454514671\\
};
\addplot [color=mycolor2, line width=1.0pt, forget plot]
  table[row sep=crcr]{%
9.99999999962142e-06	0.0152627586614766\\
0.03449275862069	0.0703056138741385\\
0.0689755172413795	0.09939594617504\\
0.103458275862069	0.121599910460104\\
0.137941034482759	0.140116785715403\\
0.172423793103448	0.156261494521033\\
0.206906551724138	0.170727849586094\\
0.241389310344828	0.183917119627383\\
0.275872068965517	0.196068995954898\\
0.310354827586207	0.207344277419402\\
0.344837586206896	0.217877124028063\\
0.413803103448276	0.237170241570996\\
0.482768620689655	0.254536332771913\\
0.551734137931034	0.270261253839187\\
0.620699655172414	0.284715964385884\\
0.689665172413793	0.298059206929405\\
0.758630689655172	0.310375583863849\\
0.827596206896552	0.321890772731774\\
0.896561724137931	0.332625779889257\\
1.00001	0.347460874298488\\
1.31035482758621	0.384670317957303\\
1.62069965517241	0.413588970876126\\
1.93104448275862	0.43635130725917\\
2.24138931034483	0.454489133783612\\
2.55173413793104	0.46893615479454\\
2.86207896551724	0.480466245559374\\
3.17242379310345	0.489709240703216\\
3.48276862068966	0.49704189914176\\
3.79311344827586	0.502926249695954\\
4.10345827586207	0.507608855247888\\
4.41380310344828	0.51131487856538\\
4.72414793103448	0.514332473841297\\
5.03449275862069	0.516712537651043\\
};
\addplot [color=mycolor3, line width=1.0pt, forget plot]
  table[row sep=crcr]{%
9.99999999962142e-06	0.0155760657043986\\
0.03449275862069	0.0669035722325182\\
0.0689755172413795	0.0944259281683735\\
0.103458275862069	0.115858637225038\\
0.137941034482759	0.133833761462243\\
0.172423793103448	0.149241448013498\\
0.206906551724138	0.163038530932689\\
0.241389310344828	0.175914684871846\\
0.275872068965517	0.187786120211364\\
0.310354827586207	0.198515335159517\\
0.379320344827586	0.218329244991778\\
0.413803103448276	0.227553751424888\\
0.482768620689655	0.244022277873417\\
0.551734137931034	0.259590509243998\\
0.655182413793104	0.279974543793113\\
0.689665172413793	0.286433534015104\\
0.793113448275862	0.303886663578351\\
0.862078965517242	0.314579583563982\\
1.00001	0.333969316700539\\
1.31035482758621	0.369599735444396\\
1.62069965517241	0.397417628544189\\
1.93104448275862	0.419236318311905\\
2.24138931034483	0.436224437309393\\
2.55173413793104	0.44984912413318\\
2.86207896551724	0.46110529583386\\
3.17242379310345	0.470130209855688\\
3.48276862068966	0.477235239891502\\
3.79311344827586	0.483229773132446\\
4.10345827586207	0.488535251459721\\
4.41380310344828	0.492950183717222\\
5.03449275862069	0.499847804150084\\
};
\addplot [color=mycolor4, line width=1.0pt, forget plot]
  table[row sep=crcr]{%
9.99999999962142e-06	0.012837854922191\\
0.03449275862069	0.0606574059579401\\
0.0689755172413795	0.0862951876547244\\
0.103458275862069	0.106112751276181\\
0.137941034482759	0.122811038813775\\
0.172423793103448	0.137461181447255\\
0.206906551724138	0.150624771904728\\
0.241389310344828	0.162658448707303\\
0.275872068965517	0.173796174436642\\
0.310354827586207	0.184179036866936\\
0.344837586206896	0.193900945226773\\
0.413803103448276	0.211721025406869\\
0.482768620689655	0.227812686985256\\
0.551734137931034	0.242425102295829\\
0.620699655172414	0.255868683671753\\
0.689665172413793	0.268254483627484\\
0.758630689655172	0.27974983974523\\
0.827596206896552	0.290461377960122\\
0.931044482758621	0.305203937337298\\
1.00001	0.314263771190851\\
1.31035482758621	0.348836496706091\\
1.62069965517241	0.375625774464258\\
1.93104448275862	0.396812298897789\\
2.24138931034483	0.413828799942494\\
2.55173413793104	0.427734332422544\\
2.86207896551724	0.439304756809699\\
3.17242379310345	0.44909574283414\\
3.48276862068966	0.457552315103601\\
3.79311344827586	0.464979840148094\\
4.10345827586207	0.471591906509381\\
4.41380310344828	0.477564095702431\\
5.03449275862069	0.487955440008392\\
};
\node[fill=white, left, align=right]
at (axis cs:4.95,0.2) {Varying effective density, $\rho_e$};
\end{axis}

\end{tikzpicture}%

%% file: sections/conclusions.tex
\section{Conclusions} \label{Sec:Conclusions}
A comprehensive unsteady aerodynamic theory is presented for lifting porous bodies. The aerodynamic problem is modelled as a singular Fredholm--Volterra integral equation, which is solved numerically using a new method developed in this paper. The foundation of this method relies on the Jacobi polynomial solution technique; the bound vorticity distribution is expanded as a series of weighted Jacobi polynomials whose parameters are determined by asymptotic analysis at the endpoints of the aerofoil. The numerical method therefore remains accurate at the endpoints, which is important when imposing the Kutta condition and in the computation of the leading-edge suction. The aerodynamic solution converges rapidly and is straightforward to implement for both continuous and discontinuous porosity distributions. 

The new numerical scheme enables the porous extension of the classical works by \cite{Theodorsen1935} for harmonic aerofoil motions and \cite{Sears1941} for harmonic gusts. Specifically, the impermeable surface boundary condition is relaxed to include the effects of {chordwise gradients of the flow resistance and effective fluid mass of the porous aerofoil.}
The porous extension of the Theodorsen function is investigated for a heaving flat plate, where the porous results depart from the traditional impermeable solution and {approach a finite limiting value at large reduced frequency that depends on the properties of the porous medium}. The magnitude of the quasi-steady lift is reduced by {the introduction of} porosity, although the relative change in circulatory lift depends on the porosity parameters and the reduced frequency. 
{Most notably, porosity augments the phase lag of the Theodorsen function, which can be tuned as a function of reduced frequency by the porosity parameters.}
{Porosity also plays a significant role in the reduction of unsteady lift in response to a harmonic gust, whose magnitude is also driven primarily by the quasi-steady lift.
The effects of porosity were seen to be most significant at reduced frequencies of practical interest, and the classical Sears function is recovered in the limits of large or small reduced frequency.} 
{The frequency-domain Theodorsen and Sears function solutions furnish porous extensions to the Wagner and K\"ussner indicial lift function for impulsive aerofoil motion and sharp-edges gusts, respectively. Flow resistance within the porous aerofoil dominates the long-time behaviour of both indicial functions. The effective density of the porous medium controls and diminishes the initial impulsive lift of the Wagner function, but the K\"ussner function is insensitive to this parameter.}


{An asymptotic analysis of the singular integral equation for unsteady porous aerofoils is performed in the limits of small or large reduced frequency. The low-frequency limit recovers the steady solution of \mbox{\cite{Hajian2017}}, where corrections to this result are on the order of the reduced frequency and may be found by regular perturbation techniques. Consideration of the high-frequency limit requires matched asymptotic techniques to determine the aerofoil circulation and unsteady lift coefficient, which scale as the square-root and square of the reduced frequency, respectively. The effective fluid density of the porous boundary condition is essential here to determine a solution to the associated singular perturbation problem. These scaling behaviours are confirmed by the numerical scheme.}

The analysis presented in this paper invites companion experimental studies for validation and to suggest appropriate model refinements as required. The present work restricts its attention to the {linearised unsteady dynamics of the flow within the porous medium}. Whilst this approach is valid when the Reynolds number of the flow through the pores is small, a higher-order quadratic model such as the \cite{Ergun1952} model may prove more appropriate in practice, especially near the leading edge where the pressure {jump across the aerofoil} is large. The analysis of \cite{Wegert1987} for steady flow through a cylindrical shell indicates that nonlinear porosity functions lead to a nonlinear Riemann--Hilbert problem, which can be solved using an iterative technique \citep{Wegert1990}. Future work will be devoted to the adaptation of the current study to more general, nonlinear porosity functions to improve the physical fidelity of the present model, possibly through the formulation of an appropriate nonlinear Riemann--Hilbert problem.


The numerical approach advocated in the present research is  sufficiently fast and accurate to be integrated into design optimisation routines. In particular, it is often desirable to reduce aeroacoustic emissions with a minimal aerodynamic penalty \citep{JWJ,jaworski2020aeroacoustics}. Initial assessment of this performance trade-off was explored by \cite{Weidenfeld2018} in the case of an elastic aerofoil, and optimization of elastic aerofoil effects on unsteady propulsion by \cite{Moore2015} found that the limiting case of a torsional spring at the leading edge of the wing led to optimal thrust conditions. More broadly, the inclusion of elastic effects is an important step forward in improving the physical fidelity of the mathematical modelling that may also contribute to biologically-inspired problems in unsteady flows. For example, the Jacobi polynomial approach of the present work may be adapted into the analysis of \cite{Tzezana2019} to consider porous, compliant wings. Similarly, the study of emergent motions of fliers and swimmers by  \cite{Moore2014} may also be extended to include porous planform effects by using the numerical scheme developed here.

%% file: sections/acknowledgement.tex
\section*{Acknowledgement}
P. J. B. was supported by EPSRC  grant no. 1625902 and an EPSRC Doctoral Prize from Imperial College London. J. W. J. gratefully acknowledges the financial support of the National Science Foundation under grant awards 1805692 and 1846852. R. H.  acknowledges the generous support of the David Crighton Fellowship from DAMTP, University of Cambridge. P. J. B. and J. W. J.  would like to thank the Isaac Newton Institute for Mathematical Sciences, Cambridge, for support during the
programme ``Bringing pure and applied analysis together via the Wiener--Hopf technique, its generalisations and applications" where some work on this paper was undertaken. This programme was supported by EPSRC
grant no EP/R014604/1.\\

{We also acknowledge the anonymous reviewers whose suggestions greatly improved the quality of the paper.}

\section*{Declaration of interests}
The authors report no conflicts of interest.

%% file: sections/appendix-jacobi.tex
\section{Identities for Jacobi polynomials} \label{Ap:Jacobi}
This appendix compiles some useful identities for Jacobi polynomials. All of the relations presented assume that $a, \, b>-1$. The Jacobi polynomials are normalised so that
\begin{equation}
P_0^{a, b}(x) = 	1. 
\end{equation}
Higher-order polynomials may be calculated using the recurrence relation
\begin{align}
	2n(n+a +b )(2n+a +b -2)P_{n}^{a ,b}(x)& \notag\\ =(2n+a +b -1)\Big \{(2n+a +b )&(2n+a +b -2)x+a ^{2}-b ^{2}\Big \}P_{n-1}^{a ,b}(x)\notag \\
	-2(n+&a -1)(n+b -1)(2n+a +b )P_{n-2}^{a ,b}(x), \label{Eq:recur}
\end{align}
for $n = 1,2, \cdots$, where we have used the convention that $P_{-1}^{a, b}(x)\equiv 0$.

The Jacobi polynomials satisfy the general orthogonality relation
\begin{align}
	\int_{-1}^1 P_m^{a,b}(x) P_n^{a,b}(x)  \weight{x}{a}{b} \d x &= \frac{2^{a + b +1}}{2 n + a + b + 1} \cdot \frac{G(n+ a +1) G (n + b + 1)}{n! G (n + a + b + 1)} \delta_{m n},   \label{Eq:JacobiOrthogonality}
\end{align}
where $G(\cdot)$ is the Gamma function. In particular, when $m=n=0$, we have
\begin{align}
    \int_{-1}^1 \weight{x}{a}{b} \d x &= 2^{1+ a+ b} B(1+ b, 1+ a), \label{Eq:JacobiOrthogonality0}
\end{align}
where $B(\cdot,\cdot) $ is the Beta function.

Analytic expressions of the finite Hilbert transform of the weighted Jacobi polynomials are \citep[p. 797]{Polyanin1998}
\begin{align}
\dashint_{-1}^{1}w^{a, b}(t) \cdot \frac{P_n^{a,b}(t)}{t-x}\d t =& \frac{\pi w^{a, b}(x) P_n^{a,b}(x)}{\tan(\pi a)} \notag\\
&- 2^{a + b} B(n+ b + 1, a) \cdot \pFq{n+1,-n-a - b}{1- a}{\frac{1-x}{2}}, \notag \\
\coloneqq& \ w^{a, b}(x) Q_n^{a, b}(x). \label{Eq:JacobiHilbertTransform}
\end{align}
%
We also require the finite Hilbert transform when the principal value part is not assumed. The identity for the zeroth Jacobi polynomial is \citep[Eqs. 12 \& 13]{Grosjean1986}
\begin{align}
\int_{-1}^{1} \frac{w^{a, b}(t) }{t-x} \d t =& \frac{2^{a+b+1}G(a+1)G(b+1)}{G(a+b+2)}\notag \\
&\times
\begin{dcases*}
\dfrac{1}{x-1} \cdot  \pFq{a+1,1}{a+b+2}{\dfrac{2}{1-x}}, \qquad \left|x-1\right|>2, \\
\dfrac{1}{x+1} \cdot  \pFq{b+1,1}{a+b+2}{\dfrac{2}{1+x}},  \qquad \left|x+1\right|>2,
\end{dcases*} \label{Eq:jacQ2}
\end{align}
and the corresponding results for the higher-order polynomials can be obtained through the recurrence relation \eqref{Eq:recur}.

Finally, the indefinite integral of a weighted Jacobi polynomial is \citep[Eq. 18.17.1]{DLMF}
\begin{align}
        \int_{-1}^x \weight{y}{a}{b}  P_{n}^{a,b} (y)\d y &= 
        \begin{dcases}- \frac{\weight{x}{a+1}{b+1}}{2 n} P_{n-1}^{a+1,b+1} (x), & n > 0, \\
       2^{1+a+ b} B\left(\frac{1+x}{2};1+ b,1+a\right), & n=0 ,
        \end{dcases} \notag\\
        &\coloneqq I_n^{a,b}(x) \label{Eq:WeightedJacobiIntegral}
\end{align}
where $B(\,\cdot\,\,;\,\cdot\,,\, \cdot\,)$ is the incomplete Beta function.

%% file: sections/large-frequency.tex
\section{Asymptotic analysis}
\label{Sec:asymp}
We now consider the asymptotic regimes of low and high frequency aerofoil motion or disturbance.
The low-frequency regime yields a regular perturbation problem that is straightforward to analyse.
In contrast, the high-frequency regime is asymptotically singular, which requires the method of matched asymptotic expansions to {resolve the region} near the trailing edge.
\subsection{Low-frequency regime}%
%
At leading order, the low-frequency problem reduces to that of the steady problem considered by \cite{Hajian2017}.
The solution is given in \eqref{Eq:qsSol} with $k=0$.
Accordingly, the effective density $\rho_{\rm e}$ vanishes and therefore does not play a significant role at low frequencies.
Additionally, the circulation and lift coefficient satisfy
\begin{align}
	C_L = -2 \hat{\Gamma} + {\it O}(k),
\end{align}
where the leading-order approximation of $\hat{\Gamma}$ is given by integrating \eqref{Eq:qsSol}:
\begin{align}
	\hat{\Gamma} = \int_{-1}^1 \left(\frac{-2}{1+(\psi(x,0))^2} \Bigg\{\psi(x,0) 
			\frac{\d \hat{y}_{\rm a}}{\d x} (x) + \frac{Z(x)}{\pi} \dashint_{-1}^1  
\frac{\d \hat{y}_{\rm a}}{\d x} (\xi) \frac{\d \xi }{Z(\xi)(\xi - x)} \Bigg\} \right) \d x,
\end{align}
which is independent of $k$.
As such, when $\d \hat{y}_{\rm a} / \d x\not \equiv 0$, the circulation and lift tend to finite values as $k \rightarrow 0$.
If $\d \hat{y}_{\rm a} /\d x \equiv 0$ then the aerofoil is an almost-stationary flat plate, so the circulation
and lift scale like $k$.
Since the perturbation problem is regular, it is simple to derive higher-order corrections to the leading-order solution,
though that is not our focus here.
\subsection{High-frequency regime}%
\label{sub:high_frequency_regime}
The high-frequency ($k\gg 1$) regime represents a singular perturbation problem;
at leading-order, the SF--VIE becomes
\begin{align}
	 -\frac{2}{\rho_{\rm e}(x)} \int_{-1}^x \hat{\gamma}_{\rm a}(\xi) \d \xi
	+ \frac{1}{\pi} \dashint_{-1}^1 \frac{\hat{\gamma}_{\rm a}(\xi)}{\xi - x}\d \xi
	= 2 \i k \frac{\d \hat{y}_{\rm a}}{\d x}(x) + \hat{\Gamma} f_{\rm w}(x).
	\label{Eq:hfSIE}
\end{align}
The singular nature of this perturbation problem stems from the forcing from the wake, $f_{\rm w}$,
and imposing the Kutta condition results in {a distinct inner region near the trailing edge.}
The Riemann--Lebesgue lemma shows that
\begin{align}
	f_{\rm w}(x) &= \frac{1}{\pi(x - 1)}+ o(1), &
	\textrm{as } k \rightarrow \infty \textrm{ with } x \textrm{ fixed},
	\label{Eq:fw1}
\end{align}
so $f_{\rm w}(x)=o(k)$.
However, this scaling is not uniformly valid throughout the entire region $-1<x<1$, 
as evidenced by the simple pole at $x=1$ in \eqref{Eq:fw1}.
In particular, expanding $f_{\rm w}$ near the trailing edge instead yields
\begin{align}
f_{\rm w}(1+x^+/k) &= \frac{\i k }{\pi} 
	\int_0^\infty \frac{\e^{-\i \xi^+}}{\xi^+ - x^+}  \d \xi^+
	+ o(k),
    & 
	\textrm{as } k \rightarrow \infty \textrm{ with } x^+ \textrm{ fixed},
	\label{Eq:fw2}
\end{align}
so $f_{\rm w}(x) = {\it O}(k)$ in this region.
Therefore, there exists a inner region of length ${\it O}(1/k)$ near the trailing edge inside which the solution behaviour is distinct from that in the outer region.
Additionally, the forcing due to the aerofoil motion $f_{\rm a}={\it O}(k)$, so $\hat{\gamma}$ itself must also be ${\it O}(k)$.
These observations motivate two separate asymptotic expansions valid in the outer and inner regions respectively:
\begin{align}
	\hat{\gamma}_{\rm a}(x) &= \sum_{j = -1}^n u_j(x) + o\left(k^{-n}\right), &
	\textrm{as } k \rightarrow \infty \textrm{ with } x \textrm{ fixed},
\end{align}
and
\begin{align}
	\hat{\gamma}_{\rm a}(1+ x^+/k) &= \sum_{j = -1}^m \tilde{u}_j(x^+) + {o}\left(k^{-m}\right), & 
	\textrm{as } k \rightarrow \infty \textrm{ with } x^+ \textrm{ fixed}.
\end{align}
These solutions are not unique since the circulation must still be specified.
The circulation is determined by matching the solutions across an intermediate region where they overlap.
The standard tool of choice for matching is the rule given by \cite{VanDyke1964}.
In words, Van Dyke's matching rule states that \emph{the inner representation of the 
outer representation must equal the outer representation of the inner representation}.
In mathematics, the matching rule states that
if 
\begin{align}
	u_j(1 + x^+/k)  &= S_j^m(x^+) + {o}\left(k^{-m}\right), & \textrm{as } k \rightarrow \infty \textrm{ with } x^+ \textrm{ fixed},
\end{align}
and
\begin{align}
	\tilde{u}_j\left(k (x - 1)\right) &= \tilde{S}_j^n(x) + {o}\left(k^{-n}\right), &
	\textrm{as } k \rightarrow \infty \textrm{ with } x \textrm{ fixed},
\end{align}
then 
\begin{align}
\sum_{j=-1}^n S_j^m \left( k(x - 1)\right)
	= \sum_{j=-1}^m \tilde{S}_j^n (x).
\end{align}
Combining the inner and outer solutions, and removing the overlapping contribution, generates
a solution that is  valid throughout the entire domain:
\begin{align}
	\hat{\gamma}_a(x) = \sum_{j=-1}^n \left( u_j(x) + \tilde{u}_j(x^+) - S_j^n(x^+) \right) + o(k^{-n}),
\end{align}
where $x^+ = k(x-1)$.

Therefore, the circulation has the asymptotic expansion
\begin{align}
	\hat{\Gamma} = \int_{-1}^1 \sum_{j=-1}^{n} \left( u_j(\xi) + \tilde{u}_j(\xi^+) - S_j^n(\xi^+) \right) \d \xi^+
	+ o (k^{-n}).
\end{align}
%
\subsection{Preliminaries of the leading-order solution}
At leading order, the composite solution is
\begin{align}
	\hat{\gamma}_{\rm a}(x)  = u_{-1}(x) + \tilde{u}_{-1}(x^+) - S_{-1}^{-1} (x^+) + o(k).
	\label{Eq:loS}
\end{align}
Note that the superscript in $S_{-1}^{-1}$ refers to the truncation of the outer solutions to
 ${\it O}(k^{-1})$, not the inverse of $S_{-1}$.
The leading-order contribution to the circulation is then
\begin{align}
	\hat{\Gamma}_{-1}
		      &= \int_{-1}^1 \left( u_{-1}(\xi) + \tilde{u}_{-1}(\xi^+) - S_{-1}^{-1}(\xi^+)\right) \d \xi
\end{align}
where $\xi^+ = k(\xi - 1).$
The last two terms in the integral may be rewritten as
\begin{align}
	\int_{-1}^1 \left(\tilde{u}_{-1}(\xi^+) - S_{-1}^{-1}(\xi^+) \right)\d \xi 
	&= \frac{1}{k} \int_{-\infty}^0 \left(\tilde{u}_{-1}(\xi^+) - S_{-1}^{-1}(\xi^+) \right)\d \xi^+
	+ {\it O}(1),
\end{align}
such that the leading-order contribution to the circulation is
\begin{align}
	\hat{\Gamma}_{-1}
		      &=  \int_{-1}^1 u_{-1}(\xi) \d \xi
		      + \frac{1}{k}  \int_{-\infty}^0 \left(\tilde{u}_{-1}(\xi^+) - S_{-1}^{-1}(\xi^+) \right)\d \xi^+ .
		      \label{Eq:circLO}
\end{align}
Furthermore, we can show that $\hat{\Gamma} = o(k)$.
In inner variables, the Cauchy principal value becomes
\begin{align}
	\dashint_{-1}^1 \frac{u_{-1}(\xi) - S_{-1}^{-1}(\xi^+)
	+ \tilde{u}_{-1}(\xi^+)}{\xi - (1 + x^+/k)} \d \xi   = 
	\int_{-1}^1 \frac{u_{-1}(\xi)}{\xi - 1}\d \xi 
	+\dashint_{-\infty}^0 \frac{\tilde{u}_{-1}(\xi^+) - S_{-1}^{-1}(\xi^+)}
	{\xi^+ - x^+} \d \xi^+ +o\left(k\right),
\end{align}%
where we have used
\begin{align}
	\dashint_{-1}^1 \frac{\tilde{u}_{-1}(\xi^+) - S_{-1}^{-1}(\xi^+)}{\xi - (1 -x^+/k)} \d \xi &=
	\dashint_{-2k}^0 \frac{\tilde{u}_{-1}(\xi^+) - S_{-1}^{-1}(\xi^+)}{\xi^+ - x^+} \d \xi^+ \nonumber \\
	&=
	\dashint_{-\infty}^0 \frac{\tilde{u}_{-1}(\xi^+) - S_{-1}^{-1}(\xi^+)}{\xi^+ - x^+} \d \xi^+ 
	+o(k).
\end{align}
Now, the SF--VIE for $\tilde{u}_{-1}$ becomes, at leading order,
\begin{align}
\frac{1}{\pi}	\dashint_{-\infty}^0 \frac{\tilde{u}_{-1}(\xi^+)}{\xi^+ - x^+} \d \xi^+
	&=
	\frac{1}{\pi} \dashint_{-\infty}^0 \frac{{S}_{-1}^{-1}(\xi^+)}{\xi^+ - x^+} \d \xi^+
	+\frac{2}{\rho_{\rm e}(x)} \int_{-1}^1 u_0(\xi) \d \xi -
\frac{1}{\pi}	\int_{-1}^1 \frac{u_{-1}(\xi)}{1 - \xi} \d \xi \notag \\
	&+ 2 \i k y_{\rm a}(1) +  \hat{\Gamma}_{-1} \frac{\i k}{\pi} \int_{0}^\infty 
	\frac{\e^{-\i\xi^+}}{\xi^+ - x^+} \d \xi^+.
	\label{Eq:sieInner}
\end{align}
The above expression implies that $\hat{\Gamma}_{-1}= o(k)$, which we will now demonstrate by contradiction.
Supposing that, instead, $\hat{\Gamma}_{-1} = {\it O}(k)$,
then the final term on the right side of \eqref{Eq:sieInner} would be ${\it O}(k^2)$.
It would follow that $\tilde{u}_{-1} = {\it O}(k^2)$, which is an order
of magnitude larger than the outer solution since $u_{-1} = {\it O}(k)$.
It is not possible to match solutions of differing integer orders of magnitude, and we therefore have a contradiction.
Accordingly, $\hat{\Gamma}_{-1} = o(k$). 
Later, we will refine this assertion and show that $\hat{\Gamma} = {\it O}(k^{1/2})$ independently of the 
flow resistivity and effective mass of the porous material.

Combining this observation with \eqref{Eq:circLO} shows that
\begin{align}
	\int_{-1}^1 u_{-1}(\xi) \d \xi = 0.
\end{align}
In other words, $u_{-1}$ is the non-circulatory solution to the high-frequency SF--VIE \eqref{Eq:hfSIE}.

We now present the solutions to the outer and inner problems, and then match these solutions.
\subsection{Outer problem}%
Noting that $\tilde{u}_{-1}(x^+) \sim S_{-1}^{-1}(x^+)$ as $k \rightarrow \infty$
with $x$ fixed, the principal value integral becomes
\begin{align}
	\dashint_{-1}^1 \frac{u_{-1}(\xi) + \tilde{u}_{-1}(\xi^+)  - S_{-1}^{-1}(\xi^+)}{\xi - x} \d \xi  
	&= \dashint_{-1}^1 \frac{u_{-1}(\xi)}{\xi - x} \d \xi + o(k).
\end{align}
Accordingly, at leading order, the SF--VIE becomes
\begin{align}
	-\frac{2}{\rho_{\rm e}(x)} \int_{-1}^x u_{-1}(\xi) \d \xi 
+\frac{1}{\pi} \dashint_{-1}^1 \frac{u_{-1}(\xi)}{\xi - x} \d \xi &=  
2 \i k \hat{y}_{\rm a}(x).
\label{Eq:outerSIE}
\end{align}
Similarly to the full problem, the presence of the Volterra term in the above expression
precludes the possibility of an analytic solution.
However, using similar arguments to \S\ref{Sec:nc}, we see that $u_{-1}$ has a square-root singularity
at the trailing edge since it corresponds to the non-circulatory solution.
As such, $u_{-1}$ may be expressed as 
\begin{align}
	u_{-1}(x) &= \frac{u_{-1}^\ast(x)}{\sqrt{1-x}} ,
	\label{Eq:genSol1}
\end{align}
where $u_{-1}^\ast(x)$ is bounded as $x \rightarrow 1$ and $u_{-1}^\ast = {\it O}(k)$.
For example, in the impermeable case we have
\begin{align}
	u_{-1}^\ast(x) &= \frac{2\i k}{\pi \sqrt{1+x}}  \dashint_{-1}^1 \sqrt{1 - \xi^2} 
	\frac{\hat{y}_a(\xi)}{\xi - x} \d \xi.
\end{align}
Expressing the general solution \label{Eq:genSol1} in terms of inner variables and expanding yields
%
%
\begin{align}
	u_{-1}(1 + x^+/k) &= \sqrt{ \frac{k}{-x^+}} u_{-1}^\ast (1)
+ o (k)
\end{align}
Accordingly, we have
\begin{align}
	S_{-1}^{-1}(x^+) = \sqrt{ \frac{k}{-x^+}} u_{-1}^\ast (1)
	\label{Eq:S1}
\end{align}
The constant $u_{-1}^\ast(1)$ must be determined numerically,
but it is a function only of the aerofoil shape and the effective mass of the porous material.

\subsection{Inner solution}
Noting that
\begin{align}
	\dashint_{-\infty}^0 \frac{\d \xi^+}{\sqrt{\xi^+} ( \xi^+ - x^+)} = 
	\begin{dcases}
		0 & \textnormal{if } x^+<0,\\
		\frac{\i \pi}{\sqrt{x^+}} & \textnormal{if } x^+>0,
	\end{dcases}
	\label{Eq:sqrtInt}
\end{align}
and using \eqref{Eq:outerSIE} allows
the inner problem \eqref{Eq:sieInner} to be simplified to
\begin{align}
\frac{1}{\pi} \dashint_{-\infty}^0 \frac{\tilde{u}_{-1}(\xi^+)}{\xi^+ - x^+} \d \xi^+
	&=
	\hat{\Gamma}_{-1} \frac{\i k}{\pi} \int_{0}^\infty 
	\frac{\e^{-\i\xi^+}}{\xi^+ - x^+} \d \xi^+.
	\label{Eq:sieInner2}
\end{align}
The solution to \eqref{Eq:sieInner2} that is bounded at $x^+=0$ is \citep{NIM}
\begin{align}
	\tilde{u}_{-1}(x^+) &= \frac{\hat{\Gamma}_{-1} \i k\sqrt{-x^+}}{\pi^2} 
\dashint_{-\infty}^0 
	\frac{1}{\sqrt{-\xi^+}(\xi^+ - x^+)} 
	\left(\int_0^\infty \frac{\e^{-\i t^+}}{t^+ - \xi^+} \d t^+\right)
	\d \xi^+.
\end{align}
%
%
%
Using \eqref{Eq:sqrtInt} allows 
the inner solution to be expressed in terms of the error function as 
\begin{align}
	\tilde{u}_{-1}(x^+) &=- \i k \hat{\Gamma}_{-1} \e^{\i x^+}
\textnormal{erfc} \left(\e^{-\i \pi/4} \sqrt{-x^+} \right) .
\end{align}
From this form it is clear to see that the Kutta condition is satisfied, as 
$\tilde{u}_{-1}(0) = - \i k \hat{\Gamma}_{-1}$, so $\Delta p(1) = 0$.

Expressing the inner solution in terms of outer variables and expanding yields
\begin{align}
	\tilde{u}_{-1}(k(x-1)) &= - \i \hat{\Gamma}_{-1}\sqrt{\frac{k}{\pi (x-1)}} \e^{\i \pi /4} + o(1)
\end{align}
so
\begin{align}
	\tilde{S}_{-1}^{-1}(x) = 
- \i \hat{\Gamma}_{-1}\sqrt{\frac{k}{\pi (x-1)}} \e^{\i \pi /4}.	\label{Eq:S2}
\end{align}
\subsection{Matching procedure}
We now match the inner and outer solutions according to Van Dyke's matching rule.
At leading order, the matching rule reads
\begin{align}
	S_{-1}^{-1}(k(x-1)) &= \tilde{S}_{-1}^{-1}(x).
\end{align}
Substituting \eqref{Eq:S1} and \eqref{Eq:S2} then yields the expression for the circulation:
\begin{align}
	\hat{\Gamma}_{-1} =  \i \sqrt{\pi} \e^{-\i \pi/4} \frac{u_{-1}^\ast(1)}{\sqrt{k}}.
\end{align}
Since $u_{-1}^\ast(1) = {\it O}(k)$, it follows that $\hat{\Gamma}_{-1} ={\it O}(\sqrt{k})$.
Accordingly, the composite vorticity distribution may be expressed as 
\begin{align}
	\hat{\gamma}_{\rm a}(x)_{\rm} = u_{-1}(x) - \i k \hat{\Gamma}_{-1} \left( \frac{-\e^{\i \pi/4}}{\sqrt{-\pi x^+}} 
		+\e^{\i x^+}
			\textnormal{erfc} \left( \e^{-\i \pi/4} \sqrt{-x^+} \right) \right).
	\label{eq:gamma_comp}
\end{align}
Therefore, for large $k$, the bound vorticity scales as ${\it O}(k)$ per the dominant contribution from the $u_{-1}(x)$ term. This result hold across the aerofoil except in the region immediately local to the trailing edge where the matching term cancels the singularity of the outer solution and the inner solution behaves like ${\it O}(k^{3/2})$. 

Using \eqref{Eq:pres2a} and integrating \eqref{eq:gamma_comp} over the aerofoil shows that the leading-order contribution of to the pressure jump is given by the non-circulatory solution
\begin{align}
	\Delta p(x) &= \Delta p_{\rm NC}(x) + o(k^2)
\end{align}
and the inner region at the trailing edge does not contribute at this order.
Instead, the pressure has an inner region at the leading edge that can be resolved using a similar method.
Per \eqref{Eq:pres2a}, the noncirculatory pressure distribution $\Delta p_{\rm NC}$ scales like ${\it O}(k^2)$, whereas the remaining term is subdominant for large $k$. We note that these scaling behaviours with respect to reduced frequency $k$ are \emph{identical} for both porous and nonporous unsteady aerofoils, where the only difference in the end results will be the scaling coefficients.

%
%